\documentclass[usenames,dvipsnames,twoside,12pt]{article}
\usepackage[table,dvipsnames]{xcolor}

\usepackage[inline,nomargin,marginclue,index]{fixme}
\makeatletter
\renewcommand*\FXLayoutInline[3]{%
  {\@fxuseface{inline} \ignorespaces[#3 \fxnotename{#1}: #2]}}
\makeatother
\fxsetup{theme=color,mode=multiuser} \FXRegisterAuthor{Javi}{ame}{\color{red}Javi}
\fxsetup{theme=color,mode=multiuser} \FXRegisterAuthor{todo}{am}{\color{green}todo}
\fxsetup{theme=color,mode=multiuser} \FXRegisterAuthor{Igor}{a}{\color{red}Igor}

\usepackage{tikz}
\usetikzlibrary{patterns}
\usetikzlibrary{plotmarks}
\usetikzlibrary{external}
\usepackage{soul}
\usepackage{booktabs}
\usepackage{xfrac}
\usepackage{bm}
\usepackage{rotating}

\usepackage{cite}
\usepackage[colorlinks=true,linkcolor=blue,urlcolor=magenta,citecolor=blue,pagebackref=true]{hyperref}
\usepackage{graphicx}
\usepackage{epsfig}
\usepackage{amssymb}
\usepackage{amsmath}
\usepackage{epstopdf}
\usepackage[latin1]{inputenc}
\usepackage{url}
\usepackage{slashed}
\usepackage{mathtools}
\usepackage[makeroom]{cancel}
\usepackage{dsfont}
\DeclareMathAlphabet{\mathpzc}{OT1}{pzc}{m}{it}

\usepackage[weather]{ifsym}
\usepackage{authblk}
\usepackage{pdflscape}

\numberwithin{equation}{section}

\newcommand{\lsim}{\mathrel{\mathop{\kern 0pt \rlap
  {\raise.2ex\hbox{$<$}}}
  \lower.9ex\hbox{\kern-.190em $\sim$}}}
\newcommand{\gsim}{\mathrel{\mathop{\kern 0pt \rlap
  {\raise.2ex\hbox{$>$}}}
  \lower.9ex\hbox{\kern-.190em $\sim$}}}
\newcommand{\alt}{\mathrel{\mathop{\kern 0pt \rlap
  {\raise.2ex\hbox{$<$}}}
  \lower.9ex\hbox{\kern-.190em $\sim$}}}
\newcommand{\agt}{\mathrel{\mathop{\kern 0pt \rlap
  {\raise.2ex\hbox{$>$}}}
  \lower.9ex\hbox{\kern-.190em $\sim$}}}

\newcommand{\gagamma}{g_{a\gamma}}
\newcommand{\ckcs}{counts keV$^{-1}$~cm$^{-2}$~s$^{-1}$ }

\newcommand{\Igor}[1]{#1}

\newcommand{\exclude}[1]{}
\newcommand{\additionalinfo}[1]{}
\newcommand{\modified}[1]{#1} 

\newcommand{\X}{X}
\newcommand{\WW}{{_{\rm W}}  }
\newcommand{\cosbeta}{\cos\hspace{-2pt}\beta}
\newcommand{\sinbeta}{\sin\hspace{-2pt}\beta}
\newcommand{\cosdosbeta}{\cos^2\hspace{-3pt}\beta}
\newcommand{\sindosbeta}{\sin^2\hspace{-3pt}\beta}
\newcommand{\mpl}{m_{\rm P}}

\newcommand{\DFSZ}{{\sf{DFSZ }}}
\newcommand{\PQWW}{{\sf{PQWW }}}
\newcommand{\KSVZ}{{\sf{KSVZ }}}
\newcommand{\KLPY}{{\sf{KLPY} }}
\newcommand{\PQ}{{\rm{PQ}}}

\newcommand{\A}{A}

\newcommand{\thetalp}{\theta_a}

\newcommand{\KQ}{{\sf Q}}
\newcommand{\KS}{{\sf S}}

\newcommand{\fB}{{\cal B}_a}

\newcommand{\V}{ {\cal V} }

\newcommand{\VA}{ {\cal V}_{\sf QCD} }

\newcommand{\gl}{{\tiny $g$-loop}}
\newcommand{\gal}{{\tiny $\gamma$-loop}}
\newcommand{\nl}{{\tiny $\nu$-loop}}

\newcommand{\mPl}{m_{\rm P}}
\renewcommand\vec[1]{{\bf #1}}

\newcommand{\SF}{{\cal R}}

\newcommand{\sigmaDW}{\sigma_{\rm DW}}

\newcommand{\vvv}[3]{ \left( \begin{array}{cccc}   #1 \\ #2 \\ #3 \end{array} \right)}

\newcommand{\n}{{\mathpzc{n}}}

\newcommand{\dmden}{\varrho}
\newcommand{\dmdensun}{\varrho}

\newcommand{\admfrac}{{\tilde \varrho_a}}
\newcommand{\dmsigmav}{\sigma_v}

\newcommand{\lvol}{V_\odot}

\newcommand{\la}{\ell_a}

\newcommand{\lambdaa}{\lambda_a}

\newcommand{\osci}{\thickapprox}

\newcommand{\geo}{{\cal G}}
\newcommand{\Trh}{T_{\rm RH}}
\newcommand{\proba}{{\cal P}}
\newcommand{\area}{{\mathcal {A}}}

\newcommand{\Higgs}{Higgs }
\renewcommand\({\left(}
\renewcommand\){\right)}
\renewcommand\[{\left[}
\renewcommand\]{\right]}

\newcommand{\be}{\begin{equation}}
\newcommand{\ee}{\end{equation}}
\newcommand{\bea}{\begin{eqnarray}}
\newcommand{\eea}{\end{eqnarray}}

\begin{document}

\title{ \vspace{1cm} New experimental approaches in the search for axion-like particles}
\author[1,2]{Igor G. Irastorza}
\author[1,3]{Javier Redondo}
\affil[1]{\normalsize Departamento de F\'{i}sica Te\'{o}rica,  
Universidad de Zaragoza, 50009 Zaragoza, Spain}
\affil[2]{\normalsize Laboratorio Subterr\'{a}neo de Canfranc, 
22880 Canfranc Estaci\'{o}n, Spain}
\affil[3]{\normalsize  Max-Planck-Institut f\"{u}r Physik,  
80805 M\"{u}nchen, Germany} 
\date{\today}
\maketitle
\begin{abstract}

Axions and other very light axion-like particles appear in many extensions of the Standard Model, and are leading candidates to compose part or all of the missing matter of the Universe. They also appear in models of inflation, dark radiation, or even dark energy, and could solve some long-standing astrophysical anomalies. The physics case of these particles has been considerably developed in recent years, and there are now useful guidelines and powerful motivations to attempt experimental detection. Admittedly, the lack of a positive signal of new physics at the high energy frontier, and in underground detectors searching for weakly interacting massive particles, is also contributing to the increase of interest in axion searches. The experimental landscape is rapidly evolving, with many novel detection concepts and new experimental proposals. An updated account of those initiatives is lacking in the literature. In this review we attempt to provide \modified{such an update.} We will focus on the new experimental approaches and their complementarity, but will also review the most relevant recent results from the consolidated strategies and the prospects of new generation experiments under consideration in the field. We will also briefly review the latest developments of the theory, cosmology and astrophysics of axions and we will discuss the prospects to probe a large fraction of relevant parameter space in the coming decade.

\end{abstract}
\newpage 
\tableofcontents

\newpage
\section{Introduction}

The 20th century witnessed a spectacular revolution in our understanding of the fundamental laws of nature,
that culminated with the establishment of the Standard Model (SM) of particle physics,
the theory that describes with accuracy (at least as far as our experimental and computational accuracy goes) the results of every experiment performed so far in particle physics.
There are however many reasons to believe the SM is not an ultimate theory of nature.
Some decades ago it could have been argued that the SM does not include the gravitational interactions --so successfully described at the classical level by Einstein's theory of general relativity-- and so it has to be extended or embedded in a more complete theory. Nowadays we can count on a few other striking observations. Perhaps the most pressing come from cosmology, 
which seems to be also extremely well described by a classical solution of Einstein's gravity equations,
a homogeneously expanding 
Universe with some primordial inhomogeneities seeded by tiny quantum fluctuations during an exponential expansion phase, so-called primordial inflation. And this excellent description requires a few ingredients that are nowhere to be found in the SM: Dark Matter (DM) --a substance that behaves under gravity as cold gas of non-baryonic weakly interacting particles, Dark Energy (DE), which gravitates as Einstein's famous cosmological constant, and at least a new field (not necessarily a fundamental field) whose potential energy drives inflation for some time and then transforms somehow into the radiation that will dominate the energy density of the Universe during Big Bang Nucleosynthesis.
Amongst these three, the evidence for Cold DM is the most precious for particle physics as it is directly attributable to the existence of new species of particles, i.e. it has been convincingly proven that the majority of DM is not in the form of neutrinos or any other SM particle.

But the SM itself also provides compelling reasons to seek a more fundamental theory of nature.
Most of them 
follow the same pattern: the lack of symmetry of the SM will be alleviated as we consider physics at higher energy scales. New particles/fields are expected to appear and restore symmetries that are not altogether evident in the SM. Couplings can be all related at high energies 
and still lose this unified character at low energies because they run with the energy scale.
Electroweak and strong interactions could be two aspects of the same Grand Unified Theory (GUT) at a very high energy scale of $10^{15}$ GeV where quarks and leptons would also be different ingredients of the same multicomponent fundamental field. Other ideas consider the unification of the fermion generations into the framework of family symmetries. Finally, theories beyond the framework of quantum field theory have to be invoked to include gravity at the same quantum footing than the rest of known interactions. The most conspicuous framework in which this appears to be possible, at least in principle, is the framework of string theories in 10 dimensions.

Another avenue of
speculation about possible extensions of the SM 
is concerned with 
the hierarchy problem and the concept of naturalness.
A very well motivated scenario predicts the existence of a ``supersymmetry" (SUSY) in nature between fermions and bosons. In addition of solving the hierarchy problem, SUSY partners contribute to the running of SM gauge couplings,  providing a strong hint for GUTs.
SUSY is also widely present in string theories. 
In addition, SUSY theories usually predict a stable weakly interacting massive particle (WIMP), typically the lightest SUSY particle (LSP).
In a rather generic way, the expected relic density of LSPs thermally produced after the Big Bang falls in same ballpark as the observed DM density. This has been called the \textit{WIMP miracle}, and has constituted a major motivation to invest large efforts to search for SUSY and WIMPs as DM candidates for the last few decades.
Unfortunately, \modified{experiments at the Large Hadron Collider} have not yet found any convincing signature of SUSY or any other new physics at the TeV scale and the many underground experiments searching for WIMP-nuclear recoils have borne any unambiguous signal either. Besides, cosmic rays from DM-DM annihilation (at the core of the WIMP miracle) above the accountable astrophysical backgrounds, have not yet been found.

We have a very strong prejudice towards nature accommodating more particles/fields and more symmetries at high energy scales. If this new physics is well above the electroweak scale, there is little hope to directly reach the needed energy scale at future accelerators. However, there are mechanisms by which physics associated with high energy scales have important measurable consequences at very low energy. There are well known examples of this. Gravity is associated with physics at the Planck scale and has very appreciable effects at low energies (thanks to the graviton being massless and its effects coherently summed up over a large amount of particles). Another example is the neutrino, whose properties are better studied not by producing the highest energies possible, but the highest luminosities and the most controlled environments for their experimental detection. Neutrinos offer more analogies with the topic of this review, regarding e.g. their role in astrophysics and cosmology.
Let us mention that
the smallness of the graviton and neutrino masses compared with the electroweak scale does not pose another hierarchy problem
because quantum corrections to their mass are protected by symmetries.
This inspires us to think about low mass particles associated with symmetries present at high energies and their effects in cosmology, in
astrophysics, and in experiments at the high intensity, precision frontier. A discipline with its own taste sometimes called the {\em low energy frontier
of particle physics}~\cite{Jaeckel:2010ni}.

The paradigm of this low energy frontier is the \emph{QCD axion}, a hypothetical spin 0 particle predicted by the Peccei-Quinn mechanism~\cite{Peccei:1977hh,Peccei:1977ur} to solve dynamically the so-called \textit{strong CP problem} (the absence of CP violation in the strong interactions) by using QCD dynamics itself. The axion was identified by Weinberg~\cite{Weinberg:1977ma} and Wilczek~\cite{Wilczek:1977pj} as the pseudo Nambu-Goldstone (pNG) boson of a new spontaneously broken global symmetry that Peccei and Quinn had postulated (and that since then bears the name of PQ symmetry).
The axion is strongly related to mesons and indeed would mix with the known $\pi^0,\eta,\eta'$ obtaining a mass and featuring couplings to hadrons and two photons. Both the mass and the strength of these couplings are inversely proportional to $f_\A$, an energy scale related to the spontaneous  breaking of the PQ symmetry, so that the smaller the mass the weaker the couplings.
Indeed, the first Weinberg~\cite{Weinberg:1977ma} and Wilczek~\cite{Wilczek:1977pj} models had $f_A$ of the order of the electroweak scale and were soon ruled out~\cite{Donnelly:1978ty,Zehnder:1981qn}.
However, very soon it was realised that $f_A$ could correspond to a much higher energy scale~\cite{Kim:1979if,Dine:1981rt} which implies very low mass and weakly interacting axions. These axions were so weakly coupled that they were dubbed \emph{invisible} axions, a term which we shall not need henceforth as any other types of axions are ruled out.
The experimental constraints we discuss later on force us to consider values of $f_A \gg 10^7$ GeV, which imply very small masses $m_\A\ll $ eV. Indeed, the axion is the paradigm of the so-called weakly interacting slim particles \emph{WISP}s~\cite{Jaeckel:2010ni} and its discovery would imply the identification of a new energy scale in particle physics.

Axions as pNG bosons are very easily embedded in extensions of the SM at high energies by invoking new fields and symmetries. Other NG bosons with similar properties to the axion have been proposed, like familons~\cite{Wilczek:1982rv,Ema:2016ops,Calibbi:2016hwq} (related to family symmetries), majorons~\cite{Chikashige:1980ui,Chikashige:1980ht} (related to lepton number) or even axi-majorons (where the lepton and PQ symmetries are the same), see~\cite{Dias:2014osa,Ballesteros:2016xej} and Refs. therein. The PQ mechanism can be easily embedded also in SUSY~\cite{Nilles:1981py}, GUTs~\cite{Wise:1981ry} and most notably it is built \modified{in} string theory in a model-independent way~\cite{Witten:1984dg,Conlon:2006tq}. Indeed string theories predict the existence of many axion-like particle candidates~\cite{Arvanitaki:2009fg,Cicoli:2012sz} one of which would be the QCD axion, but the rest could still play a similar phenomenological role~\cite{Anselm:1981aw}.
\modified{In this review we will call axion-like particle (ALP) to any such low mass pNG with weak interactions to SM particles and denote it with the letter $a$. The particular ALP solving the strong CP problem by the PQ mechanism is called QCD axion and shown as $A$, being granted the uppercase distinction by the Review of Particle Physics~\cite{AxionsPDG:2016xqp}. }

The phenomenology of \modified{axions and ALPs} is determined by their low mass and very weak interactions. ALPs (and other WISPs) could affect stellar evolution~\cite{Raffelt:1996wa,Raffelt:2006cw} and cosmology~\cite{Archidiacono:2015mda} in a similar way to thermal neutrinos. These effects are responsible for the constraint $f_A \gg 10^7$ GeV mentioned above.
Fortunately, the analogy with our standard WISPs (gravitons and neutrinos) does not stop here as axions could be discovered by experiments at the low-energy high-intensity frontier. This is because axions could mediate new long range forces~\cite{Moody:1984ba}, allow rare decays, appear after thick walls in beam-dump experiments (leading to the fascinating \textit{light-shining-through-walls} (LSW) experiment~\cite{Redondo:2010dp}), and be thermally produced in copious amounts in the Sun to be detected on Earth~\cite{Sikivie:1983ip}.
The analogy stops, however, when we realise that axions are excellent DM candidates.
Being very weakly interacting, their main production mechanisms in the early Universe are non-thermal: the
vacuum realignment mechanism~\cite{Preskill:1982cy,Dine:1982ah,Abbott:1982af} and the decay of topological defects (axion strings and domain walls)~\cite{Davis:1986xc,Lyth:1991bb}. Therefore, these axions are produced with extremely small velocity dispersion and are therefore very \emph{cold} DM, which perfectly fits the needs of the $\Lambda$CDM model of the Universe that so well describes the large scale structure of the Universe.

For these reasons, axions have been searched for in dedicated laboratory experiments since their proposal.
The techniques employed for their detection are by no means common in the particle physics community, focused to a large extent on accelerators and high energy collisions.
\modified{After the experimental exclusion of the first electroweak-scale axion models, the high values of $f_A$ needed to evade the astrophysical and cosmological bounds threatened to make the axion impossible to find.} In 1983, Pierre Sikivie came up with a seminal paper in which he proposed two of the most fruitful techniques to search for invisible axions~\cite{Sikivie:1983ip}: the axion \emph{helioscope} to detect the copious flux of axions emitted from the Sun and the axion \emph{haloscope} to detect axions from the hypothetical DM galactic halo. One idea is behind all these experiments: to use coherent effects over macroscopic distances/long times to boost the axion production or detection.
As will be seen throughout this paper, this concept is crucial for the detectability of the axion,
but Sikivie's proposal had another very important point, we can use natural sources of axions which are extremely efficient, the Sun and the Big Bang, and concentrate the searches in the detection part. Because of the extremely large fluxes of natural axions, helioscopes and haloscopes are typically much more sensitive to axions and ALPs than their purely laboratory competitors, although their luminosities are also subject to larger uncertainties, especially in the case of DM axions.

Since those early days, there has been a small but continuous experimental activity attempting the detection of axions. Relevant pioneering experimental results in the 90s include: the Brookhaven-Rochester-Fermilab collaboration implementing the first haloscope~\cite{DePanfilis:1987dk,Wuensch:1989sa} and helioscope~\cite{Lazarus:1992ry} setups with moderate sensitivity as well as, together with the Trieste group, also the first LSW setup~\cite{Semertzidis:1990qc}; the axion haloscope setup in Florida U.~\cite{Hagmann:1990tj} (later to become ADMX~\cite{ADMXweb}); a competing haloscope in Japan (CARRACK)~\cite{CARRACKwebNEW} focused on R\&D in photon counting~\cite{Fukuda:1988yw}; the first polarization experiment~\cite{Cameron:1989bm}, precursor of PVLAS~\cite{Bakalov:1994hr}; the Tokyo helioscope~\cite{Moriyama:1998kd} and, towards the end of the decade, the start of the CAST helioscope at CERN.
 
\modified{During the last two decades, the efforts and size of the community have been steadily growing but the few last years are witnessing a real blooming phase. Many new groups have entered the field, new exciting detection concepts have been proposed and several demonstrative small-scale setups have been commissioned. Moreover, well  established techniques appear now as clearly consolidated and face the upgrade to large scale experiments,  entering the radar of more formal particle-physics roadmaps.} The reason for this is, on the one hand, the development of theoretical and phenomenological aspects of axions (like their potential cosmological or astrophysical roles), that has helped sharpening their physics case and yielding further motivation and guidance for detection, and, on the other, the fact that detection technologies have reached levels that allow entering unexplored territory beyond current constraints. The lack of positive detection of SUSY at LHC and WIMPs in underground detectors has further contributed to the increased interest in axions too.

The proof that the experimental landscape is rapidly changing is that relatively recent reviews on the matter have become already obsolete~\cite{Essig:2013lka,Graham:2015ouw}. In spite of the risk of it being soon outdated too, we attempt here to provide a complete review of the experimental landscape that is lacking at the moment in the literature. We will describe the different detection strategies and their complementarity, with a focus on the novel concepts recently \modified{proposed, and a review of the future plans and prospects for the consolidated research lines.} We start by presenting the theoretical motivations of the axion and ALPs in section~\ref{sec:theory}. We follow with a short update on the cosmology and astrophysics of these particles in section~\ref{sec:cosmoastro}, explaining their potential role as DM candidate, and including an account of the status of their astrophysical hints. We then proceed with the experimental review. \modified{Section~\ref{sec:sources} provides a small bridge between the theory and the experiment in which we describe the relevant features of the natural sources as well as the most theoretical elements of the detection. The experimental part is organized, as it is customary, in three sections according to the source of axions considered: laboratory (section~\ref{sec:laboratory}), solar (section~\ref{sec:solar}) and dark matter (section~\ref{sec:DMexps}) axions.}  We finish with our discussion and conclusions in sections~\ref{sec:discussion} and \ref{sec:conclusions} respectively.

\section{Theoretical motivation to search for axions}
\label{sec:theory}

\subsection{The strong CP problem and axions}

When we consider the Lagrangian of the SM at energies below electroweak symmetry breaking (EWSSB), we
find two possible terms that violate parity (P) and time-reversal (T) without changing quark flavour,
\be
{\cal L}_{\cancel{\rm  CP}} = - ( \bar {\bf q}_L m_q e^{i \theta_Y}  {\bf q}_R + {\rm h.c.}) - \frac{\alpha_s}{8\pi}G^a_{\mu\nu}\widetilde G_a^{\mu\nu}\theta_{\rm QCD},
\ee
where ${\mathbf q}=(u,d,...)$ is a vector of quark flavours, 
$\alpha_s$ is the QCD equivalent of the fine-structure constant, $G_{\mu\nu}^a$ is the gluon field-strength tensor and $\tilde G^{\mu\nu^a}=\epsilon^{\mu\nu\alpha\beta}G^a_{\alpha\beta}/2$ its dual and $\theta_{\rm QCD}$ is the angle determining the gauge-invariant QCD vacuum.
Here, $m_q$ is an already diagonalised mass matrix and $\theta_Y$ is a common phase for all the quark masses. Note that
a chiral phase redefinition of one quark,
\be
\label{chiredef}
q_L \to e^{i \alpha/2}q_L \quad , \quad q_R \to e^{-i\alpha/2}q_R ,
\ee
shifts its mass term to $m_q e^{i(\theta_Y-\alpha)}$ and\footnote{
The shift's sign depends on conventions for $\epsilon^{0123}$ and $\gamma^5$. We use 
$\epsilon^{0123}=+1$ and  $\gamma^5=i \gamma^0\gamma^1\gamma^2\gamma^3$ as in~\cite{Peskin:1995ev}. } 
\be
\label{anomalyshift}
\theta_{\rm QCD}\to \theta_{\rm QCD}+\alpha \, .
\ee 
These shifts are a consequence of the fact that such axial U(1)$_A$ transformation of the quark fields is not a symmetry. It is violated both by non-zero quark masses at the tree-level and by the colour anomaly at the 1-loop level\footnote{In principle there would be a shift in the equivalent electroweak term $W_{\mu\nu}\widetilde W^{\mu\nu}$ but this term is not relevant, see~\cite{Perez:2014fja}.  }. 
This last shift can be also understood as a non-trivial Jacobian from the Path-integral measure, regularised to preserve gauge-invariance~\cite{Fujikawa:1979ay}. 
Therefore, while we can absorb all phase \emph{differences} between different flavours without changing $\theta_{\rm QCD}$ --a combination of these phases leading to the CP-violating phase of the CKM matrix-- the common phase has to be absorbed by a combined shift of all quarks, which will shift the CP violating phases from the Yukawa terms to the $G\widetilde G$ term, 
\be
{\cal L}_{\cancel{\rm  CP}}(\theta_Y,\theta_{\rm QCD}) \to {\cal L}_{\rm \cancel{\rm CP}}(\theta_Y-\alpha,\theta_{\rm QCD}+N_f\alpha)  ,
\ee
where $N_f$ is the number of quark flavours. 
Therefore, this particular type of P and T violation is proportional just to one particular combination,
\be
\theta = \theta_{\rm QCD} + N _f\theta_Y .
\ee
Note that the first term arises from the definition of the QCD vacuum while the second comes from quark mass matrices.
In the SM, the latter originate from the Yukawa couplings of the Higgs to quarks, which also lead to the CKM
CP violating phase ($\gamma\sim 60$ degrees in the usual parametrisation). So our naive expectation is that $\theta$ can be easily of the order 1.

The most relevant observational consequence of this P/T violation phase is the existence of electric dipole moments (EDMs) for hadrons, in particular for the neutron,
\be
\label{nedm}
{\cal L}_{\rm EDM} = -\frac{d_n}{2} (\bar \psi_n i \gamma^5 \sigma^{\mu\nu}\psi_n) F_{\mu\nu} ,
\ee
where the magnitude of the neutron EDM has been calculated to be, 
\be
\label{dn_sumrule}
d_n  =   (2.4\pm 1.0)\, \theta\, \times 10^{-3}\, {\rm e\, fm} ,
\ee
with QCD sum rules~\cite{Pospelov:1999mv}. \modified{The result agrees very well with the chiral loop estimate~\cite{Crewther:1979pi} when the CP-violating pion-proton coupling $g_0$ is extracted from the strong part of the proton-neutron mass splitting~\cite{Mereghetti:2010kp,deVries:2015una} instead of the octet baryon mass splitting, which has NLO corrections and is 100\% uncertain.}
\additionalinfo{One gets $dn = (2.1 +- 0.3) \cdot 10^-3 \theta e fm$ $dp = - (2.4 +- 0.3) \cdot 10^-3 \theta e fm$ \url{http://inspirehep.net/record/873543}, in excellent agreement with \eqref{dn_sumrule} pointing at the fact that indeed the chiral loop dominates. }
Ultimately, we expect that the uncertainty can be reduced in lattice QCD calculations but so far
the errors are large. The latest results~\cite{Guo:2015tla,Alexandrou:2015spa,Shintani:2015vsx} still have errors somewhat larger than the above estimate (note a recent correction in these results~\cite{Abramczyk:2017oxr}) but improvements are to be expected over the next few years.
The experimental search for a neutron EDM has however consistently given null results from the early 60s,
and the current most stringent upper bound~\cite{Baker:2006ts,Afach:2015sja} is $|d_n|<3.0 \times 10^{-13} {\rm e\, fm}$ ,
which imposes the spectacular restriction, 
\be
\label{thetabound}
|\theta|<1.3 \times 10^{-10}.
\ee
But, why is $\theta$ so small if composed of two a-priori-arbitrary phases of completely unrelated origin?
This is the essence of the so-called strong CP problem\footnote{The $\theta$ phase violates P {\rm and} T and conserve C, which by conservation of CPT, implies violation of CP.}.

Different solutions to this issue have been discussed in the literature, see for instance~\cite{Peccei:2006as,Pospelov:2005pr}. \todonote{More references needed eventually}
Interestingly, if one of the quark masses would be zero, for instance $m_u=0$, $\theta$ would be unphysical as it can be rotated away into the up-quark phase~\cite{Kaplan:1986ru,Banks:1994yg}. 
In other words, if the up quark is massless, the U(1)$^u_A$ phase transformations of the up quark
are a symmetry at the classical Lagrangian level, but are violated at the quantum level by the colour anomaly term, $G\tilde G$. 
We can use these transformations as redefinitions to show that $\theta$ is unphysical.  
Current lattice QCD estimates converge to $m_u/m_d=0.48(3)$, strongly disfavouring such a possibility\footnote{\modified{A possible loophole pointed out in~\cite{Dvali:2016eay} happens if the measured up-quark mass is not a hard mass from the Higgs Yukawa interaction but is an effective CP-conserving mass given by additional physics, like a neutrino condensate~\cite{Dvali:2016eay}. 
In such case the $\eta'$ meson develops a VEV that cancels the effects of $\theta$, just as the axion does.} }, see~\cite{diCortona:2015ldu} and references therein. 

However, having a U(1)$_A$ symmetry only violated by the colour anomaly would clear the strong CP problem, inspired Roberto Peccei and Helen Quinn. They proposed that such a symmetry indeed could exist in nature without the concomitant massless quark if it was spontaneously broken at a high energy scale~\cite{Peccei:1977hh,Peccei:1977ur}. Such a spontaneously broken global axial symmetry, exact at classical level and \emph{only} violated by the colour anomaly has been henceforth called Peccei-Quinn (PQ) symmetry.
Later, S.~Weinberg and F.~Wilczek independently realised that such an spontaneously broken global symmetry implied a new pNG boson, which Wilczek called the ``axion" after a famous detergent for its virtue of cleansing the SM out of its strong CP stain~\cite{Wilczek:1991jgb}.

The axion solution to the strong CP problem can be understood by writing the
low energy effective theory of the NG boson of the PQ symmetry,
\be
\label{L_axion0}
{\cal L}_a = \frac{1}{2}(\partial_\mu \A)(\partial^\mu \A)
- \frac{\alpha_s}{8\pi}G^a_{\mu\nu}\widetilde G_a^{\mu\nu}\frac{\A}{f_A} + ({\rm derivative \,\, couplings\,\, ...})
\ee
where $\A$ is the axion field and $f_A$ is an energy scale called the ``axion decay constant", related to the scale of spontaneous breaking of the PQ symmetry. The particulars of the implementation of the PQ symmetry define the model-dependent derivative couplings but the colour anomaly is the defining element of the PQ symmetry and implies the coupling of the axion field to $G\widetilde G$. 
\exclude{The interaction terms that respect the PQ symmetry below EWSSB, and that typically arise in many axion models are 
\be 
\label{axioninteractions1}
- \frac{\alpha_s}{8\pi}\frac{E}{N}F^a_{\mu\nu}\widetilde F_a^{\mu\nu}\frac{\A}{f_A} + \frac{\partial_\mu \A}{2 f_\A}\sum_f C_{\A f} \bar f  \gamma^\mu\gamma^5 f
\ee } 
The PQ symmetry reflects here as a shift symmetry $\A\to \A+\beta f_A$ that is respected by the kinetic and interaction terms, except for the anomalous coupling to gluons.
Indeed, thanks to this non-invariance we can now reabsorb $\theta$ into a redefinition of $\A$.
After such redefinition, the axion field plays now the role of a dynamical theta phase $\theta\to \theta(t,x)=\A(t,x)/f_A$,
and indeed henceforth when we refer to $\theta$ in the context of axions we will be speaking of $\A/f_A$.
\modified{Now we can} assign $\theta$ a sign change under P and T transformations $\theta\to -\theta$, so that the $\theta G\widetilde G$ term is now P and T invariant.

But the PQ mechanism is much more than merely rendering a concrete value of $\theta$ unphysical. Note that a vacuum expectation value (VEV) of the axion field has exactly the same powers to induce P,T violation as the original $\theta$ did in the SM.
The most important point of the PQ mechanism comes from the Vafa-Witten theorem stating that, in absence of other CP-violation sources, the QCD vacuum energy has an absolute minimum at $\theta = 0$~\cite{Vafa:1984xg}. Indeed, the tree-level potential is calculable in \modified{chiral perturbation theory (ChPT)}~\cite{DiVecchia:1980yfw},
\be
\label{potential}
\VA(\theta) = -(m_\pi f_\pi)^2\sqrt{1-4\frac{m_u m_d}{(m_u+m_d)^2}\sin^2\(\frac{\theta}{2}\)} ,
\ee
\modified{ where $m_u,m_d,m_\pi$ are the up, down quark and neutral pion masses and $f_\pi\simeq 93$ MeV the pion decay constant. }
In other words, the energy of the QCD vacuum depends on $\theta$ and it is minimum at the CP conserving value. In the SM, $\theta$ is a constant and this energy behaves like a cosmological constant, but once we promote $\theta$ to be a dynamical field, it can now evolve and settle into the minimum. Indeed, given enough time and regardless of  initial conditions\footnote{With a caveat, related with the domain wall number, explained later.} it will relax to $\theta = 0$.  QCD dynamics \emph{itself} solves the strong CP problem!

\subsubsection{Main axion properties}
The axion solution has some model-independent aspects worth noting already. First of all, the strong CP problem is solved regardless of the value of $f_A$, the only model parameter shown in the low-energy effective Lagrangian \eqref{L_axion0}. In this sense, the value of $f_A$ is only relevant in deciding how the axion VEV will evolve towards $\langle \A\rangle=0$. Second, if $f_A$ is very large, the axion will be a very weakly coupled field and the VEV evolves classically towards zero from its initial conditions in the early Universe. In this case, the axion field overshoots the minimum and performs damped oscillations around it, behaving like a condensate of axions, which has the exact properties of cold DM~\cite{Preskill:1982cy,Dine:1982ah,Abbott:1982af}.
Third, the model-independent axion coupling $G\tilde G \A$ will mix the axion with the $\eta'$ at energy scales below QCD confinement and through it with $\pi^0$ and the rest of the mesons.
Through this mixing, the axion acquires a mass given by $m_\A^2f_\A^2 =\partial^2 \VA/\partial{\theta^2}\equiv\chi$. Recently, the value of the topological susceptibility, $\chi$, has been calculated in ChPT~\cite{diCortona:2015ldu} with NLO corrections and in an epic lattice QCD effort~\cite{Borsanyi:2016ksw}. Both results are compatible within the errors, so we quote the one with the smallest uncertainty $\chi = (75.5(5) \rm MeV)^4$. The axion mass becomes, 
\be
\label{axionmass}
m_\A = 5.70(7) \mu{\rm eV}\(\frac{10^{12}{\rm GeV}}{f_\A}\).
\ee
The same mixing with mesons induces also a model-independent coupling to protons, neutrons and two photons, see below.

\exclude{Note that this model-independent coupling to neutrons is compatible with $0$ within the uncertainties.
Additional model dependencies can change these couplings by O(1) amounts and even cause relatively large cancellations or, (in the case of the neutron, uncancellations), see below.}

Fourth, since the axion plays the role of a dynamical $\theta$ in the SM, it also acquires a coupling to the neutron and proton EDMs, given by \eqref{nedm} with the substitution $\theta\to \A/f_A$ (the coupling to protons has the opposite sign).

Lastly, the minimum of the QCD potential is at $\theta=0$ in absence of other sources of CP violation but we know for sure that these do exist. Therefore, the axion will not be able to provide a perfect cancellation of the neutron EDM.  
In the SM the effect is quite small. If physics beyond the SM bring larger sources of CP violation, the magnitude of these couplings can be much larger but it should not shift the minimum of the axion from zero further than the bound \eqref{thetabound} and this restricts very severely these couplings.

\additionalinfo{
The CKM matrix sources CP violation and is expected therefore to shift somewhat the minimum of the potential from $\theta_0=0$ to $\theta_0\neq 0$ and induce CP-violating axion couplings. 
The estimated leading effects in the SM~\cite{Georgi:1986kr} come from integrating out the $c$ quark in 6-quark operators proportional to $V_{ud}V^*_{cd}V_{cs}V^*_{us} G_F^2f_\pi^4$, where $V_{ij}$ is the CKM matrix, $G_F$ is Fermi's constant and $f_\pi\sim 93$ MeV the pion decay constant. The imaginary part of that combination is $w=3\times 10^{-5}\times 10^{-14}=3\times 10^{-19}$ and would lead to values of $\theta_0\sim w$ and CP violating axion couplings to nucleons $N$ of the order,
\be
\label{CPviolatingproto}
{\cal L}_{a-} \sim \frac{w m_u m_d}{(m_u+m_d)f_\A}  \bar N N \A.
\ee
Reference~\cite{Moody:1984ba} mentions another calculation where the effective value of $\theta_0\sim 10^{-15}$.
}

\subsubsection{Axion models}

We can learn more about what couplings of the axion to expect, and the general relation between axion phenomenology at low energies and the high energy theory by studying some axion models.
A general U(1) symmetry acting on the coloured fermions $\{\psi_f\}$ and scalars $\{H\}$ of any extension of the SM is $\psi_f\to e^{i \X_f \alpha}\psi_f, H\to e^{i \X_H \alpha} H$ and can be represented by a vector of charges $\vec \X=\{\X_f,\X_H\}$.
The symmetry has a colour (and/or electromagnetic) anomaly if the triangle anomaly diagrams weighted by the charges are non-zero.
The coefficients of the \modified{anomalies are respectively, }
\be
N =  \sum_f  (-1)^a \X_f S_f \quad , \quad  E = 2\sum_f  (-1)^a \X_f Q_f^2
\ee
\fxnote[noinline,nomarginclue]{Check notation; change a for something more creative}
where $S_f$ is the index of the SU$_c$(3) representation of $\psi_f$  (quarks, i.e. colour triplets have $S_q=1$ with our normalisation) and $a=0,1$ for L and RH fields. \todonote{notation horror, a ALP}
Invariance of any of the Yukawa couplings with the form $\bar f_i H_j f_k$ imposes a constraint $-\X_{f_i}+\X_{H_j}+\X_{f_k}=0$
in the vector space of possible U(1) transformations, while having a non-zero colour anomaly requires
the inequality $\sum_f  (-1)^a\X_f S_f \neq 0$. In the SM, if we restrict ourselves to flavour-conserving symmetries,
we have $q_L, u_R, d_R$ quarks and the \Higgs field spanning a 4-dimensional vector space. Invariance of the
Yukawa couplings that gives mass to the up and down quarks restricts it to a 2D surface, which turns out to be
orthogonal to the vector of the anomaly, $\vec \X\cdot \vec S=0$. One of the two remaining directions corresponds to the Hypercharge gauge symmetry and the last direction is essentially baryon number (which is not chiral, and is not spontaneously broken).
Moreover, in the SM there are not enough degrees of freedom to implement the PQ symmetry because
the 3 possible Goldstone directions of the \Higgs are eaten by the $W,Z$ bosons to become
massive.

In order to implement a PQ symmetry in an extension of the SM, we need extra degrees of freedom and a colour-anomalous direction in $\X$-space. Regarding the former, there are two options: adding new complex scalars or new fermions that condense (composite or dynamical~\cite{Choi:1985cb,Kim:1984pt,Redi:2016esr} axions).
Regarding the direction in $\X$-space we can distinguish two options: the PQ symmetry involves the SM quarks (like in the original \PQWW or the \DFSZ models), or invokes new quarks (like in \KSVZ), enlarging the available $\X$-space.

{\bf PQWW model:} The first axion model considered only the SM fermions but enlarged $\X$-space by invoking 2 Higgs doublets, $H_u,H_d$ with the Yukawa interactions,
\be
{\cal L}_Y \ni -\lambda_u  (\bar q_L H_u) u_R - \lambda_d (\bar q_L H_d ) d_R  -\lambda_e (\bar L H_d) e_R +h.c.
\ee
so that up and down-type fermions get their masses through the VEV of the neutral components of $H_u,H_d$, respectively. A PQ symmetry appears if we forbid terms containing the SM-invariant $H^\dagger_d \epsilon H^*_u$ in the scalar potential, see for instance~\cite{Branco:2011iw}.
Writting the neutral Higgs components as $H_{d0}=e^{i\theta_d}\rho_d/\sqrt{2}, H_{u0}=e^{i\theta_u}\rho_u/\sqrt{2}$,
the kinetic part of the lagrangian gives, after EWSSB,
\be
\label{PQWWkinetic}
|D_\mu H_u|^2+|D_\mu H_d|^2 = \frac{1}{2}v_u^2 (\partial_\mu \theta_u)^2+\frac{1}{2}v_d^2 (\partial_\mu \theta_d)^2
+\frac{m_Z}{v_F}Z_\mu \partial^\mu(v_u^2 \theta_u-v_d^2 \theta_d) + ...
\ee
where $D_\mu$ is the covariant derivative, $v_u,v_d$ are the VEVs of $\rho_u,\rho_d$, $v_F=\sqrt{v_u^2+v_d^2}\simeq 247 $GeV plays the role of the SM Higgs VEV, and we used the hypercharges $Y_{H_d}=-Y_{H_u}=1/2$. In the space spanned by the two possible Goldstone directions $a_d=v_d\theta_d$ and $a_u=v_u\theta_u$, the ``gauge'' direction eaten by the $Z$ boson to become massive corresponds to $(v_u^2 \theta_u-v_d^2 \theta_d)=v_u a_u-v_d a_d$ and the \PQWW axion to its orthogonal direction $\propto v_d a_u+v_u a_d$.
A rotation in $a_u,a_d$ space of angle $\tan\beta = v_u/v_d$ gives a canonically normalised kinetic term ${\cal L}_\phi = (1/2)(\partial_\mu \A_{\WW})^2$ from \eqref{PQWWkinetic}
for the \PQWW axion,  
\be
\A_{\WW} = \modified{a_u \cosbeta  + a_d \sinbeta}  = (\theta_u+\theta_d) \frac{v_u v_d}{v_F} .
\ee
In this model $\theta_u+\theta_d$ is periodic in the interval $(0,2\pi)$ and so is the axion field $\A_{\WW} \to \A_{\WW} + 2\pi n v_{\WW}$ with $n\in \mathbb{Z}$ and $v_{\WW}=v_uv_d/v_F$.
The resulting Lagrangian contains the axion interactions in the Yukawa Lagrangian in terms like $m_u \bar u_L e^{i \theta_u} u_R$ and does not explicitly contain the axion-$G\tilde G$ interaction. However, a chiral redefinition of fermion fields 
\exclude{$u\to e^{-i\gamma^5\theta_u/2}u$ }
$u_L\to e^{i\theta_u/2}u_L, u_R\to e^{-i\theta_u/2}u_R$
(also for down and $e$-type), can be used to reabsorb the $\A_{\WW}$-component of the $u,d$ phase fields, which are (setting the gauge part to 0),
\be
\theta_u= \cosdosbeta\frac{A_{\WW}}{v_{\WW}} ,  \quad
\theta_d = \sindosbeta \frac{A_{\WW}}{v_{\WW}} ,
\ee
producing the required $G\widetilde G \A_{\WW}$ term and derivative interactions (from the fermion kinetic terms). Since quarks and leptons have electric charge, the redefinition also generates a coupling to photons $F\widetilde F \A_{\WW}$.  Redefining the $u_L,u_R, d_L,d_R,e_L,e_R$ fields for the three families we generate the following Lagrangian interactions for the \PQWW axion,
\bea
\label{pqwwGG}
-\frac{\alpha_s}{8\pi} G^a_{\mu\nu}\widetilde G_a^{\mu\nu} (\theta_u+\theta_d)\times 3  &&=
     -\, \frac{\alpha_s}{8\pi} G^a_{\mu\nu}\widetilde G_a^{\mu\nu} \frac{\A_\WW}{v_\WW/3}\\
-\frac{\alpha}{8\pi} F_{\mu\nu}\widetilde F^{\mu\nu} 2
\[3\(\frac{2}{3}\)^2 \theta_u+3\(\frac{-1}{3}\)^2\theta_d+(-1)^2 \theta_d\]\times 3
 &&= -\, \frac{\alpha}{8\pi} F_{\mu\nu}\widetilde F^{\mu\nu} \frac{8}{3} \frac{\A_\WW}{v_\WW/3}\\ 
+\bar u \gamma^\mu\gamma^5 u \frac{\partial_\mu \theta_u}{2}
&&= \cosdosbeta\,\, \bar u \gamma^\mu\gamma^5 u
 \frac{\partial_\mu \A_\WW}{2 v_\WW} \\
 \label{pqwwCde}
+\bar u \gamma^\mu\gamma^5 u \frac{\partial_\mu \theta_u}{2}
+\bar d \gamma^\mu\gamma^5 d  \frac{\partial_\mu \theta_d}{2}
+\bar e \gamma^\mu\gamma^5 e  \frac{\partial_\mu \theta_d}{2}
&&= \sindosbeta\,\,\, \(\bar d \gamma^\mu\gamma^5 d
+\bar e \gamma^\mu\gamma^5 e \) \frac{\partial_\mu \A_\WW}{2 v_\WW} \hspace{1cm}
\eea

The gluonic and photonic anomaly terms have the same origin as the shift in \eqref{anomalyshift}, while the fermion couplings come from the fermion kinetic terms $i \bar u \gamma^\mu \partial_\mu u$, etc. 
Note that our convention is $\gamma^5\psi_{^L_R}=\mp \psi_{^L_R}$. 
If we define the interactions of a generic flavour-conserving axion as 
\be
\label{couplingshighT}
{\cal L}_\A = \frac{1}{2}(\partial_\mu \A)(\partial^\mu \A) 
 - \frac{\alpha_s}{8\pi} G^a_{\mu\nu}\widetilde G_a^{\mu\nu} \frac{\A}{f_\A}
- \frac{\alpha}{8\pi} \frac{E}{N}F_{\mu\nu}\widetilde F^{\mu\nu} \frac{\A}{f_A} + \frac{\partial_\mu \A}{2f_A} \sum_{\psi}  C_{\A\psi}(\bar \psi \gamma^\mu\gamma^5 \psi) , 
\ee
we can now read from \eqref{pqwwGG}-\eqref{pqwwCde} the decay constant, $f_A$, and the couplings to down-type, up-type quarks and charged leptons as  
\be
f_\A = \frac{v_\WW}{3} \quad , \quad
\frac{E}{N} = \frac{8}{3}\quad , \quad 
 C_{\A u}=\frac{\cos^2\beta}{3}\quad , \quad
C_{\A d}=C_{\A e}=\frac{\sin^2\beta}{3} 
\quad (\PQWW).
\ee
Note that the physical range of the axion is $\frac{\A_\WW}{v_\WW}\in(0,2\pi)$ but the QCD potential is
periodic in the quantity $\frac{\A_\WW}{f_\A}=3\frac{a_\WW}{v_\WW}$ so this model has three physically different CP-conserving vacua: $0,2\pi/3, 4\pi/3$. The degeneracy of vacua plays a dramatic role in cosmology, because of the possible cosmologically-stable domain walls, to be reviewed later.
The \PQWW  axion is periodic with 3 vacua, and thus $N_{\rm DW}=3$.
With a convenient normalisation $N_{\rm DW}=N$.
For a generic periodic QCD axion model, we will define $\theta=\A/f_A$ and the axion periodicity as $\A/v_\PQ \in (-\pi,\pi)$ with $N_{\rm DW}=v_\PQ/f_A$ an integer.

{\bf DFSZ: } The \PQWW axion has interactions related to the EW scale and was quickly excluded by a number of experimental constraints~\cite{Donnelly:1978ty,Zehnder:1981qn}. {So were some of its flavour violating ``variants'', for instance by the absence of the rare decay $K^+\to \pi^++\A_\WW$, \cite{Bardeen:1986yb,Asano:1981nh}.}
However, the \PQWW model can be tweaked to become ``invisible", i.e. to escape such constraints, by including a new complex SM singlet scalar, $S=(v_S+\rho_S) e^{i\theta_S}/\sqrt{2}$
with a coupling $\propto H^\dagger_d \epsilon H^*_u S^t$. Two variants are possible with $t=2,1$.
This term violates the original \PQWW symmetry, but respects a new PQ symmetry where the $S$ field transforms with $t \X_S=\X_{H_u}+\X_{H_d}$. It is actually responsible for giving mass to a combination of the \PQWW would-be axion, $\A_\WW$, and the new Goldstone direction, $a_S=\theta_S v_S$.
In particular, \modified{in the spontaneously broken phase, when both Higgs have VEVs, we find,} 
\be
H^\dagger_d \epsilon H^*_u S^t + {\rm h.c.} \to
\modified{
\frac{1}{ 2^\frac{2+t}{2} }
}v_dv_u v_S^t e^{i(t\theta_S-\theta_u-\theta_d)} + {\rm h.c.}
= \modified{\frac{1}{2^\frac{t}{2}} } 
v_dv_u v_S^t \cos(t\theta_S-\theta_u-\theta_d)
\ee
so that the direction $t (a_S/v_S) + (\A_\WW/v_{\WW})$ becomes massive.
The orthogonal direction becomes the so-called \DFSZ axion~\cite{Dine:1981rt,Zhitnitsky:1980tq},
\be
\A =  a_S \cos\gamma - \A_\WW \sin\gamma  \quad {\rm with} \quad \tan\gamma = \frac{t v_{\WW}}{v_S}
\ee
The model is ``invisible'' when $v_S\gg v_F$, which corresponds to a very small mixing $\tan\gamma\ll 1$ and the axion being mostly along the new direction $a_S$. This $S$ field is a SM singlet and has therefore no interactions with the SM.
The \DFSZ axion therefore, acquires all couplings to the SM from the mixing with the $\A_\WW$ direction. This has the interesting implication that all couplings are proportional to the couplings of the \PQWW model. Indeed we can get them from the \PQWW low energy Lagrangian through the simple substitution\footnote{The ellipsis stands for the massive pseudoscalar absent in the low energy theory.}
$\A_{\WW} = -\A \sin\gamma + ...$. Since $f_A$ is defined as the energy scale that divides the axion $\A$ in the $G\widetilde G$ term, $\sin\gamma$ is by definition the ratio of the decay constants,
\be
\label{PQWWtoDFSZ}
\A_{\WW} \to  \frac{f^{\WW}_A}{f_A} \A =\frac{t v_{\WW}}{\sqrt{v_S^2+(t v_{\WW})^2}} \A =
\frac{3 t f_A^{\WW}}{\sqrt{v_S^2+ (t v_{\WW})^2}} \A
\ee
so that $f_A = \sqrt{v_S^2+ (t v_{\WW})^2}/3t$. It is easy to check that $\A$ is periodic in $\sqrt{v_S^2+ (t v_{\WW})^2}$, so $N_{\rm DW} = 3 t$ can be 3 or 6.
With the redefinition \eqref{PQWWtoDFSZ} it is clear that all the \DFSZ axion couplings to quarks and leptons  have the same $C's$ as \PQWW but with the new decay constant. 
\be
\frac{E}{N}= \frac{8}{3}\quad , \quad
 C_{\A u}=\frac{\cos^2\beta}{3}\quad , \quad
C_{\A d}=C_{\A e}=\frac{\sin^2\beta}{3} \quad (\DFSZ \, \sf I).
\ee

There is a variant of \DFSZ where the lepton mass is obtained from a Yukawa coupling to $\epsilon H^*_u$ instead of $H_d$. This gives $C_{Ae}=-C_{Au}$ instead of $C_{Ae}=C_{Ad}$ and thus $E/N=2/3$. We usually call them \DFSZ {\sf I} and {\sf II}, respectively, but in the context of 2 Higgs doublet models they are
labelled type-II and type-IV (or flipped).
Perturbativity of the Yukawa couplings constrain the ratio $0.28<v_u/v_d<140$~\cite{Giannotti:2017hny}, which implies
$0.024<|C_{Ad}|<0.33332$ and $2\times 10^{-5}<|C_{Au}|<0.31$, so are not very restrictive.

{\bf KSVZ: }
The simplest invisible axion model is \KSVZ~\cite{Kim:1979if,Shifman:1979if}, in which the SM is enlarged by a new extra heavy quark $ \KQ $, SM singlet, and a new SM singlet complex scalar ${\KS}$ with the Lagrangian,
\be
{\cal L}_{\rm KSVZ} = i\bar {\KQ}\slashed{D}{\KQ } +\frac{1}{2} |\partial_\mu {\KS}|^2 -\lambda(|{\KS}|^2-v^2)^2 - (y  \bar {\KQ}_L {\KS} {\KQ}_R+\rm h.c.)
\ee
which features a PQ symmetry
\be
{\KQ}_L\to e^{-i \alpha/2}{\KQ}_L \quad , \quad {\KQ}_R\to e^{i \alpha/2}{\KQ}_R \quad , \quad {\KS} \to e^{-i \alpha} {\KS},
\ee
spontaneously broken by the VEV of $\KS$, $\langle \KS \rangle = v_\KS$. The VEV gives a mass $m_\KQ\sim y v_\KS$ to the heavy quark and the radial component of $\KS$, $m_\rho\sim \sqrt{\lambda}v_\KS$, which are supposed much larger than the electroweak scale.
The axion is the NG boson that can be extracted by redefining the scalar as $\KS(x)=(v_\KS+\rho(x)) e^{-i\theta(x)}$ and reabsorbing it in the quark ${\KQ}\to e^{i\gamma^5 \theta(x)/2}\widetilde \KQ$.
This field redefinition generates a kinetic term for the axion field from $|\partial_\mu \KS|^2/2 = v_\KS^2 (\partial_\mu \theta)^2/2 + ...$ and the anomalous coupling to $G\tilde G$ from the anomalous triangle diagram.
Thus, at low energies where ${\KQ},\rho$ are integrated out, and identifying $f_\A=v_\KS$ we find precisely the Lagrangian \eqref{L_axion0} without any further derivative coupling.

{\bf Hadronic axion models:} 
The \KSVZ model has the virtue of simplicity but brings a possible cosmological problem because $\KQ$ turns out to be cosmologically stable and the relic population~\cite{Nardi:1990ku} can easily become quite problematic~\cite{Perl:2001xi,Perl:2004qc,Chuzhoy:2008zy,Perl:2009zz}. A simple solution is to endow $\KQ$ with the same hypercharge as up or down quarks, so that a mixing term with SM quarks is allowed and $\KQ$ decays open up. This does not generate direct axion couplings to SM fermions. However, $\KQ$ gets electrically charged and the PQ symmetry   also has an electromagnetic anomaly. The low-energy theory includes a coupling to photons with $E/N=2Y^2_{\KQ}=2/3$ if the hypercharge coincides with up-type ($Y_{\KQ}=-1/3$) or $E/N=8/3$ if it coincides with down-type quarks ($Y_{\KQ}=2/3$). 
The \KSVZ model can be further generalised to include several coloured fermions (not neccesarily triplets) and scalars. A recent study has considered the former option selecting the models cosmologically unproblematic from the point of view of the stability of coloured relics and the absence of Landau poles below the Planck scale~\cite{DiLuzio:2016sbl,DiLuzio:2017pfr}. 
Under such requirements, the values of $E/N$ found in the phenomenologically reasonable models are $E/N\in (5/3,44/3)$

A radiative correction via the photon coupling produces couplings to SM fermions, most importantly to electrons, given by:
\be
C_{\A \psi} \sim   \frac{3}{4}\frac{\alpha^2}{\pi^2}\(\frac{E}{N}\log\frac{f_\A}{\Lambda}-1.92 \log\(\frac{\Lambda}{m_\psi}\)\)
\ee
where $\Lambda\sim$ GeV and the second log comes from meson mixing, see below. 
\todonote{Some issues with notation, factor of 1/2 with respect to Srednicki?}

\todonote{Exercise with scalars?}

{\bf Axi-majorons: }
Another interesting variant of \KSVZ can be obtained by using the new scalar to  break spontanously $B-L$ at a high energy scale to give mass to RH neutrinos and explain small neutrino masses within the Seesaw mechanism.  In these models, the majoron, a NG boson related to the spontanous breaking of $B-L$~\cite{Chikashige:1980ht,Chikashige:1980ui} can  also be the axion, leading to the so-called axi-Majoron models~\cite{Dias:2014osa}, first discussed as \KLPY~\cite{Kim:1981jw,Langacker:1986rj,Shin:1987xc}. A recent proposal in this direction is the SMASH model~\cite{Ballesteros:2016euj,Ballesteros:2016xej}. Another concept for an axi-majoron is the Ma-xion~\cite{Ma:2017vdv}, where the neutrino masses are generated radiatively. 
The axi-majoron has couplings to neutrinos, \modified{but they are too small for having any relevant  phenomenological consequence.} However, if the Dirac-Yukawa RH couplings with the SM Higgs are large, a loop of RH neutrinos can induce an axion coupling to electrons of order $C_{Ae}\sim 1/16\pi^2N_{\rm DW}$, potentially larger than in Hadronic models, see~\cite{Shin:1987xc,Pilaftsis:1993af} and~\cite{Ballesteros:2016xej}.

{\bf Familons: }
Further axion models can be built by relaxing the assumptions above.
Leaving our stringent requirement of flavour dependent U(1) symmetries, we encounter the
axion as a NG boson associated with the spontaneous breaking of some family/flavour symmetry, as first proposed in~ \cite{Wilczek:1982rv}. A large number of options are available in these models. 
A recent paper proposed the Minimal-Flavour-Violation axion \cite{Arias-Aragon:2017eww} where there are no flavor-changing axion interactions but flavour universality is violated, i.e. axions do not mediate transitions between different families but couple to them with different strengths. In this model, we have $E/N=8/3$ as in \PQWW or \DFSZ. Getting the Yukawa hierarchies correctly requires PQ charges that imply a large $N_{\rm DW}=9$.

A minimal flavour-violating axion model is perhaps the axi-flavon or flaxion~\cite{Calibbi:2016hwq,Ema:2016ops}.
Here, the complex scalar Froggat-Nielsen flavon field~\cite{Froggatt:1978nt} plays the role of the Yukawa couplings when taking a  VEV, while its phase provides an axion with relatively clear predictions, for instance $E/N=8/3$. The axion-meson mixing is different from the standard flavour-neutral axion case but the effective axion-photon coupling still gets a clear prediction \modified{$|C_{\A\gamma}|\in (0.5,1.1)$. } 
The strongest constraint/discovery channel in the model is the $K^+\to \pi^+ \A$ decay, which sets a bound $f_A\gtrsim 10^{10}$ GeV depending on the $sd$ flavour violating coupling.

In general, the freedom in choosing the PQ charges of SM fermions is such that one can engineer models with very surprising properties. 
\modified{For instance, ref.~\cite{DiLuzio:2017ogq} considers a non-universal \DFSZ model with a 2+1 flavour scheme (i.e. universal PQ charges for two generations) and can cancel completely the proton and neutron couplings} (nucleophobic axion, see for an old related idea~\cite{Krauss:1987ud}) and even the electron coupling. While nucleophobia can be obtained by engineering $C_{\A u}\sim 2/3, C_{\A d}\sim 1/3$,  electrophobia requires other means. 
A nucleo and electrophobic axion would be \emph{astrophobic}, for the strongest constraints on axion properties from stellar evolution rely on axion-emission from these three couplings, see Sec.~\ref{sec:astrophysics},  and thus can be largely avoided~\cite{DiLuzio:2017ogq}.

{\bf Other: }
The above list is by no means exhaustive. As mentioned in the introduction, axions can be easily embedded in SUSY~\cite{Nilles:1981py} or GUT~\cite{Wise:1981ry} models. A recent paper summarises model building constraints when the axion is built into a GUT~\cite{Ernst:2018gso}. A recent GUT SMASH-like model, has been proposed along these lines~\cite{Boucenna:2017fna}. Other models in which the axion emerges as a composite particle have also been studied by several authors~\cite{Choi:1985cb,Choi:1985hg,Kim:1984pt,Redi:2016esr}. 


{\bf The issue with gravity: }
Quantum gravity (QG) is expected to violate global symmetries as classical black holes have no global charge, and this imposes a potential harm to the axion solution to the strong CP problem~\cite{Georgi:1981pu}. See~\cite{Alonso:2017avz} for a recent review wormholes and NG bosons. 
This can be illustrated by parametrising QG effects as a series of Planck suppressed operators~\cite{Kamionkowski:1992mf,Barr:1992qq,Masso:2004cv}, which contribute to the axion potential, e.g. in \KSVZ as 
\be
-{\cal L}_{\cancel{\rm PQ}} =  \sum_{n=4...} \frac{c_n}{2 \mPl^{n-4}} \KS^n +{\rm h.c.} 
\to \V_{\cancel{PQ}} \ni \sum_{n}  \frac{|c_n|}{2 \mPl^{n-4}} |\KS|^n \cos\(\theta +\delta_n + n \theta_{\rm SM}\)
\ee 
where the SM angle $\theta_{\rm SM}=\theta_{\rm QCD}+N_f\theta_Y$ reappears here when we absorb it in the axion field and $c_n=|c_n|e^{i\delta}$. With this contribution, the minimum $\theta_0$ of the overall potential $\VA+\V_{\cancel{PQ}}$ is, in general, not $\theta_0=0$ and the ensuing nEDM could easily violate the bound \eqref{thetabound} with $\theta\to \theta_0$, unless the $c_n$ are very small $\lesssim 10^{-10} (\chi/n \mPl^4) (\mPl/f_\A)^n$. 
The $c_n$ are expected to be exponentially suppressed by the classical action so very small values could well be natural but other solutions have been proposed. Fundamental discrete symmetries could forbid the lowest order effective operators~\cite{Georgi:1981pu,Lazarides:1985bj,Dias:2014osa,Choi:2009jt}, effectively suppressing the danger. In other proposals the PQ symmetry could be protected by gauge symmetries~\cite{Holman:1992us,DiLuzio:2017tjx}. A more recent proposal suggests that there might only \modified{be \emph{one} operator in gravity capable of invalidating the axion solution, the gravitational Chern-Simons term~\cite{Dvali:2005an}. Now, if in addition to the QCD axion there is \emph{another} NG boson in nature that couples to such term, gravity will give mass only to one linear combination of the two, leaving the other to solve the strong CP problem. This NG boson could be related with the neutrino lepton number~\cite{Dvali:2013cpa}.}

\subsubsection{Axion couplings and the generalisation to ALPs}

Most of the phenomenological implications of axions and ALPs are due to their feeble interactions with SM
particles at the relatively low energies of laboratory setups below QCD confining temperatures. The most relevant particles are thus protons, neutrons, electrons and, most importantly two photons. These are the relevant couplings also in the relatively hotter stellar interiors and in the Big Bang shortly after the QCD phase transition at a temperature $\sim$150~MeV. The Lagrangian density capturing the relevant interactions for axions at those energies can be divided into CP conserving and CP-violating interactions. The former are given by
\bea
\label{couplingslowT}
{\cal L}^{\rm CP}_\A &=& \frac{1}{2}(\partial_\mu \A)(\partial^\mu \A) - \modified{\VA}(\theta)- C_{A\gamma}\frac{\alpha}{8\pi} F_{\mu\nu}\widetilde F^{\mu\nu} \frac{\A}{f_A} + \frac{\partial_\mu \A}{2f_A} \sum_{\psi}  C_{A\psi}(\bar \psi \gamma^\mu\gamma^5 \psi) + ...
\eea
where the low-energy couplings are obtained from the quark couplings $C_{\A q}$ and the model-independent contributions from meson mixing as

\bea
C_{Ap}&=&-0.47(3) + 0.88(3) C_{Au} - 0.39(2)C_{Ad}-K_{Ah}\\
C_{An}&=&-0.02(3) - 0.39(2)C_{Au} + 0.88(3) C_{Ad}-K_{Ah}\\
K_{Ah} &=& 0.038(5)C_{As}+0.012(5)C_{Ac}+0.009(2)C_{Ab}+0.0035C_{At} \\
C_{A\gamma} &=& \frac{E}{N}-1.92(4) 
\eea
where the bracketed figure corresponds to the experimental error from quark mass estimations and NLO corrections from\footnote{Our conventions are essentially the same as~\cite{diCortona:2015ldu} so we can directly take their results. Note only that we use $\epsilon^{0123}=+1$ where~\cite{diCortona:2015ldu} uses $\epsilon^{0123}=-1$, so we have written ${\cal L}\in -G\widetilde G,-F\widetilde F$ terms with an explicit minus sign in the Lagrangian, while~\cite{diCortona:2015ldu} shows a + sign.}~\cite{diCortona:2015ldu}.
Note that the model-independent coupling to neutrons is compatible with $0$ within the uncertainties.
Additional model dependencies can change these couplings by O(1) amounts and even cause relatively large cancellations~\cite{DiLuzio:2017ogq}.
The coupling to charged leptons does not change when matching above and below $\Lambda_{\rm QCD}$.
The axion also develops couplings to pions and mesons, relevant for flavour changing searches in the context of familons and cosmology. \todonote{We have chosen not to review them?}
At dimension 6 other axion couplings arise. 
Phenomenologically,  the most relevant is the one leading to the nEDM, 
\be
\label{AxionnEDMcoupling}
-  F_{\mu\nu}\frac{\A}{f_\A}\sum_{\psi=p,n} \frac{ {C}_{\A \psi \gamma}}{2 m_\psi} (i\bar \psi \sigma^{\mu\nu} \gamma^5 \psi) + ... ,
\ee
where the couplings are estimated as $C_{\A n\gamma}=-C_{\A p\gamma}\simeq 0.011 e$ from \eqref{dn_sumrule}. 

All the above couplings conserve P and T if the axion field is pseudoscalar, i.e. if it transforms $\A\to -\A$ under a parity or time reversal transformation. However, any other source of CP violation in the theory will propagate and produce CP-violating couplings of the axion. Phenomenologically, the most relevant are perhaps the scalar Yukawa couplings, 
\be
\label{CPviolatingproto}
{\cal L}^{\cancel{\rm CP}}  = -\frac{\A}{f_\A}\sum_{\psi=p,n} {\bar C}_{\A \psi} m_\psi \bar \psi \psi+ ...
\ee
where the values of the coefficients ${\bar C}_{\A\psi}$ depend on CP violation in the SM 
and physics beyond it. The couplings \eqref{CPviolatingproto} can be generated \emph{directly} by radiative corrections and but also \emph{indirectly} though higher order CP-conserving operators like $\A^2 \bar N N$ 
because the CP-violation will in general shift the minimum of the axion potential \eqref{potential} from $\theta_0=0$ to some value $\theta_0\neq 0$, generating interactions like $2\theta_0\A \bar N N$.  
In the SM, CP-violation comes from the CKM phase and the leading effects are though to come from integrating out the $c$ quark in 6-quark operators proportional to $V_{ud}V^*_{cd}V_{cs}V^*_{us} G_F^2f_\pi^4$~\cite{Georgi:1986kr}, where $V_{ij}$ is the CKM matrix, $G_F$ is Fermi's constant and $f_\pi\sim 93$ MeV the pion decay constant. The imaginary part of that combination is $w=3\times 10^{-5}\times 10^{-14}=3\times 10^{-19}$ and would lead to values of $\theta_0\sim w$ and CP violating axion couplings to nucleons $N$ of the order $|{\bar C}_{AN}|\sim w m_u m_d/((m_u+m_d) m_N)$ in~\cite{Georgi:1986kr}. The contribution from any vacuum shifting CP-violation contribution is  $|{\bar C}_{AN}|\sim \theta_0 2m_u m_d \sigma_{N\pi} /m_N (m_u+m_d)^2$ where $\sigma_{N\pi}\simeq 60$ MeV~\cite{Moody:1984ba}. Reference~\cite{Moody:1984ba} estimates the SM CKM contribution as $\theta_0\sim 10^{-15}$.  
Note the contribution to the couplings could be correlated with the neutron EDM $d_n\propto \theta_0$, and therefore constrained to $|{\bar C}_{\A N}|<3.7\times 10^{-12}$. However, in the most general case, CP-violation effects  shift $\theta_0$ and also contribute to ${\bar C}_{\A N}$ directly so the constraint is valid barring cancelations, see~\cite{Pospelov:2005pr} for a review on EDMs and examples. 
On another note, the axion field is expected to have a VEV for dynamical reasons (see section~\ref{sec:cosmology}), although it features damped oscillations. We are talking about the axion DM field itself. Locally, it will produce CP violating effects that oscillate in time with a characteristic frequency (given by the axion mass) and will be searched for experimentally. The neutron EDM is perhaps the most interesting example but there are others~\cite{Pospelov:2008gg}.

In this review, we \modified{define axion-like particles as those} having interactions similar to the axion, whose origin is expected to be similar, even if some of the $C$ coefficients are zero at tree-level and the mass and decay constant hold a different relation than the QCD axion does.  We will denote them as $a$ and reserve the symbol $\A$ for the QCD axion.
It is convenient to define a set of generic CP-even and CP-odd ALP interactions to photons and fermions $f$,
\bea
\label{ALPinteractions}
{\cal L}_{\rm ALP-int.} &=& -\frac{\gagamma}{4}F_{\mu\nu}\widetilde F^{\mu\nu} a
- a  \sum_{\psi}  g_{a\psi}(i \bar \psi \gamma^5 \psi)
- a  F_{\mu\nu}\sum_{\psi}  \frac{ g_{a \psi\gamma }}{2} (i\bar \psi \sigma^{\mu\nu} \gamma^5 \psi) + ... \\
\label{ALPCPoddinteractions}
&& - \frac{\bar g_{a\gamma}}{4}F_{\mu\nu} F^{\mu\nu} a  - a  \sum_{\psi}  \bar g_{a\psi} (\bar \psi \psi)
\eea
\todonote{notation collision, $f, f_a, f_\A$ ... flavour changing}

Here, we have used the fermion equations of motion to put the couplings in Yukawa-form by integrating by parts 
\be
(\partial_\mu a) (\bar \psi \gamma^\mu \gamma^5 \psi) \equiv 
- a 2 m_i (i\bar \psi  \gamma^5 \psi) . 
\ee
showing that the CP conserving axial couplings are proportional to the fermion masses. 
\todonote{Note that it is not always equivalent. PDG reference}

Historically, fermion couplings, $g_{af}\propto m_f/f_a$ with some ALP decay constant $f_a$, have been defined dimensionless and the photon couplings $\gagamma$ with dimensions of inverse energy from $f_a^{-1}$.
For QCD axions, the mass and couplings are related by
\bea
g_{\A f}&\equiv\frac{C_{\A f}m_f}{f_\A}&=1.75 \times 10^{-13}C_{\A f}\frac{m_f}{\rm GeV}\frac{m_\A}{\mu\rm eV}, \\
\label{axionphotoncoupling}
g_{\A \gamma}&\equiv\frac{\alpha}{2\pi}\frac{C_{\A \gamma}}{f_\A}& =2.0 \times 10^{-16}C_{\A \gamma}\frac{m_\A}{\mu\rm eV} \, {\rm GeV}^{-1},  \\ 
\label{axionEDMcoupling}
g_{\A\gamma n} &\equiv
\frac{C_{\A\gamma n} }{m_n f_\A} &= 6.4\times  10^{-16} \frac{m_\A}{\mu\rm eV}{\rm GeV}^{-2},
\eea
note that with the last definition, an axion VEV implies a neutron EDM
$d_n= g_{\A\gamma n} \langle A\rangle $, $d_n=  g_{\A\gamma n} \langle \theta \rangle f_\A $.
Therefore, QCD axions models occupy bands in the phenomenological $(m_a,g_{a})$ planes.
We usually define ``QCD axion bands'' by bracketing the $C's$ on a reasonable set of models.
The photon coupling has been the most studied. Early works\cite{Kaplan:1985dv,Cheng:1995fd,Kim:1998va} have been recently completed by a  more phenomenological approach~\cite{DiLuzio:2016sbl,DiLuzio:2017pfr}, that we follow to define the band on the ($\gagamma$,$m_a$) space in the plots of this review. For the $g_{ae}$ coupling we follow the lower and upper bounds on $C_{ae}$ derived from the perturbativity of the Yukawa couplings presented above.

\todonote{Schizons~\cite{Hill:1988bu}}

\subsection{Beyond the QCD axion,  ALPs and other WISPs}

The theory and phenomenology of axions is to a large extent shared with any other pNG bosons
which have a low mass and very weak couplings coming from a spontaneously broken symmetry at very high energy scales. However, these \emph{axion-like particles}~\cite{Masso:1995tw,Masso:2002ip,Ringwald:2012hr} will in general have nothing to do with the PQ mechanism and will not get their masses from QCD effects but from some other dynamics that break explicitly the global symmetry.
Thus, while for the QCD axion we have $m_\A f_\A\sim m_\pi f_\pi$, the quantity $m_af_a$ can be larger or smaller for other ALPs.
As already mentioned, string theory predicts a rich spectrum of ALPs (including the axion itself)~\cite{Arvanitaki:2009fg,Cicoli:2012sz,Ringwald:2012cu}.

Most of the experiments that search for the effects of axions coupling to some SM particles (photons, electrons, nucleons) can be sensitive to ALPs with a comparatively smaller value of $\Lambda$ because for the same mass the couplings $\propto 1/f_a$ will be larger. Thus axion experiments have the potential to discover ALPs other than the axion, which as we have mentioned arise quite naturally in extensions of the SM.
Unfortunately, because of this possible degeneracy it might be complicated to attribute a possible discovery signal in one experiment to the existence of purely QCD axions, i.e. the ones involved in the solution of the strong CP problem.
In such a case, we would need as many different signals as possible to discriminate between axions and ALPs.
Nevertheless, the current variety of experiments and the rate at which new ones are proposed lead us to think that it might be possible, at least in certain cases, to identify a purely QCD axion discovery.

\begin{sidewaystable}[p]
   \centering
   \begin{tabular}{@{} c c cccc cccc @{}} 
      \toprule 
                  &    & \multicolumn{4}{c}{High-E couplings} & \multicolumn{4}{c}{Low-E couplings} \\
      \cmidrule(r){3-6}       \cmidrule(r){7-10}               
      Model    &  $\modified{N_{\rm DW}}$ & $E/N$ & $C_{\A u}$ & $C_{\A d}$ & $C_{\A e}$ & $C_{\A \gamma}$ & $C_{\A p}$ & $C_{\A n}$ & $C_{\A e}$ \\
      \midrule
      \PQWW &  3 & $\sfrac{8}{3}$     & $\sfrac{c_\beta^2}{3}$ & $\sfrac{s_\beta^2}{3}$ & $\sfrac{s_\beta^2}{3}$ 
      		   & 0.75 & ... & ... & ... 	 \\
      \DFSZ I &  6,3  & $\sfrac{8}{3}$     & $\sfrac{c_\beta^2}{3}$ & $\sfrac{s_\beta^2}{3}$ & $\sfrac{s_\beta^2}{3}$ 
      		   & 0.75 & (-0.2,-0.6) & (-0.16,0.26) & (0.024,\sfrac{1}{3}) 	 \\
      \DFSZ II & 6,3 &  $\sfrac{2}{3}$     & $\sfrac{c_\beta^2}{3}$ & $\sfrac{s_\beta^2}{3}$ & $-\sfrac{c_\beta^2}{3}$ 
      		   & -1.25 & (-0.2,-0.6) & (-0.16,0.26) & (-\sfrac{1}{3},0) 	 \\
      \KSVZ   &  1 &  0     & \gl & \gl & 0
      		   & -1.92 & -0.47 & -0.02(3) & $\sim 2\times 10^{-4}$ 	 \\
  Hadronic $1{\cal Q}$~\cite{DiLuzio:2016sbl}   &  $1...20$ &  $\sfrac{1}{6}...\sfrac{44}{3}$ & \gl & \gl & \gal  
      		   & \modified{-0.25 ... 12.7$^\dagger$} & -0.47 & -0.02(3) & (0.05 ... 5) $\times10^{-3}$	 \\	
      {\sf SMASH}~\cite{Ballesteros:2016xej}  &  1 &  $\sfrac{8}{3},\sfrac{2}{3}$     & \gl & \gl & \nl
      		   & 0.75,-1.25 & -0.47 & -0.02(3) & $(-0.16,0.16)$ 	 \\  
\midrule
MFVA~\cite{Arias-Aragon:2017eww} &  9 &  $\sfrac{2}{3},\sfrac{8}{3}$ & $0 $ & $\sfrac{1}{3} $ & $\sfrac{1}{3} $
      		   & $0.75,-1.25$ & $\sim -0.6$ & $\sim -0.26$ & $\sim \sfrac{1}{3}$ 	 \\  
Flaxion/Axi-flavon~\cite{Calibbi:2016hwq,Ema:2016ops} &  - &  $\sfrac{8}{3}$ & $\modified{\sim O(1)}$ & $\modified{\sim O(1)}$ & $\modified{\sim O(1)}$
      		   & \modified{(0.5,1.1)} & - & - & - 	 \\  
Astrophobic M1,2~\cite{DiLuzio:2017ogq} &  1,2 &  $\sfrac{2}{3},\sfrac{8}{3}$ & $\sim \sfrac{2}{3}$ & $\sim \sfrac{1}{3}$ & $\sim 0 $
      		   & -1.25,0.75 & $\sim 10^{-2}$ & $\sim 10^{-2}$ & $\sim 0$ 	 \\  
\modified{Astrophobic M3,4~\cite{DiLuzio:2017ogq} &  1,2 &  $\sfrac{-4}{3},\sfrac{14}{3}$  & $\sim \sfrac{2}{3}$ & $\sim \sfrac{1}{3}$ & $\sim 0$           & -3.3,2.7 & $\sim 10^{-2}$ & $\sim 10^{-2}$ & $\sim 0$} 	 \\  

      \bottomrule
   \end{tabular}
   \caption{CP and flavour conserving QCD Axion couplings in selected models. 
   Some ranges are limited by perturbativity of the Yukawa couplings. 
   In hadronic models, fermion couplings arise at the loop level in the high-energy theory, while the low energy couplings to nucleons include meson mixing. A tilde indicates model variability uncertain to the authors (of this review). \hspace{\textwidth}
   \modified{ $^\dagger$ Only minimum and maximum values of $E/N$ are quoted. For $C_{\A\gamma}$'s, we quote the ones giving the minimum and maximum of its absolute value (the relevant for detection).} 
   }
   \label{axionmodellist}
\end{sidewaystable}

\todonote{in table: Say N=NDW as it pertains}

Finally, it is worth mentioning that other low-mass particles, very weakly coupled to known SM particles
share many of the theoretical and phenomenological aspects of axions and ALPs. In particular, they can be often searched for with the same experiments. Moreover,  their study can also give us hints of fundamental properties of nature and/or new high energy scales.
Many of the axion experiments covered in this review are sensitive to other bosonic particles even if very weakly coupled.
A prime example are hidden photons, (also called paraphotons, dark photons, $Z's$~\cite{Okun:1982xi,Galison:1983pa,Holdom:1986eq,Ahlers:2007qf,Bjorken:2009mm}), gauge bosons of a new U$_h$(1) gauge symmetry of a hidden sector, i.e. a set of fields uncharged under the SM gauge interactions.
Even very heavy mediator fields coupling to the SM and this hidden sector can induce kinetic mixing between
the hidden photon and the standard photon~\cite{Galison:1983pa,Holdom:1985ag}. The kinetic mixing angle between the hidden and standard photon is naturally small --values $\chi_h \sim 10^{-13}-10^{-2}$ are typically predicted in the literature~\cite{Dienes:1996zr,Lukas:1999nh,Abel:2003ue,Blumenhagen:2005ga,Abel:2006qt,Abel:2008ai,Goodsell:2009xc,Bullimore:2010aj,Goodsell:2010ie,Heckman:2010fh,Cicoli:2011yh,Goodsell:2011wn}. The mixing leads to a number of observable consequences in laboratory experiments~\cite{Jaeckel:2007ch,Povey:2011ak,Schwarz:2011gu,Parker:2013fxa,An:2013yua,Mizumoto:2013jy,Dobrich:2014kda,Graham:2014sha,An:2014twa,Suzuki:2015sza,Arias:2014ela,Villalba-Chavez:2014nya,Chaudhuri:2014dla,Schwarz:2015lqa,Hochberg:2016ajh,Jaeckel:2015kea,Arias:2016vxn}, astrophysics~\cite{Redondo:2008aa,Lobanov:2012pt,Redondo:2013lna,Kazanas:2014mca,Redondo:2015iea,Ayala:2015juy,Vinyoles:2015aba,Chang:2016ntp,Hardy:2016kme} and cosmology~\cite{Pospelov:2008jk,Jaeckel:2008fi,Redondo:2008ec,Mirizzi:2009iz,Nelson:2011sf,Blennow:2012de,Arias:2012az,Vogel:2015bna,Kunze:2015noa,Graham:2015rva,Choi:2017kzp}, which are very similar to those of an ALP coupled to photons.  Recent reviews can be found in~\cite{Jaeckel:2013ija,Hewett:2012ns}.
Hidden photons can easily arise in the context of string theory, where their mass and the size of kinetic mixing are correlated with properties of the extra dimensions and the string coupling, see for instance~\cite{Goodsell:2009xc,Cicoli:2011yh}.

It is worth mentioning that low mass particles charged under the U$_h$(1) hidden group, will appear to have a small electric charge of size $g_h \chi_h/e$ where $g_h$ is the gauge coupling associated to U$_h$(1).
These minicharged particles can have similar implications in astrophysics and cosmology~\cite{Okun:1982xi,Holdom:1985ag,Davidson:1993sj,Vinyoles:2015khy} than axions and ALPs, particularly in laser experiments~\cite{Ahlers:2007qf}.

Scalar fields invoked to explain the identity of dark energy, such as quintessence fields~\cite{Wetterich:1987fm,Peebles:1987ek,Frieman:1995pm,Caldwell:1997ii}, chameleons~\cite{Khoury:2003rn,Khoury:2003aq,Burrage:2016bwy}, galileons~\cite{Nicolis:2008in}, symmetrons~\cite{Hinterbichler:2010es} also share some of the peculiar signatures of axions and ALPs.
However, contrary to axions, which have naturally pseudoscalar couplings to fermions, these have scalar couplings to matter and can mediate spin-independent long range forces that easily compete with gravity. Indeed some of these models, use non-linearities to reduce the long-range forces and evade the strong constraints, often implying violations of equivalence principle~\cite{Khoury:2010xi}. These scalar couplings can also lead to variations of the fundamental constants, subject to further constraints.

In this review we will be mostly concerned in reviewing experimental approaches for the search of axions and axion-like particles, but we will highlight whenever these searches are also competitive in the quest of identifying other types of WISPs.

\section{Axion cosmology and astrophysics: constraints and hints}
\label{sec:cosmoastro}

In this section we aim at a concise review of the current {\rm indirect} constraints on axions and ALPs from astrophysics and cosmology. Laboratory searches are the main subject of this review and will be discussed at length in the body of the paper.
The impact of axions and ALPs in astrophysics and cosmology is very often model-dependent and we will make special emphasis on the assumptions of each constraint.
We discuss the different arguments from the early big bang, inflation, dark matter and dark radiation, physics of the cosmic microwave background (CMB), the dark ages, reionisation, structure formation, stellar evolution and some other stellar probes, and ending with cosmic rays.
Most importantly, sometimes the effects of axions/ALPs actually improve the matching of observations with the theoretical expectations. In this case, we speak of \emph{hints} towards the existence of an axion/ALP with certain characteristics. The case is that observational errors and systematics are difficult to ascertain and thus we cannot take these hints as discoveries. However, if many hints point at the same type of axion with a consistent set of couplings the significance increases. We will also describe the most interesting regions of parameter space where some of these hints could be due to the same and only particle.
For the sake of generality we will refer to ALPs in this section. When we talk about the axion we will mean the QCD axion solving the strong CP problem. An comprehensive review on axion and ALP cosmology to largely complement our discussion can be found in~\cite{Marsh:2015xka}. 

\subsection{Axions and ALPs in cosmology}
\label{sec:cosmology}

\begin{figure}[t!]
\begin{center}
\tikzsetnextfilename{constraints_ALPs}
\resizebox{0.9\linewidth}{!}{\input{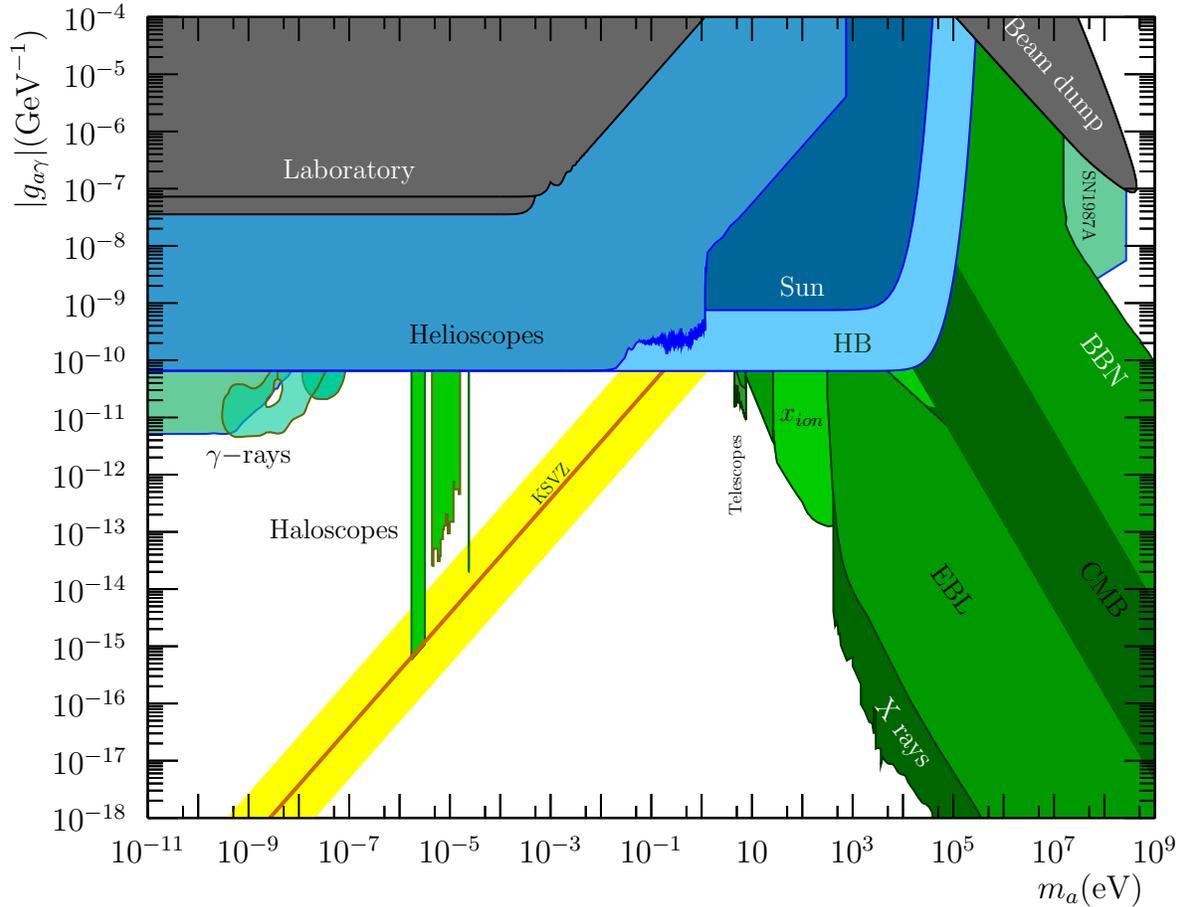}}
\caption{Review of current constraints in the overall ($\gagamma$, $m_a$) plane. We introduce the color criteria for the rest of plots of this review: black/gray for purely laboratory results, bluish colors for helioscope experiments or bounds depending on stellar physics, and greenish for haloscopes or cosmology-dependent arguments. Yellow/orange are reserved for hinted regions of the parameter space, like, in this case, the QCD axion band. In this plot we present only current bounds, for future prospects we refer to following plots later on. We refer to the text for explanation of each region. 
Adapted and updated from~\cite{Cadamuro:2010cz}.}
\label{fig:ALPs_constraints}
\end{center}
\end{figure}

\begin{figure}[t!]
\begin{center}
\tikzsetnextfilename{constraints_gae}
\resizebox{0.7\linewidth}{!}{\input{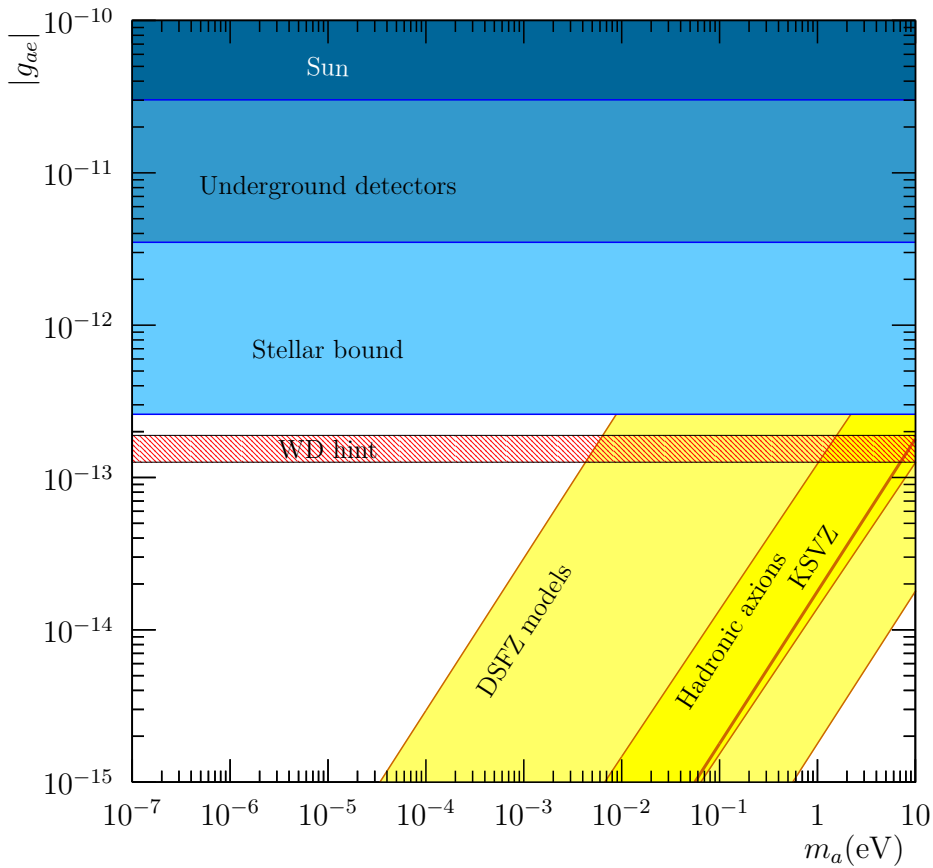}}
\caption{Review of current constraints on an ALP coupled to electrons in the mass-coupling plane, ($m_a$, $g_{ae}$). Yellow regions contain \DFSZ models respecting unitarity constraints on the SM Yukawa couplings and Hadronic axion models --like \KSVZ-- where the electron coupling arises from the 2-photon-loop and is proportional to $E/N$. The $g_{ae}$ coupling is bounded from stellar physics, solar physics and searches for solar axions via $g_{ae}$-channel in underground experiments. See text for details. The hint of $g_{ae}=1.6^{+0.29}_{-0.34}\times 10^{-13}$~\cite{Giannotti:2017hny} from stellar physics (see section~\ref{sec:astrophysics}) is also shown.}
\label{fig:ALPgae_constraints}
\end{center}
\end{figure}

Axions and ALPs appear in non-renormalisable theories below a certain energy scale, associated with $f_a$. Above such a scale the axion is typically not an adequate degree of freedom to describe physics, and in particular the vacuum of a theory. For this reason, it is expected that the axion ``appears'' after a phase transition in the early Universe at very high energies/temperatures $\sim f_a$. This phase transition is controlled by an order parameter, typically the VEV of some other field, e.g. the VEV of $\KS$ in \KSVZ. When this order parameter takes the low temperature value, the PQ symmetry becomes spontaneously broken and a ``flat'' direction appears in the potential, the QCD axion. The same scheme applies for an ALP.  
Deviations from flatness are related with explicit violations of the global shift symmetry, most importantly the axion/ALP mass, which is responsible for ALPs becoming DM. 
We assume that these deviations are irrelevant at the scale $f_a$ by considering $m_a\ll f_a$. 
In the QCD axion case this is even more so because the axion potential arises from the $G\widetilde G A$ interaction and QCD instantons, which are extremely suppressed at high temperatures. 
Thus, at the phase transition, the ALP field is effectively massless and will take different vacuum expectation values VEVs in regions of the Universe that are causally disconnected, i.e. of size comparable with the causal horizon $\sim t$ (cosmic time). One speaks of different ``patches''  of the scale of the horizon, but of course they are smoothly patched together. At the phase transition, ALP interactions with the thermal bath are not expected to be small, at least with the particles closely related to them. Therefore some thermal ALP population is expected on top of this smooth field. The wavelength related to the VEV is of horizon size, $\sim t$, while the typical wavelength for thermal fluctuations is $\sim t( T/\mPl)\ll 1$ with $T$ temperature and $\mPl=1/\sqrt{G_N}=1.22\times 10^{19}$ GeV the Planck mass. Usually one has $T\sim f_a \ll \mPl$ so there is a huge separation of scales between the long-wavelength ALPs and their thermal counterpart.
If the ALP is a periodic field, like in the cases where it corresponds to the phase of a new scalar field (\PQWW, \KSVZ, \DFSZ...) or fermion condensate, this patchy distribution is accompanied by a network of cosmic strings formed by the Kibble mechanism~\cite{Kibble:1980mv}. In the string cores, of typical size $\sim {\cal O}(1/f_a)$ the order parameter cannot break the PQ symmetry restricted by topology and a huge energy density $\sim f_a^4$ is stored. As the network evolves, the overall string length decreases by straightening and collapsing loops, some energy is radiated in the form of low-momentum axions. 

There are therefore at least two ALP populations, cold and hot. The first will become cold dark matter and the second dark radiation or hot dark matter (depending on the exact value of the ALP mass and cosmological history). We discuss them in order. 
The effects of these populations are quite diverse and so are the ensuing constraints. In particular, some constraints can depend on the effects of these populations through the ALP interactions with SM particles and will disappear at low values of the ALP couplings, but others will depend only on their gravitational effects as dark matter and will depend on $m_a,f_a$ and some extra cosmological assumption.  

Our working definitions are axion-inspired but generic for ALPs. We define our ALP, $a$ to be defined in a circle of radius $v$, $a/v\in (-\pi,\pi)$, and having a relatively small potential energy $\V(\thetalp)=\V(\thetalp+2\pi)$ with a different periodicity expressed as $\thetalp = a/f_a$ with  $f_a=v/N_{\rm DW}$.

\subsubsection{Cold dark matter}
\label{sec:axions_as_DM}

The impact of ALPs in cosmology is strongly affected by the moment where cosmic inflation happens, if it does at all.
If the PQ-like symmetry breaks spontaneously before inflation and is never restored afterwards we speak of the \emph{pre-inflation scenario}. Inflation stretches a tiny patch of the Universe until it is so large that our current Universe fits inside. In order to explain the isotropy of the CMB temperature and the homogeneity of the large scale structure of the Universe, the size of the original patch must be much smaller than the horizon at that time. Therefore the axion field inside was also essentially homogeneous and it gets even more so during inflation. The further evolution of this ALP ``zero-mode'' in our Universe follows the simple dynamics of a decoupled scalar field,
\be
\ddot a + 3 H \dot a +\frac{\partial \V(a)}{\partial a} = 0,
\ee
where the Hubble expansion rate $H=\dot \SF / \SF$ is the logarithmic derivative of the scale factor $\SF(t)$ of the assumed Friedmann-Lema\^itre-Robertson-Walker metric describing our expanding Universe. The solution is a constant initial value of the field, $a_i$, until a characteristic time 
\be
t_1\sim 1/\sqrt{\left.\partial^2 \V/\partial a^2\right|_{i}} ,
\ee
at which the ALP rolls down its potential towards its minimum and performs damped oscillations. This state corresponds to a coherent state of non-relativistic ALPs that behaves as \emph{cold dark matter} at time scales longer than the period of oscillations. If the initial condition is not far from the minimum, we can always approximate the potential as a harmonic one, $\V\sim m_a^2a^2/2$ and obtain the density\footnote{We use $\rho$ for any energy density and $\dmden$ exclusively for cold dark matter. $\Omega's$ are cosmological averaged energy densities normalised to the critical density $\rho_{\rm cr}  = 10.5\,h^2$  keV/cm$^3$ and $h=H(t_0)$ in units of km/(Mpc s). }  today~\cite{Piazza:2010ye,Arias:2012az},
\be
\label{ALPabundance}
\exclude{\rho^{\rm pre}_{ a} \sim 0.17 \frac{\rm keV}{\rm cm^3} \sqrt{\frac{m_a}{\rm eV}} \sqrt{\frac{m_a}{m_a(t_1)}}\(\frac{a_i}{10^{11} \rm GeV}\)^2 {\cal F}
}
\Omega^{\rm pre}_ah^2 \sim 0.016 \sqrt{\frac{m_a}{\rm eV}} \sqrt{\frac{m_a}{m_a(t_1)}}\(\frac{a_i}{10^{11} \rm GeV}\)^2 {\cal F}_g
\ee
where the parameter ${\cal F}_g$ is a smooth function of the temperature at $t_1$, $T_1=T(t_1)$~\cite{Arias:2012az} taking values $\in (1,0.3)$ when $T_1\in (T_0,200 {\rm GeV})$. 
The above equation is also approximately valid even if the mass depends slowly on time, i.e. it is different at the onset of oscillations and today $m_a(t_1)\neq m_a$.
This is precisely the case of the QCD axion, where one finds $\VA(\A)\sim \chi(T) (1-\cos\theta)$ at sufficiently high temperatures --the previously quoted \eqref{potential} is valid only sufficiently below the QCD cross over at $T_{\rm QCD}\sim 150$ MeV.
 The topological susceptibility, $\chi(T)$, strongly suppressed by thermal effects above the QCD confinement scale  is, because of its smallness and relation to topology, extremely difficult to compute in lattice QCD. A recent burst of interest~\cite{Buchoff:2013nra,Bonati:2015vqz,Borsanyi:2015cka} 
has resulted in the first calculation including the relevant quarks up to $2$ GeV~\cite{Borsanyi:2016ksw}. \todonote{Describe? plot?}
Departures from the harmonic approximation can be computed numerically and depend on the concrete ALP potential (The low-$T$ QCD potential~\eqref{potential} is an explicit example).

In the \emph{post-inflation scenario}, the PQ-like phase transition happens after inflation and therefore the Universe has patches of different ALP VEVs. The correlation length increases as $\sim t$ because fluctuations of wavelength shorter than the horizon decay as radiation and because the cosmic strings (which force all values of $\theta_a\in (0,2\pi)$ around them) can only annihilate by collapsing loops of size smaller than the causal horizon and straightening long strings on those length scales~\cite{Vilenkin:2000jqa}. At $t_1$, the ALP potential becomes effective in differentiating values of $\theta_a$ and the VEV in all the Universe is revealed to be misaligned and starts moving towards the closest minimum (there might be more than one!). The dominating long fluctuations have wavelengths the size of horizon at $t_1$, which is precisely $t_1$ and upon oscillating around the minimum behave as cold dark matter, with a larger velocity dispersion than the pre-inflation scenario. There are three main differences with the previous scenario. First, different patches have different values of $a_i$ so the dark matter density is inhomogeneous. The size of these patches is $t_1\sim 1/m_a(t_1)$, and today they have stretched to
$t_1 \SF(t_0)/\SF(t_1)\sim t_1 (\sqrt{m_a(t_1) \mPl}/T_0)\sim 30\sqrt{10^{-14}{\rm eV}/m_a(t_1)}$ pc, which can be much smaller than the cosmological probes of large scale structure even for tiny masses. Overdense regions will collapse to small and dense DM clumps called axion/ALP miniclusters~\cite{Hogan:1988mp,Kolb:1993zz,Zurek:2006sy,Hardy:2016mns} .
Second, if our Universe contains sufficiently many patches with random initial conditions the average density is given by calculable statistical average, and the uncertainty about $\theta_i$ disappears~\cite{Turner:1985si}. In the quadratic approximation for the potential one finds $\langle \theta_a^2\rangle=\pi^2/3$ for $\theta_a\in(-\pi,\pi)$ for the \modified{axions coming from the realignment mechanism.}
 This scenario can be much more predictive than the pre-inflation. However, the predictivity is spoiled because in general the string and domain wall network decay producing axions, which also contribute to DM. 
 See below for the discussion on the QCD axion mass needed to get the right DM density.
Third, in a general ALP potential we can have several local minima or even degenerate minima, \modified{which implies a critical issue to be discussed next.}

{\bf Domain wall problem: }
The QCD axion can be easily afflicted by having more than one CP-conserving minimum. By definition, the axion gets its potential from QCD instantons and the $G\widetilde G(\A/f_\A)$-coupling, which is $2\pi-$periodic in $\A/f_\A$.
But an axion model can define the NG boson in a physical region $N_{\rm DW}$-times larger than that. This was the case in PQWW where $\theta_{u}+\theta_{d}=\A_{\WW}/v_{\WW}$ is $2\pi$-periodic but $f^{\WW}_\A =v_{\rm PQ}/3$, therefore $\A_{\WW}/f^{\WW}_\A=3(\theta_{u}+\theta_{d})$ can turn $N_{\rm DW}=3$ times around the periodic QCD potential before reaching a physically equivalent value, and thus it will have physically-different minima at $\A_{\WW}/f^{\WW}_\A=0,2\pi,4\pi$, all of them CP conserving. This is certainly the case in the \DFSZ models, where we have $N_{\rm DW}=6,3$ for $t=2,1$, respectively, and can be the case in hadronic models with many extra new colored fermions ${\KQ}$, see Tab.~\ref{axionmodellist}.
In the post-inflation case there are patches with all values of the ALP field and different regions will tend to different minima at $t_1$. A domain wall (DW) is the \modified{zone that appears at the physical boundary between regions that chose a different minimum. Across the DW, the axion field interpolates between these minima, reaching very large densities $\VA\sim \chi$ in its centre, where $\theta_\A$ takes the value $\pi$ (modulo $2\pi$).} Given enough time, the energy stored in DWs can dominate over radiation and matter and lead to a Universe very different from observed~\cite{Zeldovich:1974uw}. This is certainly the case for axions with $N_{\rm DW}>1$~\cite{Sikivie:1982qv}, see also~\cite{Sikivie:2006ni}.
Assuming a scaling solution for the average DW energy density $\rho_{\rm DW}\sim \sigmaDW/t$, with $\sigmaDW$  the energy per surface of the order $\sim \V_{\rm max}/m_a$ and $\V_{\rm max}\sim {\cal O}((m_a f_a)^2)$, the network would not dominate today's energy density for $m_a<10^{-8} {\,\rm eV} \left(10^6 {\rm GeV}/f_a\right)^2$. 
This is never satisfied for the QCD axion, but can be the case for ALP models\todonote{further constraints apply!}.
In the context of axions, several ways to solve the PQ mechanism and axions have been proposed: inflation after the PQ-phase transtition (i.e. the previous scenario), $N_{\rm DW}=1$ where only one physical vacuum exist (see also~\cite{Choi:1985iv}) and DWs are unstable, and models where the degeneracy of different vacua is not exact~\cite{Sikivie:1982qv,Holdom:1982ew}. \todonote{More references to DW contribution to DM and observational constraints} 
\todonote{elaborate more on effects of soft violations of symmetries and their cosmological role} 
The latter is very generic if the global symmetry is just accidental since at least it should be violated by quantum gravity~\cite{Georgi:1981pu}, see also~\cite{Alonso:2017avz} and refs. therein. 
Discrete symmetries can be invoked to control the magnitude of the breaking making  
succesful cosmological scenarios for ALPs and axions~\cite{Ringwald:2015dsf}.
A difference in vacuum energy $\Delta \V_0$ between vacua causes an acceleration $\Delta \V_0/\sigmaDW$  that precipitates their annihilation in a time $\sim \sigmaDW/\Delta \V_0$.
The annihilation of the DW network solves a cosmological disaster and provides another source of potential dark matter~\cite{Hiramatsu:2012sc}, dark radiation ALPs and gravitational waves~\cite{Chang:1998tb,Sikivie:2006ni,Saikawa:2017hiv}.

{\bf QCD axion/ALP mass for the right DM density: }
In the \emph{pre-inflation} models, \eqref{ALPabundance} univocally links $m_a$ with $\Omega_a$ for a given initial value of the field $a_i = \theta_{ai}/f_a$. If one requests that $\Omega_a$ equals the observed DM density, this would give a prescription on $m_a$, if it were not for the unknown value of $\theta_{ai}$. In Fig.~\ref{fig:DR} we show the required value of $m_a$ (assumed constant) as a function of $f_a$ for $\theta_{ai}=1$ as a black line. Values in the ballpark $f_a\sim 10^{12}-10^{14}$ GeV are required for a broad range of ALP masses $m_a\sim 10^{-10}-10^{-2}$ eV. The required value of $f_a$ increases as we decrease the assumed $\theta_{ai}$ and viceversa. 
For the QCD axion case, and for the unsophisticated value of $\theta_{\A i} \sim 1$, this argument suggests $m_A\sim$ few $\mu$eV, see Fig.~\ref{fig:DR}. But different initial values, e.g. $\theta_{\A i} \in (0.3,3)$, correspond to a wider approximate range  $m_A \in (10^{-6}-10^{-4})$ eV. Even lower (or higher) finetuned values of $\theta_{\A i}$, something that could be justified by anthropic reasons~\cite{Tegmark:2005dy}, could lead to arbitrarily low values of $m_A$ (or as high as 10$^{-3}$ eV). The low mass scenario in sometimes called \textit{anthropic window}. 
We have indicated these DM-motivated mass ranges in blue in the top scale of {Fig.~\ref{fig:halo_sens}}. 
In the \textit{post-inflation} case, the uncertainty of an unknown $\theta_{\A i}$ is averaged away but the contribution of topological defects to axion DM must be taken into account, and their calculation suffers from important computational uncertainties. Early attempts to estimate the relative importance with respect the standard misalignment effect provided conflicting answers from different studies. Some authors argued that the contribution was of the same order as the one from the misalignment effect~\cite{Hagmann:2000ja}, while others~\cite{Wantz:2009it} found it considerably larger. Recent studies have shed more light on the issue, although considerable uncertainty remains. A recent computation of the decay of topological  defect and their contribution to the axion relic DM density in the post-inflation scenario predicts a range for the $m_\A \sim (0.6-1.5)\times 10^{-4}$~eV~\cite{Hiramatsu:2012gg,Kawasaki:2014sqa}. Another study claims a more definite and lower prediction $m_a = 26.5 \pm 3.4$~$\mu$eV~\cite{Klaer:2017ond}. The challenge behind these computations arise from the difficulty in understanding the energy loss process of topological defects and analyzing the spectrum of axions produced from them in numerical simulations with a relatively small dynamical range. 
The $N_{\rm DW} =1$ mass range indicated in red in the topscale of Fig.~\ref{fig:halo_sens}  arbitrarily encompasses the last two results quoted. As mentioned above, models with $N_{\rm DW} >1$ are cosmologically problematic. However, those models can be made viable if the degeneracy between the $N_{\rm DW}$ vacua is explicitly broken. In those models the topological defects live longer and produce larger amount of axions, and therefore they can lead to the same relic density with substantially larger values of $m_\A$. More specifically models with $N_{\rm DW} = 9$ or 10 evade the contraints imposed by the argument that the breaking term should not spoil the solution to the strong CP problem, while potentially giving the right DM density for a wide range $m_\A \in (5 \times 10^{-4} - 0.1)$~eV~\cite{Kawasaki:2014sqa}. The generic ALP case has not been studied in detail but the contribution from topological defects will typically increase $\Omega_a$, thus decreasing the required $f_a$ for a given $m_a$ or increasing $m_a$ for a given $f_a$. 
Let us stress again that the values of $m_a$ obtained with any of the above prescriptions correspond to a $\Omega_a$  equal to the total observed DM density, and given the approximately inverse proportionality of $\Omega_a$ with $m_a$ (common for all of the axion production mechanisms discussed), lower values of $m_a$ would overproduce DM while higher masses would lead to subdominant amount of DM.

{\bf Model dependencies on the relic density: }
The ALP relic density \eqref{ALPabundance} assumes that $t_1$ happens during radiation domination (RD) and from that moment on the oscillations are adiabatic, i.e. $\dot m_a/m_a\ll H$, and the expansion is adiabatic too. If a heavy particle decouples and decays later, significant entropy can be produced, and if  injected after $t_1$ the ALP abundance decreases by the ratio of the comoving entropies before and after the injection~\cite{Yamamoto:1985mb,Lazarides:1987zf}.
No significant injection can happen after neutrino decoupling $T_{\nu}\sim 5$ MeV or the accurate predictions of BBN would be spoiled~\cite{Hannestad:2004px,deSalas:2015glj} so \eqref{ALPabundance} is safe for $m_a\lesssim 1/t_\nu\sim 1.66T_\nu^2/\mPl\sim 10^{-14}$ eV but could be overestimated above it. 
The consequences of dropping the assumption of radiation domination during $t_1$ have been studied in some references. 
Reference~\cite{Visinelli:2009kt} studied the low-$T$ reheating scenatio and kination, where $\Omega_a$ decreases or increases, respectively, with respect to the standard radiation dominated scenario. More general kinetic terms including singularities have been considered in~\cite{Alvarez:2017kar}. Another interesting case, recently discussed in~\cite{Hoof:2017ibo} considers a second (shorter) period of inflation after cosmic inflation, where large entropy dilution factors can arise, but a little enhancement is also possible.

Perhaps it is interesting to mention in this context the relation between baryogenesis and the QCD axion.
A VEV of the QCD axion induces CP violation, which combined with baryon number violating interactions during an out-of-equilibrium period could produce the baryon asymmetry of the Universe. The power of QCD to violate CP resides in the $G\widetilde G$ operator, which only receives contributions from QCD instantons, which are highly suppressed at the temperatures where B-non-conserving interactions are active.
However, an ultra-low inflationary period with a delayed EW phase transition can potentially do the trick~\cite{Servant:2014bla,vonHarling:2017yew}.

{\bf Generic constraints on ALP DM: }
The CDM model requires dark matter to be present at matter-radiation equality, corresponding to redshift $z_{\rm eq}\sim 3000$, and any component should not exceed the measured value
\additionalinfo{ $\rho_{\rm cdm}=1.3\, $ keV/cm$^3$} $\Omega_ch^2=0.12$.
This last overproduction argument has been widely used to constraint the axion decay constant $f_a\lesssim 10^{12}$ GeV by assuming $a_i\sim O(1) f_a$, i.e. by imposing no fine-tuning on the ``misalignment angle'' $\theta_i$ and can be easily generalised to any ALP~\cite{Arias:2012az}.
However, we find these bounds extremely misleading.
In the \emph{pre-inflation scenario}, the initial misalignment angle in our Universe is a randomly chosen parameter which cannot be assumed to be of ${\cal O}(1)$ because it turns out to lead to anthropic selection ~\cite{Tegmark:2005dy,Linde:1987bx,Hertzberg:2008wr}.
In earnest, the overproduction bound only applies to QCD axions in the \emph{post-inflation scenario} where the uncertainty on the initial conditions of the axion field is averaged out.

If the ALP mass is too small, it might behave as a cosmological constant during CMB decoupling before starting to  oscillate and behave as DM.
A careful analysis of the CMB constraints $\Omega_a/\Omega_{\rm c}<0.1$ for $m_a<10^{-25}$ eV, above which the bounds relax notably~\cite{Hlozek:2014lca}.
Ultralight-axions (ULAs) have received a lot of attention recently as peculiar dark matter candidates. Their Compton wavelength spans astronomical distances and their gravitational clustering differs from the ordinary pressureless dark matter. See~\cite{Hui:2016ltb} for a recent review on this sometimes called fuzzy dark matter scenario.
Currently the most relevant constraints at those low masses come from the non-observation of a decrease in the power spectrum of density fluctuations at the small scales probed by the Lyman-$\alpha$ forest~\cite{Irsic:2017yje,Hui:2016ltb}.

For $m_a\lesssim 10^{-32}$ eV, $t_1>t_0\sim 13.9$ Gy and the ALP energy behaves as a  cosmological constant even today. In the string axiverse~\cite{Arvanitaki:2009fg}, some axions are expected to play exactly this role.

{\bf Isocurvature constraints:} In the \emph{pre-inflation scenario}, the ALP field exists during inflation and its small quantum fluctuations with $\langle a^2\rangle \sim H_I^2$ are stretched to cosmological sizes. \modified{If ALPs are a significant contribution to the DM of the Universe, their density fluctuations contribute to the temperature inhomogeneities. The characteristic imprint to the latter is an isocurvature contribution~\cite{Linde:1985yf,Seckel:1985tj} that has been searched for in CMB and not} yet found~\cite{Ade:2015lrj}. The latest analysis of the Planck collaboration gives a constraint~\cite{Kobayashi:2013nva}:
\be
\frac{\Omega_{a}}{\Omega_{\rm cdm}} \frac{d \ln \Omega_{a}}{d\theta_i}\frac{H_I}{2\pi f_a}\lesssim 10^{-5}.
\ee
The expansion rate during inflation, $H_I$, is currently unknown but it can be measured directly if the next generation of CMB polarisation experiments detects B-modes from the primordial graviational waves produced during inflation. The LiteBIRD~\cite{Matsumura:2013aja} and PRISM~\cite{Andre:2013afa} missions could measure down to
\exclude{PRISM+delensing r~3 10^-4, but } $H_I\sim 4\times 10^{12}$ GeV.
Since values of $f_a>\mpl$ are unrealistic, such a discovery without the corresponding isocurvature signal would very easily exclude axions within this pre-inflationary scenario (as it was thought to happen in the BICEP2 incident~\cite{Visinelli:2014twa}) unless they contribute a minute fraction to the total DM,
${\Omega_{ a}}/{\Omega_{\rm cdm}}\ll 1$.
There are however, a number of models that avoid this constraint~\cite{Folkerts:2013tua,Kawasaki:2014una}.

{\bf Miniclusters and axion/ALP ``stars'' (oscillatons, droplets, pseudo-breathers, axitons...): }
In the \emph{post-inflation scenario} the DM distribution is inhomogeneous at comoving scales related to $t_1$ when the ALP field starts to oscillate and behave as DM. 
In this scenario, an $\sim {\cal O}(1)$ fraction of the DM is placed in regions of large density contrast $\delta(x) = (\dmden(x)-\bar\dmden)/\bar\dmden \gtrsim 1$, where $\bar\dmden$ is the density
averaged over a comoving volume much larger than $(t_1\SF/\SF_1)^3$. 
A region with overdensity $\delta\gtrsim 1$ enters into matter-domination epoch earlier than the Universe on average, at a scale factor $\SF_\delta\sim \SF_{\rm eq}/\delta$, where $\SF_{\rm eq}\simeq 1/3400$ is the scale factor of matter-radiation equality~\cite{Ade:2015xua}. Around that time, gravitational instability becomes stronger, and the DM inside becomes gravitationally bound relatively fast, decoupling from the Hubble expansion and becoming an axion/ALP minicluster. Miniclusters were first considered in the context of axions~\cite{Hogan:1988mp,Kolb:1993zz} and recently for ALPs in the context of the string axiverse~\cite{Hardy:2016mns}. 
Assuming no significant accretion or merging, their typical size after collapse and virialisation stays $\sim$ constant as $0.5\, t_1 \SF_{\delta}/\SF_1$. Compared with the average DM density $\bar \dmden$ that redshifts as $1/\SF^3$, at matter radiation equality the minicluster is a factor of $\sim \delta^4$ denser and another factor $\sim 1/\SF_{\rm eq}^3$ is accrued until today. 
We assumed that $\SF_1<\SF_\delta$, i.e. ALPs become DM before miniclusters collapse, but this might not be the case for very small ALP masses. 
QCD axion miniclusters\footnote{In the \emph{post-inflationary scenario} we have $f_\A<10^{12}$ GeV for which $T_1\sim 1.5$ GeV, $\SF_1\sim 10^{-13}$, $t_1\sim 10^{-7} {\rm s}\sim 40$ m. A typical virialised axion minicluster radius is $10^{12}$ cm. } have very low masses but very dense cores, 
\be
\label{axionminicluster}
M_{\rm mc}\sim 10^{-12} M_\odot \quad , \quad  \dmden_\A^{\rm mc}\sim 10^4 \frac{\rm GeV}{{\rm cm}^3},
\ee
($M_\odot=3\times 10^{30}$ kg is the solar mass)
and do not get easily tidally disrupted in the galaxy by tidal interactions with stars and fluffier objects~\cite{Sikivie:2006ni,Dokuchaev:2017psd}. 
Their density profile roughly corresponds to a cored NFW~\cite{Zurek:2006sy}. 
They shall suffer from some hierarchical structure formation to form larger gravitational bound systems but at the moment is not clear whether they become homogeneously virialised objects or clusters of miniclusters\cite{Fairbairn:2017dmf}. A Press-Schecter formalism has been used to predict the minicluster halo mass function~\cite{Fairbairn:2017sil,Fairbairn:2017dmf}. A recent analytical effort unaccounting for axion self-interactions can be found in~\cite{Enander:2017ogx}. 
One can search for these large overdensities in the DM distribution with femto and pico-lensing \cite{Kolb:1995bu,Zurek:2006sy}. Current femtolensing searches of $M_{\rm mc}\sim 10^{-15}  M_\odot $ primordial black holes~\cite{Barnacka:2012bm} constrain their DM fraction to be smaller than $\sim 10\%$ but have not been analysed in terms of miniclusters. Since typical Einstein radii are larger than the cluster radius, this study would not very constraining.  More recently, very strong constraints have been devised using microlensing~\cite{Fairbairn:2017dmf,Fairbairn:2017sil}. Under some assumptions on the minicluster mass fraction in our galaxy and its neighbourhood, the EROS survey of the large magellanic cloud~\cite{Tisserand:2006zx} and Subaru's HSC survery of Andromeda (M31)~\cite{Niikura:2017zjd} (originally intended as primordial black hole searches) can be used to set strong constraints on the amount of DM in miniclusters~\cite{Fairbairn:2017dmf,Fairbairn:2017sil}. More dedicated surveys could improve the constraints or even lead to a potential discovery but more analytical and simulation work is required in order to interpret non-observations in terms of constraints~\cite{Fairbairn:2017sil}. 
\todonote{This topic would benefit from a figure, with bounds, perhaps}

Axions and ALPs as scalar fields store energy in their gradients, which leads to a particular type of pressure (usually called \modified{quantum, kinetic or gradient pressure}).   
The gravitational pull on a coherent superposition of axions can be balanced by this pressure, leading to stable coherent lumps of DM~\cite{Ruffini:1969qy}, called oscillatons, boson-stars, droplets or dilute axion stars in our context. The solution with zero-angular momentum has a radius-mass relation $R^0_{\rm das}\simeq 9.9/(G_F M_{\rm das} m_a^2)$~\cite{Ruffini:1969qy}. 
Negative self-interactions in the potential, like axions and ALPs must have to some level, can change dramatically these solutions~\cite{Colpi:1986ye}. A negative quartic term $\V\sim -\lambda_4 a^4/4!$ behaves like a negative pressure and drives a collapsing instability to these objects above a critical mass $M\sim 11 \sqrt{(f_a/m_a)^2/\lambda_4 G_N}$. The instability is delayed for higher angular momenta configurations, although perhaps not indefinitely~\cite{Davidson:2016uok}. 
In the QCD axion case, the instability drives at a very early time $t\sim t_1$ the so-called axitons found in~\cite{Kolb:1993hw}, collapsing regions that converge into a pseudo-breather, which oscillates for a few times with large amplitudes $\theta\sim {\cal O}(2\pi)$ and radiates away barely-relativistic axions. The same phenomenon was recently found in a simulation of a collapsing over-critical axion star~\cite{Levkov:2016rkk}. For the collapse to lead to a blackhole, values of $f_a$ close to $\mPl$ are needed, see~\cite{Helfer:2016ljl} for a discussion and state of the art simulations. 
A new type of equilibrium configurations, dense axion stars, where gravity is equilibrated by repulsive self-interactions was  proposed in a recent paper~\cite{Braaten:2015eeu} (see also \cite{Chavanis:2017loo}) but it has been shown that the solution found was inconsistent and the closest object to a dense axion star is the axiton/pseudo-breather, which is so short lively to be inconsequential cosmologically~\cite{Visinelli:2017ooc,Schiappacasse:2017ham}, see also~\cite{Eby:2015hyx}.   
Although a positive self-interaction like $+\lambda_6 a^6$ would lead to a repulsive force able to counterbalance gravity and create a stable configuration~\cite{Eby:2016cnq}, this does not seem to apply when all orders of a $\V\propto (1-\cos(a/f_a))$ or the axion $\VA$ are included. 
The radiation of relativistic axions due to self-interactions and photons from axion stars have been studied in~\cite{Eby:2015hyx,Braaten:2016kzc,Braaten:2016dlp,Eby:2017teq,Eby:2017xrr}. 
A recent speculation that fast-radio-bursts (FRB) could be due to the axions in a dilute axion star falling into a compact object with a large $B-$field and converting into photons~\cite{Iwazaki:2014wka,Tkachev:2014dpa,Iwazaki:2014wta,Raby:2016deh,Iwazaki:2017rtb} is most likely not viable for QCD axions~\cite{Pshirkov:2016bjr}, particularly since it is clear that dense axion stars are not stable and some FRBs are repeating. Most of the work done for QCD axions directly applies to general ALPs, where perhaps the FRB interpretation is still possible.

{\bf Axion dark matter field as background medium in the early Universe:}
Nuclear properties like masses, binding energies etc. depend on the value of $\theta$. In the SM or its extensions where $\theta$ is a constant, the stringent bound from the nEDM~\eqref{thetabound} constrains most of the effects of a non-zero $\theta$ to be unnoticeable. 
For instance, Ubaldi studied the effects of $\theta$ in two relatively fine-tuned nuclear quantities, the deuteron binding energy and the triple-$\alpha$ nuclear reaction cross-section, where a misadjustment could lead to spectacular consequences for Big Bang and stellar nucleosynthesis~\cite{Ubaldi:2008nf}. 
He concluded that $\theta\lesssim 6\times 10^{-3}$ was enough to ensure unnoticeable changes and proposed it as an anthropic constraint. 
However, axion DM consists precisely of an oscillating $\theta$ field whose amplitude decreases with time. If today QCD axions saturate the observed DM, 
their energy density is $\bar \dmden\sim m^2_\A \A^2_\osci/2=1.3\,  {\rm keV/cm}^3$ with $\A_\osci$ the typical oscillation amplitude giving $\theta_\osci\sim 8\times 10^{-22}$. The DM redshifts as $1/\SF^3$ as long as it is decoupled, so one can retrieve the typical amplitude at proton-neutron decoupling $\theta_{T\sim \rm MeV}\sim 3.6\times 10^{-7}$. During BBN it is even smaller and thus axion DM is safe, as found in~\cite{Blum:2014vsa}. Reference~\cite{Blum:2014vsa} quotes bounds on $\bar g_{\A \gamma n}$ based on direct effects from $\theta$ on nuclear properties affecting BBN yields and therefore are not valid for a generic ALP. In principle, one could use the effects of an ALP-coupled non-zero nEDM $d_n$ to get BBN bounds, and are most likely weaker than those shown for axions in~\cite{Blum:2014vsa}.

{\bf Axion/ALP dark matter field as optical medium:}
The DM ALP field coupled to two photons behaves like an active optical medium with a characteristic frequency $m_a$. The active medium changes the dispersion relation of photons imposing forbidden bands that we interpret as instability bands where photon occupation numbers can grow absorbing ALPs.
The dispersion relations and some consequences on photon propagation and cosmic rays have been worked out in detail in~\cite{Andrianov:2009kj,Espriu:2010ip,Espriu:2011vj,Espriu:2014lma}. \todonote{expand}
The birefringence effect on the CMB can cause B-mode polarisation in the CMB~\cite{Pospelov:2008gg}.

The CP violating coupling $\bar g_{a\gamma} FF a$ renormalises the electric charge in the ALP DM background with very interesting phenomenological consequences in cosmology, astrophysics and opens up interesting laboratory searches. The value of this coupling for QCD axions is extremely small and so are the ensuing effects. For reasons of space we have decided not to cover this important topic here. See~\cite{Martins:2017yxk,Stadnik:2017mid} for recent reviews. 

\todonote{Little section of BEC as DM ~\cite{Erken:2011dz}~\cite{Sikivie:2001fg}
\exclude{Sikive galaxy, caustics, cosmological, Neff, Kawasaki, Sacha, Jaeckel, Guth, Hertzberg, Dvali}
}
\todonote{supersymmetry and axions~\cite{Baer:2011uz,Bae:2013pxa}}

\todonote{Q-balls}

\subsubsection{Hot dark matter and dark radiation}

Axions and ALPs with thermal energies would have been produced by interactions with SM particles in the early Universe. The interaction set \eqref{ALPinteractions} corresponds to dimension-5 operators that give energy-independent cross sections $\sim 1/f^2_a$ for reactions involving 2 SM particles producing one ALP. Production rates in a thermal bath are therefore $\Gamma \propto T^3/f_a^2$ and so are most efficient at high temperatures, $\Gamma/H\propto T \mPl /f_a^2$, during radiation domination (RD). Hence, ALPs are most efficiently produced at high-temperatures. \todonote{DOF as caligraphic g}
If their rates are strong enough they thermalise with the SM bath and when $\Gamma/H$ drops below $\sim 1$ their abundance freezes-out. Assuming that this happens during RD and the further expansion is adiabatic, one can compute the relic density. Compared to today's CMB photon density we have $n_a/n_{\gamma,\rm CMB}= 1.95/g_S(T_d)$ where $g_S(T_d)$ is the effective number of effective degrees of freedom (d.o.f.) at the decoupling temperature (a measure of the entropy of the bath), recently computed with accuracy even around the QCD confining epoch in~\cite{Borsanyi:2016ksw}.
If an ALP has different couplings to SM particles, only the latest decoupling counts. The impact of this radiation depends very much on the ALP mass, and to a lesser extent on the decoupling temperature, which also influences the exact velocity distribution. ALP masses above $\sim $ 100 keV behave as cold DM (like the misalignement mechanism axions but with much larger velocity dispersion), around $\sim$10 keV as warm DM, for 10 keV$\sim$1 eV as hot DM and for lower masses as dark radiation. The fraction of hot DM affects structure formation and can be constrained with the CMB anisotropies. Hadronic QCD axions have been analysed in~\cite{Hannestad:2010yi,Cadamuro:2010cz,Archidiacono:2013cha,DiValentino:2015wba}, where a constraint $m_a < 0.53$ eV at 95\% C.L. is set from a combination of Planck anisotropies and other large-scale-structure data. For these masses, the relevant processes at decoupling are meson interactions $\pi^\pm \pi^0 \leftrightarrow  \pi^\pm a$ ($T_d\sim 100$ MeV) and axions still behave as DM during CMB decoupling times. For the lower masses we are mostly interested in here, ALPs behave as dark radiation.

\begin{figure}[t] 
   \centering
   \includegraphics[width=3.5in]{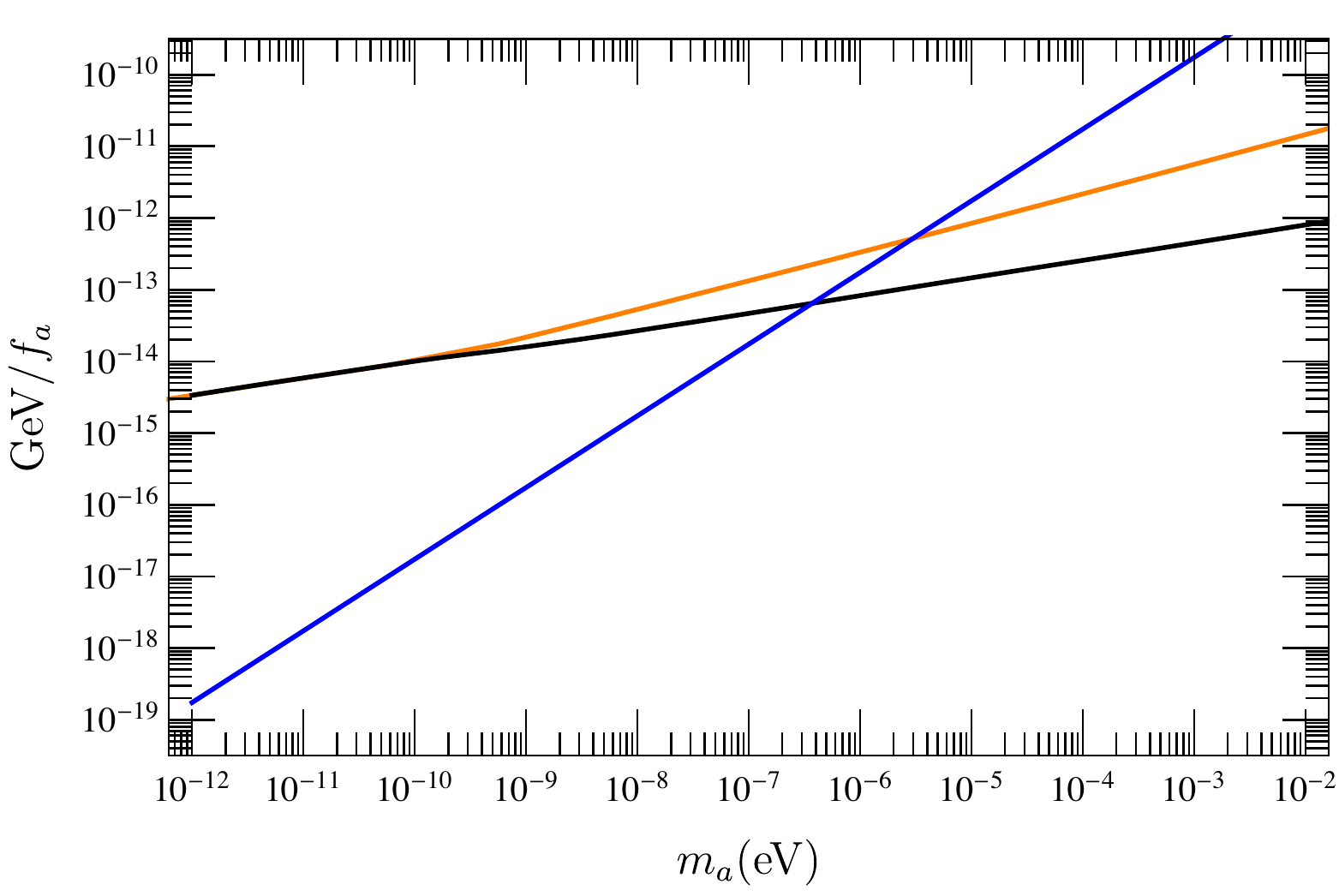}
   \includegraphics[width=3.5in]{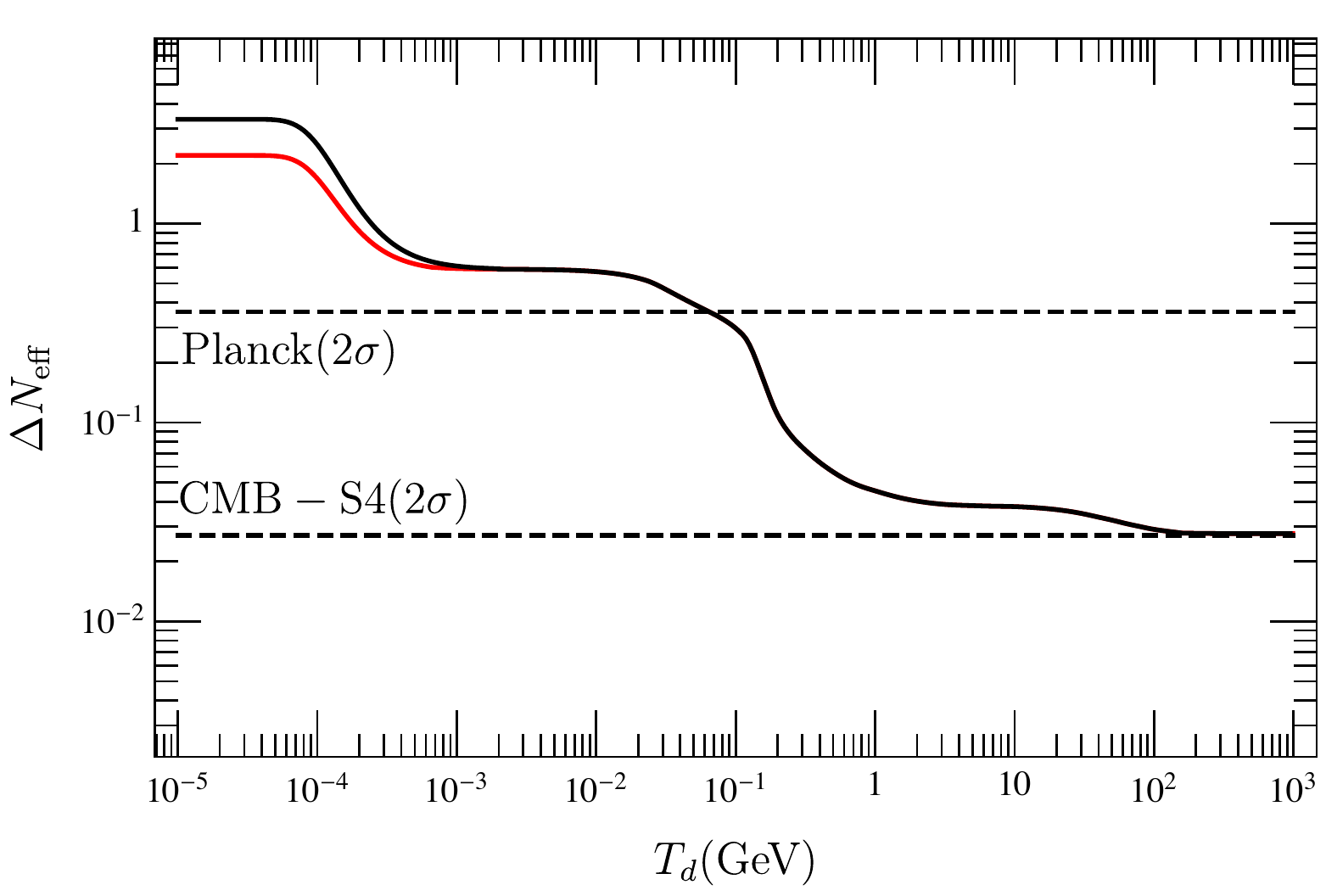}
   \caption{\emph{Left:} Value of $1/f_a$ for axions/ALPs to have the observed relic DM abundance with $\theta_a\sim 1$ for a constant $m_a$ (black) and for the case where the temperature dependence is as extreme as the QCD axion (orange).  QCD axions, however exist only in the blue band and tend to overproduce DM at low mass. The ratio between the blue and orange line corresponds to the required value of $\theta_i$ to match observations (if the ratio is smaller than 1, otherwise the relation is non-linear). 
\emph{Right:}   Dark radiation due to a generic thermal ALP relic as a function of the decoupling temperature. }
   \label{fig:DR}
\end{figure}

Any exotic contribution to the DR density $\rho_{\rm DR}$ is often expressed as an effective number of neutrino species $\Delta N_{\rm eff}=\rho_{\rm DR}/\rho^{\rm std}_{1\nu}$, where
the density corresponding to one standard neutrino d.o.f. is $\rho^{\rm std}_{1\nu}=\sfrac{7}{8}\left(\sfrac{4}{11}\right)^{\sfrac{4}{3}}\rho_\gamma$. The amount of DR due to a thermal ALP that decoupled at $T_d$ is shown in Fig.~\ref{fig:DR}, updated from~\cite{Baumann:2015rya} with the QCD equation of state of~\cite{Borsanyi:2016ksw}. Currently, the most optimistic analysis of cosmological data constrains it to be $ N_{\rm eff}=3.04\pm 0.18$~\cite{Ade:2015xua}, disfavouring $T_d<0.1$ GeV. There is no direct evidence that the early Universe was ever at a temperature higher than that, so the sensitivity to DR is extremely cosmological-model-dependent.  Nevertheless, DR searches can be a powerful tool. The sensitivity of the next generation of CMB probes can reach down to $\Delta N_{\rm eff}\sim 0.01$~\cite{Abazajian:2013oma,Wu:2014hta}. This would be enough to discover or rule out an ALP decoupling whose abundance is suppressed by all the d.o.f. of the SM. 
\modified{This has the spectacular implication than if ALPs were \emph{ever} in thermal equilibrium they could be discovered in $\Delta N_{\rm eff}$, if no d.o.f. beyond the SM is present. Tuning the argument around, if an ALP is discovered experimentally we would learn about the total number of d.o.f. in nature. 
The thermal production of axions was pioneered in~\cite{Turner:1986tb}. The relation between ALP couplings and the decoupling temperature has been worked out for  photons~\cite{Bolz:2000fu,Cadamuro:2011fd}, pions~\cite{Berezhiani:1992rk,Chang:1993gm}, gluons~\cite{Masso:2002np,Graf:2010tv}, leptons and quarks~\cite{Salvio:2013iaa,Ferreira:2018vjj}, see~\cite{Baumann:2015rya,Ferreira:2018vjj} for a recent reassessment. 
All these calculations assume RD and are therefore cosmological-model dependent. 
In particular, if the reheating temperature $T_{\rm RH}$ is smaller than $T_{d}$ the predictions get considerably diminished.  }
\todonote{Check all references;}

Relativistic ALPs can be also produced ``non-thermally". In particular by the decay of some heavy scalar field~\cite{Hasenkamp:2012ii} like the inflaton or by parametric resonance of the inflaton itself~\cite{Ballesteros:2016xej,Co:2017mop}. Depending on the ALP mass, they can become DR or even warm or hot DM~\cite{Co:2017mop}.
The former case is generic in the context of string theory where a large number of moduli fields, massive but with small (gravitational-size) couplings can be long lived and have decays to light species (axions and SM particles) with very similar branching ratios producing a late reheating and DR~\cite{Higaki:2012ar,Cicoli:2012aq}. The amount of DR is generically larger than in the case of thermal production and therefore this scenario will be very strongly constrained by CMB-S4 experiments. These ALPs can potentially affect BBN~\cite{Conlon:2013isa} and make galaxy clusters glow in soft-X-rays by converting into photons in the cluster magnetic fields~\cite{Conlon:2013txa,Angus:2013sua} for photon couplings of the order of $\gagamma\sim 10^{-12}$ GeV$^{-1}$.  

{\bf Dark matter decays:}
Axions and ALPs with masses below the electron-positron threshold decay mostly into two photons with a rate $\Gamma=\gagamma^2m_a^2/64\pi$. We are mostly interested in sub-eV masses and $\gagamma<0.66 \times 10^{-10}$ GeV$^{-1}$ as implied by the CAST helioscope (see Sec.~\ref{sec:helioscopes}) where the lifetime of DM ALPs is $\Gamma^{-1}> 3.0\times 10^{25}$ s, larger than the age of the Universe by a factor $\sim 10^8$.
For completeness, let us point that shorter DM ALP decays can have a large variety of consequences. If the decay happens earlier than the CMB decoupling, the photon injection can affect the BBN yields and distort the CMB spectrum. Photons from diffuse DM decays after recombination would be seen as a monochromatic line broadened by cosmic expansion, which can be constrained by the flux of extragalactic background light (EBL)~\cite{Cadamuro:2010cz} or direct line searches in the X-ray and gamma-ray region~\cite{Cadamuro:2010cz}. \todonote{Pospelovs: The search for DM decay photon lines is key for other DM candidates, specially sterile neutrinos, that have motivated a large number of searches~\cite{}}. In the UV region the cross section of H photoionisation is so large that these photons can alter the history of the Hydrogen ionisation fraction ($x_{\rm ion}$)~\cite{Cadamuro:2010cz}.
When we assume a sufficiently large reheating temperature to make ALPs thermalise, 
these arguments lead to the constraints shown in Fig.~\ref{fig:ALPs_constraints} as BBN, CMB, EBL, X-rays and $x_{\rm ion}$. 

Today, the signal will be larger in DM rich galaxies. Searches for decaying DM axions in a few sample galaxies have set strong constraints in the eV-mass range~\cite{Hlozek:2014lca}. Assuming ALPs are all the cold DM, the ensuing constraints are shown in  Fig.~\ref{fig:ALPs_constraints} labelled as ``Telescopes''. 
A search in the microwave region was conducted in~\cite{Blout:2000uc} but the sensitivity is not better than
the $\gagamma \sim 10^{-10}$ GeV$^{-1}$ benchmark.
The presence of slowly varying galactic $B$-fields leads to the induced decay or inverse-Primakoff conversion $a+\gamma^*\to \gamma$ that will also lead to a monochromatic signal. A recent study~\cite{Sigl:2017sew} concludes that the sensitivity of present and future radio telescopes is unfortunately not enough to detect the benchmark QCD axion but it will be sensitive to ALP DM in the mass range $0.1-100 \, \mu$eV for couplings $\gagamma\gtrsim 10^{-13}$ GeV$^{-1}$.
This paper corrected a previous claim where QCD sensitivity was envisaged with SKA sensitivity~\cite{Kelley:2017vaa}.

ALP DM decay was proposed as an explanation of the still unexplained 3.55 keV line in the Perseus cluster and M31 ~\cite{Jaeckel:2014qea} but the absence of signal from dwarf spheroidal galaxies (dSph)~\cite{Malyshev:2014xqa,Jeltema:2015mee} disfavours this interpretation. However, it could be attributable to DM decay into two ALPs, which then convert into monochromatic photons in intercluster B-fields~\cite{Conlon:2014xsa}, which are much weaker in dSph's. 
This scenario predicts a completely different signal morphology with peculiar testable signatures~\cite{Conlon:2014wna,Alvarez:2014gua,Jennings:2017gev}. The signal is proportional to the DM abundance of the heavy DM species, the DM decay rate into ALPs and the conversion probability $\propto \gagamma^2$. The degeneracy between these three parameters and uncertainties in the DM and $B-$field distributions preclude precise predictions of $\gagamma$ but the authors point to exemplary values of $\gagamma\sim 10^{-13}$ GeV$^{-1}$ and $m_a\lesssim 10^{-12}$ eV.

\subsubsection{ALPs and inflation}

The flat ALP potentials have the nice feature of being in principle protected against radiative corrections and offer excellent candidates for the inflaton field itself. These so called natural-inflation models~\cite{Freese:1990rb} have been thoroughly studied, for instance with the ubiquitous cosine potential $\Lambda^4(1-\cos \theta)$. Needless to say, string theory and its axiverse has a number of excellent candidates for such fields~\cite{Arvanitaki:2009fg,Martin:2013tda}. Although the simplest potential seems to be slightly in tension with data~\cite{Ade:2015lrj}, recent sophistications like monodromy~\cite{Silverstein:2008sg,McAllister:2008hb} can still provide satisfactory models. Unfortunately, the low-energy phenomenology is almost never correlated with the physics of inflation. Perhaps one exception is the so-called ALP-miracle model of~\cite{Daido:2017tbr,Daido:2017wwb} where an ALP with two cosine potentials giving something like a  Hilltop quartic potential can drive inflation, reheat through the coupling to photons $g_{a\gamma}\sim 10^{-11} {\rm GeV}^{-1}$ and provide the DM of the Universe through the realignment mechanism. This happens for an ALP mass in the range $\sim 0.01-1$ eV. The extreme tuning of the two potentials to give a sufficiently flat potential for inflation is correlated with the low mass needed for DM, making it more appealing than naively could be expected. Although the most appealing aspect is that this model is testable as we shall see.

\subsection{Astrophysics}
\label{sec:astrophysics}

Owing to their coupling to electrons, photons or nucleons, ALPs can be thermally produced in the hot and dense stellar interiors. The weakness of their couplings ensures a small emission rate per unit mass, but they can be emitted from the whole volume of the core and can compete with photon surface emission and even with thermal neutrino emission in certain cases. A classical textbook reviews generic ALP limits and specific implications for axions~\cite{Raffelt:1996wa}. Here we update on the latest developments.
The ALP-photon coupling is best constrained by the ratio of horizontal branch (HB) to red giants in globular clusters (GCs). The ratio decreases as the HB phase is accelerated by the increasingly faster He-burning, due to the hotter temperature implied by axion emission through the Primakoff effect $\gamma+Z\to a+Z$. The latest stacked study of 39 GCs~\cite{Ayala:2014pea,Straniero:2015nvc} gives the constraint $\gagamma< 0.66 \times 10^{-10}$ GeV$^{-1}$ (labeled as ``HB'' in Fig.~\ref{fig:ALPs_constraints}) although the data are better fit by a small axion cooling.
The photon coupling can also be constrained by a combined fit of solar data (neutrino flux and helioseismology)~\cite{Vinyoles:2015aba} (labelled Sun in Fig.~\ref{fig:ALPs_constraints}), but the constraint is not as strong. An ALP in this ballpark can produce other interesting effects in more massive stars \cite{Friedland:2012hj,Aoyama:2015asa,Dominguez:2017yhy} but data are too scarce to set reliable exclusions. 
The bounds are typically applicable for $m_a\lesssim 10$ keV \modified{(representative of typical stellar core particle temperature and particle energies).} 
Above this benchmark few constraints survive, with the SN1987A neutrino pulse duration reaching the highest masses~\cite{Masso:1995tw}, see Fig.~\ref{fig:ALPs_constraints}. The same constraints apply to the CP violating coupling $\bar g_{a\gamma}$ because the polarisation-averaged non-relativistic Primakoff cross section is the same for the $F\widetilde F a$ and $FFa$ couplings.

The coupling to electrons is very efficient in producing axions by the ABC processes (Axiorecombination, Bremsstrahlung and Compton)~\cite{Redondo:2013wwa} and is severely constrained by a number of low mass star observations. Axion bremsstrahlung in the degenerate cores of red-giants delays the He-flash enhancing the luminosity of the brightest red giant in GCs. A study of M5 outputs a constraint $g_{ae}<4.3\times 10^{-13}$ with a slight preference for extra cooling~\cite{Viaux:2013lha}. \todonote{The study of M3 yields a similar constrain without any preference, referencia-straniero.} The slightly small HB/RG ratio in GCs could also be due to this channel~\cite{Giannotti:2015kwo}. The most stringent constraints come from different white-dwarf cooling arguments: period decreases of WD variables~\cite{Corsico:2012sh,Corsico:2012ki,Corsico:2014mpa,Battich:2016htm,Corsico:2016okh} and fits to the luminosity function~\cite{Bertolami:2014wua,Bertolami:2014noa}. Surprisingly, the agreement of theory with observations of these systems improves with a bit of axion cooling.
A recent combined analysis of the M5 tip and WD data gives $g_{ae}=1.6^{+0.29}_{-0.34}\times 10^{-13}$~\cite{Giannotti:2017hny}, implying $g_{ae}<2.6 \times 10^{-13}$ at 95\%C.L. but giving a 3$-\sigma$ hint of cooling though an ALP coupled to electrons. The constraint and best fit regions are shown in Fig.~\ref{fig:ALPgae_constraints} labelled as Stellar bound and WD hint, respectively. 

The coupling to nucleons drives ALP emission from nuclear transitions in low mass stars, but it is better constrained by the SN1987A neutrino pulse duration measured on Earth~\cite{Loredo:2001rx,Pagliaroli:2008ur} and neutron-star cooling~\cite{Iwamoto:1984ir,Sedrakian:2015krq} where axion bremsstrahlung in nucleon collisions is the most effective channel.
Concerning the SN pulse, the data is very scarce (few dozen neutrinos) and axion emissivities from a nuclear dense medium are plagued with uncertainties. Moreover, SN1987A was an atypical supernova, which has not been attempted to model in detail or calibrate to account for the latest advances in SN type-II modelling. The constraints have been loosening as these points were increasingly appreciated.
An educated guess gives $g_{ap}<0.9\times 10^{-9}$~\cite{Raffelt:2006cw} that could be actually understood as a bound on $\sqrt{g_{ap}^2+g_{an}^2}$~\cite{Giannotti:2017hny}.
During the writing of this review, we were told of a further revision of the bound, which includes, among other improvements, a recent down-revision of the axion emissivity and loosening of the above constraint by a factor $\sim 5$~\cite{Chang:2018rso}. \modified{The neutrino flux from a next galactic SN explosion could be detected much better than SN1987A allowing to improve the constrains or to produce a very strong hint~\cite{Fischer:2016cyd}.}
The direct detection by a next generation helioscope is discussed below in section~\ref{sec:helioscopes}. In the $g_{aN}\sim 10^{-9}$ ballpark, type-II SN explosions could emit ${\cal O}$(1) of the gravitational binding energy in axions, leading to a diffuse supernova axion background (DSAB) of MeV energies, discussed in \cite{Raffelt:2011ft}.

Neutron star (NS) cooling does not offer much safer probes. The most recent reference concludes $g_{an}<0.8\times 10^{-9}$~\cite{Sedrakian:2015krq} with large systematics. The cooling of CAS A remnant as been measured for 6 years and shows an amazingly fast cooling attributable to neutrino pair emission in Cooper formation. An interpretation in terms of axion emission has been put forward~\cite{Leinson:2014ioa} but less exotic hypothesis are available~\cite{Leinson:2014cja}.

The CP violating versions of the couplings are also strongly constrained by stellar evolution but have received typically less attention in the axion context as they are expected to be very small.
A recent revision~\cite{Hardy:2016kme} has brought up the importance of plasma effects and strengthening previous bounds, summarised in~\cite{Raffelt:1996wa}, giving $\bar g_{ae} < 0.7\times 10^{-15}$ from the tip of the red giant branch (RGB) argument.
\exclude{alphaphiee < 8 \times 10^{-31} alphaphiee < 4 \times 10^{-32}}
The constraints are weaker than experimental searches for 5th forces at low masses, see Sec.~\ref{sec:5thforce}.

\begin{table}[htbp]
   \footnotesize 
   \centering
   \begin{tabular}{@{} c c c c @{}} 
      \toprule 
      Coupling    &  Bound & Observable & Best fit?  \\
      \midrule
      $\gagamma,\bar g_{a\gamma}$ &  $<0.65\times 10^{-10}$ GeV$^{-1}$ (95\% C.L.)  &  HB/RG stars in 39 GCs~\cite{Straniero:2015nvc} &  $0.29 \times 10^{-10}$ GeV$^{-1}$ \\  	
      $g_{ae}$       &   $<2.6\times 10^{-13}$  (95\% C.L.)  &  WD cooling + RGB tip M5 + HB/RG in GCs~\cite{Giannotti:2017hny} &  $\sim 1.5\times 10^{-13}$ \\ 
      $g_{ap}$       &   $<0.9\times 10^{-9}$    &  SN1987A $\nu$-pulse duration~\cite{Raffelt:2006cw} &  0  \\ 
      $g_{an}$       &   $<0.8\times 10^{-9}$    &  Neutron star cooling~\cite{Sedrakian:2015krq} &  0  \\ 
      $g_{an}$       &   $<0.5\times 10^{-9}$    &  CAS A NS cooling~\cite{Leinson:2014ioa,Giannotti:2017hny} &  $\sim 0.4\times 10^{-9}$  \\ 
      $ g_{a \gamma N}$       &  $<3\times 10^{-9}$ GeV$^{-2}$& SN1987A $\nu$-pulse duration~\cite{Graham:2013gfa} &  0  \\       
      \midrule      
      $\bar g_{a e}$       &   $<0.7\times 10^{-15}$    &  Luminosity of the RGB tip~\cite{Hardy:2016kme} &  -  \\       
      $\bar g_{a N}$       &   $<1.1\times 10^{-12}$     &  Luminosity of the RGB tip~\cite{Hardy:2016kme} &  -  \\  
      \bottomrule
   \end{tabular}
   \caption{Summary of Axtrophysical bounds and hints on an ALP coupled to photons, electrons, protons and neutrons. HB and RG bounds are valid for masses $m_a\lesssim 10$ keV, WD for $m_a\lesssim 1$ keV, SN and NS require $m_a\lesssim 1$ MeV. }
   \label{tab:axtrobounds}
\end{table}

\subsubsection{Black holes}
The existence of ALPs has spectacular effects on black holes with radii comparable to their Compton wavelength. The phenomenon of black hole superradiance can radiate extremely efficiently the BH's angular momentum into ALPs~\cite{Arvanitaki:2014wva}, therefore the existence of black holes with large angular momentum can be used to strongly disfavour ALPs minimally coupled to gravity. This argument excludes ALPs in the band $6\times 10^{-13} {\rm eV}<m_a<2\times 10^{-11} {\rm eV}$~\cite{Arvanitaki:2016qwi}.
The ALP cloud around the blackhole can become so dense that ALPs can annihilate into monochromatic gravitational waves that can be detected by gravitational wave interferometers~\cite{Arvanitaki:2016qwi}. The bounds are robust against uncertainties in the blackhole surrounding medium but not against strong self-interactions that can make the cloud collapse for $f_\A\sim 10^{15}$ GeV. 
Next generation experiments like eLISA will be sensitive down to $10^{-19}$ eV~\cite{Arvanitaki:2016qwi}, see also~\cite{Cardoso:2018tly}.  

\todonote{It is embarrassing to speak so little of this, not that I am not used to embarrassing}

\subsubsection{Effects on photon propagation}
\label{sec:transparency}

Photons traveling long astronomical distances and traversing galactic or inter-galactic magnetic fields could convert into ALPs and lead to observable effects. 
The most important one is an effective reduction in the $\gamma$-ray opacity of the intergalactic medium, possibly leading to the observation of distant sources that could be obscured otherwise, i.e. a LSW experiment at intergalactic scale. Indeed the observation of very high energy (VHE) photons from distant active galactic nuclei (AGNs) by imaging atmospheric Cherenkov telescopes (IACT) like HESS~\cite{Aharonian:2005gh} or MAGIC~\cite{Aliu:2008ay} triggered the first studies of this effect. It is relevant at very high energies (i.e., with energies $\gtrsim$ 100 GeV), for which the probability to interact via $e^+e^-$ pair production with the background photons permeating the Universe -- the extragalactic background light -- is important at the distances involved. Several authors have claimed that current IACT data show redshift-dependent spectral hardenings which are in tension with EBL models, requiring either a too low EBL density, or anomalously hard spectra at origin. Diverse ALP-photon oscillation schemes have been suggested to alleviate the problem~\cite{Csaki:2003ef,Mirizzi:2007hr,Hooper:2007bq,DeAngelis:2008sk,Roncadelli:2008zz,Simet:2007sa,Fairbairn:2009zi,Albuquerque:2010rq,
Avgoustidis:2010ju,DeAngelis:2011id,Horns:2012fx,Horns:2012kw,Meyer:2013pny,Tavecchio:2014yoa,Troitsky:2015nxa,Vogel:2017fmc}. Although conventional solutions to the problem cannot be excluded (e.g. unexpected effects at source origin or in EBL models), they do not seem very plausible. EBL density is now measured by its imprint in blazar spectra by both HESS\cite{Abramowski:2012ry} and Fermi~\cite{Ackermann:2012sza}, and found in agreement with models. The ALP-photon hypothesis is not absent of problems, mostly related with the uncertainties of the intervening $B$-fields. A recent account of the observational situation can be found in ~\cite{Meyer:2016xve}.

The photon-ALP conversion could take place in the intergalactic magnetic
field, or in the local magnetic fields at origin (at the AGN itself, or in the case of objects
belonging to galactic clusters, the cluster magnetic field) and in the Milky Way. Depending on the scenario, different values and uncertainties on the strength, coherence length and structure (coherent or turbulent) of the $B$ field apply. For some of the cases studied, the proponents have drawn approximate ranges for the values of $\gagamma$ and $m_a$ that seem to fit best the data. Most of them coincide roughly in requiring
very small ALP masses $m_a\lesssim10^{-10}$--$10^{-7}$ eV (to maintain coherence over sufficiently large magnetic lengths)
and a $\gagamma$ coupling in the ballpark of $\gagamma \sim 10^{-11}$--$10^{-10}$~GeV$^{-1}$. These values cannot correspond to a QCD axion, but as more generic ALPs they lie just beyond current experimental limits on $\gagamma$. A better defined region in the ($\gagamma$,$m_a$) plane is obtained in~\cite{Meyer:2013pny} from the analysis of a large sample of VHE gamma-ray spectra. Another analysis~\cite{Kohri:2017ljt} using recent observational data of the Cosmic IR background radiation also draws a hinted region largely overlapping the previous one. Both regions are shown in Figs.~\ref{fig:LSW_plot} and \ref{fig:helio_sens} as yellow regions labeled ``T-hints''.

The turbulent character of extragalactic $B$-fields introduces some randomness in the spectral distortion produced by the photon-ALP conversion~\cite{Mirizzi:2009aj}, leading to spectral irregularities that have also been used to constraint ALP parameters~\cite{Wouters:2012qd,Meyer:2014epa}. Both the HESS~\cite{Abramowski:2013oea} and the Fermi-LAT~\cite{TheFermi-LAT:2016zue} collaborations, using data from two particular sources, have relevant exclusions in the mass range $10^{-11} - 10^{-7}$ ~eV.
Both regions are shown in Fig.~\ref{fig:helio_sens} (note the latter has a small non-excluded part inside) and exclude part of the hinted regions. Sensitivity studies for the future CTA have also been carried out~\cite{Meyer:2014gta}.  Similar analysis but at X-ray energies with Chandra data~\cite{Wouters:2013hua,Berg:2016ese} produced relevant constraints at much lower masses.
\additionalinfo{The same method is used with X-ray data from the Hydra galaxy cluster~\cite{Wouters:2013hua} constraining $\gagamma < 8\times 10^{-12}\;\rm GeV^{-1}$ for ALP masses $<10^{-11}$~eV.
Still in the X-ray band, some luminosity relations of active galactic nuclei were recently shown to have precisely this particular scatter~\cite{Burrage:2009mj} although this claim is still controversial~\cite{Pettinari:2010ay}.}
\Igor{A very recent work~\cite{Majumdar:2018sbv} claims to observe significant energy-dependent modulations in high-energy gamma-ray spectra in a sample of Galactic pulsars from Fermi-LAT data, selected to have a line of sight crossing spiral arms of the Milky Way. The modulations are compatible with photon-ALP mixing in the galactic magnetic field, with ALP 
parameters $m_a = 3.6^{+0.5_{\rm stat}}_{-0.2_{\rm stat}} \pm 0.2_{\rm syst}$~neV and 
$\gagamma = (2.3 ^{+ 0.3 _{\rm stat}}_{-0.4_{\rm stat}}\pm 0.4_{\rm syst})\times 10^{-10}$~GeV$^{-1}$, 
ranges much rarrower than, and compatible with, the aforementioned hinted regions, although in slight tension with astrophysical and helioscope limits. Finally,} distortions of the CMB spectrum in primordial $B$-fields have been studied in~\cite{Mirizzi:2009nq} but they rely on the unknown $B$-fields at high redshift.  

The lack of $\gamma$-ray emission from the SN1987A supernova has allowed a constraint of $\gagamma \lesssim 5.3 \times 10^{-12}$~GeV$^{-1}$, for $m_a \lesssim 4.4 \times 10^{-10}$~eV~\cite{Grifols:1996id,Payez:2014xsa}, see Fig.~\ref{fig:helio_sens} (although see~\cite{Galanti:2017yzs} for a recent critic). 
The possible observation of a nearby supernova explosion by Fermi-LAT would allow to probe much smaller $\gagamma$ \modified{values~\cite{Payez:2014xsa,Meyer:2016wrm}} down to the dashed red line indicated in Fig.~\ref{fig:helio_sens}. Also limits on $\gagamma$ for massive (10~keV-100~MeV) ALPs have been derived from the absence of a delayed and diffuse burst accompanying the SN1987A explosion, expected from their subsequent decay into photons~\cite{Jaeckel:2017tud}.

Photon-ALP conversion in the $B$-field of single compact stellar objects has also been studied \cite{Chelouche:2008ta,Jimenez:2011pg,Perna:2012wn,Berenji:2016jji}. 
In particular, the observation of a soft X-ray modulated excess in XMM data has been attributed to solar ALPs conversion in the Earth's magnetic field~\cite{Fraser:2014wja}, however this interpretation has been shown to be problematic~\cite{Roncadelli:2014lsa}.  

Finally, photon-ALP mixing is polarisation dependent, and therefore it could leave an imprint in the polarisation pattern of a variety of sources,
(e.g. in UV photons from AGNs in radio galaxies \cite{Horns:2012pp}, in magnetic white dwarfs~\cite{Gill:2011yp} or in GRBs \cite{Bassan:2010ya})
that could be used to constraint ALP properties. It could explain long-distance correlations of quasar polarisation~\cite{Payez:2008pm}, although this possibility is nonetheless challenged by the absence of significant circular polarisation~\cite{Payez:2012vf,Payez:2012rc}. For most of these, the ALP mass needed to give visible effects is considerably lower than the previous stated ranges.

\additionalinfo{
}

\section{Sources of axions and their detection}
\label{sec:sources}

\begin{figure}[t] 
   \centering
   \includegraphics[width=5in]{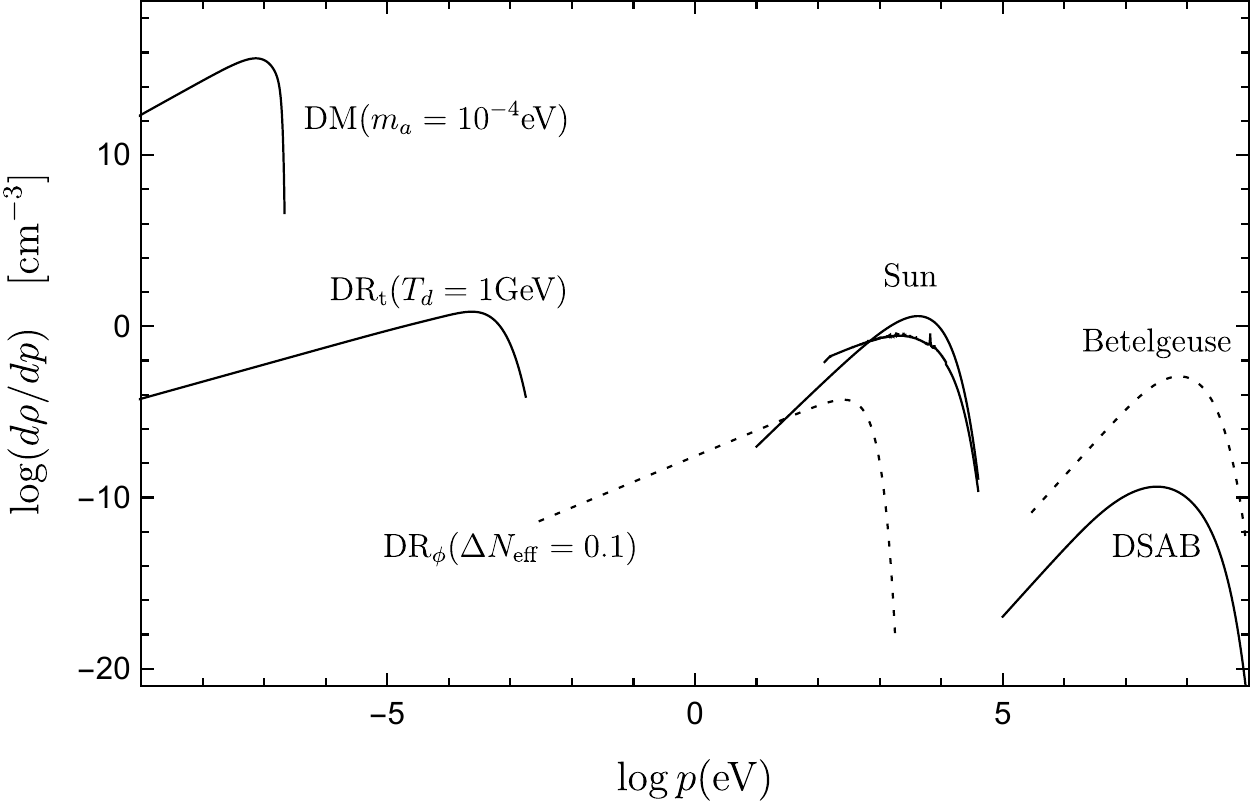}
   \caption{Energy spectrum of natural axions/ALPs as function of momentum at the Earth position. Galactic DM with $m_a=10^{-4}$ eV, thermal DR (DR$_t$) and from modulus decay (DR$_\phi$), solar Primakoff and ABC axions saturating the astrophysical bounds (from HB and WDs respectively) and maximum diffuse supernova axion background (DSAB) and axion pulse from Betelgeuse (50\% of SN energy into axions).}
   \label{fig:guap}
\end{figure}

In the previous section we have outlined a few possible natural sources of axions/ALPs, which could in principle be used for their direct detection in terrestrial experiments. Fig.~\ref{fig:guap} is an illustrative attempt at a grand unified ALP spectrum to realise the magnitude of all the components at the Earth's position. 
We discuss them in more detail in subsection~\ref{sec:natural}. 
These natural sources of axions (in particular DM and solar axions) offer excellent opportunity for detection and in fact much of the material that follows involves either a search for DM (section~\ref{sec:DMexps}) or solar (section~\ref{sec:solar}) axions. 

To the natural sources one has to add man-made sources of axions, leading to purely laboratory experiments that will be presented in section~\ref{sec:laboratory}. These are the most robust strategy in terms of model assumptions, as both production and detection would happen in the laboratory. Unfortunately, this typically comes with a penalty in sensitivity. 
We discuss the most relevant features of man-made ALP sources in subsection \ref{sec:labalps}, where we also compare them to the natural counterparts.

\subsection{Natural sources}\label{sec:natural}

The spectral energy density as a function of axion's momentum, $d \rho/dp$, of different natural sources of axions/ALPs is 
schematically shown in Fig.~\ref{fig:guap}. 
It encompasses a Galactic cold DM population of ALPs at very low momentum (from misalignment or topological defect decay), hot DM axions and/or DR (from thermal reactions, parametric resonance or moduli decay), stellar axions (at the keV energies typical of stellar interiors), for which the brighter source in the sky is our Sun, and supernova axions at MeV energies. 
We have not computed the fluxes from any particular model, but plotted approximate observational upper limits. 

{\bf Dark radiation:} 
The thermal dark radiation depicted has been chosen to correspond to decoupling at $T_d\sim$ GeV contributing $\Delta N_{\rm eff}\sim 0.04$. Assuming that ALPs were in thermal equilibrium at some point in the thermal history of the Universe does not change very much the distribution and its mean momentum, but if the reheating temperature, $\Trh$, becomes smaller than the ALP decoupling temperature, the amplitude becomes suppressed by $(T_d/\Trh)^2$ and can be much smaller.  

Much ``hotter'' dark radiation, DR$_\phi$, could come from a heavy particle decay. Here we have depicted a population giving $\Delta N_{\rm eff}=0.1$, which is below observational constraints but not much off. The typical momentum today $p_0$ is related to the CMB temperature and the heavy particle mass, $m_h$, by $p_0\sim T_0\sqrt{\mPl/m_h}$. 
We have chosen their average momentum to be $p_0=300$ eV as typical from the moduli masses $\sim 10^6-10^7$ GeV motivated in scenarios~\cite{Higaki:2012ar,Cicoli:2012aq}. The decay spectrum can be nicely approximated by $d \rho/dp \propto p^{3/2}e^{-(p/p_0)^2}$.   

{\bf Solar ALPs:} 
ALPs with a two-photon coupling, $\gagamma$, are produced through the Primakoff process $\gamma + \gamma^* \to a$ in stellar interiors. Here $\gamma$ is an ambient photon and $\gamma^*$ represents a virtual photon from the electric-field of a plasma species. Electrons, protons and other ionised or partially ionised species typically contribute. Virtual photons of the Coulomb field have negligible energy (but have momentum) and thus the ALPs inherit their energy from the ambient photons in the solar core, of $\sim $ 3 keV energies. We have depicted the flux of solar core ALPs (smooth) in the overall plot of Fig.~\ref{fig:guap}, which overshadows that of other stars, calculated by \cite{Barth:2013sma} with the value of $\gagamma$ given by the upper limit allowed by stellar constraints, see Tab.~\ref{tab:axtrobounds}. 

A useful analytic approximation to the differential flux of Primakoff solar ALPs at Earth, accurate to less than 1\% in the 1--11 keV range,  is given by~\cite{Andriamonje:2007ew}:

\be \label{bestfit}
  \frac{{\rm d}\Phi_{\rm a}}{{\rm d}E}=6.02\times 10^{10} 
  \(\frac{\gagamma}{10^{-10}\rm GeV^{-1}}\)^2
  \,E^{2.481}e^{-E/1.205}\,
  \frac{1}{ {\rm cm}^{2}~{\rm s}~{\rm keV}}\ . 
\ee

\noindent where $E$ is the axion energy expressed in keV. This Primakoff spectrum is shown in Fig. \ref{fig:axion_flux} (left). As seen, it peaks at $\sim$3 keV and exponentially decreases for higher energies. Since the solar interior is well known, the only sensible uncertainly on this flux around the peak is the overall normalisation due to $\gagamma$. At energies below $\sim 100$ eV the situation is less clear as other processes can contribute.


ALPs with coupling to electrons $g_{ae}$, like the case of non-hadronic axions, have additional production channels at the Sun: Axio-recombination, Bremsstrahlung and Compton (the ABC processes), see~\cite{Redondo:2013wwa}. Bremsstrahlung and Compton produce a continuum spectrum and dominate at low and high frequencies, respectively. The spectra of axions emitted in electron atomic transitions of free-bound or boud-bound type (Axio-recombination) have lower thresholds or are monochromatic. Again the most visible stellar source is the Sun. The ABC spectrum is included in Fig.~\ref{fig:guap}, although its spectral features, like a set of Fe lines at $\sim 7$ keV, are better seem in the zoomed version shown on the right of Fig. \ref{fig:axion_flux}. We have chosen the upper value of $g_{ae}$ allowed by other stellar bounds, see Tab.~\ref{tab:axtrobounds}. The spectral features are not known precisely due to difficulties on modelling atomic physics in the solar plasma, but completely different approaches were known to agree qualitatively very well~\cite{Redondo:2013wwa}. On the quantitative level the spectrum has small systematic errors if a spectral average is performed to average over detailed atomic physics.   
Finally, axions can also be produced in solar atomic transitions by virtue of axion-nucleon couplings.

{\bf Supernova ALPs:} 
ALPs can be emitted by the protoneutron star MeV plasma of type-II supernovae by a variety of processes, contributing to its cooling.  
In the case of QCD axions, the most relevant is nucleon-nucleon bremsstrahlung. 
The spectrum is not well known but educated guesses are available. 
The time-integrated energy flux is limited by the gravitational binding energy of the pre-collapse Iron core. 
The maximum diffuse supernova axion background is therefore constrained by the case in which proto-neutron stars are colloed exclusively by ALP emission and not to neutrino emission. 
We have depicted a DSAB flux taken from~\cite{Raffelt:2011ft} where 50\% of the core collapse energy is radiated in axions. 
For comparison, we have also depicted the peak flux obtained from the core-collapse of Betelgeuse assuming a cooling time of 10 seconds with the same 50\% assumption.  

{\bf Dark Matter: }
The galactic DM density at the Earth position, $\dmdensun$, is perhaps the most uncertain of all natural sources of axions discussed.   
ALP DM consists of small amplitude oscillations of the ALP field around the minimum of its potential. 
At large scales, ALP DM behaves as standard cold dark matter but small scale structure can be very different. 
In the \emph{post-inflationary} scenario we expect a large fraction of the DM to be in miniclusters to start with. 
In some cases, we can also have long-lived ALP domain-walls until today. 
Finally, in the limit of fuzzy dark matter we can have a large suppression of small scale structure. 
In the typical QCD axion scenario, the only potential complication is a large fraction of mass in dense objects 
because domain walls are either short-lived or leading-to-disaster and the mass $m_\A$ from \eqref{axionmass} is never large enough to become fuzzy DM if we impose $f_\A < \mPl$. For Fig.~\ref{fig:guap} we have assumed that the ALP field has negligible large-scale fluctuations of its own (the isocurvature fluctuations mentioned before) and it only inherits standard adiabatic fluctuations from the inflaton. 
Local measurements of the local DM distribution point to a local DM density of $\dmdensun \sim 0.2-0.56$ GeV/cm$^3$~\cite{Read:2014qva} while fitting kinematic data with a Milky Way DM halo with well motivated density profiles (NFW, Einasto) gives $\dmdensun \sim 0.42$ GeV/cm$^3$ with somewhat smaller errors~\cite{Bovy:2012tw}.  The local DM \emph{velocity} distribution can be read from N-body simulations of galaxy formation or estimated from orbits of very old (low metalicity) stars~\cite{Herzog-Arbeitman:2017zbm,Herzog-Arbeitman:2017fte}. A recent determination from N-body simulations with Smoothed-Particle Hydrodynamics focused on the DM axion signal gives \todonote{Notation clash: we have too many $f$'s decay constants, fermions and now distributions... }
\exclude{
\be
f^{\rm dm,\odot}(v)\propto  \(\frac{v^2}{2T}\)^\alpha \exp\(\frac{v^2}{2T}\)^\beta
\ee}
\be
\label{Lentzdistribution}
f_v\propto  \(v^2\)^\gamma \exp\(-\frac{v^2}{2\dmsigmav^2}\)^\beta
\ee
with $\gamma=0.36\pm 0.13, \beta=1.39\pm 0.28,\dmsigmav^2=(4.7\pm 1.9)\times 10^{-7}$~\cite{Lentz:2017aay}. 
This corresponds to a typical velocity dispersion at the Earth's position of  
\be
\label{veldispersion}
\dmsigmav\sim 10^{-3} . 
\ee
We must note that the above distribution implies some temporal and spatial average over a large scale and that it does not necessarily represent with accuracy the velocity distribution relevant for DM direct detection at Earth (see section~\ref{sec:directional}).

If low-mass ALPs make a large fraction of the DM, their occupation number per phase-space cell is huge, 
\be
\frac{n_{a}}{\frac{4\pi }{3} p^3/(2\pi)^2} \sim \frac{\dmdensun_a/m_a}{\frac{4\pi }{3} m_a^3 {\dmsigmav}^3} \sim 
\admfrac \times 10^{23} \(\frac{m_a}{\mu\rm eV}\)^{-4}\(\frac{\dmsigmav}{300 \, \rm km/s}\)^{-3}
\ee
where $\admfrac$ is the fraction of axion DM over the total of cold DM at the position of the solar system. 
The ALP DM field is born as a classical field and stays with large occupation numbers until today around our local vicinity as long as $m_a\lesssim $ eV,  $\admfrac$ is not too small and substructure is negligible.  
We can perform a Fourier decomposition of the field in a large volume $\lvol$ comprising our local neighbourhood, 
\be
\label{alpfield}
a(t,\vec x) = \frac{1}{\sqrt{\lvol}}\sum_{\vec p} \frac{1}{2}\(e^{i(\vec p \cdot \vec x-\omega t )} \tilde a_{\vec p}+
\tilde a^*_{\vec p}e^{i(\vec p \cdot \vec x+\omega t )} \) \equiv  
\frac{1}{\sqrt{\lvol}}\sum_{\vec p} |\tilde a_{\vec p}|\cos (\vec p \cdot \vec x-\omega t +\alpha_{\vec p}) 
, \ee
where $\omega = \sqrt{ |{\vec p}|^2+m_a^2}$ is the dispersion relation for free ALP waves and 
$\tilde a_{\vec p}=|\tilde a_{\vec p}|e^{i \alpha_{\vec p}}$. 
The group velocity for ALP wave packets\footnote{We denote vectors as boldface, e.g. velocity $\vec v$, and their modulus either with standard $|\vec v|$
notation or simply by roman versions, i.e. $|\vec v|=v$.} 
\todonote{Include in an appendix section on notations and conventions?} is $v=\partial \omega/\partial p =p/\omega$. 

In the non-relativistic limit that applies to the galactic DM, $v\ll 1$, 
\be
\omega\simeq m_a\(1+ \frac{v^2}{2} + ...\) . 
\ee
The averaged energy density can be expressed in Fourier modes and therefore we have   
\be
\label{ALPpowerFourier}
\dmdensun_a = \frac{1}{\sqrt{\lvol}}\sum_{\vec p} \frac{1}{2}\omega^2  |\tilde a_{\vec p} |^2 \to  
\int \frac{d^3\vec p}{(2\pi)^3} \frac{1}{2}\omega^2  |\tilde a_{\vec p} |^2 , 
\ee
where for DM $\vec p \sim m_a \vec v$ and we have a relation between the velocity distribution 
and the power in Fourier modes $f(\vec v) \sim  |\tilde a_{\vec p}|^2$. In the last approximate equality we have used 
$dp=\omega d v/(1-v^2)\sim m_a d v$ by dropping very small ${\cal O}(v^2)$ corrections. 
The average density also assumes a formal time average over a period longer than the natural ALP cycle $T_a = 2\pi/m_a$. The relation between the velocity distribution and the power spectrum implies that that the DM ALP field 
retains a certain level of time and space coherence. 
This can be better understood from the field decomposition in terms of a sum of cosine functions \eqref{alpfield} by noting that $|a_{\vec p}|$ is 
exponentially suppressed for values of $|\vec p|\gtrsim \dmsigmav m_a$, corresponding to 
frequencies $\omega \gtrsim m_a+m_a \dmsigmav^2/2$. 
With this maximum momentum, points separated by a distance longer than 
\be
\label{coherencelength}
\lambda_c \lesssim \frac{\pi/2}{m_a \dmsigmav}\sim 10^3 \la, 
\ee 
where $\la$ is the ALP Compton wavelength
\be
\label{Comptonwavelength}
\la=\frac{1}{m_a}=0.2 \times {\rm cm} \(\frac{m_a}{\rm meV}\)^{-1},
\ee 
have each mode in the sum \eqref{alpfield} in phase and thus the same field value to a good approximation. 
Likewise, the field at one given point, oscillates at a frequency $\sim m_a$, and it takes 
a time of order 
\be
\label{coherencetime}
t_c\sim \frac{\pi}{m_a \dmsigmav^2/2} \sim 10^6 T_a , 
\ee
for the highest modes in the distributions to dephase. We say that the axion field has a 
coherence length and coherence time given by \eqref{coherencelength} and \eqref{coherencetime}, 
respectively. 
Inside a coherence length and time we can take the axion field to be 
$\sim a_\osci \cos(\omega t+{\vec k}\cdot \vec x + \alpha)$ with some oscillation amplitude, principal frequency and momentum. 
The typical such amplitude will be on average related to the local DM density. We can write   
\be
\frac{1}{2}m_a^2 a_\osci^2 = \admfrac \dmdensun \simeq \admfrac \frac{0.4\, \rm GeV}{{\rm cm}^3} , 
\ee
and therefore  
\be
\label{ALPDMamplitude}
a_\osci \sim {\rm eV} \(\frac{m_a}{\rm meV}\)^{-1} \sqrt{\admfrac}.  
\ee
This equation gives us a sort of a maximum (time-averaged) field value to expect as a DM field. 

In the case of the QCD axion, there are two interesting points. First, the amplitude of the axion field behaves as a dynamical theta-angle, $\A = \theta f_\A$, so $\theta$ will also feature oscillations of amplitude $\theta_\osci = \A_\osci/f_\A$. 
Remarkably, the amplitude of the $\theta$-oscillations does not depend on $f_\A, m_\A$ independently, only on their product, $f_\A m_\A$, which is determined by the known topological susceptibility of QCD $\chi$, see \eqref{axionmass}.  
Therefore we have 
$m^2 \A^2_\osci = (m_\A f_\A)^2 \theta_\osci^2 = \chi \theta_\osci^2$, which implies   
\be
\label{thetaosci}
\theta_{\osci} = \sqrt{\frac{2 \dmdensun \admfrac}{\chi}}\simeq  4.3 \sqrt{\admfrac} \times 10^{{-19}},  
\ee
This amplitude is much smaller that the experimental bound \eqref{thetabound}, showing some naive consistency of the axion DM hypothesis\footnote{Naive because laboratory searches are designed to detect a constant neutron EDM, not an oscillating one. 
Interestingly, the results of these experiments can also analysed to constrain the oscillating hypothesis~\cite{Abel:2017rtm}.}. 
On the other hand, note that the QCD axion field is defined with respect to the CP-conserving value $\theta=0$, and recall that CP-violation effects will shift the minimum of the potential to some value $\theta_{0}$, so we would rather write 
\be
\theta = \theta_0+\theta_\osci\cos(m_{a}t) . 
\ee
It is curious that we have a much stronger constraint on $\theta_\osci$, which would saturate the local DM abundance for values larger than $4.3\times 10^{{-19}}$, than on $\theta_{0}$.

In Fig.~\ref{fig:guap} we have depicted a normalised distribution $d \rho/dp \propto f(|v|) $ from \eqref{Lentzdistribution} for $m_a=10^{-4}$ eV and $\admfrac=1$ as an exemplary value. For other values of $m_a$, the characteristic momentum scales as $m_a \dmsigmav$ and the normalisation as $d \rho/d p\propto 1/m_a$. 

During an encounter with a typical Axion minicluster given by \eqref{axionminicluster} the density can increase 
by a factor $\sim 10^{5}$ during a time $2R_{\rm mc}/v_\odot \sim 1$ day, but the encounter rate,   
\be
\Gamma\sim v_\odot (\pi R^{2}_{{\A},\rm mc}) \frac{\dmdensun_{\A}^{\rm mc}}{M_{{\A},\rm mc}}\sim \frac{1}{ 10^{5}\, \rm years} , 
\ee
where $v_\odot\sim 0.7\times 10^{-3}$ is the solar orbital velocity in the Galaxy, 
\modified{is quite small for experimental purposes~\cite{Sikivie:2006ni}.} 
The velocity dispersion of a self-gravitation object, like a minicluster is naturally given by 
\be
\sigma^{\rm mc}_{v}\sim \sqrt{\frac{G_N M_{\rm mc}}{R_{\rm mc}}}\sim 10^{-9}\(\frac{M_{\rm mc}}{10^{-12}\, M_\odot}\)^{1/2}
\(\frac{R_{\rm mc}}{10^{12}\rm cm}\)^{-1/2}, 
\ee
and is typically much smaller than the galactic value. In Fig.~\ref{fig:guap} a minicluster encounter would show as a sharp and very high peak. 
Encounters with \modified{tidal streams from miniclusters} last longer but bring less enhancement~\cite{Dokuchaev:2017psd}.  

\subsection{Producing ALPs in the lab}\label{sec:labalps}

The equations of motion for an axion/ALP field can be read from \eqref{ALPinteractions} and give 
\be
\label{ALPKG}
(\square +m_a^2)a = \gagamma (\vec E \cdot \vec B) 
- \sum_\psi (g_{a\psi} j^5_{\psi} + \bar g_{a\psi} j^0_{\psi}) + ...
\ee
where we have used $F_{\mu\nu}\widetilde F^{\mu\nu}=-4(\vec E \cdot \vec B)$, with $\vec E,\vec B$ the ordinary electric and magnetic fields, and $j_\psi^0=\langle \bar \psi\psi\rangle$, $j^5_\psi=\langle i\bar \psi \gamma^5 \psi \rangle$ correspond to the number and spin density of fermions of species $\psi$ in the non-relativistic limit. Therefore, we can create laboratory ALP fields by creating large $(\vec E \cdot \vec B)$ configurations or by large densities of particles, but the generated ALP field is always suppressed by a small coupling $g$. If we produce ALP waves with time-varying electromagnetic fields or matter/spin distributions, the power emitted in ALPs is proportional to $\propto a^2$ and thus to couplings squared ${\cal O}(g^2)$. If we add that we still have to measure the effects of the ALP field, we come to effects that are of order ${\cal O}(g^2)$ if only the field is involved or ${\cal O}(g^4)$ if we try to transfer the power from lab produced ALPs to a detectable signature. 
To be clear, this is no argument to favour natural ALP sources until we explore the magnitude of the fields achievable in the lab. 

Therefore, here we consider two of the most important representative examples that are actually used to search for ALPs. First, the ALP field created by a loaded cavity $\vec E_c$-field in the background strong static magnetic field $\vec B_e$. Second, the static ALP field created by a macroscopic body. 

\subsubsection{Axions from photons (two-photon $\gagamma$ coupling) }\label{sec:axionphoton}
The formal solution for the ALP generated by a source can be written using the retarded Green's function. Using $\gagamma \vec E_c \cdot \vec B_e$ as source, we assume a separable monochromatic electric field $\vec E_c =e^{-i\omega t}E_0 {\vec {\cal E}}(\vec x)$ with amplitude $E_0$ and normalised mode function $ {\vec {\cal E}_c}(\vec x)$ and a time independent $\vec B_e(\vec x)=B_0 {\cal B}(\vec x)$ where $B_0$ is the average value in the electric-field region $V$ and ${\cal B}(\vec x)$ denotes the spatial variation. We have\footnote{See for instance~\cite{Hoogeveen:1992nq,Jaeckel:2007ch}.} 
\be
\label{genericALPfieldsourcedEB}
a(t,\vec x')= e^{-i\omega t} \gagamma E_0 B_0
\int_V d^3 x \frac{e^{i k_a |\vec x'-\vec x|}}{4\pi |\vec x'-\vec x|}  \vec {\cal E}_c(\vec x) \cdot \vec {\cal B}(\vec x) ,  
\ee
where $k_a=\sqrt{\omega^2-m_a^2}$ would be the wavenumber/momentum of a propagating ALP. 
The interpretation is of course that the ALP field is the coherent sum of infinitesimal waves produced at every point in the source (here $\vec {\cal E}_c \cdot \vec {\cal B}$) with a phase difference due to the different paths to the point $\vec r$. 
If the mode function $\vec {\cal E}_c(\vec x)$ and the magnetic variation $\vec {\cal B}(\vec x)$ are arranged 
in such a way to cancel the phase accrued by $e^{i k_a |\vec r-\vec x|}$ we say that the ALP is produced \emph{coherently} and the ALP field gets potentially enhanced by a volume factor $V^{2/3}$ near the source, at least naively.  

In sec.~\ref{sec:LSW} we will review light-shining-though-walls (LSW) experiments that produce their ALPs with a long optical cavity (a Fabry-Perot resonator) in a constant $B$-field produced by a long array of particle accelerator dipole magnets. Assuming the beam width is sufficiently large, the radial integrals are trivial and we are left with the integral along the cavity's length, $L_c$, which on axis is  
\be
\label{perturbative}
a(t,z') \sim e^{-i\omega t} \gagamma E_0 B_0 \int_{0}^{L_c} dz  \frac{e^{i k_a(z'-z)}}{-i k_a}
{\cal E}_c(z) \sim e^{-i(\omega t-k_a z')}  \frac{\gagamma E_0 B_0}{2 k_a q}(e^{i q L_c}-1)
\ee
where we have used only the right-moving wave in the cavity ${\cal E}_c(z)\sim e^{i k_\gamma z}$ and defined $q=k_\gamma-k_a$ as the difference in wavenumber/momentum of the ALP propagation and the cavity mode along the long, $z$-direction. Note that it is trivial to include modulations of the $B_e$ field along the $z$ direction.  
The result is a right-moving ALP wave that we can understand as a coherent superposition of ALPs (an ALP beam). The coherence level of the ALP production is captured by the last factor. In the limit of small momentum transfer $q L_c \ll 1$, the factor $|e^{i q L_c}-1|/q$ is maximal and reaches the asymptotic value $L_c$, which reflects maximum coherence. One can imagine that in this limit, the ALPs created at one point in the cavity move in phase with the photons as they propagate through the cavity sourcing more ALPs always in phase. 

The 1-D case is however understood very nicely in terms of photon-ALP oscillations~\cite{Raffelt:1987im}. 
In a background $B$-field neither photons nor ALPs correspond to freely propagating particles because the $\vec E\cdot \vec B_e a$ interaction acts as a non-diagonal mass term and quantum mechanically mixes ALPs with photons with polarisation along $\vec B_e$. 
To understand this we consider the ALP equation of motion \eqref{ALPKG} together with Maxwell's equations in the presence of axions\footnote{
The derivation with our notation can be found in~\cite{Millar:2016cjp}. }, which are 
\bea
\label{Gauss}		\vec \nabla \cdot \vec E &=& \rho_Q -\gagamma \vec {B}\cdot \vec \nabla a \\ 
\label{Ampere} 		\vec \nabla \times \vec B - \dot {\vec  E} &=& \vec J+\gagamma (\vec B \dot a-\vec{E} \times \vec \nabla a) \\
\label{GaussB} 		\vec \nabla \cdot \vec B &=&  0 \\ 
\label{Faraday}		\vec \nabla \times \vec E +\dot {\vec B} &=&  0
\eea
where $\rho_Q$ is the electric charge density and $\vec J$ the current. In a strong static and homogeneous background $B_e$-field, we can take $\vec B\sim \vec B_e$ in the source terms, neglecting the $B$-field of purely photonic waves and $\vec E\times \vec \nabla a$ compared to $\vec B_e \cdot a$. In this way, the ALP-Maxwell's equations are all linear and can be solved by plane waves propagating along an arbitrary direction with respect to $\vec B_e$. The solutions for waves of frequency $\omega$ propagating perpendicular to the $\vec B_e$ direction are the most useful phenomenologically, 
\bea
\label{ALPlike}
\vvv{a}{i A_{||}}{i A_\perp} &\propto& \vvv{\cos\vartheta}{-\sin\vartheta}{0} 
e^{-i(\omega t - k'_{a} \hat {\vec n} \cdot \vec x)}\quad ({\rm ALP-like})\\
\label{photonlike}
\vvv{a}{i A_{||}}{i A_\perp} &\propto& \vvv{\sin\vartheta}{\cos\vartheta}{0} e^{-i(\omega t - k'_{\gamma} \hat {\vec n} \cdot \vec x)}\quad ({\rm photon-like}), \\
\vvv{a}{i A_{||}}{i A_\perp} &\propto& \vvv{0}{0}{1} e^{-i(\omega t - k_\gamma \hat {\vec n} \cdot \vec x)}\quad ({\rm photon}), 
\eea
where we have used the notation $\vec A \equiv \vec E/i\omega$ and  labeled the components: $A_{||}$ as the polarisation perpendicular to $\hat {\vec n}$ with overlap with $\vec B_e$ direction, and $A_{\perp}$ as the polarisation perpendicular both to $\hat {\vec n}$ and $\vec B_e$ (which therefore does not mix at all with ALPs)\footnote{$A_l$ would be the longitudinal polarisation component (along $\hat {\vec n}$), which is actually carried by ALP-waves but is zero in this configuration. The general case is described in~\cite{Millar:2017eoc}.}.  
The mixing phenomenon implies that the mixed waves, the real propagation eigenstates of the problem,  carry both ALP and electromagnetic (photonic) fields and have different dispersion relations~\cite{Raffelt:1987im}. The effects are controlled by the mixing parameter $\vartheta$ and the modified squared wavenumbers, 
\exclude{
\bea
 \tan\vartheta &=& \frac{2\gagamma B_e \omega}{\Delta k^2 + \Delta k'^2} \\
  k'^2_{^a_\gamma} &=& \frac{1}{2}\( k_\gamma^2+k_a^2 \mp \Delta k'^2 \) \\
  \Delta k'^2 &=& k'^2_\gamma-k'^2_a = \sqrt{(\Delta k^2)^2+(2\gagamma B_e \omega)^2} 
\eea}
\bea
 \tan\vartheta &=& \frac{2\gagamma B_e \omega}{\Delta k^2 + \sqrt{(\Delta k^2)^2+(2\gagamma B_e \omega)^2} } \\
  k'^2_{^a_\gamma} &=& \frac{1}{2}\( k_\gamma^2+k_a^2 \mp \sqrt{(\Delta k^2)^2+(2\gagamma B_e \omega)^2}  \) 
\eea
where $\Delta k^2=k_\gamma^2-k_a^2$ is the unperturbed difference. Recall that $k_\gamma, k_a$ are the wavenumbers of a pure electromagnetic or ALP wave of frequency $\omega$, $k_a=\sqrt{\omega^2-m_a^2}$, $k_\gamma=\n \omega$ where $\n$ is the refractive index of the medium. 
As an important case, in vacuum we have $\Delta k^2=m_a^2$. 

This implies that a purely electromagnetic right-moving wave at a position $z=0$ polarised along a transverse $B_e$-field must be indeed interpreted as a superposition of a photon-like and an ALP-like wave, with relative weights $\cos\vartheta,-\sin\vartheta$, respectively, to cancel the ALP component. 
Since the waves have different wavenumbers, they necessarily therefore become out of phase after some distance. In other words, their ALP components will not cancel completely and the photonic components will not  add up completely coherently, appearing as if they would decrease,  
\exclude{
\bea
E_{||}(t,z) &=& E_0 e^{-i\omega t}\[e^{ik'_\gamma z}\cos^2\vartheta+e^{ik'_a z}\sin^2\vartheta\],  \\
\label{alpwave}
a(t,z)  &=&  \frac{E_0}{\omega} e^{-i\omega t}\sin\vartheta\cos\vartheta \[e^{ik'_\gamma z}-e^{ik'_a z}\] .
\eea}
\bea
\label{Ewave}
E_{||}(t,z) &=& E_{||,0} e^{-i(\omega t-k'_\gamma z)}\[\cos^2\vartheta+e^{-iq' z}\sin^2\vartheta\],  \\
\label{alpwave}
a(t,z)  &=&  \frac{E_0}{\omega} e^{-i(\omega t-k'_a z)}\sin\vartheta\cos\vartheta \[e^{iq' z}-1\] .
\eea
where clearly $q'=k'_\gamma-k'_a$ measures the phase difference and controls the beating of the two waves. Clearly, any electromagnetic component polarised perpendicular to the external field propagates as a standard EM wave with $E$-field 
\be
\label{Ewave2}
E_\perp(t,z) = E_{\perp,0} e^{-i(\omega t-k_\gamma z)}.
\ee 

Note that the obtained expression for the ALP wave is extremely evocative of~\eqref{perturbative}, although one obtains both with different physical pictures and have some slight conceptual differences. 
The most important difference is that \eqref{alpwave} in principle is valid only in the region where $EB\neq 0$, which would be the resonant cavity in the case of~\eqref{perturbative}. However, the ALP wave essentially free-streams unperturbed out of the cavity, with the amplitude given by the value at the end of it, $z=L_c$.  
Indeed, the radiated ALP field coincides exactly in the small mixing case, 
\be
\vartheta \simeq \frac{\gagamma B_e \omega}{\Delta k^2} \ll 1 \quad ,  \quad 
k_{^a_\gamma}'^2 = k_{^a_\gamma}^2 \mp \vartheta^2 \Delta k^2+...  
\quad ,  \quad q' = q(1 + \vartheta^2)
\ee 
which corresponds to the lowest order in the photon-ALP coupling, in which the perturbative calculation~\eqref{perturbative} shall be valid. Note that \eqref{perturbative} does not include the feedback to the $E$-field as \eqref{Ewave} does.
  
\exclude{The main difference is that~\eqref{perturbative} is valid outside the cavity ($z>Z_c$) while \eqref{alpwave} would be valid inside (we can take gain the point $z=0$ as the beginning of a cavity, and there should not be right-moving ALP fields so the assumption that lead to \eqref{alpwave} applies). However, if we put a mirror at $Z_c$, the $E$-field is reflected but the ALP wave can free stream across it, so \eqref{alpwave} is valid for $z>Z_c$ as long as we set $z=Z_c$ in the last factor. }

We can now interpret the energy densities in these fields divided by a quantum of size $\omega$ as particle number densities: $|E_x|^2/2\omega$ and $\omega |a|^2/2$ would then correspond to the number density of photons and axions. 
As we move away from $z=0$ the densities change, but the total flux remains constant, and we can now speak of photon-ALP `flavour' oscillations as we do in the case of neutrinos. 
The ``conversion probability'' after some distance $L$ is 
\bea \nonumber
\proba(\gamma\to a)(L) = \frac{|a(z)|^2}{|A_x(0)|^2} &=& \sin^2(2\vartheta)\sin^2\(\frac{q' L}{2}\) \\ \nonumber
+\, {\rm small\, mixing} &\to& \(\frac{2\gagamma B_e \omega}{\Delta k^2}\)^2\sin^2\(\frac{q L}{2}\)\\ \label{oscillationprob}
+\, {\rm relativistic}       &\to& \(\frac{\gagamma B_e }{q}\)^2\sin^2\(\frac{q L}{2}\)
\eea
As long as $q' L \ll 1$ the photon ALP conversion is coherent and $\proba\propto L^2$. 
In the relativistic limit we have taken $q\sim \Delta k^2/2\omega$. 
Note that if we impose that the $E$-field is zero at $z=0$, the ALP-like and photon-like waves are by symmetry given by 
\eqref{Ewave} and \eqref{alpwave} by $\vartheta\to -\vartheta$. Therefore, the photon to ALP oscillation probability after a length $L$ \eqref{oscillationprob} is the same than the ALP-photon oscillation probability , 
\be
\label{oscillationprob2}
\proba(\gamma\to a)(L) = \proba(a \to \gamma)(L) . 
\ee

We can now compare man-made axions with the maximum local density of solar axions from Fig.~\ref{fig:guap}, which is $\sim 1/({\rm cm}^3)$. 
The LIGO cavity stores $\sim$ MWatt power of $1064$ nm ($\omega\sim 1.2$ eV) light in a 4 km resonant cavity mode of $\sim 1$ cm beam radius, leading to a photon density of $10^{13}/{\rm cm}^3$. 
If we could build an array of 9-T magnets like those curving the proton beam at the LHC around the cavity, and $\gagamma$ would saturate the astrophysical bound $<0.65\times 10^{-10}$ GeV$^{-1}$ (see table~\ref{tab:axtrobounds}) the probability could reach up to
\be
\proba(\gamma\to a) \sim \(\frac{\gagamma B_e L_c}{2}\)^2\sim 10^{-12} \(\frac{\gagamma}{0.65 \times 10^{-10}{\rm GeV}^{-1}}\)^2\(\frac{B_e}{9\, \rm T}\)^2\(\frac{L}{4\, \rm km}\)^2, 
\ee
at the cavity end. In such an extreme case, both densities would be comparable. 
Naturally, solar axions have keV energies and can lead to potentially larger signals in lower background detectors, but the fact that a laser beam can be focused and especially that is highly monochromatic can compensate. Indeed, as we will see in sec.~\ref{sec:LSW}, the ALPS collaboration wants to employ a very similar cavity as an ALP source in its experiment. Naturally, the coherent enhancement can only be obtained when $q L \lesssim 1$. In vacuum, we have $q\sim m_a^2/2\omega$ which for 4 km restricts the coherent conversion to $m_a\lesssim 10^{-5}$ eV. 

To close this section let us mention a few related issues.
{
The LIGO laser is not the most intense on Earth, but excels in intensity times $L^2$, which 
would be the relevant parameter for an ALP source. Pulsed lasers can achieve TeraWatt intensities and the required $B_e$ fields can be provided by counter-propagating fields reaching $B\sim 10^6$ T. However, the spatial dimensions at which these values fields are achieved are comparable to the laser wavelength $L\sim 1-10\mu$m. 
The quantity, $EBL$, that controls the ALP production could be competitive with long optical cavities~\cite{Dobrich:2010hi} so some works have proposed searches for ALPs with high-intensity lasers~\cite{Dobrich:2010hi,Fujii:2010is,Homma:2014vsa,Gies:2014jia,Homma:2014rja,Paredes:2014oxa,Villalba-Chavez:2013goa,Hasebe:2015jxa,Burton:2017bxi,Homma:2017cpa,Nakamiya:2015pde}.
}
Throughout the section and in most of this review we tend to focus on the CP conserving ALP-photon coupling \eqref{ALPinteractions}, $\sim \gagamma {\vec E}\cdot {\vec B}a$, but most of the phenomena studied also apply to the CP-violating version~\eqref{ALPCPoddinteractions}, which is 
$\propto \bar g_{a\gamma} (|{\vec E}|^2-|{\vec B}|^2)a$. With such an interaction and in a background $B$-field, it is the $B$-field of a photon what interacts with the ALP field. 
This is to be compared with the ${\vec E}\cdot {\vec B}$ interaction, in which the photon interacts with ALPs proportionally to its $E$-field. Therefore, with the CP-violating coupling, the  $E_{\perp}$ polarisation, and not $E_{||}$, will be affected by the propagation in a $B$-field. In natural units, the electric and magnetic fields of a photon have the same magnitude and thus, besides the polarisation issue, the magnitudes of all the effects are the same. The case where both couplings are present has been studied in~\cite{Redondo:2008tq}. 

\subsubsection{ALP fields from macroscopic bodies (fermionic couplings $g_{a\psi},\bar g_{a\psi}$)}\label{sec:axionmacro}

The ALP couplings to fermions~\eqref{ALPinteractions} are responsible for macroscopic bodies sourcing an ALP field. 
A non-relativistic fermion sources an ALP field in a similar way to charged particles sourcing an electrostatic field, 
and can be easily computed from~\eqref{ALPKG} by using the retarded massive Green's function. 
At a position $\vec r$ away from a fermion $\psi$ with spin polarisation $\bm \sigma$ we have
\be
a(\vec r) = - \frac{e^{- m_a r }}{4\pi r}\( \bar g_{a\psi} - g_{a\psi} \frac{\vec r \cdot \bm \sigma}{r} \frac{1+m_a r}{2 m_\psi r}\) .
\ee
The CP-conserving axial coupling, $g_{a\psi}$, sources a dipole field proportionally to the spin of the fermion. 
For sub eV mass ALPs, the factor $(1+m_a r)/2 m_\psi r$ is always small at macroscopic distances $r \gg m_f$ 
and thus typically the field created is parametrically smaller than the one created by the CP-violating coupling. 
In both cases, the sourced field is exponentially suppressed beyond the ALP Compton wavelength 
\eqref{Comptonwavelength}.  
\exclude{ \be \la=\frac{1}{m_a}=0.2 \times {\rm cm} \(\frac{m_a}{\rm meV}\)^{-1}.  \ee }

Atoms source ALP fields proportionally to the ALP couplings and spin of their constituents. 
We can define effective couplings $\bar g_{aS},g_{aS}$ for a particular composite species as, 
\be
\bar g_{aS} = \sum_{\psi\in S} \bar g_{a\psi} \quad , \quad 
\bar g_{aS}\frac{\bm \sigma}{m_S} = \sum_{\psi\in S}  
g_{af} \frac{\bm \sigma_\psi}{m_\psi} \, . 
\ee
It would be considered a tuned case to have the $\bar g$ couplings to protons, neutrons and electrons  adjusted in a way as to cancel the effective $\bar g_{aS}$. On the other hand, cancellations in  $g_{aS}$ can be due to two identical fermions with opposite spin. For instance, the $^3$He nucleus has paired protons and electrons, and only neutrons would couple. 

Macroscopic bodies will of course source coherently through the $\bar g_{aS}$ coupling and, if spin polarised, through $g_{aS}$. 
One would ideally have a high-density object of dimensions larger than $\la^3$ to have fields of order, 
\be
a_S \sim \bar g_{aS} n_S (\la)^3\frac{1}{\la}   \sim  10^{-2} {\rm eV} \frac{\bar g_{aS}}{10^{-19}}\(\frac{n_S}{10^{25}{\rm /cm}^3}\) \(\frac{m_a}{\rm meV}\)^{-2} .   
\ee
Note that even for these tiny couplings, the amplitudes are comparable to the maximum typical amplitude of the DM field \eqref{ALPDMamplitude}. We can also compare the associated energy density, 
\be
\rho_{S}\simeq \frac{1}{2}m_a^2 a^2_S \sim \frac{5 {\rm keV}}{{\rm cm}^3} \(\frac{\bar g_{aS}}{10^{-19}}\)^2
\(\frac{n_S}{10^{25}{\rm /cm}^3}\)^2 \(\frac{m_a}{\rm meV}\)^{-2}  , 
\ee 
with the putative local DM density $\sim 0.4\, {\rm GeV/cm}^3$, which is also based on the average ALP amplitude~\eqref{ALPDMamplitude}, and thus constitutes a measure of ALP amplitude generated by a ``natural'' ALP source.  
If we were to use the coupling to nucleons allowed by astrophysical bounds, $\bar g_{aN}\sim 10^{-10}$, see tab.~\ref{tab:axtrobounds}, the allowed amplitude would largely exceed the typical value from the local DM density~\eqref{ALPDMamplitude}, at least for sub-eV masses. 
Indeed, these amplitudes are also sufficient to mediate a \emph{new type of force} between macroscopic bodies which would largely shadow gravity, see below. 
We will review experimental searches for new forces in Sec.~\ref{sec:5thforce}, but we can already anticipate that none has ever been found and the limits are very stringent. For instance, $\bar g_{aN}$ is constrained to be $\lesssim 10^{-19}$ at $m_a\sim $ meV. Indeed, when we consider the bounds on $\bar g_{aN}$ as a function of $m_a$, see Fig.~\ref{fig:5thNariadne}, the amplitude of the fields created by laboratory test masses are smaller than the DM amplitude~\eqref{ALPDMamplitude} in the range $m_a= (10^{-9},0.1)$ eV. 

\subsubsection{ALP forces}
The ALP-fermion couplings make ALPs to mediate $\la$-range forces between macroscopic bodies. The interaction energy of a non-relativistic fermion $\psi'$ in a background ALP field is 
\be
\label{Uinteraction}
U=-\int d^3x {\cal L}_{\rm ALP-int} = \bar g_{a\psi'} a+\frac{g_{a\psi'}}{2 m_\psi'}\bm \sigma' \cdot \vec \nabla  a.
\ee
The concomitant interactions between fermions mediated by ALP exchange are of three types: monopole-monopole ($\propto \bar g_{a\psi} \bar g_{a\psi'}$), monopole-dipole ($\propto \bar g_{a\psi} g_{a\psi'}$) and dipole-dipole ($\propto  g_{a\psi} g_{a\psi'}$)~\cite{Moody:1984ba}. We review searches for these types of forces in~sec.~\eqref{sec:5thforce} but it is nevertheless interesting to compare at least the monopole-monopole interactions with the gravitational force and the dipole-dipole with the magnetic interaction. 
\todonote{And the dipole-dipole with the magnetic interaction}
The ALP force between two fermions, $\psi,\psi'$ at a distance $r$ is given by
\be
\vec F_a = -\frac{dU}{dr}= -\bar g_{a \psi} \bar g_{a \psi'}\frac{e^{-m_a r}}{r^2}(1+m_a r)\frac{\vec r}{r}, 
\ee 
The new force is attractive between same fermion species but it could be repulsive between different species if they have different signs of the $\bar g_{a \psi}$ coupling, while gravity is always attractive. 
The force is species-dependent and will violate the equivalence principle in the most general case, while gravity is universal. At distances below the ALP Compton length \eqref{Comptonwavelength},  the force goes as $1/r^2$ with the distance like gravity. In this limit, the ratio of the ALP force with gravity is 
\be
\frac{|\vec F_a|}{|\vec F_G|} \sim \modified{\frac{\bar g_{a \psi}  \bar g_{a \psi'} }{G_N m_\psi m_\psi'}}= \(\frac{\mPl}{f_a}\)^2
\bar C_{a\psi} \bar C_{a\psi'} 
\ee

Although $f_a$ is expected to be below $\mPl$, and thus the first factor can be quite large, in the case of the axion, CP-violating couplings are extremely suppressed and the force turns out to be tiny. 
\modified{Indeed, if we use $\bar C_{\A \psi} \lesssim 10^{-14}$ as the best SM guess, we would need the ruled-out values of $f_\A < 10^5$ GeV to compete in intensity with gravity. Moreover, this competition would only happen at distances $\lambda_\A < $ nm, where we have never measured gravity. }
With physics beyond the SM, $\bar C_{\A \psi}\sim 10^{-10}$ might be allowed by neutron EDM searches, and axions would compete with gravity for $f_\A<10^{9}$ GeV at few $\mu$m distances. 
Dropping the assumption of the QCD axion, ALP mediated forces can be stronger than gravity at longer distances, which strongly motivates the search for new long-range forces.  

\todonote{Often wrongly stated sign of dipole-dipole force \cite{Daido:2017hsl}}

\subsection{Coherent ALP detection}\label{sec:coherent}

Although the natural and laboratory sources of ALPs can be quite efficient, the detection of any ALP signal involves again extremely small couplings and therefore requires all possible enhancements we can devise. 
Currently, the most sensitive ALP detectors use resonant techniques to exploit the time and spatial coherence of ALP fields to detect. For future reference we summarise the concepts involved in the time-coherence enhancement currently exploited by axion DM detectors, which is also at the heart of some of the enhancements proposed for LSW experiments and NMR detection techniques. 

The most important example consists of coupling a background ALP field to the electromagnetic mode of a resonant cavity in a 
strong and homogeneous magnetic field $B_e$. 
Recall that for Ampere's equation \eqref{Ampere}, the ALP field term behaves like a current density $\vec j_a = \gagamma \vec B_{\rm e} a$, which will load a cavity resonator if properly matched. 
In the previous discussion, we used electromagnetic fields or macroscopic distributions of matter particles to source the 
axion field and now we are interested in the electromagnetic fields sourced by a background ALP field. 
In the small $\gagamma$ limit, the backreaction onto the ALP field can be neglected and the ALP current $\vec j_a$ can be taken 
as a classical source. 
We take a generic ALP field oscillating at a main frequency $\omega$ with a generic position dependence, 
\be
a = a_0 e^{-i \omega t}{\cal A}(\vec x).
\ee 
\todonote{Change notation for the spatial part of the axion-area?}
Since the response we seek is linear, we can adapt the equations to multifrequency ALP excitations easily. 
We then expand the $E$-field in orthonormal normal modes $\vec E = \sum e^{-i \omega t} E_m{\vec {\cal E}}_m(\vec x)$ that solve the Poisson eigenfunction equation with suitable boundary conditions 
\be
-\nabla^2 {\vec {\cal E}}_m = \omega_m^2{\vec {\cal E}}_m \quad, \quad \int dV {\vec {\cal E}}_m \cdot {\vec {\cal E}}_n  = V\delta_{mn} ,    
\ee
where $\omega_m$ is the eigenfrequency of the mode and $V$ the volume of the cavity. 
\exclude{A set of corresponding orthonormal magnetic field functions can be defined as $\omega_m \vec {\cal B}_m(\vec x) = \nabla \times \vec {\cal E}_m(\vec x)$ from \eqref{Faraday}. }
Introducing the ALP field and mode expansion into Ampere's equation,  and projecting onto a single mode by multiplying by ${\vec {\cal E}}_m$ and integrating over the cavity volume, we obtain an equation for the mode amplitude  
\be
\label{modeequation}
(\omega^2 -\omega_m^2 +i \omega \Gamma_m  ) E_m = -\gagamma B_e \geo_m \omega^2 a_0  \, ,  
\ee
where we have introduced the geometric factor 
\be
\label{geometricfactor}
\geo_m = \frac{1}{V B_e}\int_V  d^3\vec x\,  {\vec {\cal E}}_m \cdot \vec B_e\, {\cal A}(\vec x) \,  , 
\ee
and the cavity losses are described by the decay constant $\Gamma_m$. 
\exclude{,  formally given by  $\Gamma_i\simeq \oint dS (\Igor{ {\cal E}_m}\times {\cal B}_m)/V$.} 
So defined, the decay constant of each mode $\Gamma_m$ has the correct interpretation of being the rate at which the energy density in a mode decays in the absence of a driving source. In such case, the electric field decays as $ e^{-i\omega_m t}e^{-\Gamma_m t/2}E_i $  and the energy density as $\propto |E_m|^2 \propto e^{-\Gamma_m t}$. 
The quality factor $Q_i$ of a oscillator can be defined as the energy stored divided by the power-loss in an oscillation cycle $1/\omega_i$, which corresponds to 
\be
\label{qualityfactor}
Q_m = \frac{\omega_m}{\Gamma_m}, 
\ee
and is more widely used than $\Gamma$ in the axion literature. 

The mode equation \eqref{modeequation} shows that each mode oscillates at the excitation frequency $\omega$ with a different amplitude (and phase) given by 
\be
\label{cavityEi}
E_m  =\frac{-\gagamma B_e \geo_m \omega^2 a_0}{\omega^2 -\omega^2_m +i\omega\omega_m/Q_m } .  
\ee 
If we manage to tune the cavity resonant mode $\omega_m$ to the ALP excitation frequency, $\omega$, the induced $E_m$ field 
becomes enhanced by the quality factor of the cavity. 
The power that one can extract from the cavity is however not proportional to the square $\propto Q_m^2$. The reason is that extracting power 
counts as a loss factor in \eqref{modeequation} and decreases the overall quality factor of the resonator. We usually split the losses into signal extraction losses, $\Gamma^s$, and intrinsics from the cavity, $\Gamma^c$ (due to ohmic losses from the currents in the cavity walls etc.) and define the coupling factor 
\be
\kappa = \frac{\Gamma^s_m}{\Gamma^s_m+\Gamma^c_m} .  
\ee
Formula \eqref{qualityfactor} still holds for the combined losses $\Gamma_m=\Gamma^s_m+\Gamma^c_{m}$.  
Since the signal is proportional to $\Gamma^s_m=\kappa \Gamma_m = \kappa \omega_m/Q_m$, it can only 
be enhanced by one factor of $Q_m$.

The energy stored\footnote{The energy stored oscillates between the electric and magnetic energy components. In our complex notation and on resonance the cycle-average of each component is the same and corresponds to $|E_m|^2/4$. Off resonance a factor $\omega^2+\omega_m^2/(2\omega^2)$ arises.}  in the $m$-th mode is given by $U_m=V|E_m|^2/2$ therefore, the extracted signal power is 
\be
\label{resonantregenerationD}
P^s_m = \Gamma^s_m U_m = \frac{\kappa \omega_m}{Q_m} |\geo_m|^2  V \frac{\gagamma^2 B^2_e \omega^2 }{\modified{(\omega^2 -\omega^2_m})^2  +(\omega_m \omega/Q_m)^2 }  \frac{1}{2}\omega^2 |a_0|^2, 
\ee
which on resonance is 
\be
\label{resonantregeneration}
P^s_m(\omega=\omega_i)  = \kappa \frac{Q_m}{\omega_m}  |\geo_m|^2  V \times  \gagamma^2 B^2_e\times \frac{1}{2}\omega^2 |a_0|^2 . 
\ee
Note the last term coincides with the energy density the ALP wave and it will play the role of the DM density. 
These formulas are quite general and will be used in several contexts of next section. 
In particular, one can develop the spectral response of our cavity for the general ALP Fourier spectrum 
used in \eqref{alpfield},\eqref{ALPpowerFourier} 
as 
\be
\label{resonantregenerationDD}
\frac{d P^s_m}{d\omega} = \frac{\kappa \omega_m}{Q_m} V \frac{\gagamma^2 B^2_e \omega^2 }{(\modified{\omega^2 -\omega^2_m})^2  +(\omega_m \omega/Q_m)^2 }  
\[\int d\Omega_{\vec p} \frac{k_a \omega}{(2\pi)^3} 
\frac{1}{2}\omega^2 |\tilde a_{\vec p}|^2 |\geo_m|^2 \], 
\ee
where the bracketed part is the ALP spectral energy density, except for the geometric factor, $\geo$, which is evaluated with a plane wave ${\cal A}(x)=e^{i\vec p\cdot \vec x}$. 
In the case of a conventional haloscope for DM detection, see sec.~\ref{sec:haloscopes}, the coherence length $\lambda_c$ can be much larger than the apparatus $\sim \la/2$ so $\geo$ is effectively independent of ${\vec p}$, and the power is a convolution of the axion energy spectrum with the resonator. 
However, for some configurations, a $\vec p$-dependence, even if small,  can be useful to infer information about the DM velocity distribution, see sec.~\ref{sec:directional}.  
 
\exclude{
They can very easily extrapolated to a general ALP source. 
Note that we have not used the ALP dispersion relation for ALPs and are valid for a general ALP ansatz 
$a = a_0 e^{-i(\omega t+i\vec k\cdot \vec x)}$. A general ALP field can be written as 
\be
a = \int \frac{d^4p}{(2\pi)^4} \tilde a( 
\ee
}

\todonote{NMR theory for dummies here would be veeery nice}

\subsection{Overview}

We have seen that relying on natural sources can give access to higher axion fluxes, but at the \modified{expense to} some degree of model-dependency, moderate in the case of solar ALPs, and larger for DM searches. 
While some aspects of the expected solar axion flux depend on the axion model details (e.g. whether they couple with electrons), it is a general prediction from QCD axion physics that the Sun will emit axions via, at least, the Primakoff conversion, which can be suppressed only in very contrived models, see Tab.~\ref{axionmodellist}. Finally, if DM is made mostly of ALPs, our galactic halo would be the most prolific source of these particles, as seen in Fig.~\ref{fig:guap}. 
However, although the axion being the DM is a very appealing hypothesis, it is not a necessity derived from their existence (at least not as the main DM component).

\begin{table}[t!]
  \centering
\begin{tabular}{lcccc|ccc|c|c}
  \hline
  \textbf{Detection method}
  & $\gagamma$ & $g_{ae}$ & $g_{aN}$ & $ g_{\A\gamma n}$ & $\gagamma g_{ae}$ & $\gagamma g_{aN}$ & $g_{ae }g_{aN}$    & $g_N\bar g_N$  &   \multicolumn{1}{p{3cm}}{\centering \textbf{Model dependency}} \\ \hline
  Light shining through wall & $\times$ &  & & & & & & & no \\
  Polarization experiments & $\times$ &  & & & & & & & no \\
  Spin-dependent 5th force & & & $\times$ &  & & & $\times$  & $\times$ & no \\
\hline
  Helioscopes & $\times$ &  &  & & $\times$  & $\times$ & &  & Sun \\
  Primakoff-Bragg in crystals & $\times$ &  & &  & $\times$  &  & & &  Sun \\
  Underground ion. detectors & $\times$ & $\times$ & $\times$ &  & & $\times$ & $\times$ &    & Sun$^*$ \\
\hline
  Haloscopes & $\times$ &  & & & & & & & DM \\
  Pick up coil \& LC circuit & $\times$ &  & & & & & & & DM \\
  Dish antenna \& dielectric & $\times$ &  & & & & & & & DM \\
  DM-induced EDM (NMR)    &  &  & $\times$ &  $\times$ & & &  & & DM  \\
  Spin precession in cavity & &  $\times$  & & & & & & & DM \\
  Atomic transitions & &  $\times$  & $\times$ & & & & & & DM \\\hline
  \end{tabular}
  \caption{List of the axion detection methods discussed in the review, with indication of the axion couplings (or product of couplings) that they are sensitive to, as well as whether they rely on astrophysical (axions/ALPs are produced by the Sun) or cosmological (the dark matter is made of axions/ALPs) assumptions.
  $^*$Also ``DM'' when searching for ALP DM signals, see section \ref{sec:othersolar} }
  \label{tab:methods}
\end{table}

A number of detection techniques are being considered, exploiting the diverse possible couplings of the axion to SM particles. Most of them rely on the $\gagamma$ coupling, in part because of the generality of its presence in axion models, but also because it easily leads to coherent effects at detection. However, other channels are also being probed in current and future experiments.
In order to better frame the material of the following sections, table~\ref{tab:methods} lists all the detection strategies to be reviewed, with indication of the relevant axion coupling being probed in each of them. Some experiments are sensitive to a product of couplings, when the mechanism at origin and detection are different. For some cases, more than one coupling (or product of couplings) are indicated, because more than one type of signal or mode of operation is available for those experiments.

Finally, let us mention that the search for ALPs with ``conventional'' high energy physics tools, i.e., at accelerators, is suitable only for very high masses ($\sim$MeV or more) that are largely excluded for QCD axions. Although in some ALP models those masses may not be excluded, and indeed signatures of those models in colliders (see e.g.~\cite{Bauer:2017nlg,Bauer:2017ris,Mariotti:2017vtv}) or future beam dump experiments (see e.g.~\cite{Blumlein:1990ay,Blumlein:1991xh,Dobrich:2015jyk}) are being studied, we will not review them here. Our focus is on very low mass axions and ALPs (below $\sim$~eV) traditionally considered invisible in conventional particle physics experiments, and that require novel techniques at the low energy frontier, involving coherence effects at detection and high intensity sources.

\section{Search for ALPs in the laboratory}
\label{sec:laboratory}

The existence of an axion or ALP field could result in observable effects purely in the laboratory, i.e. without relying on an extraterrestrial sources of axions. The most straightforward of these is the photon regeneration in magnetic fields, colloquially known as \textit{light-shining-through-walls} (LSW). A powerful source of photons (e.g. a laser) is used to create axions in a magnetic field. Those axions are then reconverted into photons after an optical barrier. In addition, effects on the polarisation of the laser beam (ellipticity and/or dichroism) traversing a magnetic field due to the existence of axion can also be searched for. Finally, the axion field can also give rise to short-range macroscopic forces.

In general, the magnitude of these effects is very small and, for the case of QCD axion, currently far from experimental sensitivity (a possible exception, under some assumptions, may be the ARIADNE fifth force experiment, see below). However some ALP parameter space may be within reach of the next generation of laboratory experiments.

 \subsection{Light-shining-through wall experiments}\label{sec:LSW}

\begin{figure}[t]
\centering
\includegraphics[width=\textwidth]{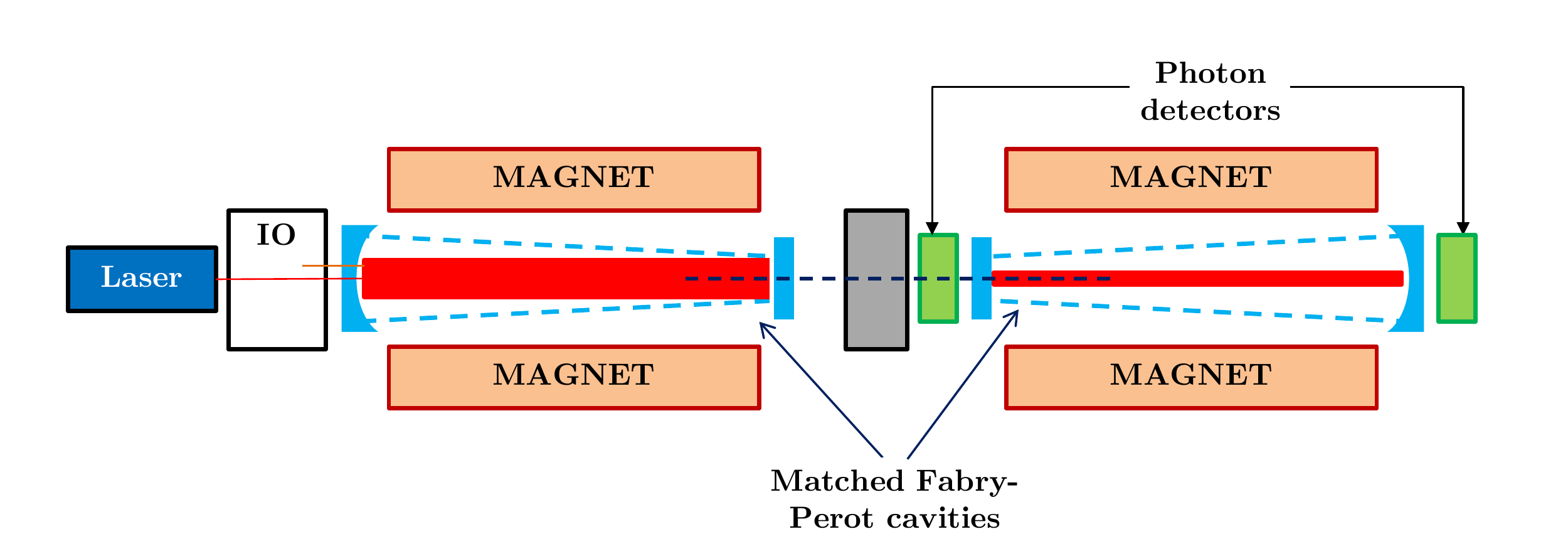}
\caption{\label{fig:LSW_sketch} The principle of photon regeneration. The Fabry-Perot cavities in both the production and regeneration regions must be actively locked in order to gain in sensitivity. Adapted with permission from~\cite{Mueller:2009wt}. }
\end{figure}

Figure \ref{fig:LSW_sketch} shows the conceptual arrangement of LSW experiments. The left half is the \textit{production} region, where photons from the source (e.g. a laser) are converted into axions.  
The right half is the \textit{reconversion} region, where axions are converted into photons, that are subsequently detected. 
The LSW probability can be written schematically as,  
\be
\proba(\gamma \to a \to \gamma) = \proba(\gamma \to a)\proba(a \to \gamma).  
\ee
In the typical 1-D approximation, the conversion probabilities are given by \eqref{oscillationprob} and \eqref{oscillationprob2}. 

The sketch shows the improved detection scheme based on resonant regeneration~\cite{FUKUDA1996363,Hoogeveen:1990vq,Mueller:2009wt}. By using resonators in both the production and the regeneration regions, the conversion probability is enhanced by a factor $\beta_P \beta_R$, where $\beta_P$ and $\beta_R$ are called the power built-up factors.
The first factor comes from the fact that inside of a cavity on resonance the laser power in the right-moving photon wave is amplified with respect to the available input laser power by the factor 
\modified{
\be
\beta_P = \frac{|E_0|^2}{|E_{\rm in}|^2} 
\sim \frac{4 T_{1}}{(T_{1}+T_{2}+R)^2+4 \Phi^2}
\xrightarrow{\text{on resonance}} \frac{4 T_{1}}{(T_{1}+T_{2}+R)^2} \sim \frac{\kappa_P Q_P}{n\pi/2}
\ee
where $T_{1},T_{2},R_{in}$ are the power transmissivities of the input (1) and auxiliary (2) mirrors forming the cavity, $R$ the round-trip losses, and $\Phi=2\omega L_P$mod$2\pi$ the round-trip phase (modulo $2\pi$), which on resonance is equal to zero, being $\omega L_P=n\pi$, an integer called the order of the resonance. 
In the last expression we have identified 
$T_1/(T_1+T_2+R)$ with the coupling factor and the quality factor as $Q_P=2\pi n/(T_1+T_2+R)$ but one 
more often talks about the \emph{finesse} of the resonator $F= Q/n$, which is independent of the resonator length. 
}
From a different more pictorial viewpoint, one can think that photons are not smashed against the middle wall, but are reflected by the mirrors and can attempt many times to convert into axions before being absorbed by internal cavity losses. 
The second factor comes from the coherent detection of the ALP field discussed in section \ref{sec:coherent}. 
The first cavity would be providing a coherent ALP field given by \eqref{perturbative}. The ALP field will excite the cavity 
modes in the regeneration cavity with amplitudes given by \eqref{cavityEi} so that one can extract a signal enhanced by the quality factor of the regeneration cavity $Q_R$, see \eqref{resonantregeneration}.  
A key consideration is that both cavities need to be mode matched and phase locked. \modified{In such 
case, the maximum $\beta_R$ enhancement of the resonant regeneration is again given by the finesse of the cavity (apart from ${\cal O}(1)$ coupling factors), i.e. independent of the order of the resonator like in the case of the production cavity. }
This can enhance the signal by many orders of magnitude, given that $\beta$ factors of several $10^4$ in the optical regime, and even several $10^5$ for microwaves, are possible.

The figure of merit of LSW experiment is best expressed in terms of the ratio of the expected power at the photon detector and the power delivered by the input laser~\cite{Hoogeveen:1992nq,Mueller:2009wt,Jaeckel:2007ch}, which can be understood as an enhanced probability of the double conversion $\gamma \rightarrow a \rightarrow \gamma$, 

\begin{equation}
\proba({\gamma \rightarrow a \rightarrow \gamma}) = 
\left( \frac{\gagamma B_e}{\omega} \right)^4  |\geo |^2 \beta_P   \beta_R, 
\end{equation}
\noindent where $B_e$ is the reference external magnetic field in the cavities and $\omega$ is the photon energy. 
The geometrical factor, $\geo$, represents the overlap of the modes resonating in the conversion and the regeneration cavities convoluted with the ALP Green's function and magnetic field spatial distribution. 
Formally it is given by \eqref{geometricfactor} by chosing ${\cal A}(\vec x)$ to represent the spatial dependence of the ALP field generated by the production cavity (essentially the factor under the integral in \eqref{genericALPfieldsourcedEB}). 

Experiments using microwave cavities have been performed with geometric dimensions of the order of the Compton wavelength $V\sim 1/\omega^3$ and production and regeneration cavities at comparable distances. In such a case, $\geo$ depends on the particular shape, position and orientation of the cavities~\cite{Caspers:2009cj}.

In the case of optical/near infrared light, Fabry-Perot resonators are used instead where the length is much greater than the wavelength $L \gg \lambda$ and the beam-waist, although typically a 1-D approximation typically suffices. 
The technique is relevant for $\omega\gg m_a$. 
In a homogeneous $B_e$-field, the ALP field generated by the production cavity is given by \eqref{perturbative}, i.e. essentially a plane wave ${\cal A}\propto e^{i k_a z}(e^{i q L_P}-1)$ that propagates along the optical axis of the production and covers the regeneration resonator. 
The geometric factor on resonance shows the typical dependency with the length $L_P, L_R$ of each cavity:
\be
\geo \propto (e^{i q L_P}-1) (e^{i q L_R}-1) 
\ee
but the relative phase between them enters as an overall phase.  
\todonote{One can derive how noise fluctuations affect the matching from the overal phase $\cal G$ }
The LSW probability can be expressed as 
\begin{equation}
\label{eqn:lsw}
\proba({\gamma \rightarrow a \rightarrow \gamma}) = 
\[ \left( \frac{\gagamma B_eL_P}{2} \right)^2 \mathcal{F}_P \beta_P\]
\[ \left( \frac{\gagamma B_eL_R}{2} \right)^2  \mathcal{F}_R \beta_R\], 
\end{equation}
where now $B_{e}=B_{e,\perp}$ is the component transverse to the photon/axion direction and we have defined for convenience the form factor,
\begin{equation}\label{eqn:lsw_F}
\mathcal{F}  = \left(\frac{2}{qL}\right)^2 \sin^2\(\frac{qL}{2}\) , 
\end{equation}
that reflects the coherence of the conversion. Recall that $q=k_\gamma-k_a\sim m_a^2/2\omega$ in the relativistic limit and in vacuum.

\begin{figure}[t!]
\begin{center}
\tikzsetnextfilename{LSW_exps}
\resizebox{0.9\linewidth}{!}{\input{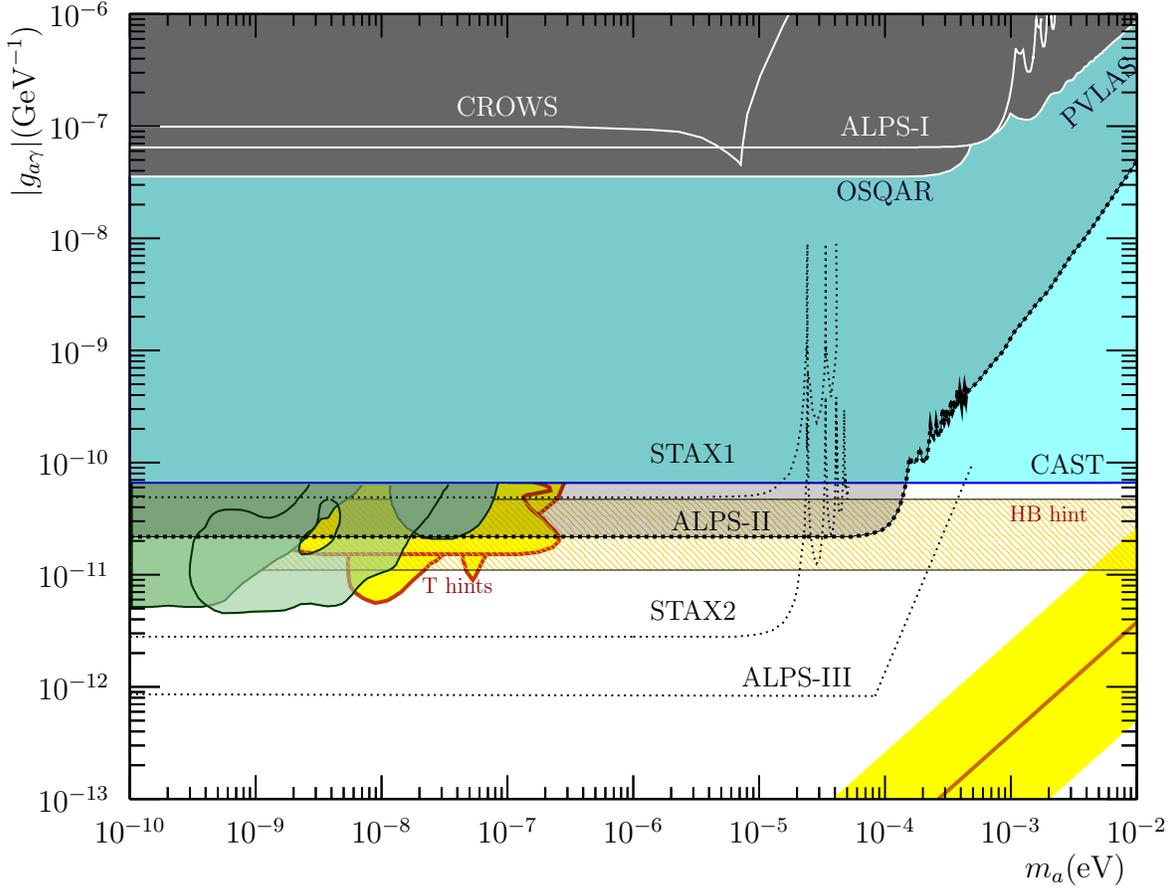}}
\caption{Exclusion regions in the ($\gagamma$, $m_a$) plane from the four most recent laboratory experiments reviewed in the text and, for LSW, in table~\ref{tab:lsw}, as well as the prospects of new generation ALPS-II experiment under preparation, that will surpass the astrophysical and CAST bounds on $\gagamma$ for $m_a \lesssim$~0.1~meV. The prospects for a more ambitious optical LSW experiment like ALPS-III are also shown. The prospects of possible future microwave LSW experiments like STAX would also surpass the CAST bound but for lower masses $m_a \lesssim 0.01$~meV. Also shown (yellow) are the regions hinted by the transparency of the universe to UHE photons, and from HB stars (orange horizontal band) as well as exclusions from several astrophysical observations (see section~\ref{sec:astrophysics} for explanations). The yellow band represents the QCD axion models, and the orange line inside it the \KSVZ model.}
\label{fig:LSW_plot}
\end{center}
\end{figure}

A number of LSW experiments have been carried out in the past: BRFT~\cite{Cameron:1993mr} at Brookhaven; BMV~\cite{Robilliard:2007bq} at Toulouse, using a short, pulsed high-field magnet and pulsed laser fields; GammeV~\cite{Chou:2007zzc} at Fermilab, using a Tevatron dipole magnet; or LIPPS~\cite{Afanasev:2008jt} at Jefferson Laboratory, using a pulsed free electron laser and two identical magnets; all of them producing limits to $\gagamma$ in the ballpark of 10$^{-6}$ -- 10$^{-7}$ GeV$^{-1}$. We refer to \cite{Redondo:2010dp} for a detailed account of past LSW experiments. Currently two active collaborations are working on LSW experiments and have produced the most competitive bounds below 10$^{-7}$ GeV$^{-1}$: The ALPS~\cite{Ehret:2010mh} experiment at DESY and the OSQAR~\cite{Ballou:2015cka} experiment at CERN, both making use of powerful accelerator dipole magnets, from HERA and LHC accelerators respectively. The main parameters of both experiments are listed in table \ref{tab:lsw} and the exclusion achieved in the $\gagamma-m_a$ space so far is shown in figure~\ref{fig:LSW_plot}. ALPS enjoys power build-up in the production region, while OSQAR has slightly higher magnet and laser parameters.

\begin{table}[b]
  \centering
\begin{tabular}{cccccccc}
  \hline
  Experiment & status & $B$ (T) & $L$ (m) & Input power (W) & $\beta_P$ & $\beta_R$  & $g_{a\gamma} [{\rm GeV}^{-1}]$  \\ \hline
  ALPS-I~\cite{Ehret:2010mh} & completed & 5 & 4.3 & 4 & 300 & 1 & 5$\times$10$^{-8}$ \\
  CROWS~\cite{Betz:2013dza} & completed & 3 & 0.15 & 50 & 10$^4$ & 10$^4$ & 9.9$\times$10$^{-8}$($^*$) \\
  OSQAR~\cite{Ballou:2015cka} & ongoing & 9 & 14.3 & 18.5 & - & - & 3.5$\times$10$^{-8}$ \\
  ALPS-II~\cite{Bahre:2013ywa} & in preparation & 5 & 100  & 30 & 5000 & 40000 & 2$\times$10$^{-11}$ \\
  ALPS-III~\cite{ALPSIII} & concept & 13 & 426  & 200 & 12500 & 10$^5$ & 10$^{-12}$ \\
  STAX1~~\cite{Capparelli:2015mxa} & concept & 15 & 0.5  & 10$^5$ & 10$^4$ & - & 5$\times$10$^{-11}$ \\
  STAX2~~\cite{Capparelli:2015mxa} & concept & 15 & 0.5  & 10$^6$ & 10$^4$ & 10$^4$ & 3$\times$10$^{-12}$ \\ \hline
\end{tabular}
  \caption{List of the most competitive recent LSW results, as well as the prospects for ALPS-II, together with future possible projects, with some key experimental parameters. The last column represents the sensitivity achieved (or expected) in terms of an upper limit on $\gagamma$ for low $m_a$. For microwave LSW (CROWS and STAX) the quality factors $Q$ are listed. $^*$ The limit is better for specific $m_a$ values, see Figure~\ref{fig:LSW_plot}}\label{tab:lsw}
\end{table}

LSW experiments with photons at frequencies other than optical have also been performed. The most relevant result comes from the CROWS experiment at CERN~\cite{Betz:2013dza}, a LSW experiment using microwaves~\cite{Hoogeveen:1992nq}. A couple of resonant RF cavities (\textit{emitting} and \textit{receiving} cavities) were placed inside a magnetic field, and a $\sim$50 W RF signal injected in the emitting one. Despite the small scale of the cavities ($\sim$15 cm diameter and height) the result is competitive with optical LSW experiments thanks to full implementation of resonant regeneration. Both cavities enjoyed $Q$ factors above 10$^4$ and they were frequency locked during the data taking time. LSW experiments have also been performed with intense X-ray beams available at synchrotron radiation sources \cite{Battesti:2010dm,Inada:2013tx}. However, due to the relative low photon numbers available and the difficulty in implementing high power built-ups at those energies,  X-ray LSW experiments do not reach the sensitivity of optical or microwave LSW.

Typically LSW limits are only valid up to an axion mass of $\sim$meV, below which $qL/2 \ll 1$ and thus $\mathcal{F} < 1$. Above this value the sensitivity drops as shown in Figure~\ref{fig:LSW_plot}. In general, past LSW bounds are considerably less restrictive than astrophysical or helioscope (see later) bounds. However, there is large margin for improvement in the current LSW figure of merit (see Table~\ref{tab:lsw}), especially in the magnetic length $L$ and in implementing resonant regeneration schemes. The ALPS-II experiment~\cite{Bahre:2013ywa}, currently in preparation at DESY, will use a string of 2$\times$10 HERA magnets (i.e. a length of 2$\times$100 m) for the production and conversion regions. It will implement resonant regeneration with power build-up factors of roughly 5$\times$10$^3$ and 4$\times$10$^4$ for the production and conversion regions respectively, as well as single photon detection capabilities with Transition Edge Sensors (TES). The HERA magnet string does not provide a completely homogeneous $B_e$ field but has $L_G \sim 1$-m long gaps of negligible field between the magnets of length $L_B\sim 8.8$ m. 
The form factor of such configuration is slightly different  
\be
{\cal F} = \(\frac{2}{qL}\)^2 \sin^2\(\frac{q L}{2N_B}\) \frac{\sin^2\(q N_B(L_B+L_G)/2\)}{\sin^2\(q (L_B+L_G)/2\)}
\ee
where $N_B$ is the number of magnets and $L=L_BN_B$ is the magnetic length~\cite{Arias:2010bh,Bahre:2013ywa}. 
The rest of parameters are shown in Table~\ref{tab:lsw}.  
Resonant regeneration imposes challenging requirements on the optical system, e.g. in regards the microroughness of the mirrors, the need to mode-match both cavities, the need to control the relative orientation and length of the cavities with extreme precision. Length fluctuations are compensated for by adjusting the laser frequency in the first cavity and the cavity length in the regeneration part by moving the mirrors with piezo-actuator down to a precision of 0.5 pm. A small frequency doubled signal is feed into the regeneration cavity for the locking phase and filtered before the detector. Mode matching is ensured by design. Without the middle wall the external mirrors would form a confocal cavity that the inner mirrors divide into two approximate concave-convex halves. 
The mode in the production cavity is a solution of the paraxial equation and can be chosen to overlap with a solution of the wall-less cavity. In the $m_a\to 0$ limit, gaussian beams are solutions of the ALP equation of motion in the paraxial approximation, so the ALP mode follows exactly the spatial distribution of the would-be mode of the whole cavity, ensuring perfect overlap.
In addition HERA magnets are slightly bent by design and need to be mechanically straightened to increase the horizontal aperture to reach specification of the optical cavities. The collaboration is progressively meeting those technical challenges in a undergoing preparatory phase~\cite{Spector:2016vwo}. Results regarding the specifications of the optical subsystems~\cite{Spector:2016ymd,Pold:2017sgm} and of the photon detectors~\cite{Bastidon:2015ifq,Dreyling-Eschweiler:2015pja,Bush:2017yuk} are already available. The full string of magnets is expected to be deployed in 2019 and first data taking will start in 2020. The expected sensitivity of ALPS-II goes down to $\gagamma < 2\times 10^{-11}$ for low $m_a$, and will be the first laboratory experiment to surpass current astrophysical and helioscope bounds on $\gagamma$ for low $m_a$, partially testing ALP models hinted by the excessive transparency of the Universe to UHE photons explained in section~\ref{sec:transparency} (see Fig.~\ref{fig:LSW_plot}).

A more ambitious ALPS-III extrapolation of this experimental technique is conceivable, for example, as a byproduct of a possible future production of a large number of dipoles like the one needed for the Future Circular Collider (FCC). Tentative experimental parameters for this future LSW are shown in Table~\ref{tab:lsw}, and contemplate a magnetic length of almost 1 km. ALPS-III would suppose a further step in sensitivity of more than one order of magnitude in $\gagamma$ with respect ALPS-II. 

In the microwave regime, resonant regeneration is less technically demanding, and larger input power is available. However, the boost factor given by the length of the cavity in the optical LSW is not available in the microwave regime. Nevertheless similar sensitivities could in principle be achieved by using extremely intense (up to 1 MW) sub-THz photon sources such as gyrotrons, as proposed by STAX~\cite{Capparelli:2015mxa}. The latter also invokes challenging single-photon detection in the RF regime. Despite those extrapolations, QCD axion sensitivity remains unattainable to LSW experiments, and such ambitious endeavors may eventually need more specific motivation, like e.g. a confirmation of possible hints produced in other experiments. In addition, increased sensitivity to a particular mass can be obtained by alternating the polarity of some magnets of the array~\cite{Arias:2010bh}, something that could be considered e.g. in the event of a determination of the ALP mass in haloscopes or helioscopes.

\subsection{Polarization experiments}

The vacuum permeated by a static $B_e$-field is birefringent and dichroic~\cite{Erber1961,Iacopini:1979ci} due to the photon-photon interactions mediated by virtual electron loops, as pointed out in the Euler-Heisenberg seminal work~\cite{Heisenberg:1935qt}. 
The index of refraction for photons polarised along the $B_e$ field or perpendicular to it are given by 
\be
\n_{||} = \frac{14 \alpha^2 B_e^2 }{45 m_e^4} \quad , \quad \n_\perp = \frac{8 \alpha^2 B_e^2 }{45 m_e^4} , 
\ee
where $m_e$ is the electron mass. 
On the other hand, the ALP photon coupling~\eqref{ALPinteractions} in an external B-field writes $- \gagamma {\bf B}\cdot {\bf E} a$,  so a laser beam with its E-field polarised at at angle $\alpha_0$ with respect to the B-field direction will have its parallel component depleted (by $\gamma\to a$ conversion) and phase delayed (due to $\gamma\to a\to \gamma$) but the perpendicular component untouched~\cite{Maiani:1986md,Raffelt:1987im}. 
Including vacuum birefringence in our example of section \ref{sec:axionphoton}, 
the laser polarisations after a length $L$ are given by \eqref{alpwave}-\eqref{Ewave2} to be 
\be
E_{||,\perp}(L) = E_{||,\perp}(0)(1-\eta_{||,\perp}-i\varphi_{||,\perp}),
\ee
with 
\be
\eta_{||} =  \vartheta^2 2\sin^2\frac{q L}{2} \quad , \quad
\varphi_{||} = \n_{||}\omega L+\vartheta^2(q L-\sin qL)  \quad , \quad
\varphi_\perp = \n_\perp \omega L, 
\ee
which produces a rotation of the laser polarisation given by $\delta\alpha = (\eta_{||}-\eta_{\perp})(\sin2\alpha_0)/2$  and an ellipticity angle $\varepsilon=(\varphi_{||}-\varphi_{\perp})(\sin2\alpha_0)/2$. 

In a $B_e=10$ Tesla, $L=10$ m field, the pure QED effect gives a maximum ellipticity angle $\varepsilon \simeq 2.4\times 10^{-14}$ for 1064 nm light ($\omega=1.2$ eV).
An ALP respecting astrophysical bounds can induce up to $\varepsilon \simeq 3.4\times 10^{-18} $ and a rotation of polarization $\delta\alpha = 5.3 \times 10^{-18}$.
In an optical resonant cavity, the effect increases linearly with the quality factor $Q$ and inside a delay line with the number of reflections~\cite{Raffelt:1987im}.
After first experiences in Brookhaven~\cite{Semertzidis:1990qc,Cameron:1993mr}, the PVLAS collaboration settled in Legnaro laboratory and instrumented a vertically oriented optical cavity with an ellipticity modulator between cross polarisers to measure the ellipticity generated by gases in B-fields (Cotton-Mouton effects) with the ultimate goal of reaching QED sensitivity. The 5 Tesla  1 m-long B-field was provided by a superconducting dipole, which rotated in the horizontal plane at a small frequency $\nu_r\sim 0.3$ Hz to modulate the signal proportional to  $\sin2 \alpha_0$ at $2\nu_r$ and thus improving the signal to noise ratio. A positive signal in rotation and ellipticity was announced in~\cite{Zavattini:2005tm} but subsequent investigation showed it was a spurious effect of unknown systematics~\cite{Zavattini:2007ee}. The PVLAS claim motivated a number of theoretical speculations to make the ALPs compatible with the astrophysical bounds and many experimental efforts to disprove it, and largely boosted the field of axion research in the years up to now.
The collaboration moved to a table-top set up in Ferrara to better understand the noise and increase the finesse. They have recently produced best results with a record\footnote{References mention the cavity's Finesse $F=7\times 10^5$} $Q=2.2\times 10^5$ and two $\sim10$ Hz rotating 2.5 T permanent magnets in Hallbach configuration giving $B^2L =10.25~{\rm T}^2$m~\cite{DellaValle:2014xoa,DellaValle:2015xxa}, reaching a sensitivity only a factor of $\sim 8$ away from the QED effect~\cite{DellaValleTrento}. The current limiting noise source is speculated to be thermal effects in the mirror's birefringence. Current plans involve cooling the mirror's and rotating magnets at higher frequencies. For comparison, note that the phase noise equivalent is $3\times 10^{-12} {\rm rad}/\sqrt{\rm Hz}$ at 10 Hz, better than the state of the art gravitational wave interferometer LIGO~\cite{TheLIGOScientific:2014jea}.
A recent review on the experimental search of non-linear QED effects can be found in~\cite{Fouche:2016qqj}.
\exclude{In terms of the index of refraction, QCD is Delta \n = 4 10^-24 / T^2, and they reached 40\pm 200 10^-24 /T^2, a factor of 50}
After a photon-regeneration experiment to disprove the ALP interpretation of PVLAS (see next section), the BMV collaboration in Toulouse, formed with the goal of measuring the QED birefringence using strong pulsed magnets from the LNCMI. Their current setup is not as sensitive as PVLAS. A noise analysis concluded their sensitivity to be limited by cavity intrinsic birefringence and R\&D continues~\cite{Hartman:2017nez}.

The signal could be increased by increasing the magnetic length using arrays of accelerator magnets like ALPS-II but these are difficult to modulate. An alternative was proposed to modulate the polarisation~\cite{Pugnat:2005nk} but it is severely limited by the mirrors intrinsic birrefringence.
A promising alternative is to introduce two corotating half-wave plates~\cite{Zavattini:2016sqz}.

There is little doubt that the QED birefringence will be measured in a purely laboratory experiment in a few years time-scale.  Looking into the future, it would be a background for future ALP searches, but
present prospects of improving over the astrophysical bounds are somewhat discouraging.

\subsection{5th force experiments}
\label{sec:5thforce}

\begin{figure}[t] 
   \centering
   \includegraphics[width=3.5in]{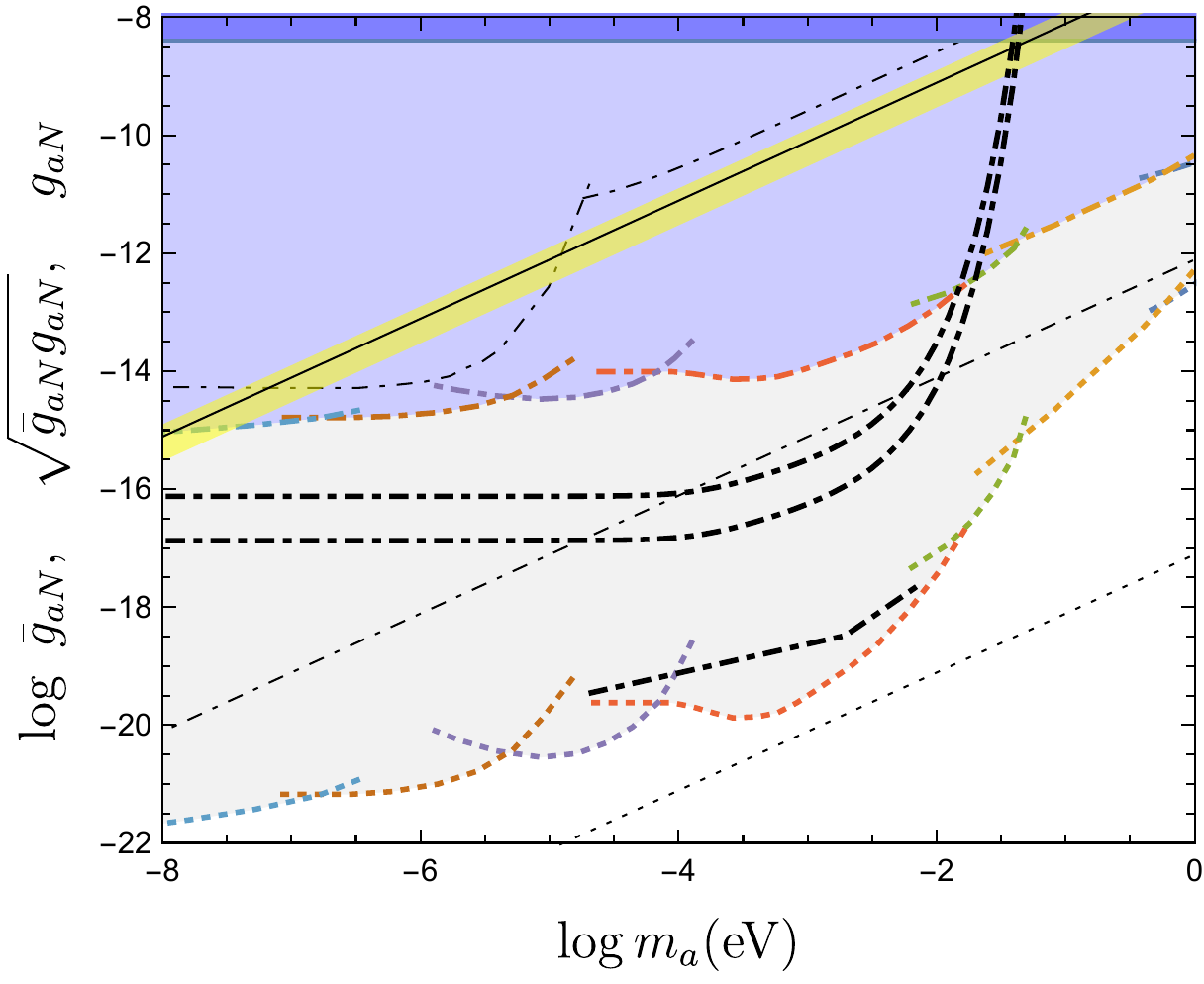} \includegraphics[width=2.8in]{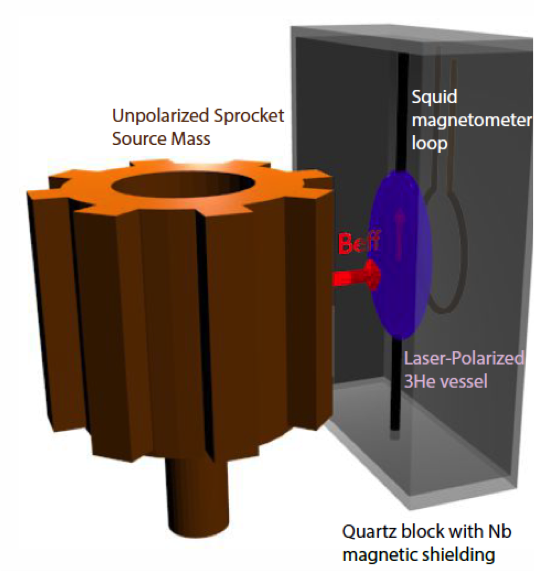}
   \caption{Left: Limits on ALP nucleon couplings: CP violating $\bar g_{aN}$ from laboratory 5th-force searches (dotted), CP conserving $g_{aN}$ from the SN1987A cooling argument (solid) 
   and their product $ \bar g_{aN}  g_{aN}$ (dot-dashed) from pure lab experiments (thin)
   and the direct product on the astro+lab constraints from~\cite{Raffelt:2012sp}.
   The yellow band represents the range for the \DFSZ QCD axion, $g^{\rm \DFSZ}_{\A p}$, and the line for \KSVZ, $g^{\rm \KSVZ}_{\A p}$, while dotdashed and dotted lines are approximate \emph{upper bounds} for the CP-violating axion couplings using $\theta_0=10^{-10}$. 
   The ARIADNE prospects for $\bar g_{aN}  g_{a n}$ are shown as black dot-dashed curves.
   Right: Sketch of the ARIADNE experiment. Credit: ARIADNE collaboration, used with permission. }
   \label{fig:5thNariadne}
\end{figure}

Axion-like particles can mediate monopole-monopole forces between baryons that compete with gravity at distances $\sim 1/m_a$ and have been constrained by precision measurements of Newton's $1/r^2$ law~\cite{Hoskins:1985tn,Kapner:2006si,Decca:2007jq,Geraci:2008hb,Sushkov:2011zz} and searches of violations of equivalence principle~\cite{Smith:1999cr,Schlamminger:2007ht}. The most sensitive technique depends on the range: micromachined oscillators for Casimir force measurements~\cite{Decca:2007jq}, torsion penduli~\cite{Sushkov:2011zz}, micro-cantilevers~\cite{Geraci:2008hb} and torsion-balance experiments~\cite{Kapner:2006si,Hoskins:1985tn,Smith:1999cr,Schlamminger:2007ht} in order of increasing range.
The difficulties in handling large quantities of polarised atoms have rendered monopole-dipole and dipole-dipole interactions traditionally much less sensitive. For instance, the combinations $\bar g_{aN}g_{ae}$, $\bar g_{aN}g_{aN}$ ($N=p,n$) are much better constrained by combining limits on monopole-monopole interactions, $\propto{\bar g_{aN}}^2$, and astrophysical bounds on $g_{ae}$ and $g_{aN}$~\cite{Raffelt:2012sp}. The direct constraints on dipole-dipole interactions are much weaker than astrophysical limits, both in the case of nucleons~\cite{Vasilakis:2008yn} and electrons~\cite{Heckel:2013ina,Terrano:2015sna}. The situation is depicted in Fig.~\ref{fig:5thNariadne}, which displays the 
upper bounds on $\bar g_{aN}$ (dotted) from gravity tests, $g_{aN}$ (solid) from astrophysics and the product $\sqrt{\bar g_{aN} g_{aN}}$ (dot-dashed). In the latter case, we show the constraints from direct laboratory tests (black) and the mentioned combination of astro and monopole-monopole laboratory experiments (colored), always as dot-dashed lines. 
For comparison we have shown the \DFSZ and \KSVZ prediction for the axion-proton coupling. 
The CP-conserving coupling $g_{\A p}$ of the \DFSZ is shown as a yellow band and of \KSVZ as a solid black line. 
The CP-violating counterpart $\bar g_{\A p}$ is not known, but has a contribution of order $\sim g_{\A p}\theta_0$, 
where $\theta_0$ is the minimum of the axionic QCD potential including CP-violation effects. 
The upper limit on $\theta_0\lesssim 10^{-10}$ from \eqref{thetabound} gives the upper bound to this contribution shown as a black dotted diagonal line, although the SM expectation is much smaller, see \eqref{CPviolatingproto}. 
The dot-dashed version is the similar upper bound for $\sqrt{g_{\A p}\bar g_{\A p}}$. 

This was indeed the situation until the recent
proposal~\cite{Arvanitaki:2014dfa} of detecting the axion field sourced by a macroscopic object by NMR techniques instead of measuring the force exerted to other body. 
Indeed, the interaction energy~\eqref{Uinteraction} of a fermion in the macroscopic ALP field 
resembles a magnetic dipole interaction with an ``equivalent'' magnetic field,
\be
\label{Bficticious}
\fB= -  \frac{g_{a f}}{m_f \gamma_f}\vec \nabla a ,
\ee
with $\gamma_f$ the gyromagnetic ratio. If we arrange a slow time-variation of the ALP field, the gradient can be detected with precision magnetometry just like any other tiny oscillating $B$-field, with the advantage that the axion field will traverse any shielding.
The ARIADNE collaboration is developing this concept in Reno U. \cite{Geraci:2017bmq}.
A sketch of the setup can be seen in Fig.~\ref{fig:5thNariadne}.
A sprocket-shaped source mass creates the $a$ field. The $\fB$ field points radially and is more intense near the ``teeth''. The detector is a laser-polarised $^3$He sample in a spheroidal quartz vessel  located inside a superconducting Nb magnetic shielding. It is optimised to lie as close as possible to the teeth of the rotating mass.
The mass rotates such that the teeth glide on the sample at the precession frequency and the transverse magnetisation is read out using a SQUID magnetometer. The proposed setup considers a $\sim$cm Tungsten cylinder  and 3\,mm\,$\times$\,3\,mm\,$\times$\,0.15\,mm $^3$He vessel shielded by a 25\,$\mu$m Nb foil screen and is limited by transverse projection noise in the sample~\cite{Arvanitaki:2014dfa}. A data integration time of $10^6$ s gives the sensitivities shown in Fig.~\ref{fig:5thNariadne} as dot-dashed black lines. The upper and lower curves consider transverse relaxation times of $T_2=1$ and 1000 s for the 3He sample, respectively. The bottom line made of two straight segments is the projected sensitivity of a scaled up version with a larger 3He cell reaching liquid density, see~\cite{Arvanitaki:2014dfa}.
Using a spin-polarised source like Xe or Fe, the scaled setup is potentially sensitive to dipole-dipole interactions.
The magnetic shielding strategy has been outlined in~\cite{Fosbinder-Elkins:2017osp} and recent progress reported in~\cite{Geraci:2017bmq}. The collaboration plans to start full construction in Summer 2018, and commissioning will occur in 2019-2020 with initial data expected by early 2021, together with  R\&D for the upgraded version.

\begin{figure}[t!] 
   \centering
	\raisebox{-0.5\height}{\includegraphics[width=4.in]{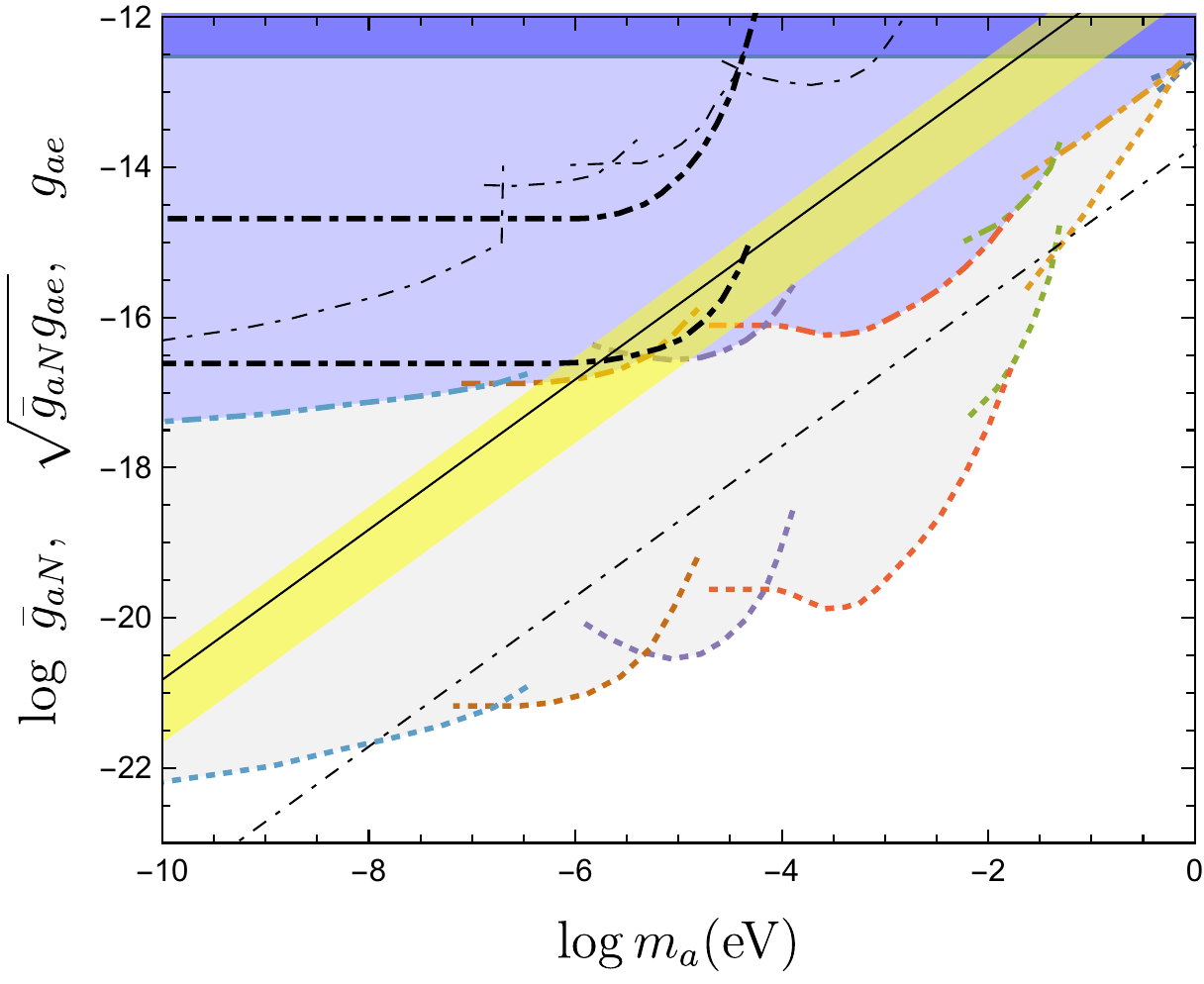}\hspace{5mm}}
   \raisebox{-0.5\height}{\includegraphics[width=1.8in]{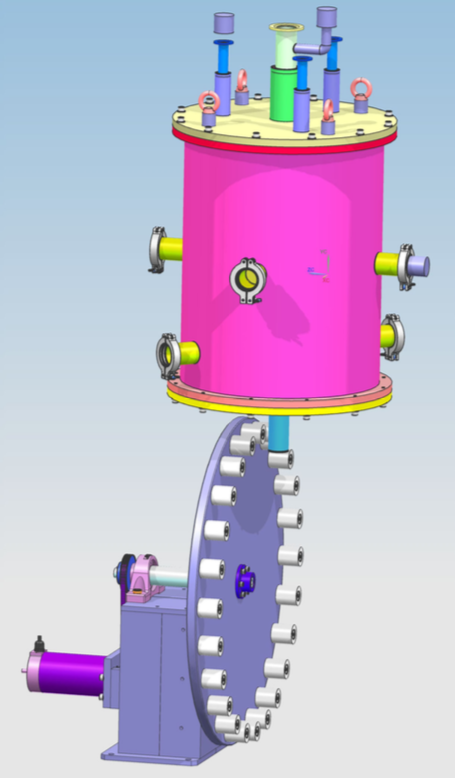}}
   \caption{Left: Limits on ALP nucleon/electron couplings: CP violating $\bar g_{aN}$ from laboratory 5th-force searches (dotted), CP conserving $g_{ae}$ from stellar cooling~\cite{Giannotti:2017hny} (solid) and
   their product $ \bar g_{aN}  g_{ae}$ (dot-dashed) from pure lab experiments (thin)
   and the direct product on the astro+lab constraints. Adapted from~\cite{Raffelt:2012sp}.
   The yellow band represents the range for the electron coupling of the QCD axion in \DFSZ I, $g^{\rm \DFSZ}_{\A p}$, and the line to a typical value with $C_{\A e}=1/6$.
  The dotdashed diagonal line is an approximate \emph{upper bound} for the QCD axion combination $\bar g_{\A N} g_{\A e}$. Black dot-dashed lines correspond to the QUAX-$g_sg_p$ limit~\cite{Crescini:2017uxs} and prospects~\cite{Crescini:2016lwj} on the same combination $\bar g_{aN} g_{ae}$. Right: Sketch of the QUAX-$g_sg_p$ experiment, taken from~\cite{Crescini:2016lwj} with permission. }
   \label{fig:5thNeQUAX}
\end{figure}

ARIADNE can improve the sensitivity of previous searches by $\sim 2$ orders of magnitude in the coupling before using the scaled up version. Moreover, it will reach the sensitivity QCD axions under the assumptions that 
\begin{itemize}
\item the CP-violating coupling to Tungsten $\bar g_{\A ^{74}W}\sim 74(\bar g_{\A p}+\bar g_{\A e})+110 g_{\A n}$ doesn't have fine cancellations between the constituent components,
\item physics beyond the SM produce a sufficiently large CP violating coupling for at least one of the constituents $\bar g_{\A f}$, for instance by making $\theta_0$ close to the experimentally excluded value $\sim 1.3\times 10^{-10}$, 
\item and the axion CP conserving axial coupling to neutrons is of natural size $C_{\A n}\sim {\cal O}(1)$, 
and not severely suppressed as in \KSVZ.
 \end{itemize}
The last reason stems from the fact that, in its current version, the experiment is only sensitive to the CP-conserving coupling to neutrons, $g_{an}$, in the detection part because the protons and electrons in 3He are paired so $g_{\A{^3}\rm He}=g_{\A n}$, recall sec. \ref{sec:axionmacro}. 
This could be unfortunate, for \DFSZ and \KSVZ-type axions have sensibly smaller couplings to neutrons than to protons (the coupling could even vanish within the uncertainties). Further research is encouraged to develop a detection scheme sensitive to $g_{ap}$.

\todonote{The CP-violating photon coupling generates radiatively a $\bar g_{ap}$ that is more constrained 
that astrophysics!~\cite{Dupays:2006dp}}

The QUAX collaboration has recently proposed a novel scheme to search for the monopole-dipole force coupled to \emph{electron}-spins $\propto \bar g_{aN} g_{a e}$ by detecting the magnetisation induced by $\fB$ in a paramagnetic material~\cite{Crescini:2016lwj}. The idea is conceptually similar to ARIADNE, see Fig.~\ref{fig:5thNeQUAX} (right).
The lead teeth of a rotating wheel are made to pass near a cryostat containing a 1 cm$^3$ sample of GSO (Gd$_2$SiO$_5$). The effective $\fB$ created by the axion field gradient (generated by the lead) induces a small magnetisation in the GSO that is modulated at the frequency $n_{\rm teeth}\omega_{\rm rotation}$ and read by a SQUID. The signal can be amplified by a RLC circuit tuned at the signal frequency.
First results of the so-called QUAX-$g_sg_p$ experiment without the boost of the RLC have been presented in~\cite{Crescini:2017uxs}, which already improve over previous experiments~\cite{Hammond:2007jm,Hoedl:2011zz} (shown as thin black dot-dashed lines in Fig.~\ref{fig:5thNeQUAX} (left)).
The prospects presented in~\cite{Crescini:2016lwj} show that the setup will be able to reach a sensitivity $\propto \bar g_{aN} g_{a e}\lesssim 10^{-33}$ for sub $\mu$eV masses, approaching Raffelt's constraint from the combination of laboratory $\bar g_{aN}$ and astrophysical constraints on $g_{ae}$, shown as coloured dot-dashed lines in Fig.~\ref{fig:5thNeQUAX} (left). 
However, this is still a factor of $10^4$ or more away from QCD axion sensitivity\footnote{Somehow, the QCD predictions in Fig.~6 of~\cite{Crescini:2016lwj} are upscaled by a factor of $\sim 10^4$.}.

\section{Detection of solar axions}
\label{sec:solar}

\begin{figure}[t]
\centering
\includegraphics[width=8cm]{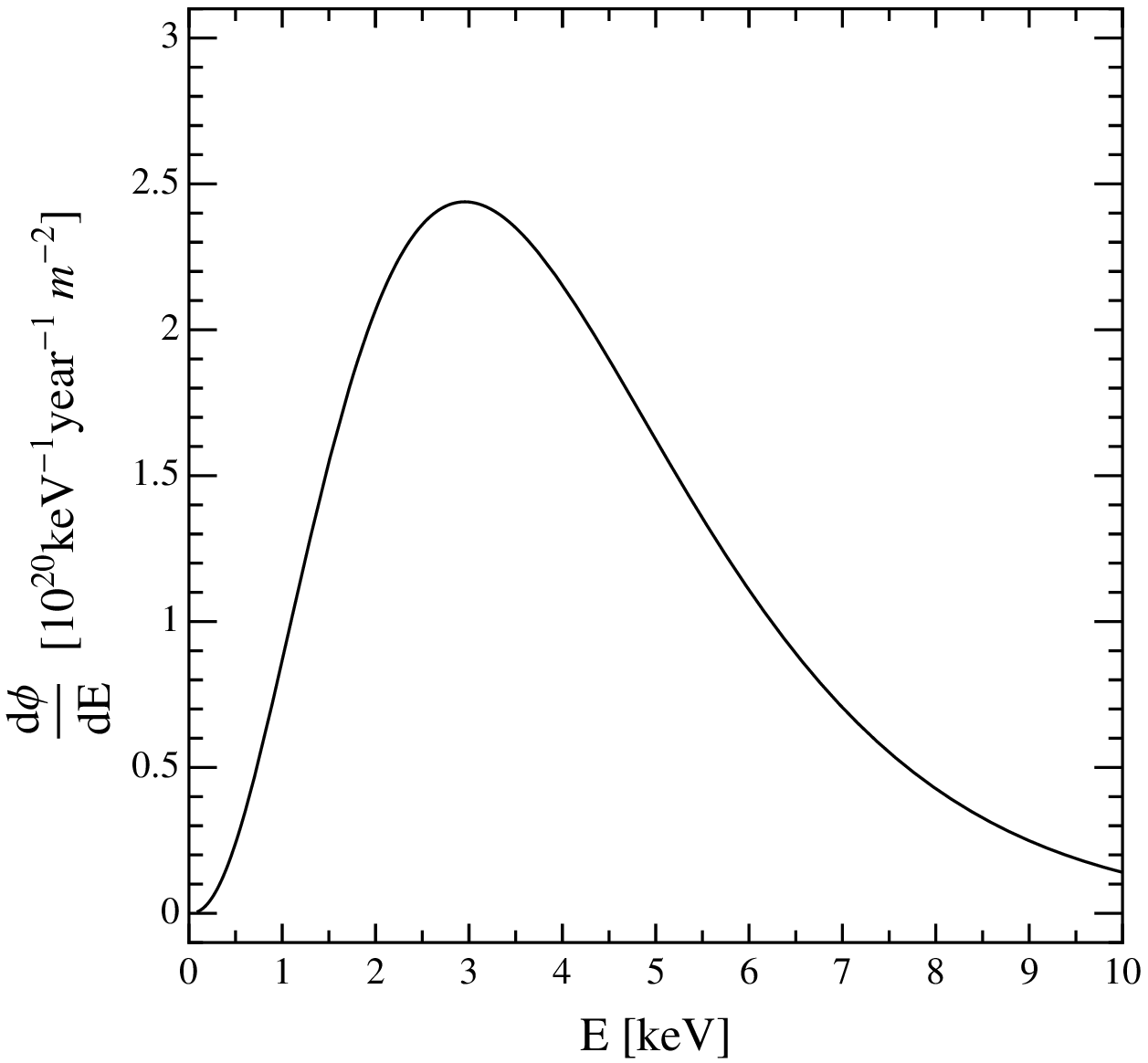}
\includegraphics[width=8cm]{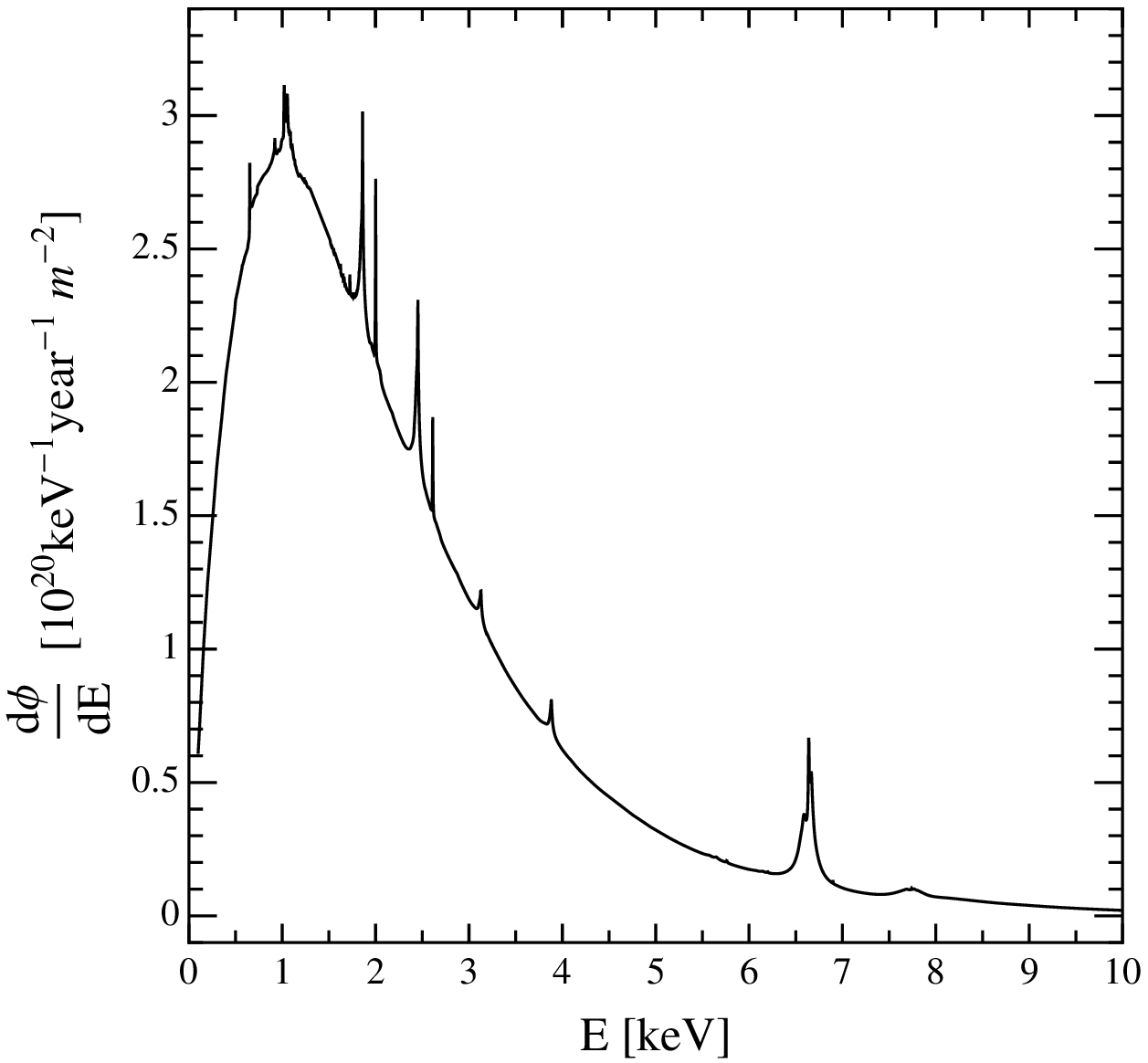}
\caption{\label{fig:axion_flux}Solar axion flux spectra at Earth by different production mechanisms. On the left, the most generic situation in which only the Primakoff conversion of plasma photons into axions is assumed. On the right the spectrum originating from processes involving electrons, bremsstrahlung, Compton and axio-recombination~\cite{Redondo:2013wwa,Barth:2013sma}. The illustrative values of the coupling constants chosen are $\gagamma = 10^{-12}$~GeV$^{-1}$ and $g_{ae} = 10^{-13}$. Plots from~\cite{Irastorza:1567109}.}
\end{figure}

ALPs can be produced in the solar interior by a number of reactions. The most relevant channel is the Primakoff conversion of plasma photons into axions in the Coulomb field of charged particles. If the ALPs couple with electrons, they can also  be produced via the ABC mechanisms discussed in section~\ref{sec:sources}. Both the Primakoff and ABC differential fluxes are shown in Fig.~\ref{fig:axion_flux}. The former peaks at 4.2 keV and exponentially decreases for higher energies. The latter results in slightly less energetic ALPs, with a maximum at $\sim$1 keV. 

The most relevant technique to search for solar axions is the axion helioscope~\cite{Sikivie:1983ip}. Axion helioscopes invoke the conversion of the axions into photons (X-rays) in strong laboratory magnets, therefore the detection relies on the $\gagamma$ coupling. The usual procedure in helioscopes considers only the Primakoff component because it maintains the broadest generality and produces relevant limits on $\gagamma$ over large mass ranges. The signals of non-Primakoff axions in helioscopes depend of the corresponding product of couplings, and typically do not compete with astrophysical limits. This however may no longer be true in the future, as projected experimental sensitivities (i.e. IAXO) will supersede astrophysical limits on $g_{ae}$, opening the possibility to probe an interesting set of non-hadronic axion models. We will review in the following the past and current efforts in the helioscope technique, while later on in \ref{sec:othersolar} we briefly review other solar axion detection techniques. We finish this section commenting the possibility of detecting axion produced by Supernovae.

\subsection{Axion helioscopes}

\label{sec:helioscopes}

By means again of the $a\gamma\gamma$ vertex, solar axions can be efficiently converted back into photons in the presence of an electromagnetic field. 
If the background field is static, the energy of the reconverted photon is equal to the incoming axion, so a flux of detectable X-rays with energies of a few keV is expected. We have already calculated the probability of axion-photon conversion in a transverse magnetic field $B_e$ over a length $L$ in \eqref{oscillationprob}, \eqref{oscillationprob2}. A reference formula for helioscopes is~ 
\cite{Sikivie:1983ip,Zioutas:2004hi,Andriamonje:2007ew}:
\be
\label{conversion_prob}
  \proba(a\to \gamma) = 2.6 \times 10^{-17} \(\frac{\gagamma}{10^{-10}\rm GeV^{-1}}\)^2 \left(\frac{B_e}{10 \mathrm{\ T}}\right)^2
  \left(\frac{L}{10 \mathrm{\ m}}\right)^2 \nonumber  \mathcal{F}(qL), 
\ee
with the homogeneous $B_e$-field form factor, ${\cal F}(qL)$, given by \eqref{eqn:lsw_F}. 
\exclude{
\begin{equation}\label{matrix_element}
    \mathcal{F}=\frac{2(1-\cos q L)}{(qL)^2}
\end{equation}
\noindent and $q$ is the momentum transfer. 
The fact that the axion is not massless, puts the axion and photon waves out of phase
after a certain length. }

Coherent conversion along the whole length gives ${\cal F} =  1$ and happens when $q L \ll 1$. 
In vacuum and for relativistic ALPs, the difference of photon and ALP wavenumbers is 
$q=k_\gamma-k_a\simeq m_a^2/2\omega$ to a good approximation. Then, the coherence 
condition for solar axion energies and a magnet length of $\sim$10 m
is satisfied for axion masses $m_a \lesssim 10^{-2}$~eV. For higher masses, $\mathcal{F}$
decreases as $(2/qL)^2\propto 1/m_a^4$, and so does the sensitivity of the experiment. 
To mitigate the loss of coherence, a buffer gas can be introduced into the magnet
beam pipes \cite{vanBibber:1988ge,Arik:2008mq} to impart an effective
mass to the photons $m_\gamma = \omega_{\rm p}$
(where $\omega_{\rm p}$ is the plasma frequency of the gas,
 $\omega_{\rm p}^2=4\pi\alpha n_e/m_e$, being $n_e$ and $m_e$ the electron density and the electron mass respectively).
If the axion mass matches the photon mass, $q=0$ and the
coherence is restored. By changing the pressure of the gas inside
the pipe in a controlled manner, the photon mass can be
systematically increased and the sensitivity of the experiment can
be extended to higher axion masses. 
In this configuration, in the event of a positive detection, helioscopes can determine the value of $m_a$. 
Even in vacuum, $m_a$ can be determined from the spectral distortion produced by the onset of ALP-photon oscillation in the helioscope of the low energy part of the spectrum, something that can be detectable for masses down to 10$^{-3}$~eV, depending of the intensity of the signal~\cite{mass_det_paper}.

\begin{figure}[t] \centering
\includegraphics[width=\textwidth]{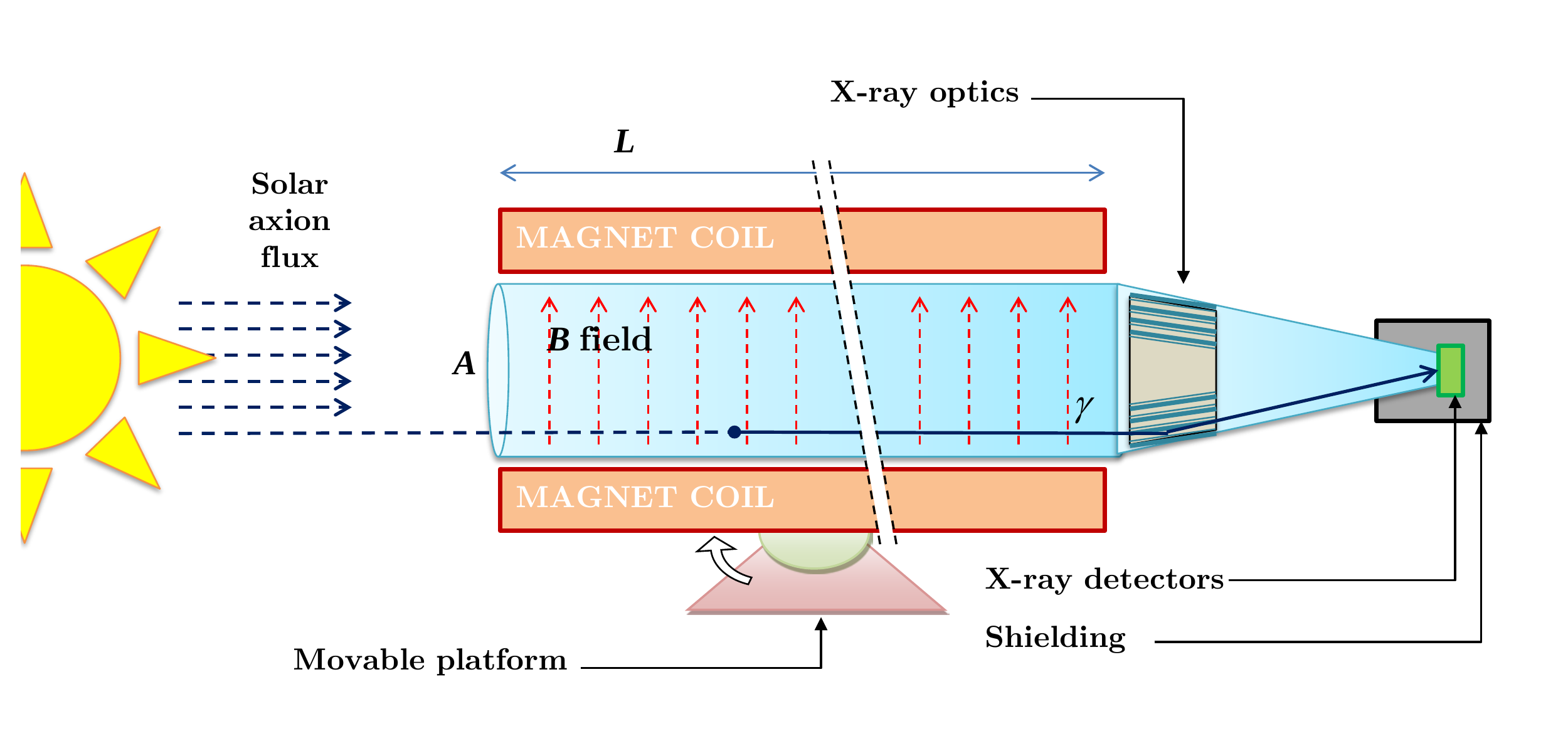}\hspace{2pc}%
\caption{\label{fig:helioscope_sketch} Conceptual arrangement
of an enhanced axion helioscope with X-ray focusing. Solar axions are converted into photons by the transverse magnetic field inside the bore of a powerful magnet. The resulting quasi-parallel beam of photons of cross sectional area $\area$ is concentrated by an appropriate X-ray optics onto a small spot area $a$ in a low background detector. Figure taken from~\cite{Irastorza:2011gs}.}
\end{figure}

The basic layout of an axion helioscope thus requires a powerful magnet
coupled to one or more X-ray detectors. In modern incarnations of the concept, as shown in figure~\ref{fig:helioscope_sketch}, an additional focusing stage is added at the end of the magnet to concentrate the signal photons and increase signal-to-noise ratio. When the magnet is aligned
with the Sun, an excess of X-rays at the detector is
expected, over the background measured at non-alignment periods. This detection concept was first experimentally realised
at Brookhaven National Laboratory (BNL) in 1992. A stationary dipole magnet with a field of $B = 2.2$~T and a length of $L = 1.8$~m was oriented towards the setting Sun~\cite{Lazarus:1992ry}. The experiment derived the upper limit $\gagamma < 3.6\times10^{-9}$ GeV$^{-1}$ for $m_a < 0.03$ eV at 99\% C.L. At the University of Tokyo, a second-generation experiment was built: the SUMICO axion heliscope. Not only did this experiment implement a dynamic tracking of the Sun but it also used a more powerful magnet ($B =4$~T, $L = 2.3$~m) than the BNL predecessor. The bore, located between the two coils of the magnet, was evacuated and higher-performance detectors were installed~\cite{Inoue:2002qy,Moriyama:1998kd,Inoue:2008zp}. This new setup resulted in an improved upper limit in the mass range up to 0.03 eV given by $\gagamma < 6.0\times 10^{-10}$ GeV$^{-1}$ (95\% C.L.). Later experimental improvements included the additional use of a buffer gas to enhance sensitivity to higher-mass axions.

\begin{figure}[t] \centering
\includegraphics[width=0.9\textwidth, trim={0 6cm 0 6cm}, clip]{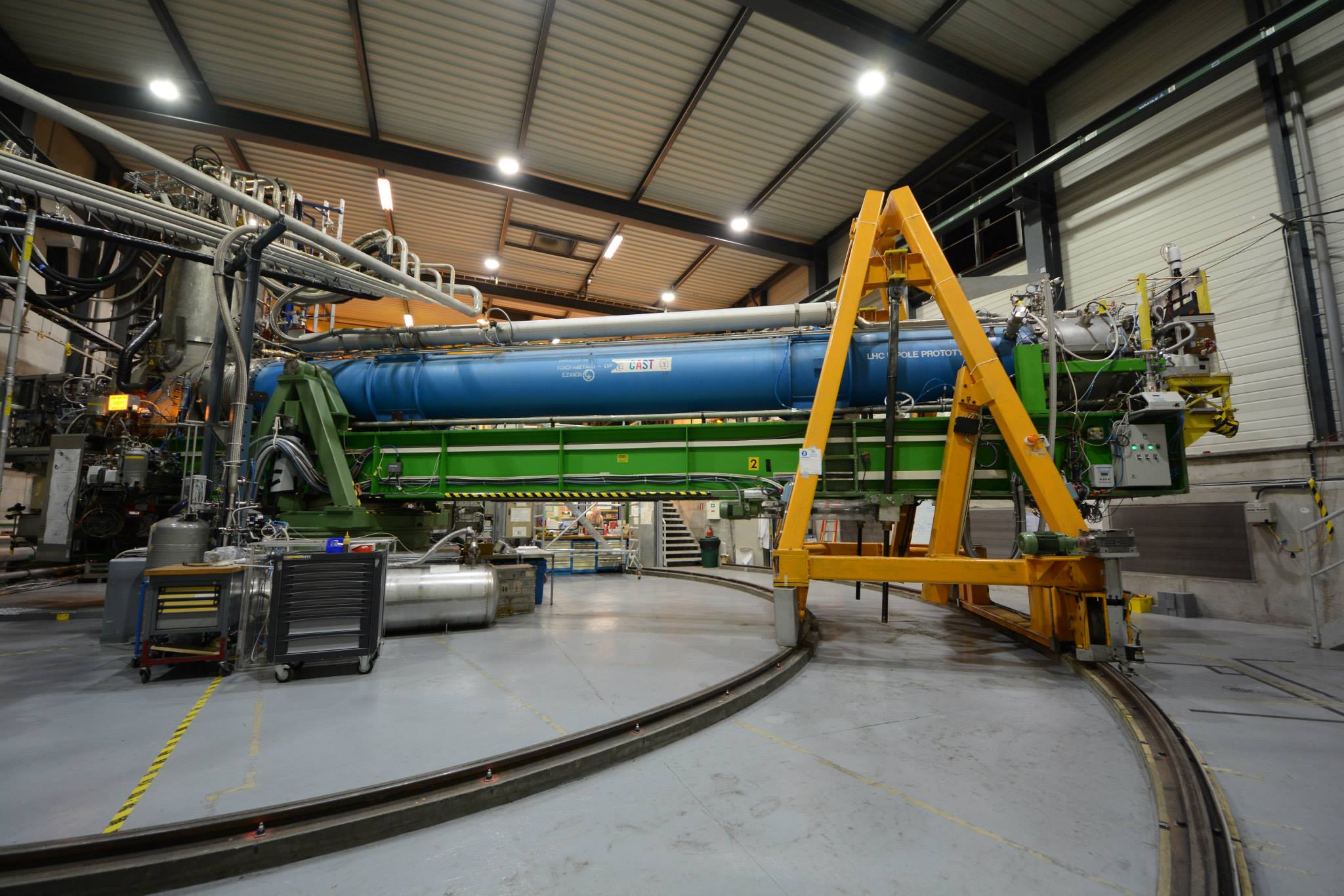}\hspace{2pc}%
\caption{\label{fig:cast} Picture of the CAST experiment at CERN. Credit: M. Rosu/CAST collaboration, CERN}
\end{figure}

A third-generation experiment, the CERN Axion Solar Telescope (CAST), began data collection in 2003. The experiment uses a LHC dipole prototype magnet with a magnetic field of up to 9 T over a length of 9.3 m~\cite{Zioutas:1998cc}. The magnet is able to track the Sun for several hours per day using a elevation and azimuth drive (see Fig.~\ref{fig:cast}). This CERN experiment has been the first helioscope to employ X-ray focusing optics for one of its four detector lines~\cite{Kuster:2007ue}, as well as low background techniques from detectors in underground laboratories~\cite{Abbon:2007ug}. During its observational program from 2003 to 2011, CAST operated first with the magnet bores in vacuum (2003--2004) to probe masses $m_a < 0.02$~eV, obtaining a first upper limit on the axion-to-photon coupling of $\gagamma< 8.8\times 10^{-11}$ GeV$^{-1}$ ~(95\%~C.L.) ~\cite{Zioutas:2004hi,Andriamonje:2007ew}. The experiment was then upgraded to be operated with $^4$He (2005--2006) and $^3$He gas (2008--2011) to obtain continuous, high sensitivity up to an axion mass of $m_a = 1.17$ eV. Data from this gas phase provide an average limit of $\gagamma \lesssim 2.3\times10^{-10}$ GeV$^{-1}$ (95\%~C.L.), for the higher mass range of 0.02 eV $< m_a <$ 0.64 eV~\cite{Arik:2008mq,Aune:2011rx} and of about $\gagamma \lesssim 3.3\times10^{-10}$ GeV$^{-1}$ (95\%~C.L.) for 0.64 eV $< m_a <$ 1.17 eV~\cite{Arik:2013nya}, with the exact value depending on the pressure setting.

CAST has more recently (2013-15) revisited the vacuum phase with improved detectors and a novel X-ray optics. These improvements are the outcome of R\&D done in preparation of the next generation axion helioscope, IAXO. In particular, one of the detection lines, dubbed IAXO pathfinder system~\cite{Aznar:2015iia}, combines for the first time both low background techniques and a new X-ray optics built purposely for this goal, and enjoys an effective background count rate of 0.003 counts per hour in the signal region. The outcome of this phase represents the most restrictive experimental limit to $\gagamma$ for masses $m_a < 0.02$~eV~\cite{Anastassopoulos:2017ftl}:
 \begin{equation}
 \gagamma  < 0.66\times 10^{-10} \mathrm{GeV}^{-1}\,  (95\%~{\rm C.L.} ).
   \end{equation}

\begin{figure}[t]
\begin{center}
\tikzsetnextfilename{Helioscopes}
\resizebox{0.9\linewidth}{!}{\input{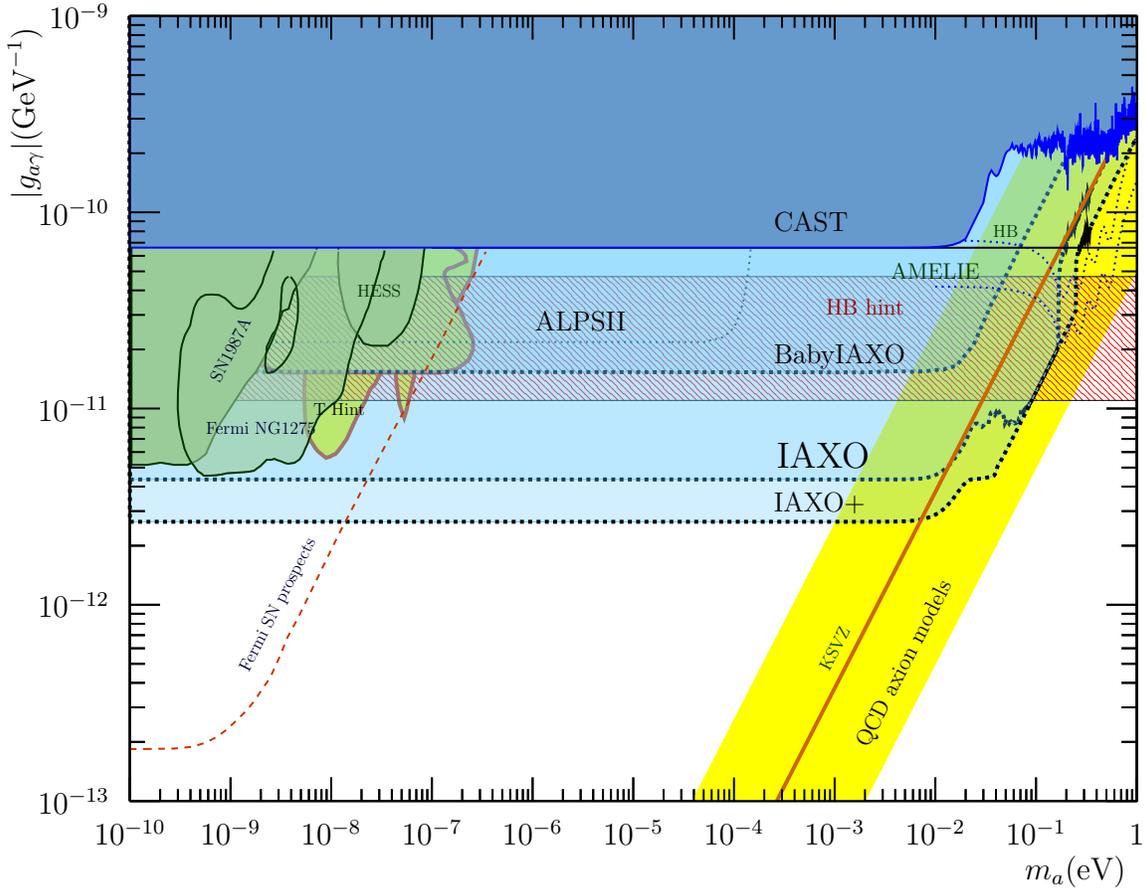}}
\caption{Latest excluded region in the ($\gagamma$, $m_a$) from helioscopes (CAST), as well as prospects from future experiments, most relevantly from BabyIAXO and IAXO. Also shown are the prospects at high mass from a possible implementation of the AMELIE concept. Also shown are prospects from the ALP-II experiment and hinted regions from astrophysics as explained in the text.}
\label{fig:helio_sens}
\end{center}
\end{figure}

CAST has been the first axion helioscope with sensitivities to $\gagamma$ values below $10^{-10}$ GeV$^{-1}$ and competing with the most stringent limits from astrophysics on this coupling, see Tab.~\ref{tab:axtrobounds}. As shown in Fig.~\ref{fig:helio_sens}, in the region of higher axion masses ($m_a \gtrsim 0.1$ eV), the experiment has entered the band of QCD axion models and excluded \KSVZ axions of for specific values of the axion mass in the range $m_a\sim$ eV.

In addition to this main result, CAST has also searched for other axion production channels in the Sun, enabled by the axion-electron or the axion-nucleon couplings. As mentioned above, in these cases helioscopes provide limits to the product of $\gagamma$ and the corresponding coupling. More specifically, CAST has provided results on the search for :
\begin{itemize}
\item 14.4 keV solar axions emitted in the M1 transition of Fe-57 nuclei~\cite{Andriamonje:2009dx}, 
\item MeV axions from $^7$Li and D($p,\gamma$)$^3$He nuclear transitions~\cite{Andriamonje:2009ar}, 
\item keV axions from the ABC processes involving the axion-electron coupling~\cite{Barth:2013sma}, 
\item more exotic ALP or WISP models, like chameleons~\cite{Brax:2011wp,Baum:2014rka}
\end{itemize}

So far each subsequent generation of axion helioscopes has resulted in an improvement in sensitivity to $\gagamma$ of about a factor of a few over its predecessor (see table~\ref{tab:helioscopes}). All helioscopes so far have largely relied on reusing existing equipment, especially the magnet. CAST in particular has enjoyed the availability of the first-class LHC test magnet. Going substantially beyond CAST sensitivity appears to be possible by designing a dedicated magnet, optimised to maximise the helioscope magnet's figure of merit $f_M  = B^2\:L^2\:\area$, where $B$, $L$ and $\area$ are the magnet's field strength, length and cross sectional aperture area, respectively. $f_M$ is defined proportional to the photon signal from converted axions. Improving the value of $f_M$ obtained by CAST can only be achieved~\cite{Irastorza:2011gs} by a completely different magnet configuration with a much larger magnet aperture $\area$, which in the case of the CAST magnet is only $3\times10^{-3}$~m$^2$. However, for this figure of merit to directly translate into a signal-to-noise ratio of the overall experiment, the entire cross sectional area of the magnet must be equipped with X-ray focusing optics. The layout of this \textit{enhanced axion helioscope}, sketched in Figure \ref{fig:helioscope_sketch}, was proposed in~\cite{Irastorza:2011gs} as the basis for IAXO, the International Axion Observatory.

\begin{table}[b]
  \centering
\begin{tabular}{cccccccc}
  \hline
  Experiment & references & status & $B$ (T) & $L$ (m) & $\area$ (cm$^2$) & focusing & $g_{10}$ \\ \hline
  Brookhaven & \cite{Lazarus:1992ry} & past & 2.2 & 1.8 & 130 & no & 36 \\
  SUMICO & \cite{Moriyama:1998kd,Inoue:2008zp} & past & 4 & 2.5 & 18 & no & 6 \\
  CAST & \cite{Zioutas:1998cc,Zioutas:2004hi,Arik:2008mq,Aune:2011rx,Arik:2013nya} & ongoing & 9 & 9.3 & 30 & partially & 0.66 \\
  TASTE & \cite{Anastassopoulos:2017kag} & concept & 3.5 & 12 & 2.8$\times10^3$ & yes & 0.2 \\
  BabyIAXO & \cite{babyiaxo} & in design & $\sim$2.5 & 10 & 2.8$\times10^3$ & yes & 0.15 \\
  IAXO & \cite{Irastorza:2011gs,Armengaud:2014gea} & in design & $\sim$2.5 & 22 & 2.3$\times10^4$ & yes & 0.04 \\ \hline
\end{tabular}
  \caption{List of past and future helioscopes with some key features. The last column represents the sensitivity achieved (or expected) in terms of an upper limit on $g_{10}=\gagamma \times 10^{10}$~GeV for low $m_a$. The numbers for the TASTE, BabyIAXO and IAXO helioscopes correspond to the design parameters considered in the quoted references.}\label{tab:helioscopes}
\end{table}

\hyphenation{tech-no-lo-gy}

\begin{figure}[!t]
\begin{center}
\raisebox{-0.5\height}{\includegraphics[width=0.54\textwidth] {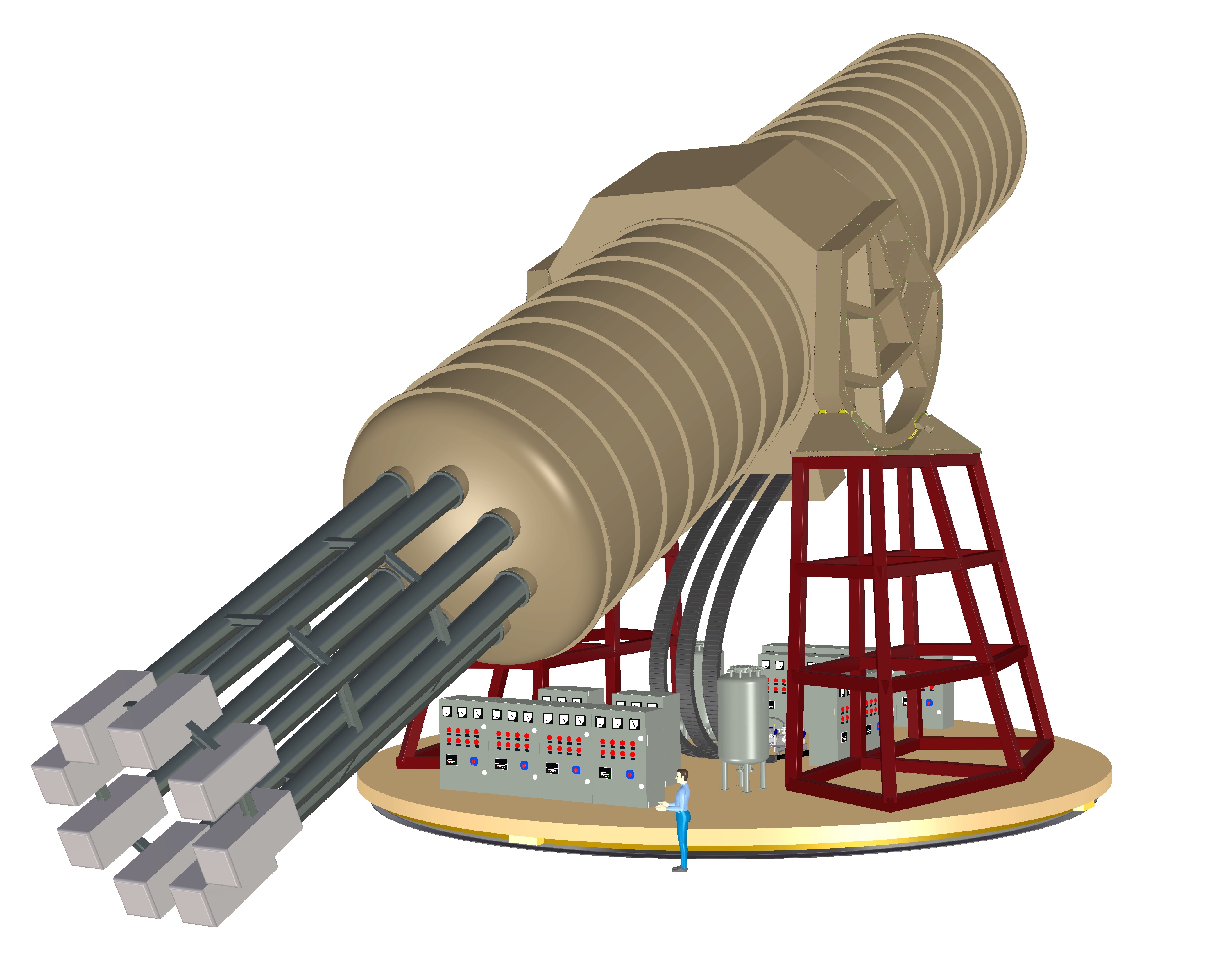} }
    \raisebox{-0.5\height}{\includegraphics[width=0.44\textwidth] {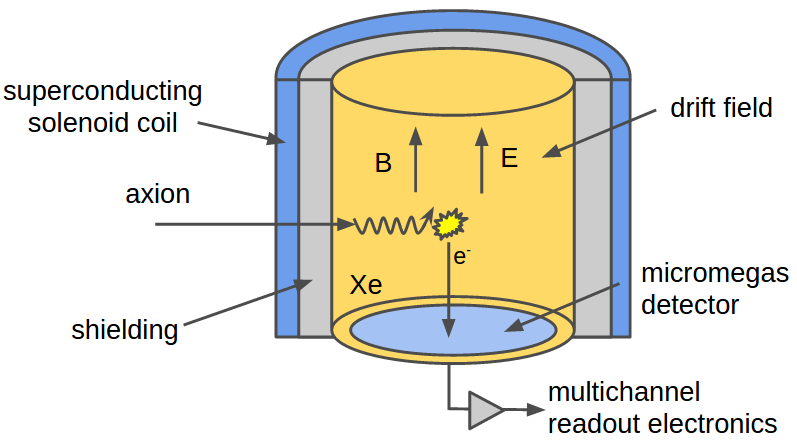}}
    \caption{Left: schematic view of IAXO. Shown are the cryostat, eight telescopes, the flexible lines guiding services into the magnet, cryogenics and powering services units, inclination system and the rotating platform for horizontal movement. The dimensions of the system can be appreciated by a comparison to the human figure positioned by the rotating table~\cite{Armengaud:2014gea}. Right: Possible implementaion of the AMELIE concept described in secion~\ref{amelie} in a cylindrical gas TPC read out by a Micromegas plane. Figure taken from~\cite{Galan:2015msa} with permission.}
    \label{fig:IAXO_cdr}
\end{center}
\end{figure}

IAXO is the next generation axion helioscope, currently at design stage. It builds upon the experience of CAST, and aims at building a new large-scale magnet optimised for an axion search and extensively implementing focusing and low background techniques. Thus the central component of IAXO is a new superconducting magnet that, contrary to previous helioscopes, will follow a toroidal multibore configuration~\cite{Shilon:2012te}, to efficiently produce an intense magnetic field over a large volume. Current design considers a 25 m long and 5.2 m diameter toroid assembled from 8 coils, and effectively generating and average (peak) 2.5 (5.1) Tesla in 8 bores of 600 mm diameter. This represents a 300 times better $f_M$ than the CAST magnet. The toroid's stored energy is 500 MJ. The design is inspired by the ATLAS barrel and end-cap toroids\cite{tenKate:1158687,tenKate:1169275}, the largest superconducting toroids built and presently in operation at CERN. Beyond the magnet, several improvements are foreseen also in the optics and detector parameters. Each of the 8 magnet bores of IAXO will be equipped with an X-ray telescope of 60 cm diameter. The collaboration envisions to build optimised optics based on thermally-formed glass substrates, similar to those used on NASA's NuSTAR~\cite{nustar2013}. The number and position of the substrates, as well as their coating, will be carefully designed to optimise reflectivity in the energies corresponding to the solar axion spectrum~\cite{doi:10.1117/12.2024476}. At the focal plane in each of the optics, there will be low background pixelated detectors able to image the focused signal, built with radiopure components and properly shielded. The baseline technology for those detectors are small gaseous chambers read by pixelised planes of microbulk Micromegas, already developed and used in CAST. The latest generation of those detectors have achieved record background levels below $\sim 10^{-6}$~\ckcs~\cite{Aune:2013pna,Aune:2013nza}, and prospects for reducing this level to $10^{-7}$ \ckcs\ or even lower appear feasible~\cite{Irastorza:2015geo}. Additional detection technologies are under consideration, like Ingrid Micromegas, Metallic Magnetic Calorimeters, Transition Edge Sensors or Silicon detectors, that could complement the former by featuring lower energy thresholds or better energy resolutions, appealing for other observations, like  the detection of the ABC solar axions emitted through the $g_{ae}$-coupling. Figure~\ref{fig:IAXO_cdr} shows the conceptual design of the overall infrastructure~\cite{Armengaud:2014gea}.

IAXO is expected to improve the CAST signal-to-noise by more than a factor 10$^4$, corresponding to more than one order of magnitude in $\gagamma$. IAXO will also feature a buffer gas phase to enhance its sensitivity at high masses. Figure~\ref{fig:helio_sens} shows the current sensitivity prospects in the ALP parameter space. At low masses, IAXO will entirely cover the region of the ALP parameter space invoked in the anomalous transparency of the Universe to UHE photons. In the $g_{ae}$ channel, IAXO will be sensitive to non-hadronic axion models that provide satisfactory fits to the stellar cooling anomaly~\cite{Giannotti:2017hny}. As a first step towards IAXO, the collaboration aims at building a scaled-down version of the setup, called BabyIAXO. The BabyIAXO magnet will be only 10 m long and will feature only one bore of 60 cm diameter. Therefore BabyIAXO will be equipped with only one optics and detector, but of similar dimensions to the final IAXO system. BabyIAXO will therefore constitute a representative prototype for the final infrastructure. But it will also provide relevant physics outcome at an intermediate level between current best CAST limit and the full IAXO prospects. Figure~\ref{fig:helio_sens} shows also current sensitivity projections for BabyIAXO. The design and operational experience with BabyIAXO, in particular with the magnet, is expected to provide relevant feedback for the technical design of the full infrastructure and enable improvements in the final figure of merit. A possible enhanced sensitivity projection for and improved IAXO is represented in Figure~\ref{fig:helio_sens}. Finally, the Russian groups in IAXO have recently proposed a medium-scale axion helioscope, called TASTE~\cite{Anastassopoulos:2017kag}, of similar scale and sensitivity reach than BabyIAXO, that could take place at INR in Russia, leveraging existing equipment and resources. 
Discussions are ongoing to explore how this project could constitute a pathfinder experiment for the future IAXO.

\subsection{Other techniques to search for solar axions}

\label{sec:othersolar}

\subsubsection{Primakoff-Bragg conversion in crystalline detectors}

Axion-photon conversion (and viceversa) can also happen in the atomic electromagnetic field inside materials. In the case of crystalline media, the periodic structure of the field imposes a Bragg condition, i.e., the conversion is coherently enhanced if the momentum of the incoming particle matches one of the  Bragg angles~\cite{Buchmuller:1989rb,Liao:2010ig} (similar Laue-like conversion in crystals is studied in~\cite{Yamaji:2017pep}). This concept has been applied to the search for solar axions with crystalline detectors~\cite{Paschos:1993yf,Creswick:1997pg}. The continuous variation of the relative incoming direction of the axions with respect to the crystal planes, due to the Earth rotation, produces very characteristic sharp energy- and time-dependent patterns in the expected signal in the detector, which can be used to effectively identify a putative signal over the detector background. This technique has the advantage that can be used as a byproduct of ongoing low-background underground detectors (e.g. those in WIMP or double-beta decay experiments) provided they have low enough threshold and the crystal plane orientation of the detectors is (at least partially) known.

After the first application of this technique with small Ge detectors by the SOLAX~\cite{Avignone:1997th} and COSME~\cite{Morales:2001we} experiments, it has been also used as byproducts of the DAMA~\cite{Bernabei:2001ny}, CDMS~\cite{Ahmed:2009ht} and EDELWEISS~\cite{Armengaud:2013rta} experiments. It is also foreseen as part of the physics program of -at least- CUORE~\cite{Li:2015tsa}, GERDA and MAJORANA~\cite{Xu:2016tap} experiments. However, in the mass range where helioscopes enjoy full coherent conversion of axions, the prospects of this technique is not competitive~\cite{Cebrian:1998mu,Avignone:2010zn}. The best result so far is that of DAMA, that sets $\gagamma < 1.7\times 10^{-9} \mathrm{GeV}^{-1}$ (90\% C.L.). Note that this is less stringent than the limit from solar physics itself, and thus these bounds are not yet self-consistent. Although this will improve in forthcoming results (e.g.~\cite{Li:2015tsa}), not even with future multiton target masses can one reach sensitivity to $\gagamma$ similar to current helioscopes~\cite{Cebrian:1998mu}. For higher masses above $\sim$1~eV, where the sensitivity of helioscopes drops, this technique does surpass the former, but this region of parameters is disfavoured by astrophysics and cosmology, as explained above.

\subsubsection{Modulation helioscope}
\label{amelie}

A variant of the helioscope technique can be realised in a large magnetised gaseous detector~\cite{Galan:2015msa}.
In this configuration the detector gaseous volume plays both the roles of buffer gas where the Primakoff conversion of solar axions takes place, and X-ray detection medium. The proposed realization is a gas time projection chamber (TPC) placed inside a strong magnet (AMELIE concept~\cite{Galan:2015msa}) like the one shown on the right of Fig.~\ref{fig:IAXO_cdr}. Contrary with standard helioscopes, in which the resulting X-rays need to cross the buffer gas to reach the detectors, here high photoabsorbtion in the gas is sought. Therefore, high pressures or high-$Z$ gases are preferred. The ALP-photon probability in a refractive and absorbing medium can be computed directly in the ALP-photon mixing picture outlined in sec.~\ref{sec:axionphoton} with a complex index of refraction, i.e. $k_\gamma\to k^*_\gamma \to \n_r \omega +i\Gamma/2$ where $\Gamma$ is the X-ray absorption coefficient.  
In the small mixing regime gives~\cite{vanBibber:1988ge}, 
\be
\proba(a\to \gamma) = |\vartheta\(e^{-i q^* L}-1\)|^2 = \frac{\gagamma^2 |B_{e\perp}|^2}{4 q^2+\Gamma^2}
\(1+e^{-\Gamma L}-2e^{-\Gamma L}\cos(q L)\)
\xrightarrow[]{\Gamma L\gg 1}  
\frac{\gagamma^2 |B_{e\perp}|^2}{4 q^2+\Gamma^2}, 
\ee 
and the probability than the ALP is absorbed along the length $L$, leading to a signal, in the $\Gamma L \gg 1$ limit~\cite{vanBibber:1988ge} tends to 
\be
P \xrightarrow[]{\Gamma L\gg 1}   \proba(a\to \gamma)\Gamma L .
\ee 
Due to the short range of the X-rays in the gas, the coherence of the conversion is lost and the sensitivity of the experiment is proportional to the  volume of detection (rather than to $L^2 \area$ as in standard helioscopes). This means that there is no privileged direction and moving the magnet to track the Sun is no longer necessary. 
Nevertheless, the signal depends on the $\vec B_e$ field component \emph{perpendicular to the ALP incident direction\footnote{In a general $B_e$ field, the ALP-like wave has an electric field along the transverse and \emph{longitudinal} components, see e.g. the discussion in~\cite{Millar:2017eoc}. In principle, the longitudinal field, which would mix with longitudinal plasmons, can also produce ionisation. However, in the relativistic limit $\omega \gg m_a$ in which this technique is relevant, the longitudinal component is suppressed by a factor $(q/\omega)^2$ and is therefore entirely negligible. }}, $\vec n_\odot$,  $B_{e\perp}=|\vec B_e \times \vec n_\odot|$, and therefore even in a stationary magnet a daily modulation of the signal is expected, which give a useful signal signature. 
Due to the loss of the coherence the technique is not competitive with standard helioscopes at low mass, although it could have some window of opportunity at higher masses $\gtrsim 0.1 $~eV where buffer gas scanning in helioscopes is increasingly difficult. The technique could profit from recent efforts in the field of TPCs with low background and relatively large volumes (e.g. \cite{Irastorza:2015geo,Iguaz:2015myh,Giomataris:2008ap}). Figure~\ref{fig:helio_sens} includes some projections of this technique, for different values assumed for the instrumented volume. \modified{The most sensitive line was drawn assuming that a magnet volume similar to that of IAXO is filled with gas TPCs. }

\subsubsection{Non-Primakoff conversions}

Axions could also interact with matter via their coupling with electrons or with nucleons. For example, solar axions could produce visible signals in ionisation detectors by virtue of the axioelectric effect~\cite{Ljubicic:2004gt,Derbin:2011gg,Derbin:2011zz,Derbin:2012yk,Bellini:2012kz}. The use of this technique to search for solar axions produced by $g_{ae}$ processes is particularly appealing, as the final signal depends only on $g_{ae}$ and robust limits on this coupling can be set. Large liquid Xe detectors, aiming at the detection of dark matter WIMPs, like XMASS~\cite{Abe:2012ut}, XENON~\cite{Aprile:2014eoa}, PANDAX-II~\cite{Fu:2017lfc} and LUX~\cite{Akerib:2017uem}, have all performed this search as a byproduct of their experiments. The latter has produced the most competitive result to date, setting an upper bound $g_{ae} < 3.5\times 10^{-12}$ (90\% C.L.). However this value is still considerably larger than the limit from astrophysics presented in section~\ref{sec:astrophysics}. Due to the mild dependency on exposure $g_{ae} \propto (MT)^{-1/4}$ (being $M$ the mass and $T$ the exposure time), even the future DARWIN detector, with a target mass of 50 ton of liquid Xe, will still fall short of reaching it~\cite{Aalbers:2016jon}.

For monochromatic solar axions emitted in M1 nuclear transitions, a reverse absorption can be invoked at the detector, provided the detector itself (or a component very close to it) contains the same nuclide. Due to Doppler broadening of the emission line, these axions should be able to excite the nucleus, whose subsequent decay can be registered. This detection strategy has been applied for a number of nuclides, like $^{57}$Fe~\cite{Moriyama:1995bz,Krcmar:1998xn}, $^7$Li~\cite{Krcmar:2001si}, $^{169}$Tm~\cite{Derbin:2009jw} or $^{83}$Kr~\cite{Derbin:2015bea,Gavrilyuk:2018jdi}. The upper limits on the nucleon couplings (and therefore to the mass) obtained by these methods are however larger than the bounds set by astrophysics.

\subsection{Axions from supernovae}
\label{sec:supernovae}

In section~\ref{sec:astrophysics} we saw how the (lack of) observation of $\gamma$-rays or neutrinos accompanying a SN explosion can indirectly constraint axion properties ($\gagamma$ and $g_{aN}$ respectively). But one could consider direct detection of axions produced in the SN at Earth. Indeed, if axion couplings are close to the upper bounds allowed by astrophysics (as the hints from stellar cooling seem to suggest) SNe would emit a significant part of their energy in the form of axions~\cite{Raffelt:2011ft}. A strong burst of axions, similar to that of neutrinos, is expected from each SN. In addition a large cosmic diffuse background flux from all past SNe, the \textit{diffuse SN axion background} (DSAB) (in analogy to the diffuse SN neutrino background, DSNB) is also expected. These DSAB axions have \modified{ ${\cal O}(100\, \rm MeV)$ } \todonote{Check other instances of MeV energies in SN axions; it is better to say O100 MeV} energies and can have energy density comparable with other diffuse stellar backgrounds like the DSNB or the EBL. Their direct detection, however, is very challenging~\cite{Raffelt:2011ft}. 
An interesting opportunity seems to be to attempt detection of the burst of a nearby SN with a helioscope equipped with MeV detectors~\cite{Raffelt:2011ft}. Assuming all SN energy is released in axions of average energy 80 MeV, Betelgeuse (distance 200 pc) provides an axion fluence at Earth of $5\times 10^{14}/$cm$^2$. Pointing the helioscope towards Betelgeuse in time for the burst is in principle possible due to the early warning ($\sim$ few days) due to the thermal neutrinos from the silicon burning phase preceding core collapse, that could be detectable by the neutrino detectors associated in the SuperNova Early Warning System (SNEWS)~\cite{Antonioli:2004zb}. 
For such energies, coherence in the axion-photon conversion of a $L= 20$ m helioscope like IAXO is achieved even for $m_a\sim$ eV. 
The expected number of events is 
\be
\sim  \frac{\area}{2\times 10^4 \, \rm cm^2}
\(\frac{\gagamma }{10^{-11} \rm GeV^{-1}}\)^2
\(\frac{B_a}{3\rm T}\)^2
\(\frac{L}{22\rm m}\)^2 . 
\ee
at essentially zero background. The discovery potential depends thus on the relation between the nucleon coupling responsible for the axion emission and the photon coupling. 
Reference~\cite{Raffelt:2011ft} used the educated guess that $g_{\A p}\sim 10^{-9}$, saturating the SN1987A neutrino pulse duration constraint, would reach such fluxes. From that point on, the signal depends on the particular axion model. For instance, interpreted as \KSVZ, $f_\A = 0.47 m_p/g_{\A p}\sim 4.4\times 10^8$ which gives $\gagamma= 0.5 \times 10^{-12}$ GeV$^{-1}$. The most optimistic case would correspond to having overestimated the axion emission from a nuclear medium (as seems to be the case~\cite{Chang:2018rso}). 
This would mean that the maximum SN axion emission would happen for larger values of $g_{\A p}$, smaller values of $f_\A$ and thus larger photon couplings. 
All in all, at this stage it does not seem implausible at all to discover axions in the $m_\A \sim 10$ meV ballpark with such a technique. Strong constraints on ALP models would ensue, otherwise.

\section{Direct detection of dark matter axions}
\label{sec:DMexps}

If our Milky Way dark matter halo is entirely composed of ALPs, we would be embedded in a sea with huge number density of about $3\times10^{14} (\mu$eV$/m_a)$~cm$^{-3}$. Despite their feeble interactions, these ALPs could lead to detectable effects in the laboratory experiments, if coherence effects are exploited. The absence of a signal in such experiments produce a limit to an ALP coupling under the assumption those particles form the dark matter, or more strictly, they produce limits to the product $g_a \sqrt{\admfrac}$, where $g_a$ is the relevant coupling for detection and $\admfrac$ is the fraction of the local ALP density, $\admfrac = \dmdensun_a/\dmdensun$, being $\dmdensun_a$ the local ALP density and $\dmdensun = 0.45$~GeV~cm$^{-3}$ the total local DM density\footnote{We shall keep in mind that the range of estimates for $\dmdensun$ is still relatively large, $\dmdensun=0.2-0.56$~GeV~cm$^{-3}$~\cite{Read:2014qva}. However, it has become customary to keep the quoted value used in axion DM exclusion plots for the sake of comparison among different results and projections. Lower $\dmdensun$ values would imply a need to scale all results accordingly}. If ALPs are a subdominant component of dark matter $\dmdensun_a < \dmdensun$, the sensitivity of these experiments to $g_a$ is correspondingly reduced.

The conventional strategy to detect axions is the \textit{axion haloscope} technique proposed by Sikivie~\cite{Sikivie:1983ip}, which invokes the conversion of axions into photons in a magnetic field and therefore is based on the coupling $\gagamma$. DM axions are highly non-relativistic and therefore almost monochromatic, because the relative spread in frequency is related to the virial velocity dispersion  in the gravitational potential well of the galaxy+DM halo, $\Delta \omega / \omega \sim \dmsigmav^2\sim  10^{-6}$ , see~\eqref{Lentzdistribution}. 
This fact can be exploited to devise a resonantly enhanced conversion using microwave cavities with high quality factor $Q$. The resonant frequency of the cavity must be matched to the unknown axion mass within an error $\sim m_a/Q$ in order to enable the resonant conversion. Given that not only the axion mass is unknown, but, as shown in section~\ref{sec:cosmology}, there is strong motivation to probe a large range of values, the experimental setup must allow tuning the resonant frequency over a range as wide as possible. 
Data taking with conventional haloscopes thus entails scanning very thin mass-slices of parameter space. The experimental implementation and the development of related technologies (high-$Q$ cavities inside magnetic fields, low noise RF detection, etc.) have been led for many years by the ADMX collaboration. ADMX has demonstrated that the haloscope technique can realistically achieve sensitivity to QCD axion models in the few $\mu$eV mass ballpark. In recent years, a number of new experimental efforts are appearing, some of them implementing variations of the haloscope concept, or altogether novel detection concepts, making this subfield one of the most rapidly changing in the axion experimental landscape. In the following subsections, we attempt a complete review of those efforts, stressing the complementarity among many of them, conceived to extend sensitivity to different axion mass ranges.

Applying the haloscope technique to frequencies considerably higher or lower than the one ADMX is targeting is challenging, for different reasons. Lower frequencies imply proportionally larger cavity volumes and thus bigger magnets. Higher frequencies imply lower volumes and correspondingly lower signals and sensitivity. Several R\&D lines are being pursued to adapt the concept at different frequency ranges, and we review them in~\ref{sec:haloscopes}. These developments are always associated with the technologies needed to go to larger and/or more intense magnetic fields, higher quality factors, cryogenics and noise reduction at detection, among others. Higher frequencies together with relatively high volume are in principle achievable with more complex extended structures resonant at high frequencies, of which several implementations are being explored. A more radical variant, in which $Q$ is (almost) given up, is followed by the \textit{dish antenna} concept and its relative, the \textit{dielectric haloscope}, presented in~\ref{sec:dielectric_haloscopes}. Other altogether different detection concepts include the use of $LC$ circuits inside magnetic fields (\ref{sec:LC}) to generate the resonance, or the use of nuclear magnetic resonance (NMR) techniques (\ref{sec:NMR}), both with promise to achieve good sensitivity at much lower masses than the conventional haloscopes, and the latter invoking the interaction of the axions with electrons or nuclei, instead of $\gagamma$. In addition, the effect of the DM axion field in atomic transitions (\ref{sec:atomic}) could lead to observable effect at much larger $m_a$ than previous techniques. Finally, some of the detection techniques offer more refined detection strategies involving low dispersion streams or sensitivity to the incoming direction of the axion, something that we briefly discuss in~\ref{sec:directional}.

Before proceeding with the rest of the section, let us mention that the possibility that the DM halo is made of more massive ALPs with coupling to electrons could produce a signal in massive ionisation detectors via the axioelectric effect. Although not a particularly motivated scenario by theory, this possibility can be explored as a byproduct of ongoing low background massive detectors developed for WIMP searches, showing up as a non-identified peak at an energy equal to the ALP mass. Experiments like CDMS~\cite{Ahmed:2009ht}, XENON100~\cite{Aprile:2014eoa} or EDELWEISS~\cite{Armengaud:2013rta} have analysed their data in search for such signal. They globally exclude values of $g_{ae} \lesssim 10^{-12}$ for ALP masses in the range 1--40 keV (always under the assumption that DM is entirely made of those particles). Next generation WIMP detectors, like DARWIN~\cite{Aalbers:2016jon}, could improve this limit by more than one order of magnitude.

Finally, let us mention that it has been claimed that DM axions can generate observable signals in resonant Josephson junctions~\cite{Beck:2011tz}, and indeed an experimental result of unknown origin~\cite{Hoffmann:2004faa} has been interpreted as a detection of a DM axion of $m_a = 0.11$ meV in~\cite{Beck:2013jha}. Although this value of $m_a$ is well inside one of the ranges expected for a DM axion as explained in section~\ref{sec:cosmology}, we must cast serious doubts on the validity of this interpretation. The formalism is based on formal similarity of the dynamical equation of the axion and the Josephson junction, but it is not clear what would be the origin of the physical coupling between both systems. At some point in~\cite{Beck:2013jha} the Primakoff conversion of the axion is invoked (with an unrealistic $\cal{O}$(1) conversion probability), however the final expression predicted for the signal strength is independent on $\gagamma$ or any other axion coupling, which means that the formalism has the strange property of predicting a signal even with vanishing axion interactions.

\subsection{Conventional haloscopes}
 \label{sec:haloscopes}

\begin{figure}[t] \centering
\includegraphics[width=\textwidth, trim={0 1cm 0 1cm}, clip ]{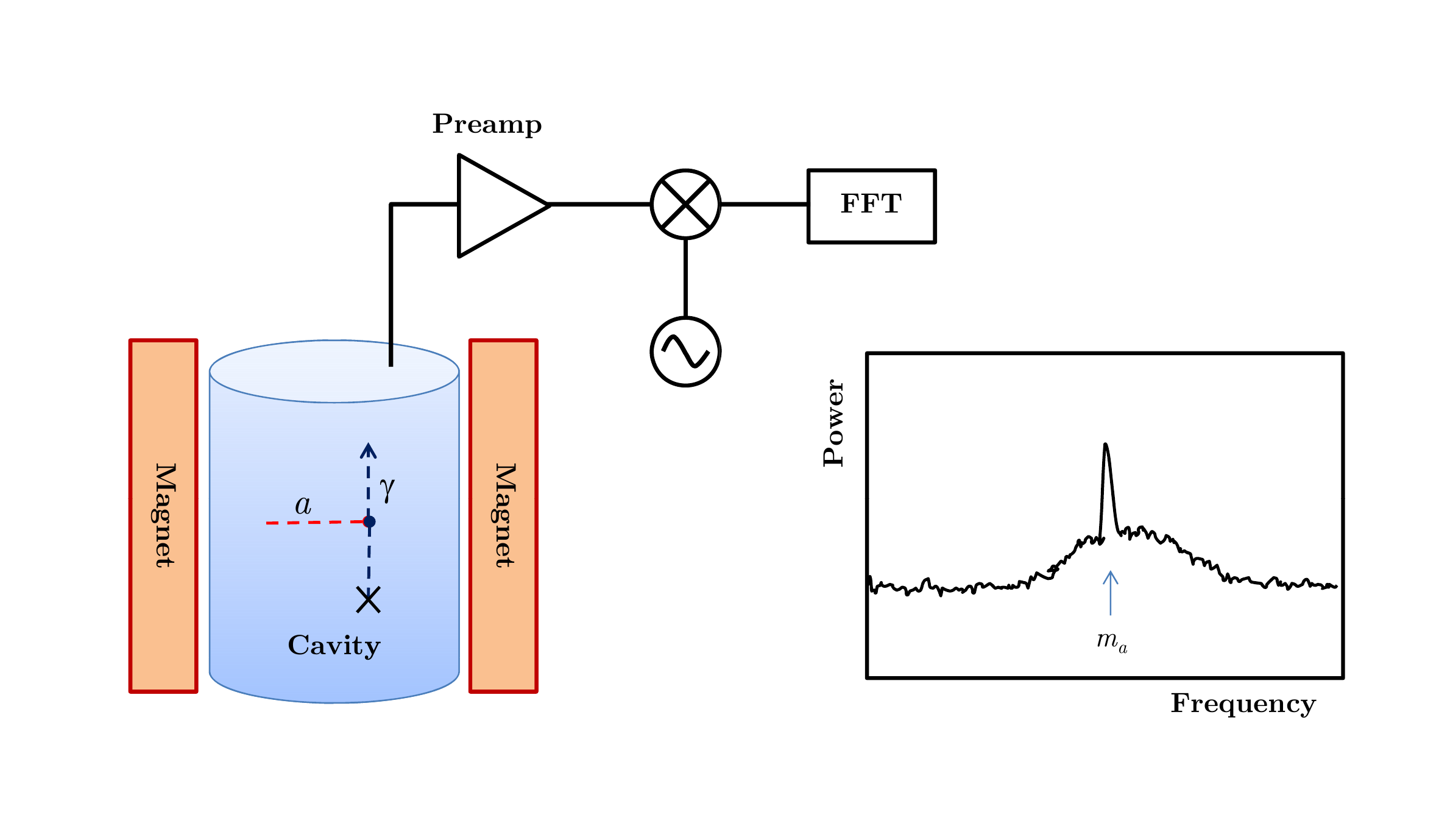}\hspace{2pc}%
\caption{\label{fig:haloscope_sketch} Conceptual arrangement
of an axion haloscope. If $m_a$ is within 1/$Q$ of the resonant frequency of the cavity, the axion will show as a narrow peak in the power spectrum extracted from the cavity.}
\end{figure}

The conventional axion haloscope technique~\cite{Sikivie:1983ip} consists of a high-$Q$ microwave cavity inside a homogeneous magnetic field of intensity $B_e$ to trigger the conversion of DM axions into photons. Figure ~\ref{fig:haloscope_sketch} shows a sketch of the haloscope concept. Being non relativistic, DM axions produce monochromatic photons of energy equal to $m_a$. For a cavity resonant frequency matching $m_a$, the conversion is enhanced by a factor proportional to the quality factor of the cavity $Q$ and the signal power in the band $m_a\pm m_a/Q$ is~\eqref{resonantregeneration}, 
\bea
P_s  &=& \kappa \frac{Q}{m_a} \gagamma^2 B^2_e |\geo_m|^2  V \dmdensun_a \\
	&=& 7.2\times 10^{-23} {\rm W}
		\(\frac{\kappa}{0.5}\) 
		\(\frac{Q}{10^5}\) 
		\(\frac{\mu\rm eV}{m_a}\)
		\(\frac{\gagamma}{2\times 10^{-16}\rm GeV^{-1}}\)^2 
		\(\frac{B_e}{8\rm T}\)^2 
		\(\frac{|\geo_m|^2}{0.69}\)  
		\frac{V}{200 \rm l}
		\admfrac
\eea
where $\geo_m$ is the geometric factor of the resonant mode, as defined in Eq.~\ref{geometricfactor}, and we have used $\dmdensun_a= m^2_a|a_\osci |^2/2$ by assuming that $\dmsigmav^2\lesssim 1/Q$, i.e. the DM bandwidth is smaller than the width of the cavity resonance and can be taken as a delta function. 
The general formula \eqref{resonantregenerationDD} can be used when $\dmsigmav^2\sim 1/Q$ or for other bandwidths. 
This resonant amplification only works for values of $m_a$ within a thin frequency window $\Delta \nu/\nu \sim Q^{-1}$ around the resonant frequency, see \eqref{resonantregenerationD}, but typically the DM signal frequency bandwidth is even smaller. One usually defines a DM quality factor $Q_a\sim 1/\dmsigmav^2\sim 10^6$ to reflect the ALP DM signal width.  
The cavity must be tuneable and the data taking is performed by subsequent measurements with the resonant frequency centred at slightly different values, scanning the ALP DM mass in small overlapping steps. 
For QCD axions, the signal is typically much smaller than noise, 
\bea
P_{n} &=& T_{sys} \Delta \nu = T_{sys} \frac{m_a}{2\pi Q_a} \\
	 &=& 3.3 \times 10^{-21} \(\frac{T_{sys}}{K}\) \(\frac{m_a}{\mu\rm eV}\) \(\frac{10^6}{Q_a}\) , 
\eea
where $T_{\rm sys}$ is the effective noise temperature of the detector (typically amplifier + thermal fluctuations). 
One hopes that measuring enough time, the signal becomes larger than noise fluctuations. The signal to noise as a function of the measurement time in a frequency bin $\Delta \nu$ is given by Dicke's radiometer equation 
\be
\frac{\rm S}{\rm N} = 
\frac{P_{s}}{T_{\rm sys}}\sqrt{\frac{\Delta t}{\Delta \nu}}, 
\ee
where $T_{\rm sys}$ is the effective noise temperature of the detector (typically amplifier + thermal fluctuations).
Therefore, given a theoretical axion signal $P_{s}$,  a time 
$\Delta t=({\rm S/N})^2(T_{\rm sys}/P_{s})^2 \Delta \nu$ is needed to achieve a given detection significance specified by a signal to noise. 
In order to scan an ALP mass interval, $d m_a$ with measurements of width $\Delta \nu = m_a/Q$, we need a number 
$(Q/Q_a)(d m_a/m_a)$ of $\Delta t$ measurements, and so the scanning rate is
\bea
\frac{d m_a}{d t} &=& \frac{Q_a}{Q}\frac{2\pi \Delta \nu}{\Delta t}=\frac{Q_a}{Q} \(\frac{\rm S}{\rm N}\)^2 \(\frac{T_{sys}}{P_s}\)^2 . 
\eea
A useful figure of merit of these experiments is proportional to the time needed to scan a fixed axion mass range~\cite{Asztalos:2001tf} down to a given S/N level  for a given value of the coupling $g_{a\gamma}$:
\begin{equation}\label{FOM}
F \sim \dmdensun_a^2 g_{a\gamma}^4 m_a^2 B_e^4 V^2 T_{\rm sys}^{-2} |\geo|^4 Q
\end{equation}

ADMX has been the major experimental effort in this category for many years, drawing on the experience of a couple of pilot small-scale experiments in the 80s, at BNL~\cite{DePanfilis:1987dk,Wuensch:1989sa} and at University of Florida~\cite{Hagmann:1990tj}. ADMX has built the largest and most competitive axion haloscope to date. It uses a NbTi superconducting solenoid with an inner cylindrical bore of 60~cm~$\times$~110~cm and producing a field of up to 8~T. The microvawe cavity, shown in Fig.~\ref{fig:haloscope_photos}, is made by electrodepositing high-purity copper on a stainless steel body, followed by annealing, and reaches a quality factor of $Q\sim 10^5$. 

For a cylindrical cavity in a solenoidal field, the TM$_{0n0}$ modes are the ones that couple with the axion, as they feature an electric field component parallel to the magnetic field. The resonance frequency, and therefore the \additionalinfo{This is the correct definition of lambdaa} wavelength of light that resonates that frequency $\lambdaa=2\pi/m_a$, is in general determined by the cavity dimensions $V\sim (\lambdaa/2)^3$. For an empty cylinder the relation is found analytically:

\begin{equation}\label{TM0n0_freq}
\omega_{{\rm TM}_{0nl}} = \sqrt{\(\frac{\xi_n}{r}\)^2 + \(\frac{l\pi}{h}\)^2}, 
\end{equation}
where $\xi_n$ is the $n$th zero of the Bessel function of order zero, and $r$ and $h$ are the radius and height of the cylinder. The fundamental TM$_{010}$ mode provides the larger form factor $|\geo_{\rm TM_{010}}|^2\sim 0.69$. In the implementation at ADMX, a set of movable rods inside the cavity slightly distorts the mode while allowing for the tuning of the resonance frequency in the approximate range of 0.5 to 2 GHz.

In a first phase (1995-2004), the experiment was cooled down to $T\sim1.5$~K, and HFET amplifiers with noise temperature $T_N\sim 1.5$~K were used, thus having a system noise temperature of $T_{\rm sys} \sim  3$~K. In this conditions, ADMX~\cite{Asztalos2010} already covered the frequency range of 460 - 890 MHz (1.9 - 3.65 $\mu$eV) with sensitivity down to \KSVZ models (see Figure~\ref{fig:halo_sens}). This has been the first axion haloscope search with sensitivity reaching the axion QCD band. Since then the collaboration has been focused on improving the setup, mostly regarding low noise detection, in order to improve sensitivity in $\gagamma$ or, alternatively, scan faster in $m_a$. In a second operational phase (2007-09), Microstrip coupled SQUID Amplifiers (MSA) were employed, with much lower noise temperature \cite{Muck:1999nra}. These devices can operate close to the \modified{quantum limit, $T_N\sim \omega$,}  but to exploit this the temperature of the cavity must be lowered substantially. In the current stage of the experiment, a dilution refrigerator has been added, that allows the cool down the cavity to a temperature $T \sim 150$~mK, reaching $T_{\rm sys} \sim 500$ mK. First data in these conditions are already being taken. A new scan of the 645-680 MHz region has been performed in the first half of 2017 with expected sensitivity \cite{Rybka_Trento} down to the \DFSZ model.

ADMX is now ready to continue to go up in frequency with relatively high scanning speed. It is expected to probe up to $\sim$2 GHz (8 $\mu$eV), with \DFSZ sensitivity, in the next 2 years (see Fig.~\ref{fig:halo_sens}). Then the collaboration is targeting going to higher masses, up to 10 GHz, in a few years time. For this the single cavity will be replaced with multiple power-combined cavities with higher resonant frequencies, but tuned
in a similar manner. Going to even higher frequencies requires new detection strategies, for which intense R\&D is ongoing, as we describe below.

Finally, let us mention that the ADMX infrastructure has also been used to search for other WISPs, as byproducts of the experiment. Chameleon particles have been searched through the ``afterglow'' effect of photon-chameleon-photon transitions~\cite{Rybka:2010ah}, while hidden photons were searched for implementing a LSW setup with a emitter cavity driven with microwave power~\cite{Wagner:2010mi}.

\subsubsection{ Lower $m_a$ }


Going to frequencies lower than the low end of the ADMX range requires larger cavity volumes. In principle, the dependence on $V$ and the fact that larger cavities easily reach higher $Q$ values should make the implementation of those haloscopes less technologically challenging. However, the construction of sufficiently large and powerful magnets to hold the cavities will require important investments that could constitute a major drawback for such implementations. Access to existing large magnets could offer appealing options to carry this type of searches. 

The WISPDMX experiment is a joint venture between DESY and Hamburg University with the aim of searching for WISPy DM with a 208-MHz resonant cavity used at the DESY HERA accelerator. 
In the first stage it is focusing on hidden photon DM search, which does not require the presence of a strong background magnetic field. The cavity has a volume of $\sim 500$ litres and the ground TM$_{010}$ mode has $Q=46000$. The signal is amplified with $T_{\rm sys}=100$~K. 
Two dielectric plungers allow the tuning of the ground and excited frequencies by a few MHz. The collaboration considers placing their cavity into the H1 solenoid magnet ($B_e=1.15$ T, $V\sim 7.2$ m$^3$) to search for ALPs~\cite{Nguyen:2015ktw}. 

The proponents of KLASH~\cite{Alesini:2017ifp} aim at using the KLOE superconducting solenoid magnet located at the National Laboratory of Frascati, INFN. This magnet could host a large cylindrical cavity of 50~m$^3$ inside a 0.6~T field. The cavity would be copper coated and could achieve a $Q$-factor of about $4.5\times 10^{5}$ at 4~K, the temperature to which the magnet is cooled down. 
A mechanical tuning similar to that of ADMX, based on movable rods, should allow to tune the resonant frequency in an approximate range of 20\% around the base frequency of 57~MHz. Finally, a low noise RF detection at 4~K is foreseen. As seen in Fig.~\ref{fig:halo_sens}, such a haloscope implementation would have a potential sensitivity of $\gagamma \sim 6 \times 10^{-17}$~GeV$^{-1}$ for $m_a$ values in the ballpark of 0.2~$\mu$eV. The large magnetic volume available, compensates the relatively low field and modest noise temperature imposed by the temperature of the magnet.

A similar large volume implementation could be conceived in (one of the bores of) the large magnet planned for IAXO (described in section~\ref{sec:helioscopes}). Using relatively conservative values for $Q$ and $T_{\rm sys}$, a very competitive sensitivity could be achieved thanks to the large $B^2 V$ available. A preliminary estimation ~\cite{redondo_patras_2014} gives sensitivity down to \DFSZ models for an approximate mass range of 0.6 -- 2 $\mu$eV.
The cavity geometry best fitting the planned bores, would be similar to the rectangular type proposed in~\cite{Baker:2011na}, which is currently being used by the CAST-CAPP and RADES searches in the CAST magnet (see below).  

A haloscope implementation in a toroidal geometry is being explored at CAPP in South Korea, called ACTION~\cite{Choi:2017hjy}. A toroidal cavity could offer some advantages over conventional cylindrical ones, like a more efficient use of the magnet conductor and a higher geometric factor, as well as lower fringe field (good for handling external sensitive equipment). Frequency tuning is achieved by moving a toroidal bar displaced parallel to the symmetry axis. First tests with a small toroidal cavity have recently been carried out. A large scale ACTION experiment
would require a large toroidal magnet with minor/major radius of 50/200~cm and a $B=$5~T, and could effectively explore the 0.8--2 $\mu$eV range. 


\subsubsection{ Higher $m_a$ }

\begin{figure}[t] \centering
\includegraphics[height=6cm]{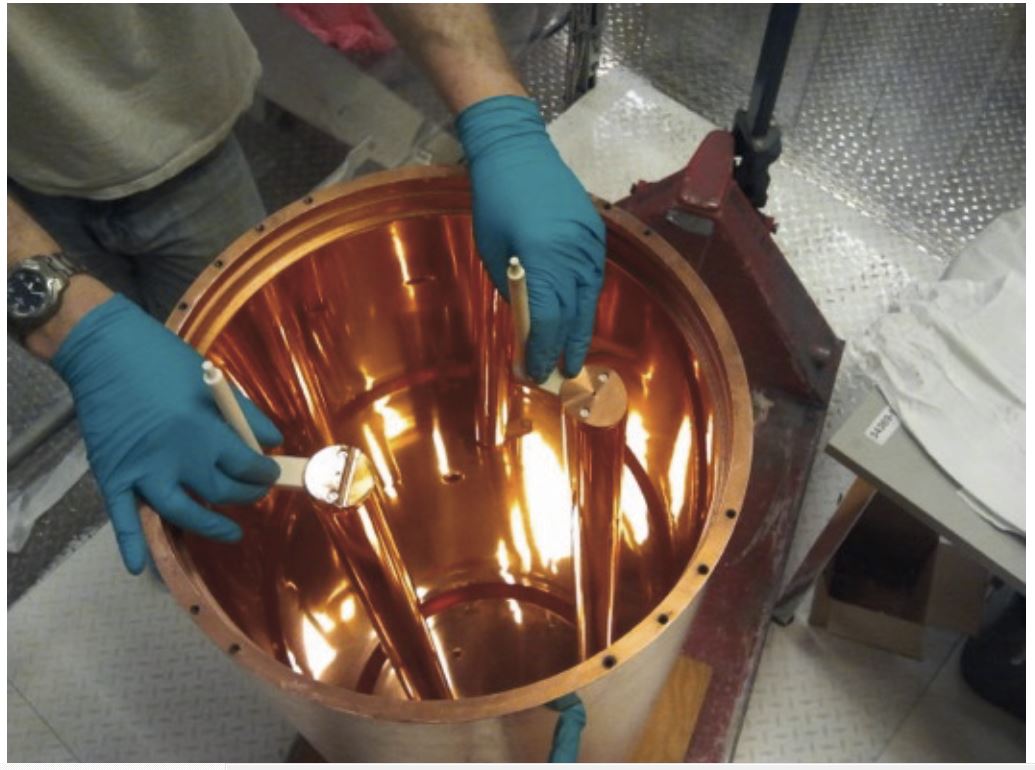} \hspace{1mm}
\includegraphics[height=6cm]{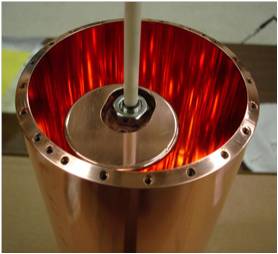}
\caption{\label{fig:haloscope_photos} Pictures of some of the microwave cavities implemented in active axion haloscopes comented in the text, with the mechanical tuning mechanism. On the left the 50~cm diameter and 1~m long cavity of the AMDX experiment (used with permission, credit ADMX Collaboration). \modified{On the right the 10.2~cm diameter and 25.4~cm} long cavity of the HAYSTAC experiment (used with permission, credit HAYSTAC Collaboration).}
\end{figure}

Pushing the haloscope technique to higher masses is challenging, for a number of reasons. First, given that the cavity size is matched to the photon wavelength, higher frequencies correspond to lower $V$, which means lower sensitivity. 
Furthermore, $Q$ typically decreases for higher frequency cavities 
\todonote{dont know which ref to put here}, and typically, the noise of microwave amplifiers increases with frequency. This is true for state of the art HEMT amplifiers, $T\sim (\nu/{\rm 2.5GHz})$ K, but also for the quantum limit $T_{N}=\omega$, achievable with SQUIDs, see e.g. for a discussion in the context of axion DM searches~\cite{Lamoreaux:2013koa}. 
The obvious exception are bolometers or photon counters that do not measure both amplitude and phase of light like linear amplifiers do. Photon counting at $\sim$ GHz frequencies is at the moment not a better option, but is not subject to the 
quantum uncertainty limitation, just to shot noise and thermal fluctuations, which at high frequencies can be strongly suppressed by cooling to $T<\omega$. In this context we lament the discontinuation of the CARRACK experiment~\cite{CARRACKwebNEW,Tada:1999tu}, which focused on developing $\sim 10\mu$eV photon counting with Rydberg atoms~\cite{Fukuda:1988yw} in a secondary coupled cavity and released some promising results~\cite{Tada2006488,Shibata2008}. 

One can in principle compensate the loss in $V$ and $Q$ by improving other parameters entering the figure of merit and indeed substantial effort in these directions is ongoing in the community. One obvious line of research is to target higher $B$-fields, piggybacking on advances in magnet technology, taking advantage on the fact that only smaller cavity volumes $V$ need to be instrumented. Alternatively (or in addition), research on low noise RF photon detection could push $T_{\rm sys}$ down to the quantum limit, or even beyond it. Higher $Q$ could be achievable using superconducting cavities, although $ Q$-values above $Q_a\sim 10^6$ do not translate to a further increase in sensitivity (except in the case of low dispersion streams, see below in~\ref{sec:directional}). 
Finally, the coherent combination of several cavities or the development of more complex resonant structures could effectively decouple $V$ from the photon wavelength and access higher $V$ values.

The ADMX collaboration has been leading several of the mentioned R\&D lines to go to higher frequencies, and relevant demonstrating results have been achieved in test benches, like e.g.: the combination of several cavities~\cite{Kinion:2001fp}, the development of multiwavelength resonators, based on photonic band-gaps or an  open resonator with $B_e$ fields alternating polarity to match a mode above the fundamental (the ORPHEUS resonator~\cite{Rybka:2014cya}, see below), or the development of novel single photon detectors to reach very low effective $T_{\rm sys}$. The latter is based on single photon manipulation hardware developed by the quantum computing community. The outcome of this R\&D program could be implemented in the future in the main ADMX setup to configure a research program beyond the one described above.
%

\begin{figure}[t] \centering
\includegraphics[height=6cm]{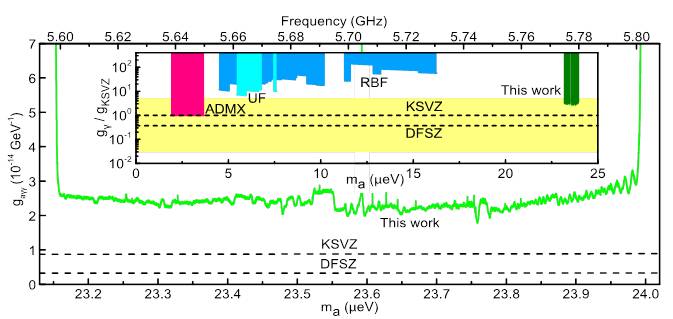}
\caption{\label{fig:haystac_plot} Zoom of the region excluded by HAYSTAC, both the 2016 as well as the new 2017 data. An axion DM local density of $\dmden_a=0.45$~cm$^{-3}$ is assumed. Courtesy of K. van Bibber / HAYSTAC team, used with permission. }
\end{figure}

The HAYSTAC experiment~\cite{Kenany:2016tta} at Yale University was born in part out of developments initiated inside the ADMX collaboration~\cite{Shokair:2014rna} to develop experiments at  higher frequencies. It is conceived as a data pathfinder and an innovation testbed in the 2.5-12 GHz (10-50 $\mu$eV) mass range, where new amplifier and cavity designs could be tested out and prepared for larger volume experiments. It has been the first setup reaching axion sensitivity in the decade of mass above ADMX. In its current experimental setup, HAYSTAC follows a geometry similar to ADMX but with scaled-down dimensions. The 1.5 liter cavity (shown in Fig.~\ref{fig:haloscope_photos}) is immersed in a superconducting 9 T magnet, and is cooled down to 127~mK with a dilution refrigerator. 
Signal amplification is done with Josephson Parametric Amplifiers (JPA), with which values of $T_{\rm sys}$ only a factor of 2 above the Standard Quantum Limit (SQL)
have been achieved. The first data-taking was performed in 2016, scanning a first frequency range of 5.7-5.8 GHz, and providing the best limits to date at these higher frequencies~\cite{Brubaker:2016ktl}, only a factor of $2.3$ in $\gagamma \sqrt{\admfrac}$ from \KSVZ models. Another data taking campaign has been carried out in 2017, soon to be published, doubling the scanned range. Figure~\ref{fig:haystac_plot} shows both the 2016 and 2017 results spanning a mass range of 5.6-5.8 GHz (approx. 23.15 - 24.0 $\mu$eV). A detailed account of the data analysis can be found in~\cite{Brubaker:2017rna}.

Current near-term plans include the development of novel squeezed-vacuum state receivers, that could surpass the SQL~\cite{Zheng:2016qjv}. This could allow a substantial improvement in the sensitivity and the scan speed of the experiment. In addition, a new cavity design optimised to efficiently use higher order modes is ongoing to push sensitivity to higher masses. These improvements are expected to provide sensitivity below \KSVZ for masses in the range 6-12 GHz (roughly 25-50~$\mu$eV). R\&D to access even higher masses is also ongoing, based on photonic band gap concepts and metamaterials.

An intense and varied experimental activity towards higher frequencies is now also taking place in the recently-established Center for Axion and Precision Physics (CAPP) in South Korea. CAPP is emerging as a major player in the axion experimental landscape, and one of its main goals is the search of DM axions in the mass range 4-40 $\mu$eV decade with sensitivity beyond \DFSZ models. The CAPP R\&D programme encompasses dedicated activity in all the technological issues already mentioned above, including ultrahigh field superconducting magnets with various bore sizes (5, 10 and 35 cm inner diameter), superconducting films (to achive high $Q$ cavities), high-gain gigahertz superconducting quantum interference device (SQUID), new cavity designs, and multi-cavity phase locking schemes.

\begin{figure}[t] \centering
\includegraphics[height=5cm, trim={0 10cm 0 10cm},clip]{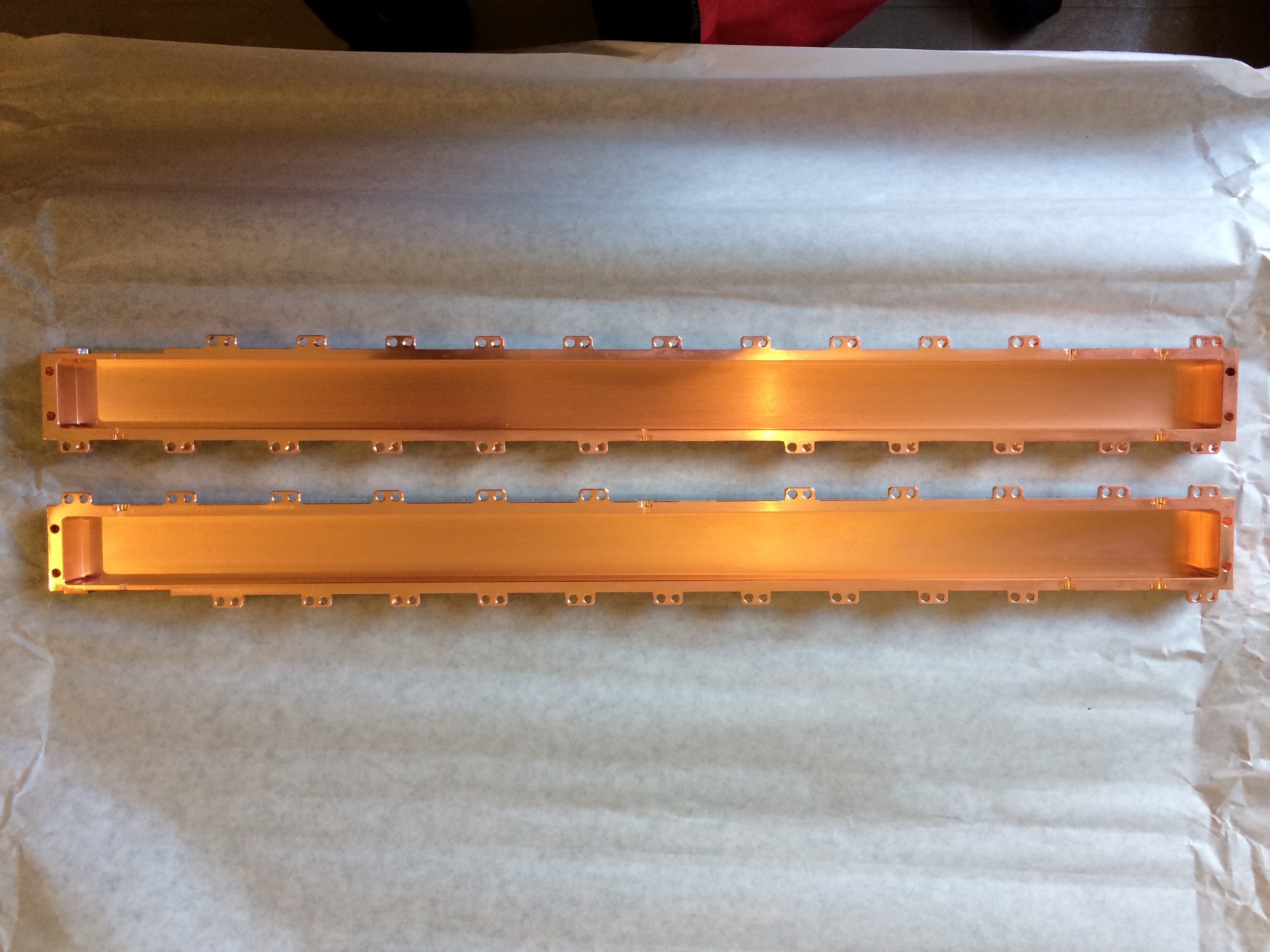} \hspace{1mm}
\includegraphics[height=5cm, trim={0 1cm 0 1cm},clip]{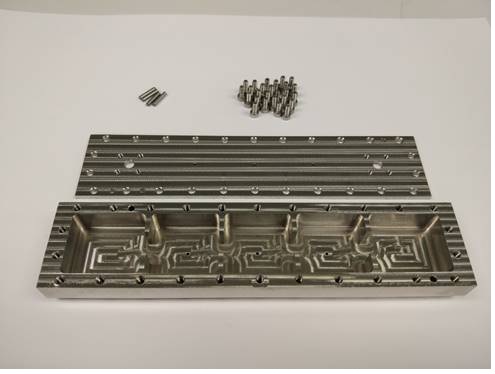}
\includegraphics[height=5cm, trim={0 0 0 0},clip]{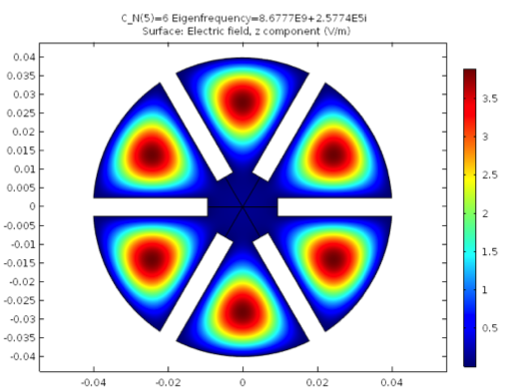}
\caption{\label{fig:long_cavities_photos} On the top, pictures of one of the CAST-CAPP 40~cm long cavities soon to be installed inside the CAST magnet. On the left, the first RADES 5-cell cavity currently in operation also in the CAST magnet. Used with permission, credit CAST-CAPP and RADES teams respectively. Right, simulation of a pizza-like sliced cylindrical cavity, taken from~\cite{Jeong:2017hqs} with permission.}
\end{figure}

\begin{figure}[!t]
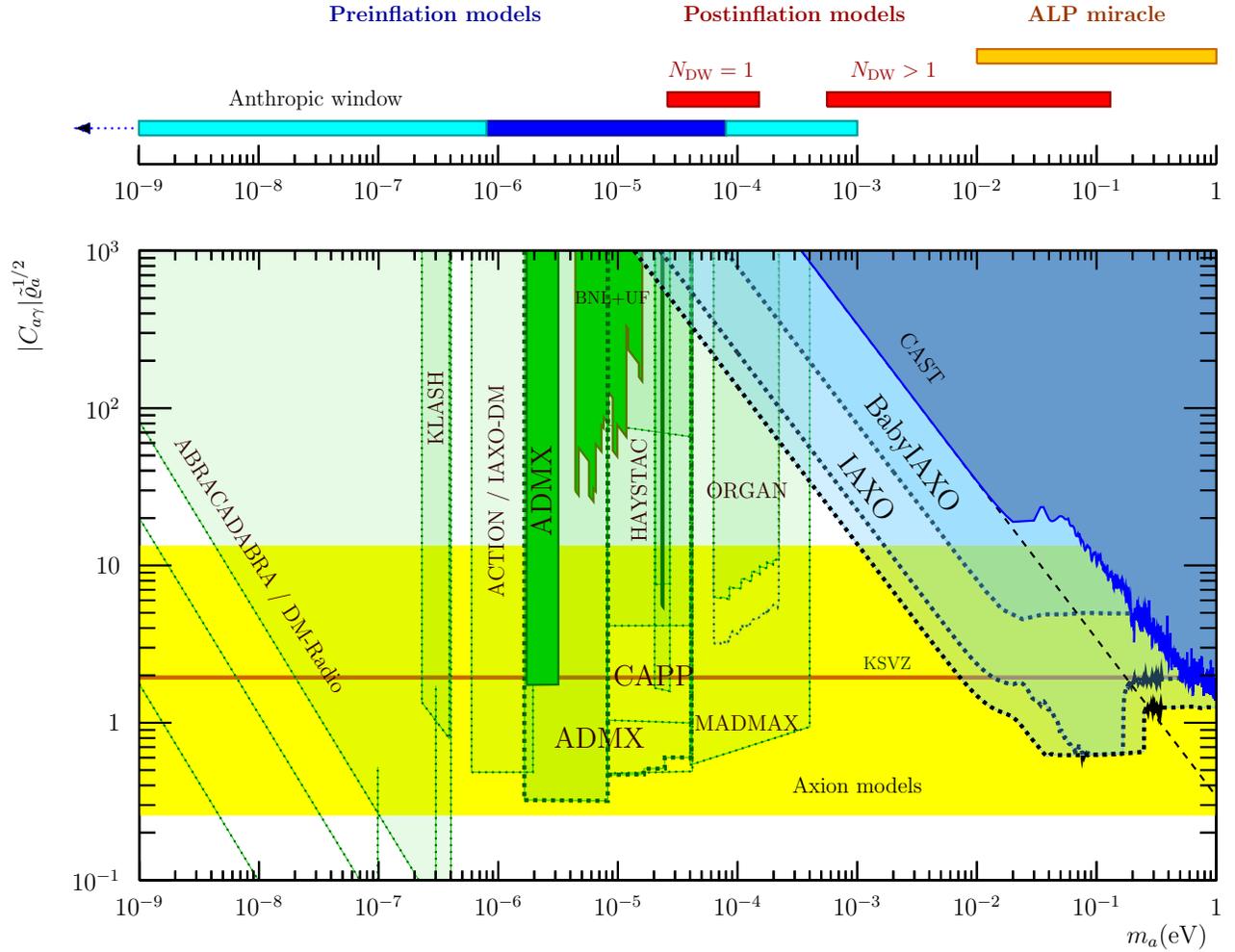

\begin{center}
\tikzsetnextfilename{DM_models}
\resizebox{\linewidth}{!}{\input{pics/DM_models.tex}}
\tikzsetnextfilename{Haloscopes_Cg}
\resizebox{\linewidth}{!}{\input{pics/Haloscopes_Cg.tex}}
\caption{Exclusion regions from haloscope searches (in green) expressed in terms of $| C_{a\gamma} | \sqrt{\admfrac}$.  
We display $C_{a\gamma}$ in the sense of $C_{a\gamma}= \gagamma f_A (2\pi/\alpha)$ from \eqref{axionphotoncoupling}
by rescaling sensitivities on $g_{a\gamma}$ by the known relation between $f_\A, m_\A$. 
Some of the regions  tentatively within reach of future experiments are indicated as semi-transparent green areas. Some of those regions are dependent, to different extents, on successful completion of R\&D on novel detection concepts, as explained in the text. Regions explored and projected by helioscopes are also shown (in blue). As usual the yellow band and orange line represent the QCD axion models and the benchmark \KSVZ model respectively. The sketch on top shows the mass ranges for which total DM density can be obtained in different models, as explained in section \ref{sec:axions_as_DM}.}
\label{fig:halo_sens}
\end{center}
\end{figure}

The main experiment at CAPP is CULTASK~\cite{Woohyun:2016hkn}, which is based on the conventional haloscope concept with cylindrical cavities. A first setup includes a copper cavity of 9~cm diameter with a sapphire tuning rod, whose lowest mode corresponds to about 2.5~GHz, and is therefore able to explore the region around $m_a\sim10~\mu$eV.  The cavity is cooled down to below 100 mK and immersed in a 8~T magnetic field, and first engineering runs are ongoing. This is expected to be the first of a series of setups with progressively stronger magnets and more ambitious parameters, results of the ongoing R\&D lines being pursued in parallel. Up to seven independent low vibration pads are being setup at CAPP to host different experimental setups, using a suite of superconducting magnets of different strengths and sizes. In particular, a 18~T magnet developed with high temperature superconductor (HTS) cables and 5~cm bore is soon to be installed. More magnets with much bigger bore sizes (35 cm and 50 cm) are also in the pipeline. Another stronger 25~T HTS magnet of 10~cm diameter is under development with Brookhaven National Laboratory (BNL). The design is relatively compact and foresees future upgrades to 35~T or even 40~T by adding external conductor coils. In combination with near-quantum limit receiver and higher-$Q$ superconducting cavities gives promise to cover the full mass range indicated in figure~\ref{fig:halo_sens} with sensitivity to $\gagamma$ that will progressively go down to \KSVZ models in the next decade.

Even higher frequencies, in the range of 15--50 GHz, corresponding to a mass range of 62-207 $\mu$eV, are targeted by the ORGAN~\cite{McAllister:2017lkb} program, recently started in the University of Western Australia in Perth. The use of a variety of thin, long resonant cavities of different dimensions, immersed in a 14~T magnet with a bore size of 65 mm and a length of 445 mm, are considered to cover all the frequency range. A first pathfinder run has already taken place~\cite{McAllister:2017lkb}, at a fixed frequency of 26.531 GHz, corresponding to  $m_a=110 \mu$eV, using the TM$_{020}$ mode of a cavity inside a 7~T magnet and an exposure of 4 days. As a result a limit $\gagamma$~(90\% CL)~$< 2.02 \times 10^{-12}$~GeV$^{-1}$ was set, over a narrow mass span of 2.5 neV defined by resonance width. The immediate plans are to carry a one-year data taking targeting the 26.1-27.1 GHz region. At a second stage, the full mentioned frequency range will be scanned in 5~GHz regions (see Fig.~\ref{fig:halo_sens}). Tunability of these cavities is achieved by movable dielectric dishes inside the cavity~\cite{McAllister:2017ern}.  The use of cavity arrays are also being considered for future implementations~\cite{GORYACHEV2017}. Bringing  the sensitivity down to QCD models will require moving to higher magnetic fields and developing squeezing techniques to go beyond the quantum noise limit.

An appealing option to go to higher $m_a$ would be to effectively increase $V$ by filling a large volume with many high-frequency resonant structures, i.e. effectively decoupling the detection volume $V$ from the volume of a cavity and the resonant frequency. Literally replicating a cavity many times and combining their output is possible in theory, although  in practice is difficult to phase match them to obtain maximum sensitivity. This kind of combination has already been performed long ago for four cavities within the ADMX R\&D~\cite{Kinion:2001fp}, but going to a much larger number of cavities has been considered not feasible in practice. Renewed efforts in this direction are being done at CAPP in South Korea~\cite{Jeong:2017xqz}. The concept of the sliced-as-a-pizza cavity consists of dividing the cylindrical cavity in sections connected by a longitudinal iris along the cylinder's axis of symmetry and has been designed to make optimal use of a solenoidal magnet~\cite{Jeong:2017hqs}. Slicing up to 8 subcavities seems to be plausible. 

The CAST-CAPP team will apply some of these efforts to combine several long-aspect-ratio rectangular (i.e. waveguide-like) cavities to be inserted in the 4~cm diameter 10~m long bores of the CAST 9~T magnet at CERN~\cite{Desch:2221945,Fischer:2289074}. Figure~\ref{fig:long_cavities_photos} shows one of them. The use of these types of cavities was proposed in~\cite{Baker:2011na} and has interesting technical advantages. The resonant frequency in these geometries is mostly determined by the smaller dimensions of the parallelepiped, and therefore $V$ can be increased (in principle, arbitrarily) by increasing its length. In practice, mode crossing and mode crowding limits the length of the cavities, but this could be overcome by phase matching several smaller cavities. Current CAST-CAPP design considers 40~cm long cavities. Tuning of these cavities can be accomplished by the use of small movable slabs inside the cavity or by having the cavity cut in two longitudinally and precisely moving the two halves. This approach should give competitive sensitivity for a small frequency range around the 6~GHz.

A variant of the above concept is being followed by the RADES~\cite{Melcon:2018dba,Desch:2221945,Fischer:2289074} team also in the CAST magnet. In this case the long-aspect-ratio cavity is realised by physically appending many smaller rectangular cavities, interconnected by irises, in what resembles a RF filter structure. The precise geometry of the cavity can be optimised to obtain maximal coupling to the axion field for a given resonant mode, or alternatively to simultaneously share it among several modes at different frequencies~\cite{Melcon:2018dba}, which opens interesting possibilities. This approach allows to (in principle, arbitrarily) enlarge the volume of the effective cavity with a much better control of mode crowding. 
Moreover, the need for external phase matching is avoided as it is guaranteed by design. At the moment, the obvious disadvantage is tuning over a range range of masses, which is to be overcome by building a large number of cavities. 
A first small-scale fixed-frequency 8.4 GHz prototype of only 5 poles, shown in Fig.~\ref{fig:long_cavities_photos}, is being now tested inside the CAST magnet, with plans to progressively instrument larger volumes. This technique seems best suited for frequencies slightly above those of CAST-CAPP and with similar sensitivity for equal instrumented $V$. 

The RADES approach can also be seen as a large $V$ cavity with an internal structure (a set of irises in this case) to appropriately shape the modes to maximize $\geo_m$. In this respect it is similar to a category of efforts in the community that try to overcome the problem of effectively coupling a large cavity to the ALP DM field for $V \gg\lambdaa^3$. It is worth noting that the basic idea was already anticipated in an early work by D. Morris and presented in an unpublished preprint~\cite{Morris:1984nu}.


The fundamental issue with a large $V$ cavity is that the ALP DM field tends to couple to a high harmonic, which does not couple effectively to a homogeneous excitation. 
This situation is nicely explained already in a 1D cavity of length $z\in (0,L)$ and TE mode functions ${\cal E}\propto \hat {\vec x}\sin(k_n z)$ with $k_n=\pi n/L$, $n\in 1,2,...$ The ALP DM field resonates with a mode when the condition $m_a=\pi n/L$ holds, and in the large volume limit this mode has a large number of nodes $n=L m_a/\pi \gg 1$ inside the volume. 
The geometric factor \eqref{geometricfactor} for a homogeneous ALP DM field and $\vec B_e$ along $\hat {\vec x}$ is,   
\be
\label{geo1D}
\geo_n = \frac{1}{B_0 L}\int_0^L  d z \,  {\vec {\cal E}}_m(z)\cdot \vec B_e(z)
=  \frac{1}{B_0 L}\int_0^L  d z \{\sqrt{2}\sin(k_n z) \} B_e(z), 
\ee
which gives $|\geo|^2 = 8/(\pi^2 n^2)$ for odd-$n$ if the $B_e$ field is constant. 
The coupling of high-$n$ modes is suppressed because 
$E$-field crests and valleys cancel by pairs in the integral \eqref{geo1D} leaving, at most, half an oscillation to couple to the axion DM field.  

\begin{figure}[htbp]
\begin{center}
\includegraphics[width=\textwidth]{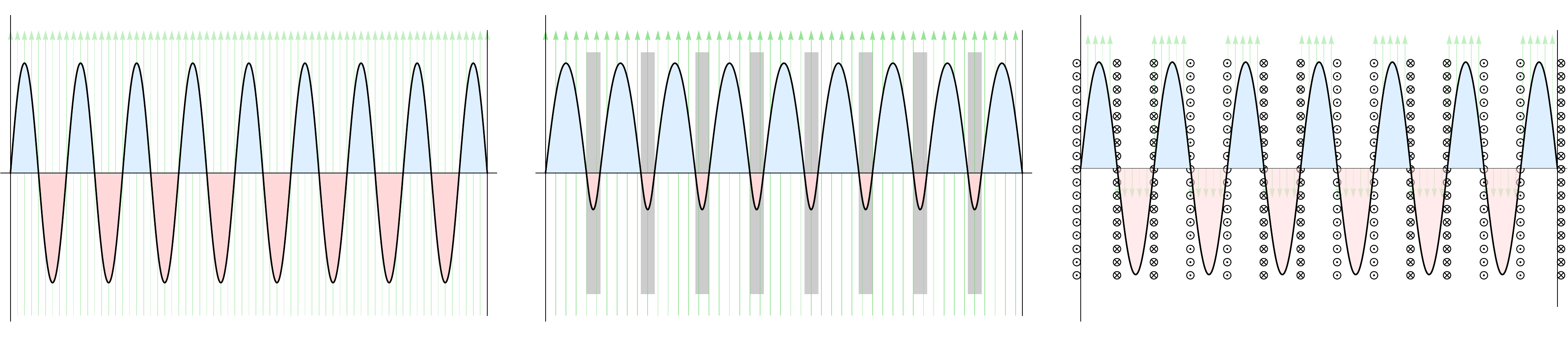}
\caption{The geometric factor of an ideal 1D cavity in a homogeneous $B$-field (green arrows) cancels between crests and valleys of a high mode (left). The cancellation can be avoided by placing high-$\n$ dielectrics --grey regions-- in the valleys (centre) or by alternating the polarity of the external $B_e$ field to track the mode variations (right). This case can be done by introducing wire planes with suitable currents~\cite{Rybka:2014cya}.}
\label{fig:highVhighQ}
\end{center}
\end{figure}

This suppression can be overcome by introducing in the cavity some structure of order $\lambda_a/2$ 
that allows the background axion field to couple effectively to high-frequency modes. 
There are two types of structures discussed: 1) periodic dielectric insets that decrease $E$-fields where located thus improving $\geo_m$~\cite{Morris:1984nu}, and 2) alternating the polarity of the B-field to adapt it to a high mode of the cavity~\cite{Sikivie:1985yu}. Schematically they are both explained in Fig.~\ref{fig:highVhighQ}. 
In the first concept one introduces dielectric slabs, ideally one covering each valley of the $E$-field, see Fig.~\ref{fig:highVhighQ} (center). 
In the dielectric regions, the $E$-field amplitude is smaller by a factor of $1/\n$ and the wavelength is shortened by the same factor so that, now, crests and valleys do not cancel perfectly, 
\be
\int_{\rm crest} dz  \sin\(\frac{n \pi}{L} z\)+\int_{\rm valley} dz \frac{1}{\n}\sin \(\n \frac{n \pi}{L}z\)=1-\frac{1}{\n^2}, 
\ee
and the geometric factor reaches a finite value at large $m$, $|\geo|^2 \sim  8/(\pi^2)(1-1/\n^2)\sim 1$. 
Note also that with the dielectrics, the resonant frequency, given by the wavelength of the vacuum oscillation, becomes smaller for the same resonator length. 
 
In the second concept, the $B_e(z)$ field polarity is modulated to follow the $E$-field $z$-dependence of the desired mode. Sikivie already suggested the use of superconducting wires embedded in a material transparent to RF radiation~\cite{Sikivie:1985yu}.
The concept has been recently revived~\cite{Rybka:2014cya} as an RF Fabry-Perot resonator with planes of superconducting wires in the nodes of a mode to alternate the polarity as sketched in Fig.~\ref{fig:highVhighQ} (right). The ORPHEUS demonstrating setup~\cite{Rybka:2014cya} consisted of a half Fabry-Perot cavity made of two 15 cm diameter Al reflectors (one flat and the other with a radius of curvature ROC$=33$ cm) and 8 frames where copper wires were wound perpendicularly to the resonator axis. Distances between the wire frames was sought to maximise the resonator $Q$, corresponding to frames placed in the nodes of the used TEM$_{00-19}$ mode. The quality factors were promising, $Q\sim 10^4$ at 18 GHz, but the $B_e$-field only of a few G. 
An interesting aspect of high-V resonators is that the quality factor can in principle increase due to the smaller ohmic losses per cycle, since effectively a photon spends more time inside the cavity and less bouncing off the walls. This suggests an increase of $Q$ proportional to the mode number $n$. 
A clear path to increase sensitivity requires increasing the $B$-fields, $Q$-factors, volume and including low-noise detectors. Magnetic fields up to 3 T could be achieved with planes of 0.4 mm spaced wires carrying 470 A and higher with a higher wire densities. Photolithographically patterned wires with 60 $\mu$m spacing and 144 A could reach 6 T reaching the critical current of NB$_3$Sn. Besides controlling the forces between wires, a scaled up version of the apparatus could be easily be afflicted by diffraction losses so much larger values of $Q$ are not guaranteed.


The original idea of a periodic dielectric loaded resonator of Morris is been revisited by ADMX in the Electric Tiger setup~\cite{electrictiger}. One of the first issues that one encounters in the conceptual design is the need to tune the distance between the dielectrics since the whole cavity is tuned to different frequencies. 
Ideally, the dielectrics have to be tuned to be equidistant with good accuracy. The Electric-Tiger setup consists of a waveguide where the dielectric plates (currently three nylon blocks) are mounted on to the central knobs of a scissor-jack, $><><>...<><>$. By actuating on the jack, the distances between dielectrics can be extended proportionally. 
The last one is placed at a $\lambda$/4 distance to the cavity wall (which we have not shown in Fig.~\ref{fig:highVhighQ}) to reduce the surface currents in the mirror and improve the quality factor~\cite{Morris:1984nu}. 
The still preliminary setup works in the 4-8 GHz range with a limited $Q\sim 10^2$ and is being used as testbench for future designs on the tuning mechanism and mode identification. Also an ALP DM search with moderate $\sim 10^{-10}$ GeV$^{-1}$ sensitivity is foreseen.    

Another problem at high masses is mode crowding, which sheds doubts on the identification of the tracked mode and its coupling to the axion DM field. A photonic band gap is an open lattice with a defect that traps some TM mode while allowing TE modes to be radiated away. Implementing this idea in high-mass resonant cavities would ease the frequency scanning. 
Currently it forms part of the set of R\&D concepts of the HAYSTAC collaboration. 
A related proposal to perform LSW experiments and DM searches with hidden photons and ALPs can be found in~\cite{Seviour:2014dqa}. 
Another idea to increase Q at high frequencies is to use distributed Bragg resonator (DBR) concepts to decrease E-fields at surfaces (like the $\lambda/4$ plate of Morris). Strategically-placed sapphire inserts have been used to achieve room temperature $Q\sim 650,000$ of a TE mode at 9.0 GHz in reference, at the expense of reducing the volume~\cite{Flory1997}. 

\todonote{MW detection section; Chou Patras, etc... Qnondemolition}

\subsection{Dish antenna and dielectric haloscopes}
\label{sec:dielectric_haloscopes}

As has been explained above, the conventional resonant cavity technique becomes increasingly challenging at higher frequencies, and this has inspired alternative proposals attempting to instrument large volumes $V \gg 1/m^3_a$, i.e. to decouple the detector volume $V$ from the frequency of operation. 
The approaches commented in the previous section represent an extrapolation of the haloscope concept in the sense that it deals with new types of cavities with large $V$ \emph{and} large $Q$. In all these approaches, the idea of a resonant cavity is kept and one seeks to obtain $Q$'s as high as possible. 
\exclude{(even if at higher frequencies and more complex structures, the achievable $Q$ could in principle increase but the adjustment/tuning of the modes becomes more critical and high values could be very difficult to achieve.
}

\subsubsection{Dish Antenna}
A second, completely radical approach to high-mass axion DM searches gives up the resonant enhancement altogether compensating with \emph{very} large volumes. Giving up a resonant structure also opens up the possibility of considering a broadband axion receiver. This concept was proposed in the very early days by P.~F.~Smith~\cite{Smith:1987kz}  and recently rediscovered in the \textit{dish-antenna} concept~\cite{Horns:2012jf}. 
Finally, a hybrid concept between the large-$V$/large-$Q$ and the huge-$V$/no-$Q$ is the \emph{dielectric haloscope} recently proposed~\cite{TheMADMAXWorkingGroup:2016hpc,Millar:2016cjp},  the main subject of the MADMAX collaboration. 

\begin{figure}[!t]
\begin{center}
\raisebox{-0.5\height}{\includegraphics[width=0.3\textwidth]{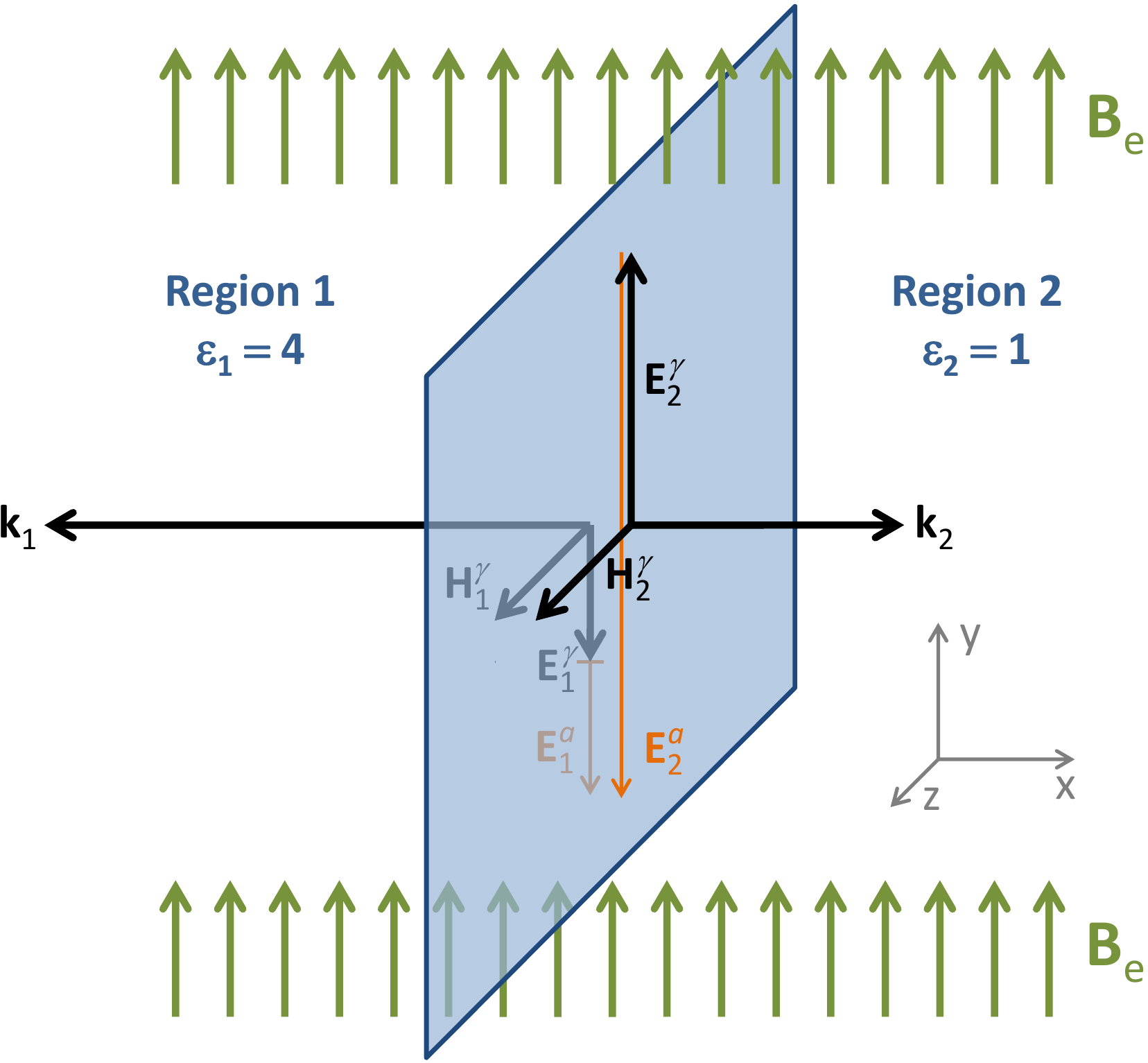} \hspace{1mm}}
\raisebox{-0.5\height}{\includegraphics[width=0.62\textwidth]{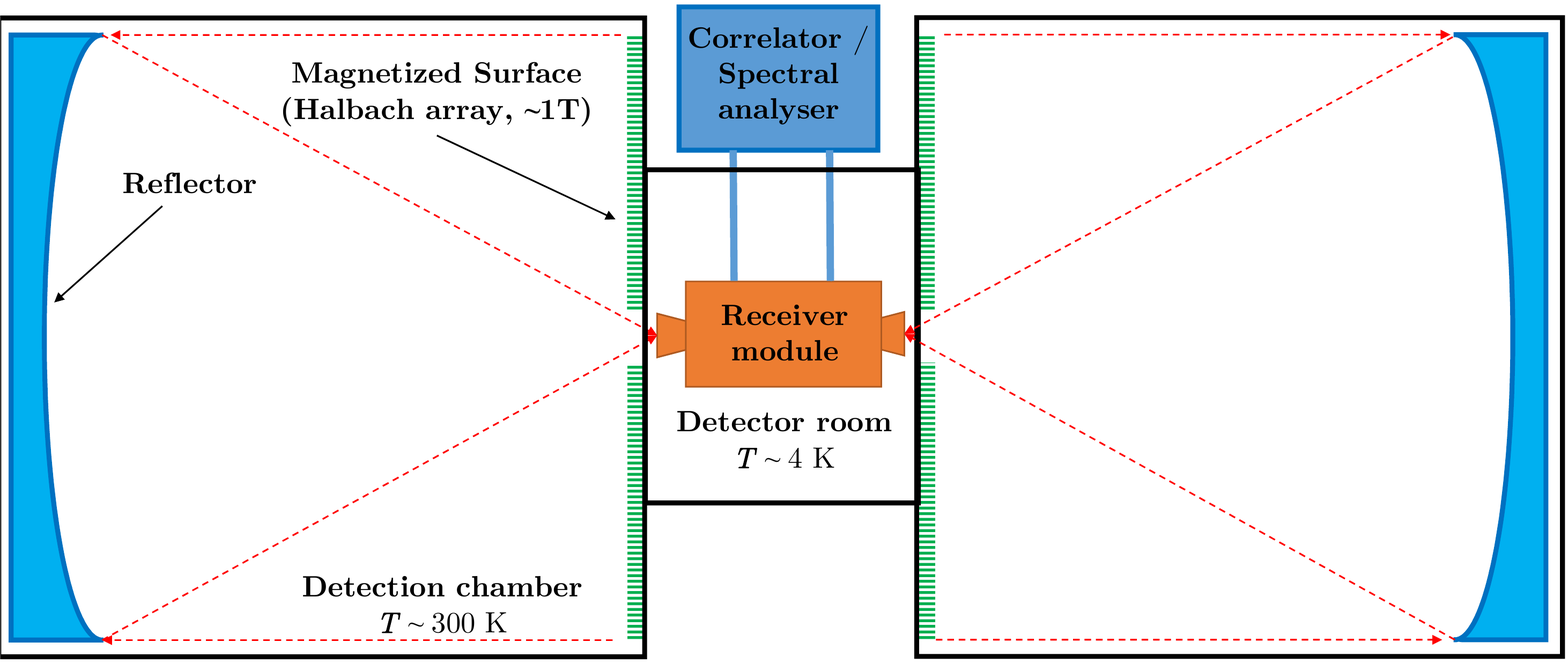}}
\vspace{-1cm}
\caption{Left: matching of the axion-induced $E_a$-fields across a boundary with outgoing EM waves, from~\cite{Millar:2016cjp}. Right: Sketch of the implementation of the dish antenna concept as followed by the BRASS experiment. Adapted from~\cite{brassweb}.}
\label{fig:dish}
\end{center}
\end{figure}

The basic picture behind the dish-antenna concept is slightly modified with respect to the standard understanding of haloscopes. The key point is to realise that photons are emitted by reflective (or refractive) surfaces embedded in a magnetic field and the DM axion field. To understand this, let us go back to the Maxwell-ALP equations and consider a locally homogeneous $\vec B_e$ field in the background of a homogeneous axion DM field, $a(t)$. An obvious solution of the Maxwell-Axion equations \eqref{Gauss}-\eqref{Faraday} is  
\be
\label{Eafield}
\vec E_a(t) = -\gagamma {\vec B}_e a(t), 
\ee
i.e. the axion DM field induces a homogeneous electric field in the $\vec B_e$ volume. Since the axion DM field oscillates at frequency $\omega\sim m_a$, so the axion-induced $E_a$ field does. 
This electric field fits very nicely in the photon-ALP mixing picture we discussed around equations \eqref{ALPlike}-\eqref{photonlike}. Indeed, in such a $B$-field it does not make sense to consider a purely ALP wave, because the freely propagating fields are just ALP-like. An ALP-like wave with no (or small) momentum, see \eqref{ALPlike}, has the electric field given by \eqref{Eafield}. The amplitude is model dependent, as it depends on the unknown combination $\gagamma a(t)$, however, for QCD axions, this combination is independent of the axion mass and only dependent on the axion DM fraction and the model-dependent O(1) coefficient $C_{\A\gamma}$ by virtue of \eqref{axionmass}, \eqref{axionphotoncoupling} and \eqref{thetaosci}, 
\be
|\vec E_\A| = \gagamma |{\vec B}_e| \A(t) = 
\frac{\alpha C_{\A\gamma}}{2\pi} |{\vec B_e}| \theta(t) \sim 1.6\times 10^{-12} \, \frac{\rm V}{\rm m}\, |C_{\A\gamma}|\, \,  \frac{|\vec B_e|}{10\, \rm T}.  
\ee
In a medium with linear dielectric or conducting properties, the induced field decreases due to polarisation and free-charge currents~\cite{Millar:2016cjp}, 
\be
\vec E_a(t) = -\frac{\gagamma {\vec B}_e a(t)}{\epsilon} , 
\ee 
where $\epsilon$ is the generally complex dielectric constant. 
Therefore, the ALP-like waves have different values of their associated $E_a$ fields in different media. 
This would lead to a discontinuity of fields at their boundary.  
However, the continuity of parallel Electric and Magnetic fields, $E_{||}, H_{||}$, across boundaries is ensured by the constraint part of Maxwells equations, Faraday \eqref{Faraday} and Magnetic Gauss law \eqref{GaussB}, which is not modified in the presence of ALPs and thus it must be respected. 
Therefore, continuity of the $E_a$ fields does not allow a pure ALP-like wave crossing the boundary between magnetised media to solve Maxwell's equation. Photon-like waves are needed to match the boundary conditions and are emitted 
from the surface towards both media. The matching of  $\vec E_a$ fields and the $\vec E,\vec B$ fields of the photon-like wave in the interface of two media with $\epsilon_1,\epsilon_2$, respectively is shown in Fig.~\ref{fig:dish} (left), from~\cite{Millar:2016cjp}. 
A first approach to this picture was outlined in~\cite{Horns:2012jf}, which discussed the emission from the boundary between a metallic mirror and vacuum where $\epsilon=1$. 
Inside the mirror $\epsilon$ would be large and imaginary so $E_a\to0$ and the continuity of $E_{||}$ at the surface is ensured by the emission of an outgoing photon-like wave of amplitude $E_a$, given by \eqref{Eafield}. The cancellation of $E_{||}$ at all times implies that the photon frequency is given by the axion's, $\omega= \omega_a=m_a+...$ Photons are emitted from magnetised mirrors! 
In a sense, the phenomenon resembles the ordinary reflection of an electromagnetic wave by a mirror, where  
a reflected wave cancels the total $E_{||}$ at the boundary. The interesting difference with the ALP DM case is that the ALP-like wave is reflected into a photon-like wave\footnote{An ALP-like wave is reflected as well but is terribly small. }, which we can detect. 
The concomitant Snell's law is slightly different and somewhat surprising. The DM ALP wave has wavenumbers much smaller than its frequency, but photons in vacuum have $\omega=k_\gamma$, therefore the photon momentum can only come from the boundary itself and photons are emitted perpendicular to it\footnote{Indeed a recent publication shows that the emission can be understood in a broad sense as transition radiation from an ALP crossing an interface~\cite{Ioannisian:2017srr}. 
The classical treatment shown here translates directly to the quantum level where the ALP/photon waves are wave functions. 
\exclude{The fact that Ionisian allows himself to call this phenomenon transition radiation might redden a distinguished ERC referee that accused one of the authors of not knowing his standard physics. We sincerely apologise, but we got the worst part.} }.
Indeed, the emitted photons will carry a small perpendicular component $\sim {\cal O}(\dmsigmav)$ because the axion DM momentum along the surface is conserved. This can cause a smearing of the signal in the centre of the sphere, but can be used to extract directional information~\cite{Jaeckel:2013sqa,Jaeckel:2015kea}, something the we comment on later in section~\ref{sec:directional}.

The region around the surface that needs to be magnetised is of order of half an oscillation $\lambdaa/2$ for the wave to be fully developed~\cite{Horns:2012jf}. Optimal coupling happens when the $\vec B_e$ direction lies along the surface. 

If the surface is a spherical mirror dish, and provided its size is much larger than the photon wavelengths, the photons are all concentrated in the centre of curvature of the dish
\exclude{, see Fig.~\ref{fig:dish}}.  
The power emitted from unit area by a generic ALP or the axion is given by 
\bea
\label{dishpower}
P_{a}/\area  &=& \frac{1}{2}E^2_a =  3.3\times 10^{-27} \frac{W}{\rm m^2} \(\frac{\gagamma}{2\times 10^{-14}\rm GeV^{-1}}\)^2
\(\frac{10^{-4}\,\rm eV}{m_a}\)^2\frac{|\vec B_e|}{10\, \rm T}^2, \\
\label{dishpower2}
P_\A/\area &=&   2.2\times 10^{-27} \frac{W}{\rm m^2} |C_{\A\gamma}|^2\(\frac{|\vec B_e|}{10\, \rm T}\)^2 . 
\eea
Comparing these numbers with the haloscope formula using $V\sim \lambda_a^3$ we find, 
\be
\label{dishvshaloscope}
\frac{P_{\rm dish}}{P_{\rm haloscope}}\propto \frac{m_a^2 \area}{Q}, 
\ee
which transparently shows the source of the enhancement in each case. Assuming $Q\sim Q_a$ could be achieved, 
a dish with a magnetised area of $\area\sim 1 $ m$^2$ would compete with the corresponding haloscope for $\la \sim $ mm, corresponding to $m_a\sim 2\times 10^{-4}$ eV. The lack of resonance can be compensated by area of the dish, $\area$. This concept is particularly appealing for very high axion masses which are very difficult to access with cavities. Unfortunately, the sensitivity of neither of them would be good enough for  QCD axions with $C_{\A\gamma}\sim 1$. 

The dish setup has several advantages. First, it is sensitive to the whole mass range at once (as long as the mirror is reflective at all frequencies) and therefore broadband searches are possible. \modified{In practice, the bandwidth will be mostly limited by the receiver. }
Furthermore, having the detector in the centre of curvature of a spherical mirror, environmental backgrounds can be 
quite suppressed as far away sources are focused onto the focal point at half ROC. 
Moreover, a perfect mirror does not radiate thermally, so the black body radiation onto the detector is supressed by 1-reflectivity, which can be quite small. In the idealised version of the setup, just the detector is able to radiate thermal photons that can later return and be detected as background. Therefore, only the detector needs to be at cryogenic temperatures in principle.

An implementation of this concept, the BRASS experiment~\cite{brassweb} will take place in the U. of Hamburg. 
Preliminary estimations~\cite{Horns:2012jf,brassweb} suggest the need of $B_e^2 \area = 100$~T$^2$m$^2$ to reach sensitivity to the most optimistic QCD axion models. Despite the simplicity of the concept, the need of embedding the dish in a magnetic field poses practical limits to its scaling. 
The current setup consists of a planar Halbach array of permanent magnets to which magnetises a large surface $\sim 8$ m in radius with a parabolic mirror to focus the radiation into a detector. The correlation length of the $B_e$-field is of the order of cm, so ALP DM with masses leading to $\lambdaa/2\gg$ cm ($m_a<0.6\times 10^{-4}$ eV) will be radiated at a smaller rate than \eqref{dishpower2}. 
The Halbached surface and mirror are in a detection chamber at room temperature. In fact, the setup is doubled, increasing the signal and allowing to study signal correlations, see Fig.~\ref{fig:dish} (right). It is worth noting that this concept, without magnetic field, has already being used to search for hidden photon dark matter in the Tokyo dish experiment~\cite{Suzuki:2015sza}, and continues to be developed by the FUNK experiment~\cite{Experiment:2017icw} in Karlsruhe.

\subsubsection{Dielectric haloscope}

\begin{figure}[!t]
\begin{center}
\raisebox{-0.5\height}{\includegraphics[width=0.45\textwidth]{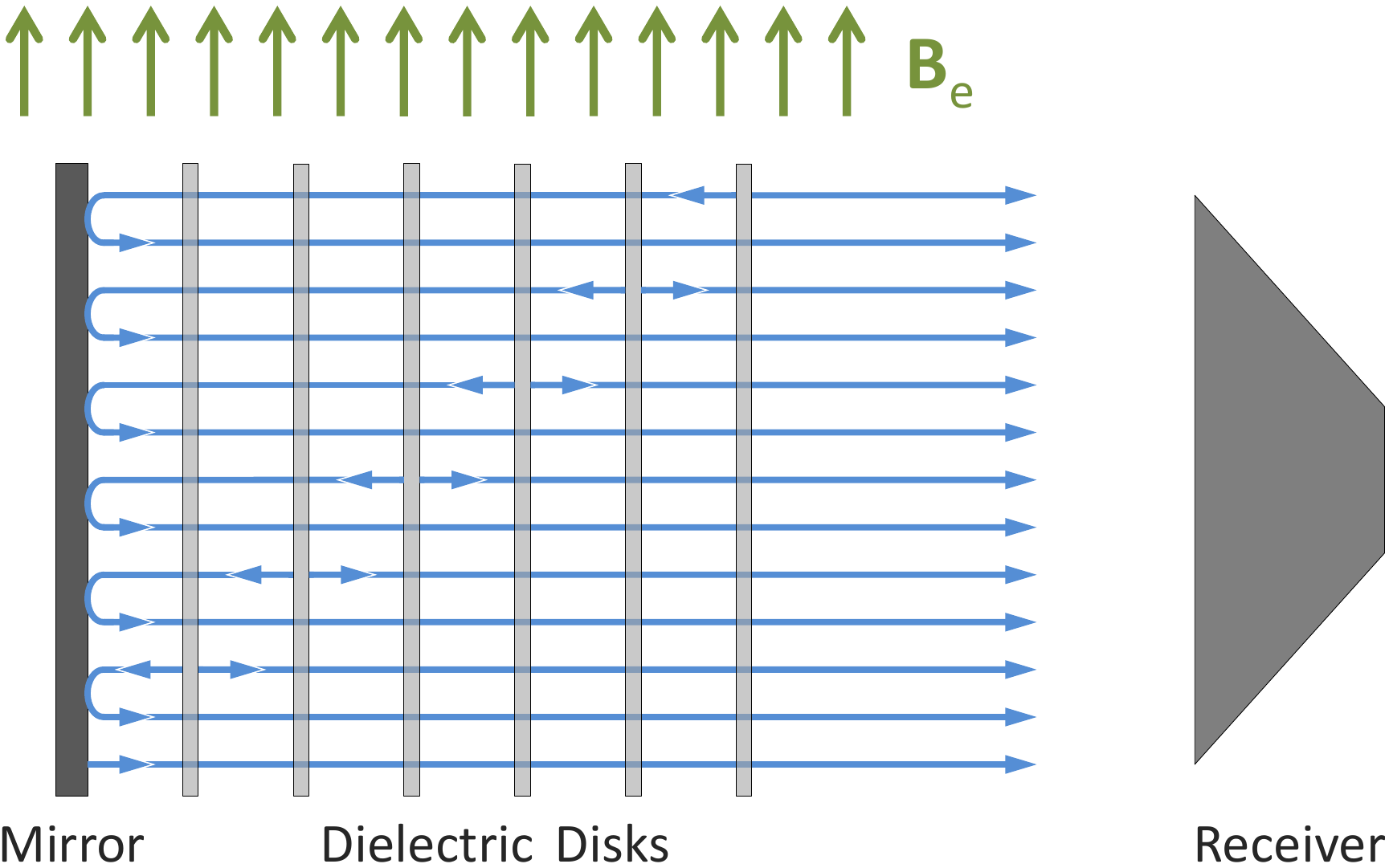}}
\raisebox{-0.5\height}{\includegraphics[width=0.5\textwidth]{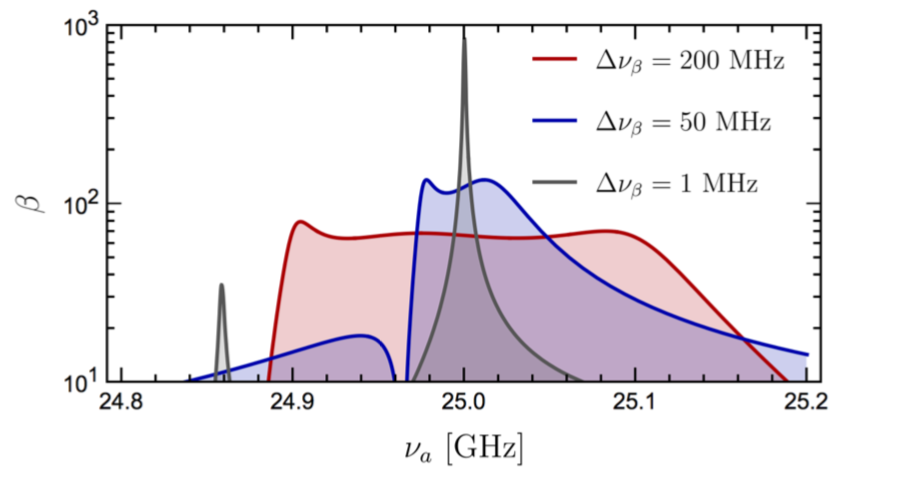}}
\caption{Left: sketch of the dielectric haloscope experiment. Photons in the $B_e$ field are emitted from the dielectric surfaces and reflected in the leftmost mirror and other surfaces to be measured coherently by a receiver, from~\cite{TheMADMAXWorkingGroup:2016hpc}. 
Right: Adjusting the distances between the layers, the frequency dependence of the boosted sensitivity can be adjusted to 
different bandwidths, from~\cite{MADMAXWP}.
}
\label{fig:dielectrichaloscope}
\end{center}
\end{figure}

The dielectric haloscope is an evolution of the dish concept with the objective of increasing even further the emitting area and being able to use a more intense $B_e$ field more efficiently. A sequence of mirrors would lead us back to resonant cavities, so instead dielectric slabs are proposed in addition to a mirror surface~\cite{TheMADMAXWorkingGroup:2016hpc,Millar:2016cjp}. A lossless dielectric slab of $\lambda/2$ is completely transparent to radiation but it still emits EM waves from axion DM if placed in a magnetic field by the same logic outlined for the mirror case~\cite{Millar:2016cjp}. In this \emph{transparent} mode, the radiated EM wave can be enhanced by a factor of $N_d^2$ where $N_d$ is the number of disks, if these are placed in such a way that the emission from the different interfaces are summed coherently. A mirror at the far end reflects the left-going waves back to interfere with the right going. 
The setup would be very similar to Fig.~\ref{fig:highVhighQ} with an open end, see Fig.~\ref{fig:dielectrichaloscope} (left).  
The transparent mode of a dielectric haloscope (with mirror) is absolutely not a resonator but the power emitted is boosted by factor
\be
\beta^2 = \frac{P_{dh}}{P_{\rm dish}} \sim 2 N_d^2 , 
\ee
which can easily cover the sensitivity boost required by the dish antenna idea to reach the QCD axion. 

This innocent idea could be thought extremely unpractical for a scanning experiment. The reason is that dielectrics become partially reflecting once used at frequencies away from $\lambda/2$. The boost factor enhancement is therefore frequency dependent. 
However, the reflections between dielectrics will generate small cavity enhancements and the final boost factor can even increase. The theory of such dielectric haloscopes has been discussed in~\cite{Millar:2016cjp} where a transfer matrix formalism is developed to easily compute the boost factor in the idealised 1D setup. 
In the same publication the authors managed to establish a clear connection between the boost factor and the overlap integral between the $E$-field distribution induced by an EM wave shone in from the open end and the external B-field, i.e. the geometric factor that we introduced in the context of closed resonators \eqref{geometricfactor}. This connection was developed even further in~\cite{Ioannisian:2017srr} where the modes of a simple open resonator are identified with Garibian wave-functions. From a practical point of view, the dielectric haloscope turns out to be quite flexible in the end. 
There is indeed a sort of area-law that considers different configurations of distances between the dielectrics that states that the area below the frequency dependent $\beta^2$ is just proportional to $N_d$. One can thus add more dielectrics to increase the boost factor or the bandwidth by playing with the distances, see Fig.~\ref{fig:dielectrichaloscope}.

The flexibility can be key to a successful experiment. Dielectric haloscopes are relatively complicated setups where a large number of dielectrics will need to be adjusted to get an optimal boost factor. It turns out that including a nonzero re-tuning time $t_R$ in the optimisation of the figure of merit, the optimum strategy is not a boost factor (quality factor) as large as possible, which would reach the required sensitivity in as small as possible $\Delta t$, but the one that corresponds to $\Delta t =t_R$~\cite{Millar:2016cjp}. 

In the computation of $\beta$, the effect of the coherent sum of layer emission and the resonant effects are mixed, but one can understand it in the language of conventional haloscopes \eqref{FOM} as a sort of $\beta^2 \sim |\geo|^2 Q$. In the transparent mode, the geometric factor is large and $Q$ is 1 but at different frequencies $Q$ increases.  
Admittedly this experimental arrangement seems very similar to the resonant cavity filled with periodic dielectric structure to allow resonance at higher frequencies~\cite{electrictiger}. 
The difference is that here the focus is on $V$ rather than on $Q$, and indeed the structure proposed in~\cite{TheMADMAXWorkingGroup:2016hpc} is a very poor resonator. In this sense, the dielectric haloscope concept is closer to the dish antenna concept than of the conventional haloscope.

\begin{figure}[!t]
\begin{center}
{\includegraphics[width=0.7\textwidth]{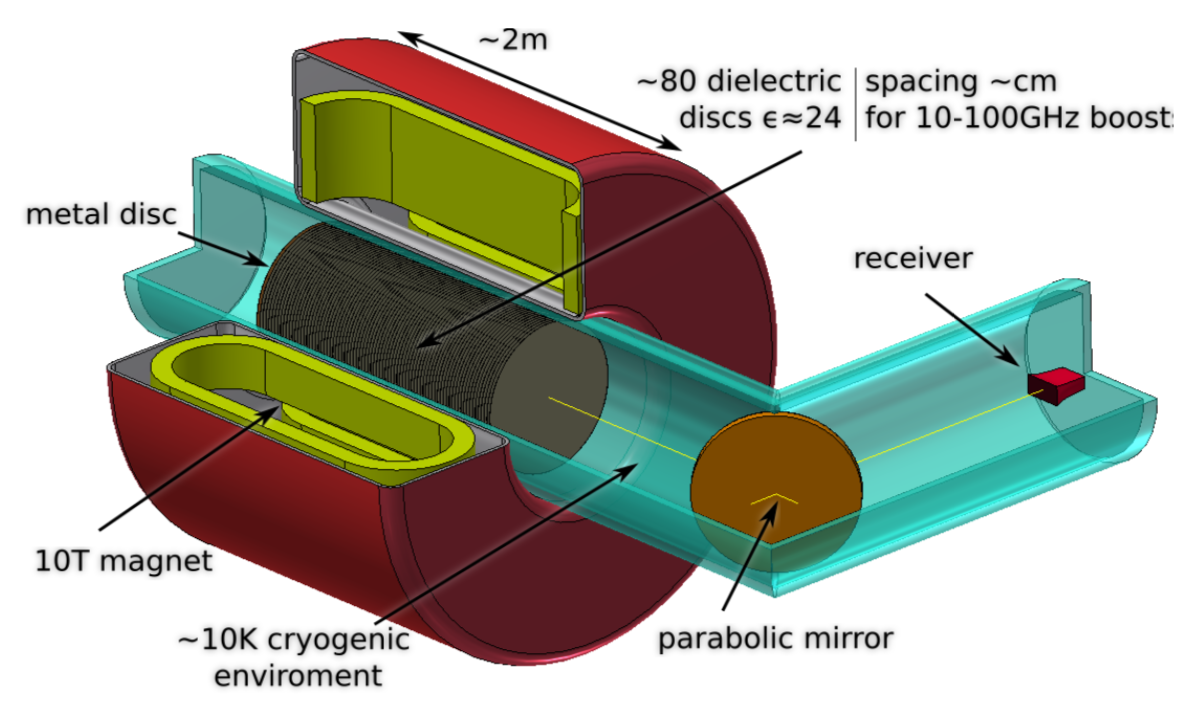}}
\caption{The concept of the MADMAX experiment, see text for details. From~\cite{MADMAXWP}.
}
\label{fig:madmax}
\end{center}
\end{figure}

The MADMAX collaboration~\cite{TheMADMAXWorkingGroup:2016hpc} aims at building a 80-disk system of $\area \sim 1$~m$^2$ inside a 10~T magnet. 
An scheme of the setup is depicted in Fig.~\ref{fig:madmax}. Photons from ALP DM conversion are emitted from the mirror-and-disk region where an intense $B_e$ field is created and then focused to a RF detector. 
With an expected $\beta \sim 5\times10^4$ and equipped with a quantum-limited receiver, should be able to scan the 40-400~$\mu$eV mass range with sensitivity down to \DFSZ axions as shown in Fig.~\ref{fig:halo_sens}. Dielectrics with high dielectric constants are preferred as they can increase the $Q$ enhancements and they emit more efficiently even in the transparent mode. Any candidate for the material must also have very small losses, should be non magnetic, mechanically stable, adequate for cryogenic environments and affordable. The baseline MADMAX choice is LaAlO$_3$, which satisfies the most critical issues of the previous list, especially a large $\epsilon\sim 24$. Sapphire is a safer option although has only $\epsilon\sim 10$. Even larger $\epsilon$ could be obtained for instance with rutile (TiO$_2$). 
Crystals of 60 cm diameter to get the required $\area \sim 1$~m$^2$ have never been grown in any of those materials so a program to study cutting and tiling smaller pieces and checking their properties in terms of losses and diffraction has started. 
The two perhaps more critical aspects of the project are the scanning by tuning the distances between the dielectrics and the design of the powerful magnet. Preliminary 1D simulations show that a precision of $\sim 20\mu{\rm m} \sqrt{100/\beta}(100\mu{\rm eV}/m_a)$ is required to avoid modifications of the boost factor at the 10\% level. 
An automatic tuning mechanism is being developed in a small demonstrative setup at the Max Planck Institute Munich with 20-cm diameter Sapphire disks.  
A first rough adjustment is performed first and and smaller corrections are performed while measuring reflectivity and phase delay of a wave send from the detector side. Distances between the mirrors are adjusted to minimise a $\chi^2$ between the measurements and a 1D lumped element circuit model. Disks are moved by precision pico motors with 100 nm precision. First results are very promising but diffraction losses into empty space are noticeable and expected to play a much smaller role in the 60 cm diameter version. The collaboration also considers optical interferometry measurements to feedback on the tuning of the apparatus. The thickness of the layers defines an O(1) frequency range where the dielectric haloscope can work because at a frequency corresponding to $\lambda$ inside the disks, the EM wave radiated by axion DM vanishes. To cover this gap in sensitivity MADMAX would surely need at least two thickness of disks to cover the whole projected mass range. The collaboration is however mostly motivated by the predictive \emph{post-inflation} scenario where $m_\A\sim 100\mu$eV seemed the best prediction for the axion mass~\cite{Ballesteros:2016xej} until the recent claims of $26$ $\mu$eV~\cite{Klaer:2017ond} so the thickness of the first set can be optimised to one of these values.  The magnet design shown in Fig.~\ref{fig:madmax} is by no means definitive. Indeed a $\area \sim 1{\rm m}^2$ aperture dipole field with $B_e\sim 10$ T has never been attempted and turns out to be quite challenging.  A design study with the participation of the magnet division of CEA/Saclay and Babcock Nell has started. On the detector side, the collaboration has already designed the acquisition chain with the required specifications for a sufficiently broadband acquisition using Low Noise Factory\footnote{\url{www.lownoisefacory.com}} HEMT amplifiers reaching $5-9$ K noise temperature in the relevant frequency range.   
The collaboration was officially created in October 2017 and plans to focus on magnet design studies and in a first intermediate setup with 20 30 cm disks within a small magnet of 3-4 T while continuing with R\&D on the LaAlO$_3$ disk option, tuning mechanics, 3D simulations of diffraction and thermal noise from side lobes. A full scale version of the experiment could happen around $2022$. 

\subsection{Low frequency resonators with LC circuits}
\label{sec:LC}

In the previous section we have explained how DM axions in an external homogeneous $\vec B_e$ field produce an electric field $\vec E_a$ oscillating with a frequency $\omega\sim m_a$. Together with the $\vec E_a$-field, at least two types of oscillating $B$-fields can be excited. First, if the axion DM wave is not completely at rest it induces a small intrinsic ${\vec B}_a$-field~\cite{Millar:2017eoc}, which can be calculated by applying Faraday's equation to the ALP-like wave \eqref{ALPlike}, 
\be
{\vec B_a}=\frac{1}{i\omega}\nabla \times {\vec E}_a= -\gagamma \vec v \times {\vec B}_e a
\ee
where $\vec v =\vec k/\omega$ is the axion DM velocity. As explained in sec.~\ref{sec:natural}, the velocity of the ALP field takes a coherence time \eqref{coherencetime} to change and sweeps values according to a velocity distribution like \eqref{Lentzdistribution} but the Sun orbital motion around the Galaxy ensures that a non-zero velocity is singled out on average, $v_{\odot}\sim 220$ km/s $\sim 0.7\times 10^{-3}$. 
This ${\vec B}_a$ field is therefore smaller than ${\vec E_a}$ by factor $\sim \dmsigmav \sim 10^{-3}$ in the DM field. 
On the other hand, the axion-induced $\vec E_a$-field can produce a current in a conductor or similar that induces a new 
$B$-field. For instance, in the dish antenna concept, EM waves are radiated off the disk and they feature $B$-fields of size $|{\vec B}|=|\vec E_a|$.  Regardless of its origin, the small oscillating magnetic field could be measured by a carefully placed pick-up coil and associated amplifying $LC$ circuit. 
The amplified signal can then be detected by a sensitive magnetometer like a SQUID. 
The first proposal by Sikivie~\cite{Sikivie:2013laa} considered measuring the small intrinsic $B$-field, while further ideas were presented to measured the secondary $\vec B$ created by ${\vec E}_a$~ \cite{Chaudhuri:2014dla,Kahn:2016aff}. 
The signal strength depends on the magnetic flux going through the pick-up coil, which, for relevant configurations, and provided the axion wavelength is much larger than the dimensions of the magnet,  is proportional to $B_eV_{B_e}$, where $V_{B_e}$ is the total volume of the magnet. This method could achieve competitive sensitivity for very low masses $m_a \lesssim 10^{-6}$~eV, if implemented in magnet volumes of few~m$^3$ volumes and few~T fields. 
\todonote{limitations? Capacities? tunability?}

\begin{figure}[t!]
   \centering
   \includegraphics[width=0.9\textwidth,trim={0 2cm 0 2cm}, clip]{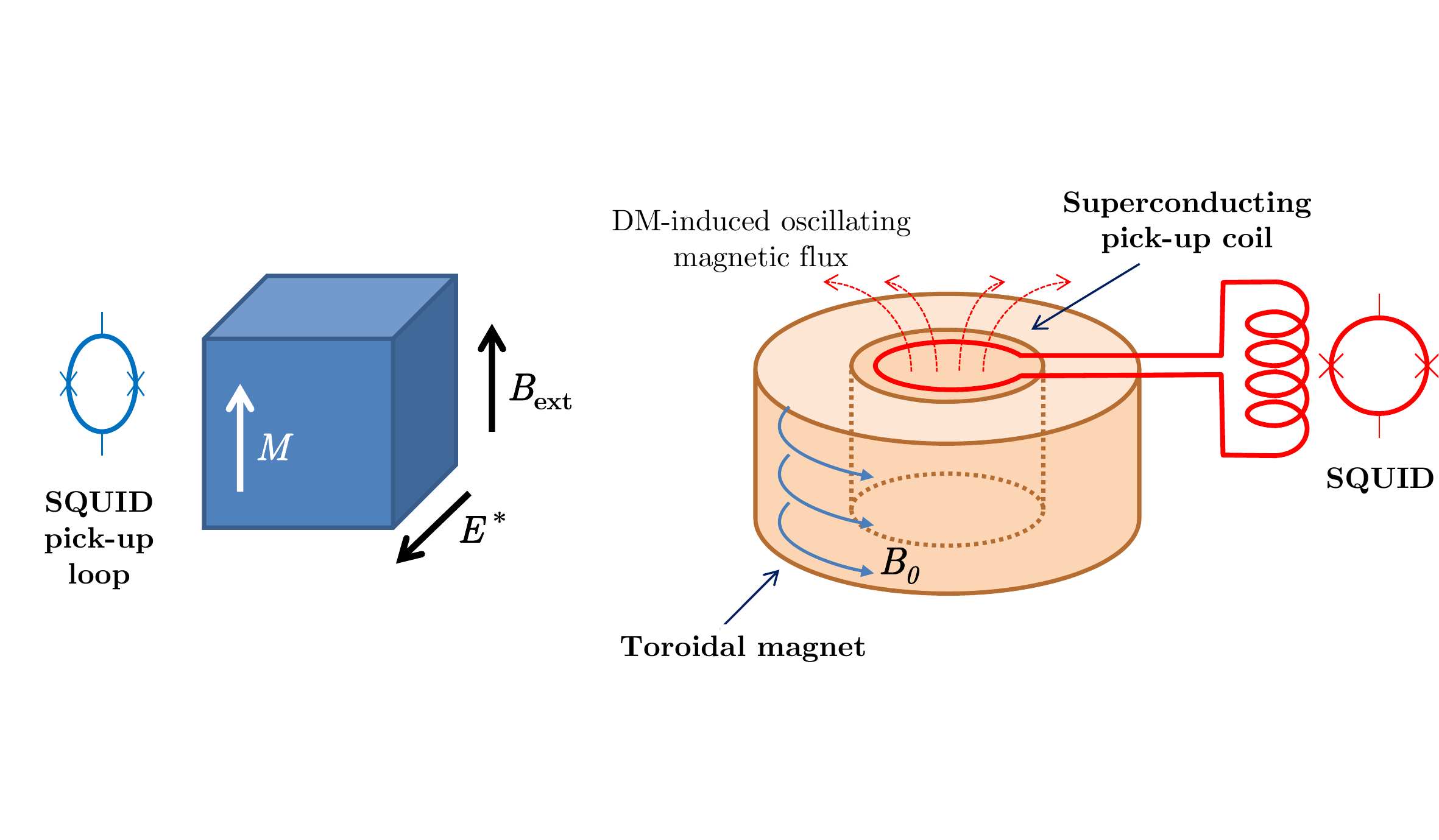}
   \caption{Left: Schematic of the detection of an oscillating nuclear EDM caused by ALP DM in the CASPEr-Electric experiment. Adapted from~\cite{Budker:2013hfa}. 
   Right: Schematic of an implementation of the pick-up coil in a toroidal magnet. The ALP DM field excites an oscillating $E_a$ field along the field lines of a static toroidal field $B_e$. The oscillating $E_a$ induces an oscillating $B_a$ field along the symmetric axis read by a pickup coil connected to a SQUID.  
    Adapted from~\cite{Kahn:2016aff}.}
   \label{fig:sketch_lc_nmr}
\end{figure}

Particularly appealing is the implementation of this concept in a toroidal magnet geometry providing a toroidal\footnote{In earnest, the solution $\vec E_a=-\gagamma {\vec B_e} a$ has been obtained for a homogeneous background $\vec B_e$ field and not valid for a toroidal field. 
It should be however a good approximation in the limit where the radius is much larger than $\la$.} oscillating ${\vec E}_a$ because the pickup coil can then be placed in the centre of the toroid, where the static background magnetic field is practically zero~\cite{Kahn:2016aff}. As proposed in~\cite{Kahn:2016aff}, the concept allows for a non-resonance (i.e. broadband) mode of operation, in which the coil is inductively coupled to the SQUID without a tunable capacitor, see Fig.~\ref{fig:sketch_lc_nmr}. This mode of operation has the advantage of probing large ranges in $m_a$ at once (something particularly useful also in the search for hidden photons~\cite{Chaudhuri:2014dla}), and is more efficient than the narrowband mode for lower axion masses. When in narrowband mode, the amplifying resonance is produced externally by the $LC$ circuit and not by mechanical modification of the cavity, which makes tuning in principle easier than in conventional haloscopes.

\additionalinfo{The paper \cite{Hong:2014vua} might contain issues}

Two teams are already developing experimental setups implementing this concept. The DM-Radio~\cite{Silva-Feaver:2016qhh} team is setting up a pathfinder experiment at Stanford University with a liter-scale detector, while preparations for a second stage with a 30L detector are ongoing. The ABRACADABRA~\cite{Kahn:2016aff} experiment at MIT is also preparing a small 10~cm prototype with plans to scale up. The sensitivity lines shown in Fig.~\ref{fig:halo_sens} to represent this technique are extrapolations to a large magnet of 5~T and 1~m$^3$, as well as more ambitious 20~T / 1~m$^3$ and 5~T / 100~m$^3$ combinations, taken from~\cite{Kahn:2016aff}. The latter suggests the implementation of this technique in a large toroidal magnet like the one foreseen for IAXO (see section~\ref{sec:helioscopes}). Of course, these projections rely on successful experience with the ongoing small scale prototypes, whose sensitivity (not shown in the figure) is not yet enough to reach QCD axions. Finally, there are plans at the University of Florida to explore the concept with the possibility of implementing this type of search in the AMDX magnet as part of the future program of the experiment~\cite{Battaglieri:2017aum}.

%

\subsection{NMR techniques}
\label{sec:NMR}

The fact that the gradient of the axion DM field couples to non-relativistic fermion spins like a ficticious magnetic field $\fB$, see \eqref{Bficticious}, which can be searched for with NMR techniques was already known since the 90s, see for instance~\cite{Krauss:1985ww}. We do not know why this idea was not pursued before, but very recently the idea resurged~\cite{Budker:2013hfa} and, what is more important, reached the ears of several groups of dashing experimentalists who have responded to the challenge. Something completely new to us is the idea of detecting the oscillations of nuclear electric-dipole-moments (EDMs) produced directly by axion DM~\cite{Graham:2011qk}, i.e. probing the same axion coupling that would reflect its relation with the Peccei-Quinn solution of the strong CP problem. This turns out to be extremely challenging, but certainly not impossible and has become a most exciting venue for testing axions.
Another novel aspect to discuss is a new experiment that would be sensitive to the axion DM coupling to the \emph{electron} spin, which was deemed impossible in a number of works including~\cite{Krauss:1985ww,Budker:2013hfa}, but not completely forgotten~\cite{Barbieri:1985cp} and is now one of the core proposals of the recently formed QUAX collaboration~\cite{Barbieri:2016vwg}.

\todonote{Oscillating electron EDM of Hill here}

\subsubsection{Oscillating EDMs}
If QCD axions comprise the DM of the Galaxy, the axion field would be oscillating as $
\theta(t)\sim \theta_0 +\theta_{\osci} \cos(m_\A t)$ where the amplitude $\theta_{\osci}\sim 4\times 10^{-19}$ is fixed by the local DM density, see sec. \ref{sec:natural}.
Protons and neutrons have EDMs proportional to $\theta$, see \eqref{dn_sumrule}, and thus they have oscillating EDMs copying the axion DM oscillations, e.g. $d_n(t)\simeq 0.0024 \times \theta_{\osci}\cos(\omega t)$ e fm. Detecting these oscillations can be much easier than detecting the static value $\propto \theta_0$ because one can use a resonant detector tuned to the axion natural frequency, $m_\A$.
Recently, P. Graham and S. Rajendran proposed to detect nEDM oscillations by searching for tiny atomic energy shifts~\cite{Graham:2011qk} in highly polarised cold molecules. The idea is that an external E-field, $E_{e}$,  can polarise molecules to develop large intrinsic $E$-fields, $E_{\rm int}$ at the position of the nucleus. The $\vec d_n(t)\cdot \vec E_{\rm int}$ interaction with the nEDM then causes the sought time-dependent small shifts.
Unfortunately, Schiff's theorem states that charges in an atom would reorganise in a way as to make the E-field vanish at the position of a charged \emph{point} particle, which would kill the effect.
The fact that nuclei have finite size allows to define a nuclear Schiff's \emph{effective} moment so that the interaction is still written as $\epsilon_S \vec d_n(t)\cdot \vec E_{\rm int}$ with a so-called Schiff's suppression factor $\epsilon_S<1$. They proposed an atomic interferometer with cold molecules including large-$Z$ atoms, RaH for instance, which have radii almost as large and negligible Schiff's supression.

In the basic setup, the nuclei have their spins polarised along one direction, given by an external B-field, $\vec B_{\rm e}$, and an external E-field setup at right angles. The spins precess at the Larmor frequency given by $B_e$.  The E-field polarises the molecules creating energy shifts $\pm \epsilon_S \vec d_n(t)\cdot \vec E_{\rm int}$ of typical size $10^{-24}$ eV. Standard NMR techniques are used to prepare a state combination of the $\pm$ shifted ones. When the Larmor frequency $ \omega_L = 2 \mu_N B_{\rm e}$ ($\mu_N$ nuclear \emph{magnetic} moment) coincides with the oscillations of the small EDM, the two states accrue a phase difference $\sim 2\epsilon_S \vec d_n(t)\cdot \vec E_{\rm int} t$ that under ideal conditions increases linearly up to the axion DM coherence time. 
Like in other haloscopes, the experiment has to be tuned to the axion mass to see a signal. Here, the magnitude of $B_{\rm e}$ is the variable element and is considerably simpler to tune than the cavity shape or dielectric positions in haloscopes.
Since $m_\A$ is unknown, the experiment has to scan over Larmor frequencies $\omega_L = 2 \mu_N B_{\rm e}\sim 2\pi {\rm MHz}(B_{\rm e}/0.1 \rm T)$ until a signal is found. 
The typical interrogation time in free-fall interferometer is limited to $\sim 1$ s, giving a tiny $\sim 10^{-10}$ phase difference, very hard to measure. A set of two interferometers separated by a distance smaller than the coherence time could be used to suppress noise. From the practical point of view, suitable nuclei have lifetimes $\sim$ day, which complicate the experiment even further.  

Although the proposed experiment seemed extremely challenging, it drove discussions that ultimately lead to other more feasible possibilities and new observables~\cite{Graham:2013gfa}. 
In two followup papers that appeared side to side~\cite{Budker:2013hfa,Graham:2013gfa} the same authors developed the idea of using NMR techniques to search for general ALP DM through ALP-fermion couplings.
They have generated several projects with the Cosmic Axion Spin Precession ExpeRiment (CASPEr) label, involving experimental groups in Mainz, LBNL and Boston U. 

The so-called CASPEr-Electric~\cite{Budker:2013hfa}, consists of a more feasible version of the cold-molecule detector of~\cite{Graham:2011qk} to detect the oscillating electric dipole moments of nuclei. A scheme of the experiment is displayed in Fig.~\ref{fig:sketch_lc_nmr}.
The high-$Z$ atoms are now set in a ferroelectric crystal, that possesses permanent electric polarisation fields $E^*$, avoiding the need for external $E$-fields to polarise the molecules ($E^*$ plays directly the role of $E_{\rm int}$ in the previous experiment). The setup considers such a crystal placed in an external B-field at right angles with $\vec E^*$ in the nuclei.
Nuclear spins are polarised along $\vec B_{ e}$, and precess at the Larmor frequency $\omega_L$. 
The interaction of the axion-induced nuclear EDM with the external $E^*$-field, $\epsilon_S \vec d_n(t)\cdot \vec E_{\rm int}$, is essentially equivalent to the interaction of the spin with a an oscillating $B$-field transverse to $B_{e}$, $\vec \mu_N \cdot \vec B_{\rm \perp}(t)$. 
The latter is well known to produce a resonant increase of the transverse magnetisation of the sample when $B_{\rm \perp}(t)$ oscillates at a frequency matching $\omega_L$, i.e. the nuclear magnetic resonance.
The same happens with the electric analogue!
The amplitude of the transverse magnetisation, $M_\perp$, will increase linearly  with time as 
\be
M_\perp \simeq (p \, n_N  \,  \mu_N) |d_n| E^* t ,
\ee
($p$ is the spin polarisation fraction, $n_N$ the spin density and $|d_n|$ is the amplitude of EDM oscillations), until some non-ideal effect cuts the resonance. This could be the axion coherence time $t_c\sim 10^6/m_a$ from \eqref{coherencetime}, the spin-lattice relaxation time $t_1$, or the transverse relaxation time $t_2$. 
The increase in $M_\perp$ at the sought frequency can be measured with a sensitive SQUID.
Compared with the previous version of the experiment, CASPEr will benefit from the coherent amplification of the signal driven by a macroscopic number of spins precessing coherently.
We emphasise that as a resonant measurement, the experiment has to scan over axion masses until a signal is found.
The maximum axion mass is limited by the largest feasible values of $B_{e}\sim 30$ T, giving $m_a\lesssim \mu$eV.
\exclude{Different crystals are identified as candidates, PbTiO$_3$, seems an excellent option, giving $E_{\rm int}\sim 3\times 10^8$V/cm, but also Pb$_5$Ge$_3$O$_{11}$, PbZr$_y$Ti$_{1-y}$O$_3$, CdTiO$_3$ and paraelectric materials like SrTiO$_3$, see~\cite{JacksonKimball:2017elr}.}
A demonstrative Phase I of the experiment~\cite{JacksonKimball:2017elr} is planed with thermally polarised $^{207}$Pb nuclear spins in lead magnesium niobate-lead titanate crystals at cryogenic temperatures, where $E^*\sim 3\times 10^8$V/cm and a value of $t_2\sim$ ms can be achieved, typically much smaller than $t_c$.
The strategy is to slowly ramp down the external magnet $B_{e}$ for the spin-lattice time $t_1\sim 1000$ s, measuring increasingly smaller masses. Then, repolarise the sample and measure again until 10$^6$ s of measurement time. Several samples with different $\vec E^*$ orientations will be used to reduce noise.
Sensitivity to QCD axion DM below $m_A\sim$ neV can be reached after several improvements, scheduled in two subsequent phases, II and III, see Fig.~\ref{fig:casper_sens} (left) adapted from~\cite{JacksonKimball:2017elr}. The benchmarks are increasing the polarisation by optical pumping and other techniques, increasing $t_2$ by decoupling protocols, implementing resonant RLC detection schemes, and increasing sample size and measurement time. The prospects assume that many technological benchmarks that have been demonstrated in independent setups can be combined together. Let us note that CASPEr could also in principle employ SERF magnetometry, which have completely different challenges but can be even more sensitive at the lowest masses~\cite{Wang:2017ixp}.

During this section we have been talking about axion DM, but indeed any ALP can have couplings to nuclear EDMs like the axion is expected to. The axion couples to $G\widetilde G$, obtaining at the same time a mass and the coupling to the nEDM given by \eqref{AxionnEDMcoupling}. 
\modified{Hence there is a relation between the two given by \eqref{axionEDMcoupling}. Within axion models it is conceivable to break this relation because the EDMs can have other contributions from axion-quark couplings but additional contributions to the mass bring back the strong CP problem very easily. \todonote{some exceptions!}
An ALP coupled to nucleon EDMs, will have a mass induced by loops that tends to bring it close to the QCD relation, even if it is not coupled to $G\widetilde G$ to start with. Therefore, although in principle it makes sense to consider all the parameter space not-excluded in Fig.~\ref{fig:casper_sens}, 
it might be in practice unnatural or even impossible to find models very far above the QCD band. Furthermore, a recent paper points out very strong phenomenological constraints on this kind of models~\cite{Hook:2017psm}.} 

\exclude{NMR EDM, heating E-field \cite{Eckel:2012aw}}

Another technique recently proposed is to search for the axion/ALP induced EDM in the future proton storage ring developed to measure the static proton EDM~\cite{Chang:2017ruk}. The range of ALP masses to which it is sensitive is very similar to the NMR techniques described here. Preliminary sensitivities are given around $|d_n|\sim 10^{-17}$ efm, a factor of $10^4$ from QCD axion expectations, but very interesting for generic ALP models featuring couplings to the proton EDM. \todonote{read new version, update mass range? answer Yannis}

It is worth noting that the most recent search for a static neutron EDM has been recently reanalysed in terms of ALP DM coupled to the nEDM~\cite{Abel:2017rtm}.  The limits surpass  astrophysical bounds and will be improved by next generation experiments but are very far from QCD axion sensitivity. 

\subsubsection{DM ALP ``wind''}

\begin{figure}[t!]
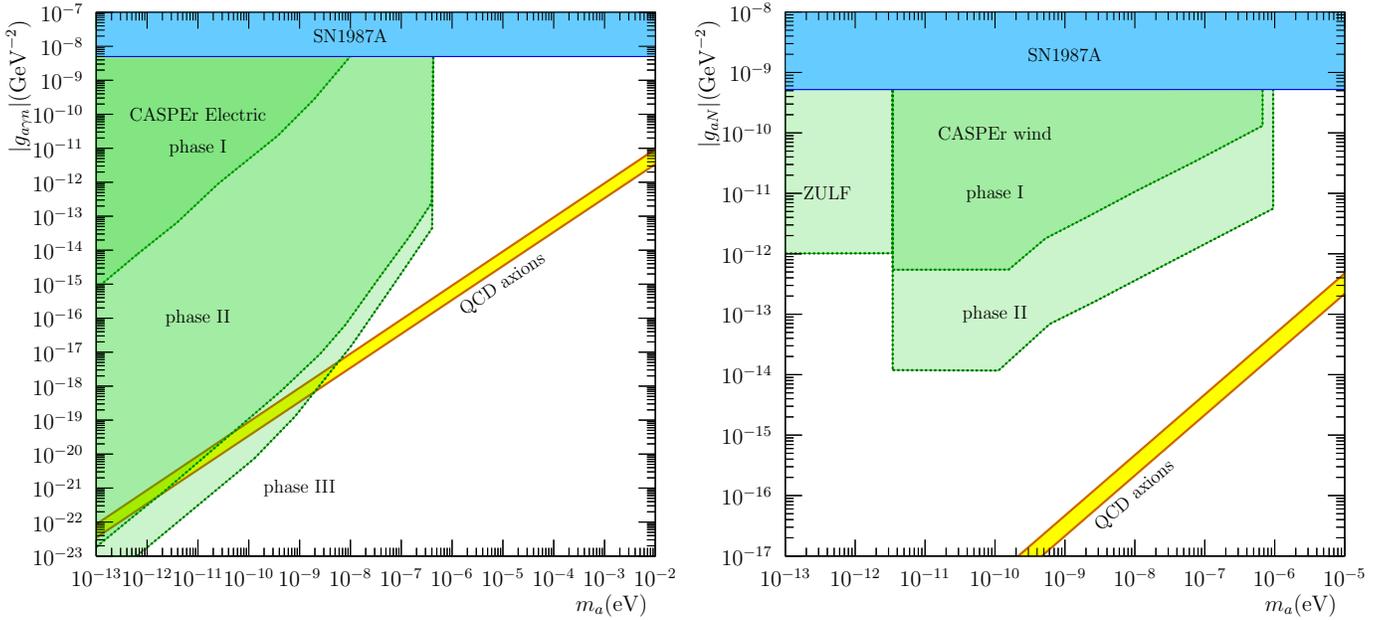

   \centering
\tikzsetnextfilename{gd_plane}
\resizebox{0.5\linewidth}{!}{\input{pics/gd_plane.tex}}
\hspace{-0.5cm}
\tikzsetnextfilename{gan_plane}
\resizebox{0.5\linewidth}{!}{\input{pics/gan_plane.tex}}
   \caption{Left: limits on $ g_{a\gamma n}$ and prospects for CASPEr-Electric. Right: Limits on $g_{aN}$ and prospects of CASPEr ``wind" experiments. From~\cite{JacksonKimball:2017elr}. }
\label{fig:casper_sens}
\end{figure}

The ALP DM field interacts directly with fermions through the couplings $g_{af},\bar g_{af}$ in \eqref{ALPinteractions}. In particular, the interaction of fermion spins with the ALP field is given by \eqref{Uinteraction} and is equivalent to their interaction with a fictitious B-field given by~\eqref{Bficticious}.
The ALP DM as a non-relativistic classical field can be represented as
\be
a \sim a_{\osci} \cos(\omega t -{\vec k}\cdot {\vec x}), 
\ee
where the wavenumber $|{\vec k}|\ll \omega$ is practically constant during a coherence time. 
The instantaneous effective field at a given point is thus $\fB \sim -  \frac{g_{a f}}{m_{f}\gamma_f} m_a{\vec v}a_{\osci}\sin (\omega t)$, but $m_a a_{\osci}$ is fixed by the local DM constraint \eqref{ALPDMamplitude} so that the magnitude only depends on $g_{af}/m_f \gamma_f$. The instantaneous velocity $\vec v$ is distributed randomly in time within the DM velocity distribution \eqref{Lentzdistribution}, which has a preferred direction in the Earth's frame  if the galactic DM distribution is relatively isotropic in the galactic rest frame. This is of course due to the fact that  the solar system orbits around the galaxy with a velocity $v_\odot \sim 220$ km/s and the Earth together with all its labs rotates with a daily period. Thus the $\fB$-field has an averaged preferred direction that is well known. This axion DM ``wind'' and its implied $\fB$ field can be detected with standard NMR techniques and is the subject of the CASPEr ``wind'' experiments.
Compared with the CASPEr electric version, the experiment is much simpler because the sample does not require large $E^*$ fields in the nuclei and thus standard NMR media like Xe or 3He can be used. The advantage of these nuclei are that they can be almost completely polarised and have very large coherence times, e.g. $t_2\sim 1000$ s. With these numbers, the measurement times are essentially limited by the coherence time $t_c$.
CASPEr-``wind'' considers three ALP mass ranges depending on the strength of the external $B_e$-field required to match the Larmor frequency of the sample to the ALP mass: a high field ($B_{e}=0.1-14$ T, $m_a\sim $neV-$\mu$eV) a low field ($B_{ e}=10^{-4}-0.1$ T corresponding to $m_a\sim $peV-neV) and a ZULF (zero-ultra-low) version $B_{e}<10^{-4}$ T. The sensitivities are shown in Fig.~\ref{fig:casper_sens}, from~\cite{JacksonKimball:2017elr}. Exploring the shown regions will take many years and yet sensitivity to QCD axions is not guaranteed. The ultra high field version is challenged by $B_e$-field homogeneity requirements, which are considered secondary in the low field range. Phase I considers sample volumes of 1 cm$^3$ and a Phase II is planned to benefit from much larger sample sizes.

\Javinote{Figure 24  right needs relabelling gan into gaN}

\todonote{Figure \ref{fig:casper_sens} left misses a hat in the g. Make axis label font larger. Include reference~\cite{Abel:2017rtm}.}
   
The QUAX experiment considers detecting ALP DM via the electron coupling using magnetic materials~\cite{Barbieri:2016vwg}.
Although conceptually the idea is very similar to those exposed above, there are certain crucial differences due to the fact that the electron mass is much smaller than a nucleon's.
For one, the Larmor frequencies are larger by a factor of $m_N/m_e\sim 2000$ and therefore much larger ALP masses are accessible. Indeed, QUAX focuses on the $m_a\sim 200\, \mu$eV ($m_a/2\pi\sim 48$ GHz) range where $B_{e}\sim 1.7$ T.
Second, electron spin precession has much shorter coherence times because electrons emit dipole radiation very efficiently. In order to overcome this difficultly, QUAX proposes to build a resonant MW cavity around the sample in the strongly coupled regime to inhibit damping. As a consequence, the detection scheme greatly changes. The spin-preccesion resonance hybridises with the electromagnetic mode of the cavity and the effect of ALP DM is to excite this hybrid mode.
The signal is picked up with an electromagnetic antenna, dominated by amplifier and thermal noise familiar from the haloscope techniques. In order to reach QCD axion sensitivity with $C_{\A e}\sim {\cal O}(1)$ like for \DFSZ models, they require sensitivity below the standard quantum noise, i.e. a single photon counter and ultralow temperatures $\sim 100$ mK are required~\cite{Barbieri:2016vwg}. Such a photon counter could be extremely interesting not only for QUAX but for every other experiment searching for axion/ALP DM coupled to photons in the multi GHz range like MADMAX, ORPHEUS and ORGAN. Moreover, it would also allow a considerale improvement of a  LSW experiment along the lines of the proposed STAX~\cite{Capparelli:2015mxa}.

\subsection{Atomic transitions}
\label{sec:atomic}

DM axions can produce atomic excitations in a target material to levels with an energy difference equal to the axion mass. This can happen via the axion couplings to the spin of electrons or the spin of the nucleus. The use of the Zeeman effect has been proposed~\cite{Sikivie:2014lha} to split the ground state of atoms to effectively create atomic transition of energy levels that are tunable to the axion mass, by changing the external magnetic field. The excited state is then efficiently brought to a higher energy level (at visible or NIR energies) by a properly tuned pump laser that is permanently shining on the target. The photon produced in the deexcitation of this state is then detected by conventional means.

According to preliminary estimates~\cite{Sikivie:2014lha}, obtaining relevant count rates requires one to instrument target materials of $\sim$kg mass cooled down to mK temperatures, reaching sensitivity to axion models (with fermion couplings) in the ballpark of $10^{-4}-10^{-3}$~eV. The technique could be extended to somewhat higher masses in suitable antiferromagnetic targets with resonant transitions. The Italian collaboration AXIOMA~\cite{axioma_web} has started feasibility studies to experimentally implement this detection concept. First results involve the exploration of suitable targets involving molecular oxygen~\cite{1367-2630-17-11-113025}, or rare-earth ions doped into solid-state crystalline materials~\cite{Braggio:2017oyt}, as well as the study of possible backgrounds to this detection method.

\subsection{Non homogeneous DM phase-space distribution }

All DM experiments so far compute their expected signals under the basic assumption of homogeneity in the DM distribution.
We now turn to more specific detection modes that contemplate the possibility of particularly non-homogeneous distribution of DM axions, whether in space (DM in the form of compact objects or \textit{miniclusters}) or momentum (DM in the form of low dispersion streams). The latter can be addressed with high resolution channels in haloscopes or devising detection strategies sensitive to the incoming axion direction. The former involves the use of a geographically distributed network of detectors.

\subsubsection{Directional detection and low dispersion streams}
\label{sec:directional}

As widely recognised and studied in the field of WIMP searches~\cite{Ahlen:2009ev}, the possibility of registering information on the momentum distribution of the incoming DM particles would provide an unmistakable signature of the extraterrestrial origin of a signal, as well as valuable information on the structure and history of the DM Galactic halo. 
Despite the current knowledge of the local distribution of DM particles presented in section~\ref{sec:sources}, it is worth stressing that the distribution determined from local kinematic data or by N-body simulations (like \eqref{Lentzdistribution}) represents an average over temporal and spatial scales that are much larger that the ones relevant for direct detection experiments. Indeed, the velocity distribution of DM particle at Earth is rather uncertain and could suffer large departures from the average one stated above. There are attempts at modelling those spreads  e.g. in the context of studies of systematic uncertainties in WIMP direct detection rates (an old but illustrative example can be found in e.g.~\cite{Belli:2002yt}, and more recent ones, already in the context of axion searches, in \cite{OHare:2017yze,Foster:2017hbq}).

\exclude{In fact, the local distribution of velocities of DM particles is not known and there is a large spread in prediction from different halo models (an old but illustrative example of the spread of models can be found in e.g.~\cite{Belli:2002yt}). The simplest one of these halo models is the isothermal sphere model, where the dark matter particles follow a spherical distribution with a flat rotation curve. Although this model is often used as a benchmark in the calculation of expected DM signal in terrestrial detectors, its underlying assumptions are not totally sustained on physics grounds or observations. Many alternative models introduce gradual deviations of this model (e.g. with some degree of non-isotropy, non-sphericity or corotation of the DM halo), which have been studied e.g. in the context of studies of systematic uncertainties in WIMP direct detection rates.
\Javinote{smoothly merge with Section 4}
}

A particularly interesting departure from the standard Maxwellian distribution are models in which a substantial fraction of the DM density is in the form of low dispersion streams. Those streams could be formed by tidal disruption of dwarf satellites or due to the late infall of dark matter onto the galaxy. They would have a much lower velocity dispersion than the main thermal component in \eqref{veldispersion}. There is no general consensus on the presence, type or importance of dark matter streams in the Milky Way and in particular which fraction of the local dark matter density is in the form of streams, however their existence could have important consequences for direct detection experiments. The particular way axions are detected in haloscopes, in which the total energy of the axion can be measured with exquisite precision, opens attractive opportunities in this regard. Axion dark matter streams would manifest themselves as peaks in frequency of much narrower width than expected for a virialised axionic component $\Delta \omega/\omega \sim 10^{-6}$. Presumably those peaks would appear on top of a standard thermal component. The needed spectral resolution to resolve them is easily accomplished with small adaptation of the standard readout of axion haloscopes, and indeed the ADMX collaboration has already implemented a high-resolution channel for the search of non-virialised dark matter peaks in their spectrum~\cite{Hoskins:2011iv}. In the absence of a positive detection, these searches can only improve the main result in a model-dependent way. In the event of a positive detection, however, the precise study of the spectrum features might quickly provide a wealth of information on the DM astrophysics~\cite{OHare:2017yze,Foster:2017hbq}.

A very appealing possibility are detection methods with sensitivity to the direction of incoming DM axions, i.e. the axionic counterpart to WIMP directional detectors~\cite{Ahlen:2009ev,Battat:2016pap}. It turns out that the standard haloscope technique develops some directional dependence for long-aspect ratio cavities with length $L$ exceeding the de Broglie wavelength of the DM axion~\cite{Irastorza:2012jq}. Due to the anisotropy of incoming velocity distribution induced by the movement of the solar system with respect the DM halo, even with a thermal Maxwellian distribution a dependence on the orientation of the cavity is expected. The effect is maximal and of O(1) for cavity lengths of $L \sim 10 \times (10^{-4} \rm{eV} / m_a)$~m. The effect is of course sharper in halo models containing low dispersion streams. This detection concept is particularly interesting in view of the implementation of long thin cavities in dipole magnets e.g. in CAST (see above).

The dish antenna concept described in section~\ref{sec:dielectric_haloscopes} also enjoys a directional effect~\cite{Jaeckel:2013sqa}. The momentum of the photons emitted (mostly) perpendicular to the surface due to axion conversion contain a small parallel component equal to the parallel component of the incoming axion momentum. This means that the focusing of photons onto the centre of curvature of the dish is in reality slightly smeared by the axion momentum distribution. If a pixelised detector is used at the focal plane of the dish antenna it will directly image the  distribution of momenta (the component parallel to the dish) of the incoming DM axions. The resolution in momentum is limited by geometric considerations (dish radius $r$ must be much smaller than curvature radius $R$) and  by diffraction (dish radius must be much larger than photon wavelength). Finally  the relative movement of the Earth with respect the DM halo will offset the image and will introduce some temporal modulation, both of which should be easy to correct.

In the event of a determination of the axion mass, the detection of axions with high signal-to-noise will be relatively easy. The concepts above promise access to a wealth of information on DM astrophysics. The discovery of the axion will be followed by a new era of axion astronomy.

\subsubsection{A global detector network for axion miniclusters and topological defects}

In some axion or ALP DM models, and in the post-inflation scenario explained in \ref{sec:cosmology}, the initial inhomogeneities of the galactic DM distribution enable gravity or self-interactions to generate minicluster of ALPs (also called ALP stars, soliton stars or Q-stars). The size and mass of these miniclusters, as well as the fraction of the DM that is in this form, depend on details of the model. The existence of compact DM objects is heavily constrained by e.g. microlensing observations. However, current bounds leave space for models with relatively light ALP ($M\lesssim 10^{-16}M_\odot$) miniclusters~\cite{JacksonKimball:2017qgk}. Interestingly, for some of the remaining parameter space, the rate of encounters with miniclusters is sufficiently frequent (at least one encounter in a typical observational campaign of 1 year). This possibility would have important consequences in the detectability of DM. Most of the experiments presented above might miss a signal because they are designed to be sensitive to a narrow mass band at a time, and use long integration times.

A similar observational problem is posed by topological defects like domain walls. As seen in section~\ref{sec:cosmology} for axion models with $N_{DW}>1$ topological defects form and are stable. This possibility has disastrous cosmological consequences and is ruled out. However, for some more generic ALP models this possibility is viable. If our galactic DM environment is populated with domain walls, there could be a non-negligible probability of having frequent encounters with the Earth. Such a crossing would produce a detectable signal in sensitive magnetometers (via spin interactions of the type discussed in \ref{sec:NMR}). However, due to the rarity of such event it could be difficult to discriminate them from other detector artifacts or noise.

The GNOME collaboration has built a network of optical magnetometers placed in geographically distinct locations and synchronised using the global positioning system (GPS). GNOME is prepared to detect transient spin-dependent interactions and, by means of correlated measurements between stations, establish whether the signal is attributable to the crossing of a compact DM object, like a minicluster or a domain wall~\cite{Pospelov:2012mt}. At present GNOME consists of six stations in continuous operation\cite{gnome_web}, plus more are in construction.

\section{Discussion}
\label{sec:discussion}

\begin{figure}[t]
\begin{center}
\tikzsetnextfilename{ALP_map}
\resizebox{0.75\linewidth}{!}{\input{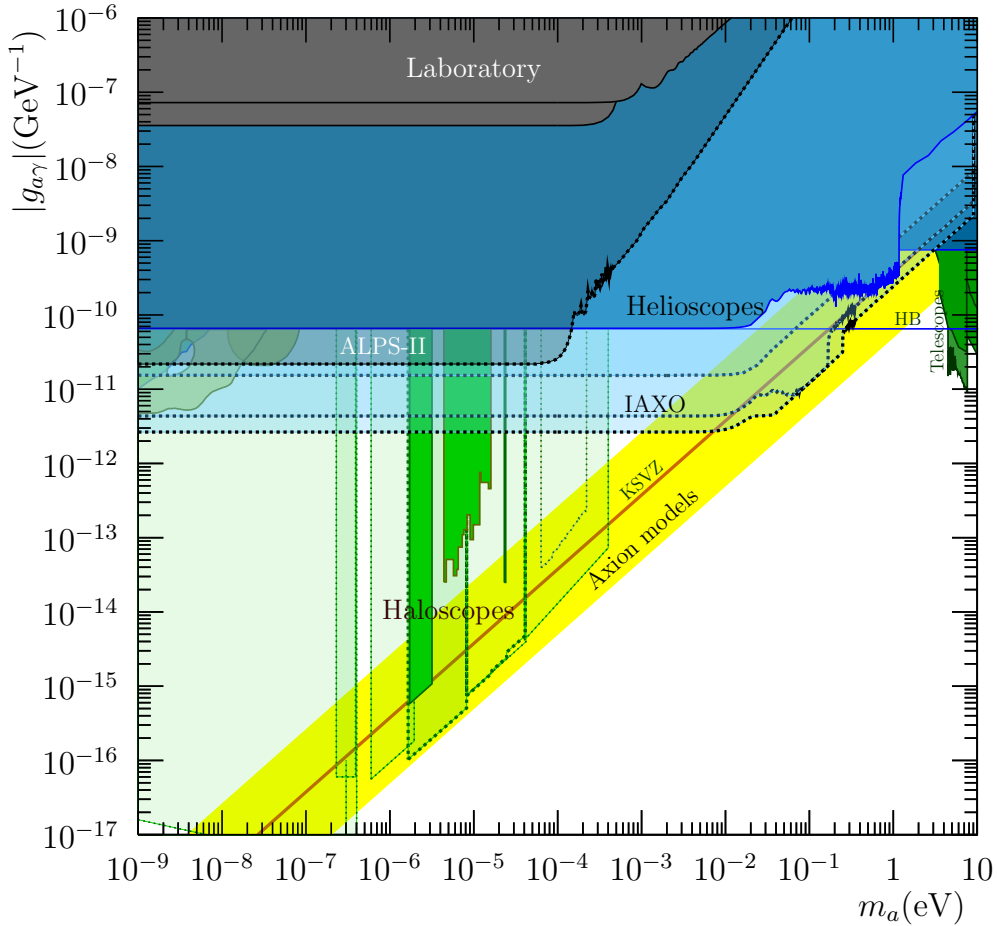}}
\caption{Overall panorama plot in the ($\gagamma$,$m_a$) plane. As usual laboratory, helioscope and haloscope areas are colored in black, blue and green respectively. Some prospect regions shown in previous plots are here collected in semi-transparent colors. }
\label{fig:overall}
\end{center}
\end{figure}

Fig.~\ref{fig:overall} is an attempt to concisely summarise most of the results and prospects detailed in previous pages. As such it is partial as it shows only the ($\gagamma$,$m_a$) parameter space and, as shown before, there are now a number of experiments active in other detection channels (see Table~\ref{tab:methods}). Nevertheless, the $\gagamma$  channel still gathers most of the experimental activity and probably remains the most promising channel for a discovery, although other channels will be crucial to identify a future putative signal as a QCD axion (or other type of ALP). Fig.~\ref{fig:overall} includes all prospect regions in the ($\gagamma$,$m_a$) plane that have been shown in previous plots without individual labels. Although many of those are still somewhat far in the future and depend on successful completion of previous R\&D, it gives a nice account of the potential of the field to collectively explore a large fraction of the allowed parameter space for axions (and ALPs) in the future.

Despite some (healthy) overlap between experiments, it is to be stressed the high degree of complementarity in the experimental landscape. Current and future DM axion searches will presumably cover the $m_a$ ranges cosmologically favoured by pre-inflation and the post-inflation $N_{DW}=1$ models, under the assumption that DM is mostly comprised of axions,  while helioscopes will be active in the higher mass range $m_a>1$~meV and access the post-inflation $N_{DW}>1$ range and the astrophysically motivated regions. Laboratory experiments with ALPS-II will soon surpass astrophysical and helioscope bounds and will be the first to partially test the transparency hint for ALPs. In the event that axions comprise only part of the DM, $\admfrac < 1$, the sensitivity of DM searches must be corrected up by a factor $\admfrac^{-1/2}$ in the plot of Fig.~\ref{fig:overall}. Therefore it is important that future experiments try to push their sensitivity in $\gagamma$ well beyond benchmark models to gain some margin. In such eventuality, theoretical predictions on $m_a$ are also moved to higher values by a factor of approximately $\admfrac^{-1}$, and so there is a strong motivation to push haloscope sensitivities to even higher masses and helioscopes to lower masses along the QCD band, and try bridge the gap between them. Although perhaps comparatively less motivated, one cannot exclude a $m_a$ of much lower values deep into the anthropic window. The LC circuit concept, and especially in its broadband mode, is an ingenious idea best suited for this mass range. We need to follow the progress on small scale prototypes by the experimental groups active there to better assess its future prospects. The same is to be said on the emerging activity on the new detection concepts involving other axion couplings like the NMR techniques, the atomic transitions, 5th forces, etc. The evolution of the ongoing demonstrating experimental activity in small test setups will be crucial to assess their future potential. The confirmation that QCD axion sensitivity is really reachable by one or more of these complementary channels would be of the utmost importance.

\section{Conclusions}
\label{sec:conclusions}

Four decades after their proposal, axions are now a focus of the utmost interest. Still the most compelling solution to the strong CP problem, axions (as well as more generic ALPs) are in addition motivated by a wealth of other arguments, considerably developed over recent years. Many plausible theoretical frameworks beyond the SM (like string theory) naturally predict them. They appear in many cosmological contexts, like in models related with inflation, dark radiation and even dark energy. Most importantly, axions and ALPs could be excellent candidates to compose all or part of the dark matter. With the persisting negative outcome of WIMP searches, axions are becoming the next most promising hypothesis to clarify the nature of DM. Contrary to WIMPs, axions and ALPs could play very relevant roles in astrophysical environments. Indeed, some astrophysical observations might already be \textit{hinting} at their existence. Axions produced by astrophysical bodies like our Sun offer particular detection possibilities, like e.g. axion helioscopes, without analogue in the field of WIMP searches.

Long considered ``invisible'', axions are now within reach of current and near-future technologies in different parts of the viable parameter space. The field is undergoing a blooming phase. As has been shown in this review, the experimental efforts to search for axions are rapidly growing in intensity and diversity. Consolidated detection techniques are now facing next-generation experiments with ambitious sensitivity goals and serious chances of impact in the field. In addition, novel and ingenious detection concepts are being proposed and new R\&D lines are starting with potential to open exploration of new ranges of parameter space, previously considered unreachable. The experimental landscape is rapidly evolving, so  relatively recent reviews are quickly becoming obsolete~\cite{Essig:2013lka,Graham:2015ouw}.

The topic of axion searches is slowly entering the formal agenda of \textit{larger scale} particle physics research, as larger infrastructures, collaborations, and institutions are needed. The most striking example of this is the recent foundation of an IBS Center for Axion and Precision Physics (CAPP) in South Korea, specifically devoted to axion physics. CAPP has as one of its goals ``to launch a state-of-the-art axion DM experiment''. In the US, ADMX has been recently selected by DOE and NSF as one of the Generation-2 DM experiments (together with LZ and SuperCDMS, both looking for WIMPs), and agency-driven events to identify novel non-standard DM detection strategies are being organized~\cite{Battaglieri:2017aum}, with relevant presence of axions. It is remarkable the success of some of the novel axion projects in attracting private funds, notable from the Heising-Simons and Simons foundations. In Europe, DESY is taking important steps to becoming a pole on axion physics, presumably hosting the future IAXO and MADMAX experiments, on top of the already ongoing preparations for the ALPS-II experiment. Although the recently issued European Astroparticle Physics APPEC roadmap~\cite{appec} fails to acknowledge the emergence of axions in the wider DM experimental landscape (something that we hope will be corrected in the next edition of the roadmap), the CERN-driven Physics Beyond Colliders (PBC) process~\cite{pbc} has a very relevant presence of axion searches. The PBC process, currently ongoing at CERN, aims at producing feedback, by the end of 2018, for the next European Strategy for Particle Physics. Finally, let us mention that steps towards a centre on axion DM in Australia are being taken~\cite{tobar}.

We want to conclude by stating that the theoretical, cosmological and astrophysical motivation for axions and ALPs is sufficiently strong so that to tackle their experimental detection as one of the major goals for particle physics nowadays. As has been shown in previous pages, the detection technologies already show promise of sensitivities sufficient to reach unexplored parameter space, or have a clear R\&D roadmap to be in such position. A diverse experimental program is strongly emerging with definite prospects of probing large fractions of viable parameter space in the coming years. If the axion exists, there is a reasonable chance for positive detection in the near future. It would be a breakthrough discovery that could reshape the subsequent evolution of particle physics, cosmology and astrophysics in the widest sense.

\section*{Acknowledgements}

We are indebted to many colleagues for providing updated information on their projects for this review, in particular to A. Lindner, K. van Bibber, G. Rybka, G. P. Carosi, D. Kimball, D. Budker, L. Miceli, Y. Semertzidis, S. Youn, M. Tobar, G. Zavattini and A. Geraci. Special thanks go to K. van Bibber and the HAYSTAC team for providing the unpublished results appearing in Figure~\ref{fig:haystac_plot}. R. Essig, S. MacDermott and J. H. Chang kindly informed us prior to publication of their revision on the SN1987A constraints, for which we are very grateful (and to G. Villadoro for the hint that they were working on it). 
We are greatly indebted to C. O'Hare, for his careful reading of the full draft. We acknowledge conversations with many colleagues of different aspects of the review as well as comments and corrections. We are very thankful to G.~Raffelt, M.~Pospelov, A.~Millar, T.~Dafni, G.~Gal\'{a}n, C.~Tamarit, C.~O'Hare, A.~Vaquero, L. Di~Luzio, J. de Vries, L.~Calibbi, R. Ziegler, G.~Villadoro, A. Payez, and many others. I.G.I acknowledges support from the Spanish MINECO under grant FPA2016-76978-C3-1-P and the European Research Council (ERC) under grant ERC-2009-StG-240054 (T-REX). J.R. is supported by the Ramon y Cajal Fellowship 2012-10597, the grant FPA2015-65745-P (MINECO/FEDER),
the EU through the ITN ``Elusives'' H2020-MSCA-ITN-2015/674896 and the Deutsche Forschungsgemeinschaft
under grant SFB-1258 as a Mercator Fellow.

\footnotesize

%

\bibliographystyle{utphys}
\bibliography{igorbib,axionDM,additional}


\newpage
\listoffixmes

\end{document}